\documentclass[twocolumn,aps,superscriptaddress,prL,floatfix,showpacs,longbibliography]{revtex4-2}

\usepackage{graphicx}
\usepackage{dcolumn}
\usepackage{bm}

\usepackage{ifthen,amsmath,amssymb}
\usepackage{hyperref}
\usepackage{float}
\usepackage{aas_macros}
\usepackage[export]{adjustbox}
\usepackage{bookmark}
\usepackage{newtxtext,newtxmath}
\usepackage[dvipsnames,table]{xcolor}

\begin{document}

\preprint{APS/123-QED}

\title{Missing baryons recovered: a measurement of the gas fraction in galaxies and groups with the kinematic Sunyaev-Zel'dovich effect and CMB lensing}

\author{Boryana Hadzhiyska}
\email{boryanah@alumni.princeton.edu}
\affiliation{Physics Division, Lawrence Berkeley National Laboratory, Berkeley, CA 94720, USA}
\affiliation{Berkeley Center for Cosmological Physics, Department of Physics, University of California, Berkeley, CA 94720, USA}
\affiliation{Institute of Astronomy, Madingley Road, Cambridge, CB3 0HA, UK}
\affiliation{Kavli Institute for Cosmology Cambridge, Madingley Road, Cambridge, CB3 0HA, UK}

\author{Simone Ferraro}
\affiliation{Physics Division, Lawrence Berkeley National Laboratory, Berkeley, CA 94720, USA}
\affiliation{Berkeley Center for Cosmological Physics, Department of Physics, University of California, Berkeley, CA 94720, USA}

\author{Gerrit~S. Farren}
\affiliation{Physics Division, Lawrence Berkeley National Laboratory, Berkeley, CA 94720, USA}
\affiliation{Berkeley Center for Cosmological Physics, Department of Physics, University of California, Berkeley, CA 94720, USA}

\author{Noah Sailer}
\affiliation{Berkeley Center for Cosmological Physics, Department of Physics, University of California, Berkeley, CA 94720, USA}
\affiliation{Physics Division, Lawrence Berkeley National Laboratory, Berkeley, CA 94720, USA}

\author{Rongpu Zhou}
\affiliation{Physics Division, Lawrence Berkeley National Laboratory, Berkeley, CA 94720, USA}
\affiliation{Berkeley Center for Cosmological Physics, Department of Physics, University of California, Berkeley, CA 94720, USA}


\date{\today}

\begin{abstract}
We present new constraints on the halo masses and matter density profiles of DESI galaxy groups by cross-correlating samples of Luminous Red Galaxies (LRGs) and Bright Galaxy Survey (BGS) galaxies with the publicly available CMB lensing convergence map from ACT DR6. This provides an independent, lensing-based calibration of halo masses, complementary to methods relying on clustering or dynamics. We derive constraints on the mean halo mass for three DESI-selected samples, finding $\log(M_{\rm halo}/(M_\odot/h)) \approx 13.18$, 13.03 and 13.02 for the Main LRG, Extended LRG, and BGS samples, respectively. Using a halo model approach, we also compare the projected galaxy-matter density profiles with previously reported gas profiles inferred from measurements of the kinematic Sunyaev-Zel’dovich (kSZ) effect. This work addresses one of the key uncertainties in interpreting kSZ signals -- the unknown host halo mass distribution -- by providing an independent and consistent mass calibration. The agreement between the gas and total mass profiles at large aperture suggests that sufficiently far from the group center (2–3 virial radii), we recover all the baryons, offering a resolution to the `missing baryon' problem. We further study the cumulative gas fractions for all galaxies as well as for the most massive galaxy groups in the sample ($\log(M_{\rm halo}/(M_\odot/h)) \approx 13.5$), finding values that are physically sensible and in agreement with previous findings using kSZ and X-ray data: compared to the TNG300 simulation, the observed gas fractions are systematically lower at fixed radius by $\gtrsim$4$\sigma$, 
providing compelling, independent evidence for stronger baryonic feedback in the real Universe. These findings highlight the power of combining CMB lensing with galaxy surveys to probe the interplay between baryons and dark matter in group-sized halos.
\end{abstract}

\maketitle


\section{Introduction}
\label{sec:intro}

The standard cosmological model, $\Lambda$CDM, has proven remarkably successful at describing the evolution of the Universe across a wide range of scales and epochs. Observations of the cosmic microwave background (CMB), baryon acoustic oscillations (BAO), and large-scale structure (LSS) have established a precise framework in which only a few parameters are needed to explain the expansion history and the growth of structure.
Nevertheless, some of the most fundamental components of this model, dark energy and dark matter, remain enigmatic. With the advent of high-precision datasets from DESI, Rubin, Euclid, and Simons Observatory, we are entering an era where new data will be sensitive to subtle departures from $\Lambda$CDM predictions. A critical challenge in leveraging this sensitivity is our incomplete understanding of baryonic physics, particularly baryonic feedback processes that redistribute gas within and beyond halos, altering the observable matter distribution on non-linear scales \citep{2011MNRAS.415.3649V,2019OJAp....2E...4C}. This manifests itself in a large deficit of observed baryon abundance in late-time galaxies and groups, known as the ``missing baryon'' problem \cite{Fukugita_2004, Cen_2006}.

Baryonic feedback, especially from active galactic nuclei (AGN) and supernovae, expels gas from the centers of halos into their outskirts, suppressing the small-scale matter power spectrum and modifying weak lensing observables \cite{Schneider:2018pfw, 2024arXiv240406098B}. This has profound implications for cosmological inference: if not properly modeled, baryonic effects can bias constraints on parameters such as $S_8$ and the total neutrino mass \citep{2019JCAP...03..020S, 2022MNRAS.516.5355A, 2025MNRAS.537.2160E}. Yet the modeling of feedback remains highly uncertain, in its amplitude, redshift and scale dependence, with state-of-the-art hydrodynamical simulations yielding divergent predictions despite matching many galaxy observables \citep{2019MNRAS.488.1652H}. This degeneracy between baryon physics and cosmology motivates the development of new empirical probes that directly constrain the gas distribution across cosmic time.

One of the most powerful tools in this regard is the kinetic Sunyaev-Zel’dovich (kSZ) effect, a secondary CMB anisotropy generated when CMB photons are Doppler shifted by scattering off free electrons with bulk line-of-sight velocities \citep{2019SSRv..215...17M}. Unlike the thermal SZ (tSZ) effect, which depends on electron temperature and is concentrated in hot cluster gas, the kSZ signal traces the momentum of all free electrons, making it sensitive to the total ionized baryon distribution. This is especially valuable in the circumgalactic medium (CGM) and the outskirts of halos ($\sim$1–4 virial radii $R_{\rm vir}$), where most of the ``missing baryons" 
have now enabled significant detections of the kSZ signal on halo mass scales down to $10^{13} M_\odot$ \citep{Schaan_2021,2024arXiv240707152H, 2024arXiv241203631H, RiedGuachalla:2025byu}. These measurements suggest that the gas distribution is significantly more extended than the dark matter, consistent with strong feedback scenarios.

Nonetheless, the interpretation of the kSZ signal is limited by uncertainties in the halo mass of the galaxy sample. Since the kSZ amplitude depends linearly on the optical depth, which in turn is proportional to the host halo mass, the comparison between the matter and gas profiles significantly benefits from external mass information;
uncertainty in the halo mass propagates directly into uncertainties on feedback strength \citep{2021ApJ...919....2M,2021PhRvD.103f3514A}. Current approaches often rely on stellar mass estimates combined with empirical stellar population or halo occupation distribution (HOD) models calibrated on clustering. However, these indirect methods introduce substantial systematic uncertainties. As a result, attempts to use kSZ to test hydrodynamical simulations or inform cosmological models are substantially limited without more accurate halo mass measurements. 

Gravitational lensing offers a natural solution. In particular, CMB lensing provides a clean, unbiased measurement of the projected mass distribution along the line of sight, without the complications of intrinsic alignments, shape noise, or shear calibration biases that affect galaxy-galaxy lensing (GGL). When cross-correlated with galaxy positions, CMB lensing can be used to measure the average halo mass and the matter density profile of a given sample \citep{2015ApJ...806..247B,2020ApJ...903L..13M}. This makes it an ideal counterpart to kSZ measurements: while kSZ traces the baryon density, lensing traces the total matter. Their combination enables a direct estimate of the gas-to-mass ratio (or gas fraction) 
as a function of scale, a critical test for both feedback models and for resolving the ``missing baryon'' problem.

An early comparison of the kSZ signal to small-scale galaxy-galaxy lensing (GGL) was performed in \cite{2021PhRvD.103f3514A} and further studied in \cite{Sunseri:2025hhj} within the halo model framework and in \cite{2025MNRAS.540..143M} with the use of state-of-the-art hydrodynamical simulations. A combined analysis of cosmic shear and the kSZ effect was conducted in \cite{2024MNRAS.534..655B}.


In this work, we perform a joint analysis of CMB lensing and kSZ measurements around DESI galaxies. Using high-resolution CMB lensing convergence maps from ACT and the Planck legacy release, we extract the small-scale projected mass profiles of photometrically selected DESI targets in multiple redshift and luminosity bins. We then compare these profiles to the corresponding kSZ measurements from \citet{2024arXiv240707152H} to assess whether the recovered gas content matches theoretical expectations. This approach allows us to: (1) empirically constrain the average host halo mass of DESI galaxies in a manner that is independent of stellar mass modeling, (2) infer their baryon fraction as a function of scale, and (3) place novel constraints on the amplitude and scale dependence of baryonic feedback. Beyond characterizing the total mass of galaxy groups and comparing it with the gas mass, this analysis has significant implications for the use of DESI galaxies in upcoming lensing and clustering cosmology. The ability to infer their halo masses directly enables new inputs for HOD-based analyses and cross checks with GGL and 2-point correlation function analyses. 

This paper is structured as follows. In Section~\ref{sec:data}, we describe the galaxy and CMB datasets used in our analysis. Section~\ref{sec:method} outlines the halo model framework used to model the lensing signal. Section~\ref{sec:results} presents our mass profile measurements, the comparison to kSZ measurements of the gas density profiles, and a comment on the observed Stellar-to-halo mass relation. We summarize our findings and discuss implications for cosmology and galaxy formation in Section~\ref{sec:conclusions}.

\section{Data}
\label{sec:data}

In this section, we summarize the galaxy and CMB observational data sets used in this study.

\subsection{Dark Energy Spectroscopic Instrument}

The Dark Energy Spectroscopic Instrument (DESI) is a robotic, fiber-fed, highly multiplexed spectrograph operating on the Mayall 4-meter telescope at Kitt Peak National Observatory \citep{2022AJ....164..207D}. It is capable of obtaining simultaneous spectra for nearly 5000 objects across a $\sim$$3^\circ$ field of view \citep{2016arXiv161100037D,2023AJ....165....9S,2023arXiv230606310M}, and is currently conducting a five-year dark energy survey covering approximately one-third of the sky \citep{2013arXiv1308.0847L}. Upon completion, the survey will have collected spectra for roughly 40 million galaxies and quasars \citep{2016arXiv161100036D}. 

In this work, we use the extended photometric sample of Luminous Red Galaxies (LRGs) \citep{2023AJ....165...58Z,Zhou:2023gji}, selected from the DESI Legacy Imaging Surveys, which combine data from three telescopes: Blanco for the Dark Energy Camera Legacy Survey (DECaLS), Mayall for the Mayall $z$-band Legacy Survey (MzLS), and Bok for the Beijing–Arizona Sky Survey (BASS). We make use of the photometric redshifts provided in \citet{Zhou:2023gji} for Data Release 9 (DR9), and the stellar mass estimates from \citep{2023AJ....165...58Z}.

We additionally include Bright Galaxy Sample (BGS) galaxies, as described in \citet{Hahn:2022dnf, Chen:2024vvk}, selected from the same imaging surveys and designed to trace the nearby galaxy distribution with high completeness.

\subsection{Atacama Cosmology Telescope}
\label{subsec:actdata}

We utilize the publicly available CMB lensing convergence ($\kappa$) maps from the Atacama Cosmology Telescope (ACT) Data Release 6 (DR6), described in \cite{ACT:2023kun, ACT:2023dou, ACT:2023ubw, ACT:2023oei}. ACT was a 6-meter telescope located at an altitude of 5,190 meters in the Atacama Desert of northern Chile. It operated from 2007 until its decommissioning in 2022, providing high-resolution measurements of the CMB over a wide range of angular scales.

The DR6 lensing data set used here leveraged multifrequency observations collected between 2017 and 2021, covering roughly a third of the sky in two frequency bands: f090 (77–112 GHz) and f150 (124–172 GHz). The lensing convergence field was reconstructed using a quadratic estimator (QE) applied to both CMB temperature and polarization maps, and included ``profile hardened'' estimators \cite{Sailer:2020lal, ACT:2023ubw} to mitigate the impact of extragalactic foregrounds.

The resulting $\kappa$ map is provided as spherical harmonic coefficients ($a_{\ell m}$'s), along with a corresponding analysis mask. The reconstruction imposes a 
low-pass filter at $\ell > 3000$ to exclude low signal-to-noise modes and suppress contamination from foregrounds and instrumental effects. 

\subsection{Measurement of convergence profiles}

We measure the CMB lensing convergence, $\kappa$, profiles around galaxies by stacking cutouts from the ACT DR6 lensing map. For each galaxy, we extract a square region centered on its coordinates using the \texttt{reproject} function from the \texttt{pixell} package, employing bilinear interpolation to project the 1-arcmin pixel $\kappa$ map into a tangential projection with a pixel size of 0.5 arcmin. These cutouts are then stacked and averaged in radial bins to construct the galaxy-convergence cross-correlation signal.

We define the radially averaged convergence profile $\hat \kappa(\theta_d)$ at angular separation $\theta_d$ as
\begin{equation}
  \hat \kappa(\theta_d) =  \mathcal{N}^{-1}(\theta_d) \int d^2\theta \, W(\theta, \theta_d)\,\kappa(\theta)\,,
    \label{eq:CAP}
\end{equation}
where $W(\theta, \theta_d)$ is a top-hat annular (ring) window centered at angular distance $\theta_d$, and $\mathcal{N}(\theta_d)$ is a normalization factor equal to the area of the annulus (i.e., $\mathcal{N}(\theta_d) = \int d^2\theta\, W(\theta, \theta_d)$). The integration is performed in 2D pixel space for each stacked image.

We bin the signal into 8 linearly spaced annuli in angular separation with $\Delta \theta = 1.5 \ {\rm arcmin}$, covering the range from $0$ to $10.5$ arcmin. This range leverages the need for high signal-to-noise on small scales with the goal of probing the outskirts of the halo and surrounding large-scale structure. The resulting $\kappa$ profile represents the average gravitational lensing signal sourced by the projected mass distribution around the selected galaxy sample. 

\subsection{Measurement of the gas profiles} 
\label{sec:kSZ}

In this work, we utilize previously published measurements of the kinetic Sunyaev-Zel'dovich (kSZ) effect obtained via stacking around DESI LRGs to draw a comparison between the total matter distribution and the gas distribution. In this section, we briefly summarize the procedure for self-consistency. 

The methodology follows that described in Refs.~\cite{Schaan_2021,2024arXiv240707152H} and is implemented in the publicly available \texttt{ThumbStack} pipeline \footnote{\url{https://github.com/EmmanuelSchaan/ThumbStack}}. It requires two ingredients: an estimate of the line-of-sight peculiar velocity for each galaxy and a measurement of the temperature modulation 
in the CMB at the location of each galaxy. The temperature signal is extracted from CMB maps using a compensated aperture photometry (CAP) filter, which subtracts the mean temperature in an annulus from that in a central disk to isolate small-scale temperature fluctuations associated with individual halos. The resulting temperature profile $T_{\rm kSZ}(\theta_d)$ is computed as a function of aperture radius $\theta_d$.

The peculiar velocity field is reconstructed by solving the linearized continuity equation in redshift space under the assumption of linear galaxy bias, using the \texttt{pyrecon} package~\cite{2015MNRAS.450.3822W}. This yields an estimate of the 3D velocity field from which only the line-of-sight component is used for the kSZ analysis. The kSZ signal $T_{\rm kSZ}(\theta_d)$ is then estimated using the velocity-weighted stacking estimator from Ref.~\cite{Schaan_2021}, which correlates the filtered CMB temperature with the reconstructed velocity field. 
We adopt a velocity cross-correlation coefficient of $r = 0.3$ for the Main DESI LRG sample \citep{2024PhRvD.109j3534H} and $r = 0.25$ for the Extended sample, based on values of the photometric redshift errors from \cite{Zhou:2023gji} and the dependence of $r$ on redshift errors from \cite{2024PhRvD.109j3533R}. The uncertainty on that quantity is about 10\%; a more precise determination of $r$ using the latest spectroscopic data is deferred to future work.
Outliers are removed and a symmetric velocity distribution imposed 
to suppress contamination from the thermal SZ effect and the Cosmic Infrared Background, ensuring a foreground-free measurement (see \cite{2024arXiv240707152H} for details). In this work, we make use of the final, stacked kSZ profiles for the Main and Extended LRGs from \cite{2024arXiv240707152H}.

\section{Methods}
\label{sec:method}

In this section, we describe the simulation-based model we adopt in our analysis for modeling the gas and dark matter distribution. 

\subsection{AbacusSummit}

\textsc{AbacusSummit} is a suite of large-volume high-resolution cosmological $N$-body simulations, which was designed to meet the Cosmological Simulation Requirements of the DESI survey \citep{2021MNRAS.508.4017M, 2022MNRAS.509.2194H}. The simulations were run with the cosmological $N$-body simulation code \textsc{Abacus} \citep{2019MNRAS.485.3370G,2021MNRAS.508..575G}, optimized for GPU architectures on the Summit supercomputer at the Oak Ridge Leadership Computing Facility. 

While the \textsc{AbacusSummit} suite spans a wide range of cosmologies and box sizes, here we make use of the \texttt{base} resolution boxes, which contain 6912$^3$ particles in a $2 \ {\rm Gpc}/h$ box, with a mass resolution of $M_{\rm part} = 2.1 \times 10^9 \ M_\odot/h$. As this analysis is performed at fixed cosmology, we employ the fiducial cosmology boxes which have cosmological parameters set to their \textit{Planck} 2018 values: $\Omega_b h^2 = 0.02237$, $\Omega_c h^2 = 0.12$, $h = 0.6736$, $A_s = 2.0830 \times 10^{-9}$, $n_s = 0.9649$, $w_0 = -1$, $w_a = 0$. 

As we are interested in the DESI LRG and BGS galaxy samples, we adopt the snapshot outputs at $z = 0.3$ (for BGS), 0.5 (for Bin 1 and 2 of LRGs), 0.8 (for Bin 3 and 4 of LRGs). We also make use of the particle outputs available for these snapshots, which feature both a 3\% and a 7\% subsample. As we are not shot noise dominated in our cross-correlation measurements between galaxies and matter, we opt to only use the `A' particle subsample, which has 3\% of all particles in the simulation. For full details on the simulation products, see \citet{2021MNRAS.508.4017M}. For some of the internal tests for validating our pipelines, we also utilize the halo light cone catalogs and weak lensing maps generated on the \texttt{huge} \textsc{AbacusSummit} simulation \cite{2021MNRAS.508.4017M}.

All halo masses quoted in this work correspond to the mass definition adopted by the AbacusSummit halo finder CompaSO \citep{2022MNRAS.509..501H}, which defines the virial mass using the spherical collapse model and the fitting formulae from \citet{1998ApJ...495...80B}.

\subsection{Halo Occupation Distribution Model}

To model the distribution of Luminous Red Galaxies (LRGs) within dark matter halos, we adopt a standard (`vanilla') five-parameter Halo Occupation Distribution (HOD) framework \citep{Zheng:2004id}. In this model, the mean number of central and satellite galaxies in a halo of mass $M$ is given by:
\begin{align}
    \langle N_{\mathrm{cen}}(M) \rangle &= \frac{1}{2} \left[1 + \mathrm{erf}\left(\frac{\log M - \log M_{\mathrm{cut}}}{2 \ \sigma_{\log M}}\right)\right], \\
    \langle N_{\mathrm{sat}}(M) \rangle &= \langle N_{\mathrm{cen}}(M) \rangle \left(\frac{M - \kappa M_{\rm cut}}{M_1}\right)^{\alpha}, \quad \text{for } M > \kappa M_{\rm cut}.
\end{align}

We adopt the \textsc{AbacusHOD} prescription within the \texttt{abacusutils} package \footnote{\url{https://github.com/abacusorg/abacusutils}}. The five free parameters are:
\begin{itemize}
    \item $M_{\mathrm{cut}}$: the characteristic halo mass at which a halo has a 50\% probability of hosting a central galaxy.
    \item $\sigma_{\log M}$: the scatter in $\log M$ describing the smooth transition of the central galaxy occupation function.
    \item $\kappa M_{\rm cut}$: the cutoff mass below which no satellites are hosted.
    \item $M_1$: the typical halo mass required to host one satellite galaxy.
    \item $\alpha$: the power-law slope governing the number of satellites in high-mass halos.
\end{itemize}
This HOD formalism provides a simple, flexible, and empirically motivated description of galaxy bias and halo occupation that we use to construct the three-dimensional galaxy-matter power spectrum $P^{gm}(k,z)$.

\subsection{Lensing observables}
\label{sec:lensing}

To predict the stacked CMB lensing signal for a given galaxy sample, we Fourier transform the galaxy-matter power spectrum into the real-space continuous convergence profile $\kappa(\theta)$ via:
\begin{equation}
\label{eq:kappa}
    \kappa(\theta) = \int \frac{\ell\, d\ell}{2\pi} J_0(\ell \theta) C^{g\kappa}_\ell,
\end{equation}
where $J_0$ is the zeroth-order Bessel function and $C^{g\kappa}_\ell$ is the angular cross-power spectrum between galaxies and convergence, computed from the simulation-measured cross-power spectrum $P_{gm}(k, z)$. In particular, under the Limber approximation, the cross-correlation angular power spectrum is given by:
\begin{equation}
   C_\ell^{g \kappa} = \int_0^{\chi_\ast} \frac{d \chi}{\chi^2} W^\kappa(\chi) W^g(\chi) P^{g m}\left(k = \frac{\ell + 0.5}{\chi}, z(\chi)\right)
\end{equation}
where $\chi$ is the comoving distance, $\chi_\ast$ the comoving distance to the last-scattering surface ($z_\ast = 1089.3$), and $P^{g m}(k, z)$ the three-dimensional cross-power spectrum. The lensing and galaxy kernels are defined as:
\begin{equation}
W^\kappa(\chi) = \frac{3 H_0^2 \Omega_m}{2 c^2} \frac{\chi}{a(\chi)}\left( 1 - \frac{\chi}{\chi_\ast} \right),
\end{equation}
\begin{equation}
W^g(\chi) = H(z(\chi)) n_g(z(\chi)),
\end{equation}
where $H(z)$ is the Hubble parameter, $H_0$ the Hubble constant, $c$ the speed of light, $\Omega_m$ the energy density of matter, and $n_g(z)$ the normalized lens galaxy redshift distribution. Since in building our model, we use the \textsc{AbacusSummit} snapshots at fixed redshift, we can substitute the lens redshift distribution with a Dirac delta function at the lens redshift, $z_{\rm l}$, $n_g(z) = \delta^D(z-z_{\rm l})$, arriving at
\begin{equation}
\label{eq:cell}
   C_\ell^{g \kappa} = \frac{1}{\chi_{\rm l}^2} W^\kappa(\chi_{\rm l}) P^{g m}\left(k = \frac{\ell + 0.5}{\chi_{\rm l}}, z_{\rm l}\right).
\end{equation}
Having converted the cross-power spectrum into an estimate of the convergence profile, $\kappa(\theta)$, we apply a low-pass filter with a cutoff at $L_{\rm max} = 3000$, consistent with the filtering applied in the ACT DR6 lensing map reconstruction \citep{ACT:2023dou}. We then bin the resulting $\kappa(\theta)$ profile into the same annular bins used in the data measurements to obtain the final binned profiles, $\hat \kappa(\theta)$. This ensures that we forward-model the lensing signal in a manner that accounts for the full measurement pipeline, enabling an accurate comparison with observations. 


\subsection{Validation}
\label{sec:validation}

A rigorous test we administered is an end-to-end validation of the modeling pipeline using the \textsc{AbacusSummit} light cone maps and catalogs. We utilized the Main and Extended LRG mock samples from \cite{Zhou:2023gji}, which have a matching redshift distribution and footprint to the DESI Legacy Survey. We first tested the validity of the approach in Section~\ref{sec:lensing} by directly comparing the stacked profiles from the light cone simulation (i.e., performing an explicit stack using the 2D $\kappa$ map and LRG-like catalogs on the DESI and ACT footprints) against the predicted $\kappa(\theta)$ from Eq.~\ref{eq:kappa} and found excellent agreement.


An essential component of our modeling pipeline is the assumption that the reconstructed convergence field from the DR6 lensing map provides an unbiased estimate of the true projected mass distribution. To test this, we create a lensed CMB temperature map by first generating a Gaussian realization of the primary CMB using \texttt{CAMB} \footnote{\url{https://camb.readthedocs.io/en/latest/}} with the \textit{Planck} 2018 cosmological parameters. This map is then lensed using the convergence map from \cite{2023MNRAS.525.4367H} via the \texttt{pixell}\footnote{\url{https://github.com/simonsobs/pixell}} package. To approximate ACT DR6 observations, we add white noise with an amplitude of 15~$\mu$K-arcmin and convolved the map with a Gaussian beam of ${\rm FWHM} = 1.6 \ {\rm arcmin}$. We then perform QE CMB lensing reconstruction on the lensed CMB map with the ACT DR6 analysis pipeline which utilizes the \texttt{so-lenspipe} \footnote{\url{https://github.com/simonsobs/so-lenspipe}} package, keeping the settings unchanged. Finally, we directly compare the reconstructed $\kappa$ to the true input convergence maps over the same scales used in this analysis. We find excellent agreement between the reconstructed and true $\kappa$ profiles across all angular scales of interest, confirming that the DR6 QE reconstruction yields unbiased measurements of $\kappa$ for our galaxy sample and modeling framework. This validates the fidelity of our lensing modeling and removes potential concerns about reconstruction bias in the small-scale regime. While we have shown here that for the masses and noise levels of interest the standard QE is appropriate to recover unbiased masses, we note that for more massive clusters and in the low-noise regime of future experiments, asymmetric estimators such as those in \cite{Hu:2007bt, ACT:2020izl} or optimal methods \cite{Baxter:2014frs, Raghunathan:2017cle, Horowitz:2017iql,Levy:2023moy} become advantageous. 

To further validate the robustness of our method, we conduct two key tests focused on recovering the mean halo mass -- our primary science result, for LRG and BGS host halos. First, we evaluate the accuracy of the mean mass inference on a mock galaxy sample generated from a randomly chosen set of HOD parameters. These parameters are drawn from within the range of our Latin Hypercube training set (see Section \ref{sec:emulator}, i.e., no extrapolation is required), but were not used during training.
For this sample, we compute the true mean halo mass as well as its corresponding $\kappa(\theta)$ profile. We then recover the best-fit parameters using a standard minimization procedure with the emulator. The recovered mean halo mass agrees with the true value to within $0.01$ dex in $\log_{10}(M/h^{-1}M_\odot)$, approximately $2\%$, demonstrating excellent internal consistency of our pipeline in the regime where the emulator is well-trained.

To provide a more stringent and conservative test, we also assess the emulator's performance on a mock sample with fixed mass ($M=10^{13.5}h^{-1}M_\odot$) and no satellites. 
This is near the upper edge of our mass range and presents a challenging test case, as the emulator has seen few examples at such high masses and was not explicitly trained on pure halo samples. Moreover, unlike realistic HOD samples, the halo-only sample has no satellites and a sharply peaked mass distribution, making it structurally different from the emulator training set. In this case, we find that the recovered mass differs from the true value by $0.05$ dex, corresponding to approximately a $10\%$ discrepancy. Given that the actual galaxy samples of interest are more similar to the HOD-based test case than to the extreme pure-halo sample, we adopt a conservative estimate of $5\%$ as the systematic uncertainty on our mean mass measurements. 

We also test for potential contamination of the lensing signal by extragalactic foregrounds using the \textsc{WebSky} simulations \citep{Stein:2020its}, which include a range of astrophysical foregrounds (e.g., tSZ, CIB, radio sources) embedded in a simulated CMB sky. As a first test, we use lensing maps reconstructed on foregrounds only (but with the weights and analysis choices appropriate for reconstruction in the presence of primary CMB), similarly to what was performed in \cite{ACT:2023ubw, ACT:2023oei, Sailer:2020lal}. We find that the maximum bias to $\kappa(\theta)$ is approximately 1.3\% for the BGS sample, and less for the others. In addition, we repeat the analysis with the ACT DR6 CIB-deprojected lensing map, finding 5\% differences at worst, noting that part of the difference is likely due to the different noise realization and level in the two lensing maps. Finally, the difference between the full minimum variance reconstruction (which includes polarization data) and temperature-only reconstruction is 3\% at worst.  

Therefore, we conservatively budget a 5\% systematic uncertainty to account for foreground biases. This is in line with the results from the profile hardened QE in \cite{Sailer:2020lal}: at the highest multipole used here ($L=3000$), the total foreground bias in cross-correlation with low-redshift galaxies is of $\mathcal{O}(5\%)$, but much lower at lower $L$. We note that smaller scales, for which the bias rapidly increases, are not used in our analysis. This uncertainty is subdominant to the other sources of uncertainty at play (especially the uncertainty on the kSZ measurements to which the lensing results are compared), and therefore, we won't explore this topic further here. 
However, we note that these biases could be further mitigated by shear-only estimators \cite{Schaan:2018tup, Qu:2022qie}, gradient cleaning methods \cite{Madhavacheril:2018bxi}, polarization-only reconstruction \cite{Sailer:2022jwt}, the use of asymmetric, or optimal estimators discussed above. Future experiments will also be able to leverage more extensive multi-frequency information to further reduce any remaining contamination \cite{Darwish_2021, Sailer_2021}.


\subsection{Gas observables}
\label{sec:gas}

Here we derive the connection between the two-dimensional gas density profile and the observed gas density signal from the kSZ.

The gas is assumed to be fully ionized with primordial abundances at the scales of interest. Under this assumption, the free electron number density is given by
\begin{equation}
    n_{\rm e} = \frac{\rho_{\rm gas}}{\mu_e m_{\rm p}} \; ,
\end{equation}
where $\mu_e = {2}/({X_{\rm H} + 1})$ is the electron chemical potential, $X_{\rm H} \approx 0.76$ is the primordial hydrogen mass fraction, and $m_{\rm p}$ denotes the atomic mass unit. Given a model for the three-dimensional gas density profile $\rho_{\rm gas}(r)$, the stacked (averaged) kinematic Sunyaev-Zel’dovich (kSZ) temperature fluctuation imprinted on the CMB is given by:

\begin{equation}
    \frac{T_{\rm kSZ}(\theta)}{T_{\rm CMB}} = \frac{v_{\rm rms}^{\rm true}}{c} \frac{\sigma_{T}}{\mu_e m_{\rm p}} \
    {\int_{\rm LoS} {dl} \, \rho_{\rm gas}\left(\sqrt{l^2 + d_{\rm A}(z)^2 \theta^2} \right)} .
\label{eq:kSZ}
\end{equation}
Here, $v_{\rm rms}^{\rm true}$ is the true (three-dimensional) root-mean-square velocity of the galaxy sample ($\approx 300 \ {\rm km}/s$), $\sigma_T$ is the Thomson scattering cross-section, and $d_{\rm A}(z)$ is the angular diameter distance to redshift $z$. The integral computes the line-of-sight projection of the gas density profile, yielding a two-dimensional projected gas density.

To enable direct comparison with the CAP-filtered temperature fluctuation maps observed by ACT, we must account for the instrumental beam response. This is done by convolving the projected gas density field with the ACT beam profile $B(\theta)$ in real space. The resulting beam-convolved temperature fluctuation is given by:
\begin{equation}
    T_{\rm kSZ}(\theta) = T_{\rm CMB} \left(\frac{v_{\rm rms}^{\rm true}}{c}\right) \frac{\sigma_{T}}{\mu_e m_{\rm p}} \left[ \rho^{\rm 2D}_{\rm gas}(\theta) * B(\theta) \right] \; .
\end{equation}
This expression gives the expected kSZ signal for the average galaxy in our sample, incorporating the effects of line-of-sight projection and instrumental resolution. Note that for ACT DR6, the beam FWHM is 1.6 arcmin. To compare with the data, we apply the same CAP filtering procedure described in Section~\ref{sec:kSZ}, which is necessary to account for the signal attenuation introduced by the filter. This ensures a consistent forward-modeling pipeline for connecting theoretical predictions to observational measurements.

\subsection{Emulator}
\label{sec:emulator}

To accelerate inference of halo occupation parameters from galaxy-galaxy lensing measurements, we developed a Gaussian Process (GP) emulator for the lensing convergence signal, $\kappa(\theta)$, over a range of angular separations $\theta$. We train this emulator at the mean redshifts for each of the lens samples considered in this study. The emulator takes as input five HOD parameters:
\begin{itemize}
    \item $\log M_{\mathrm{cut}} \in [12.24, 13.34]$
    \item $\log M_1 \in [13.38, 14.38]$
    \item $\sigma \in [0.01, 1.06]$
    \item $\alpha \in [0.35, 1.85]$
    \item $\kappa \in [0.01, 2.65]$
\end{itemize}
These bounds were sampled using Latin Hypercube Sampling (LHS) to efficiently cover the five-dimensional parameter space. A total of 1000 samples were drawn.

For each set of HOD parameters, we compute $\kappa(\theta)$ at 9 angular bins linearly spaced between $0$ and $15$ arcminutes. We train an independent GP model for each $\theta$ bin using the \texttt{scikit-learn} implementation of the GaussianProcessRegressor with an RBF kernel plus a white noise term. Inputs were scaled using a standard normalization prior to training.

To evaluate the emulator's accuracy, we held out a test set and compared predicted values against simulated truth. Across all bins, the emulator achieves excellent performance, with a total RMSE $\sim 10^{-4}$ and $R^2$ scores close to unity ($R^2 \approx 0.9998$). The per bin performance is similarly high, with the worst offender having an RMSE of 0.0001 and $R^2$ of 0.9997.

The mean fractional error across bins ranges from $0.0006$ to $0.0032$, well below typical statistical uncertainties on the CMB lensing signal, validating the emulator's fidelity for use in inference.

\subsection{Likelihood}
\label{sec:like}

To constrain the halo occupation parameters, we compare the measured galaxy-galaxy lensing signal, $\mathbf{d}_\kappa$, to the model predictions $\mathbf{m}_\kappa(\boldsymbol{\theta})$ obtained from the Gaussian Process emulator (Section~\ref{sec:emulator}). The likelihood is assumed to be Gaussian in the difference between the observed and predicted $\kappa(\theta)$ profiles:
\begin{equation}
    \ln \mathcal{L}\left[ \mathbf{d}_{\kappa} \big| \boldsymbol{\theta} \right] = -\frac{1}{2} \left(\mathbf{d}_{\kappa} - \mathbf{m}_\kappa(\boldsymbol{\theta})\right)^T \mathbf{C}^{-1}_{\kappa} \left(\mathbf{d}_{\kappa} - \mathbf{m}_\kappa(\boldsymbol{\theta})\right),
\end{equation}
where $\boldsymbol{\theta}$ denotes the set of five HOD parameters and $\mathbf{C}_\kappa$ is the covariance matrix associated with the measurement.

The covariance matrix $\mathbf{C}_\kappa$ is estimated empirically from the stacked $\kappa(\theta)$ profiles of individual galaxies. Specifically, we compute the covariance between radial bins across the galaxy sample, and normalize it by the number of galaxies $N_g$:
\begin{equation}
    \mathbf{C}_\kappa \approx \frac{1}{N_g} \hat{\mathbf{C}},
\end{equation}
where $\hat{\mathbf{C}}$ is the unnormalized empirical covariance. This method assumes the independence of individual galaxy profiles and yields an estimate accurate to approximately 10\% \citep{2024arXiv240113033C}. We find that the off-diagonal terms of the resulting $\mathbf{C}_\kappa$ are dominated by noise, which can lead to instabilities when inverting the matrix. To regularize the covariance, we perform an eigenvalue decomposition and clip the condition number by enforcing a maximum eigenvalue ratio of 100:
\begin{equation}
    \mathbf{C}_\kappa \rightarrow \mathbf{U} \, \mathbf{\Lambda}_\mathrm{clip} \, \mathbf{U}^\top,
\end{equation}
where $\mathbf{C}_\kappa = \mathbf{U} \, \mathbf{\Lambda} \, \mathbf{U}^\top$ and the eigenvalues in $\mathbf{\Lambda}$ are clipped such that $\lambda_\text{max} / \lambda_\text{min} \leq 100$. This procedure preserves the dominant structure in the covariance while mitigating the impact of noisy modes. We check that this changes the $\chi^2$ value by about 10-15\%.

We explore the posterior distribution of the HOD parameters using dynamic nested sampling, as implemented in the \texttt{dynesty} sampler \cite{2020MNRAS.493.3132S}. This approach efficiently samples multi-modal posteriors and computes the Bayesian evidence. The prior on each HOD parameter is uniform over the same range used to train the emulator (see Section~\ref{sec:emulator}). In practice, this allows us to evaluate the likelihood over the domain of validity of the emulator, ensuring accurate interpolation.

To validate our results, we also perform a direct minimization of the negative log-likelihood to identify the best-fitting point. We find good agreement between the minimizer and the peak of the posterior distribution in most cases, though the nested sampler provides additional information on the 
parameter posteriors and degeneracies.

\section{Results}
\label{sec:results}

In this section, we present our main findings regarding the masses of the DESI samples and compare the gas density profiles against the dark matter profiles derived from the measured CMB lensing convergence profiles.

\subsection{Mass estimates}

\begin{figure}[h]
    \centering
    \includegraphics[width=0.48
    \textwidth]{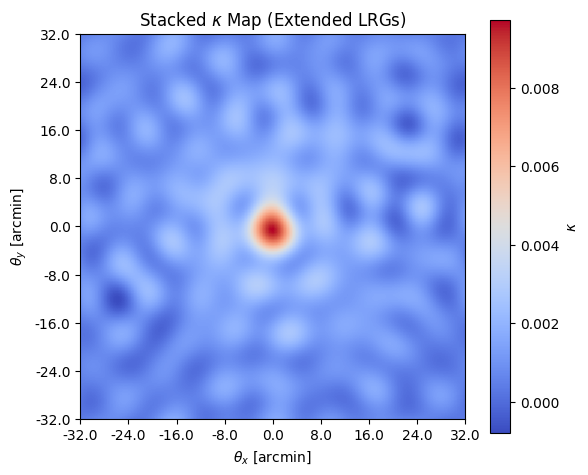}
    \caption{Stacked convergence map centered on the positions of DESI Extended LRGs using the ACT DR6 $\kappa$ map. The central feature corresponds to the mean lensing signal around DESI galaxy halo hosts. The $\kappa$ map released by ACT features a Fourier-space cutoff at $L < 3000$, which is applied to suppress small-scale noise. It introduces coherent features on few-arcminute scales and smears out the 2D profile. 
    }
    \label{fig:stack}
\end{figure}

We begin with Fig.~\ref{fig:stack}, which examines the stacked $\kappa$ map at the positions of DESI Extended LRGs using the ACT DR6 CMB lensing convergence map. This image corresponds to the average projected matter distribution around the selected galaxy sample. The central feature corresponds to the mean lensing signal around DESI galaxy halo hosts and is clearly detected. The convergence map released by ACT features a multipole cutoff of $L < 3000$, which is designed to reduce contamination from small-scale foregrounds and noise. 
At the same time, it reduces the small-scale power, causing the profile to appear smeared and broadened. This effect is also visible in the surrounding structure of the 2D stacked image, where the correlations introduced by the filtering manifest as coherent features on a few-arcminute scales. These features are a known consequence of the filtering and do not reflect the intrinsic structure of the halos.

\begin{figure}[h]
    \centering
    \includegraphics[width=0.45\textwidth]{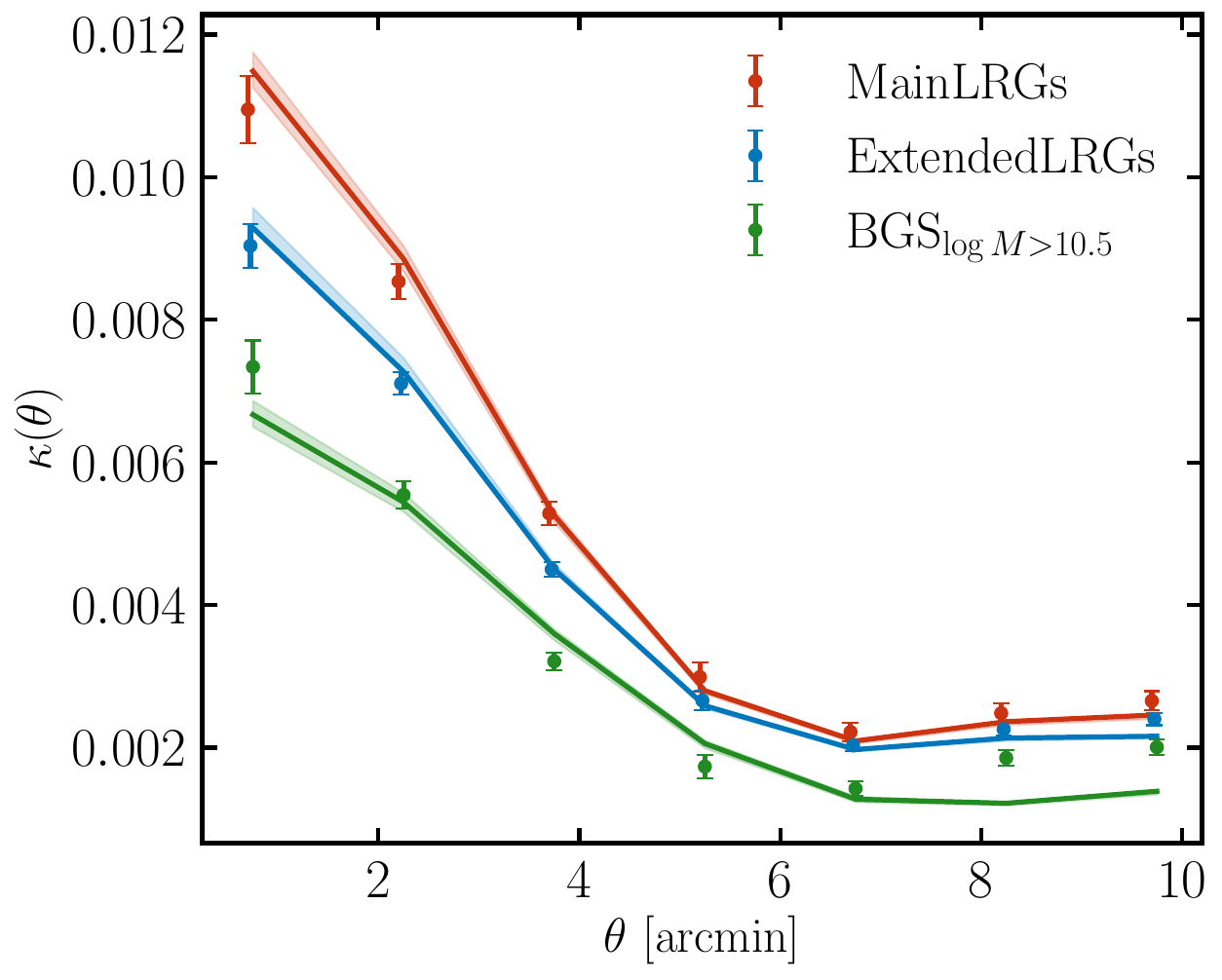}
    \caption{Lensing convergence $\kappa$ as a function of aperture angle for the three DESI galaxy samples: Main LRGs (red), Extended LRGs (blue), and BGS galaxies (orange). Points show the measured $\kappa$ signal in the DR6 map, and the solid lines show the best-fit models based on the simulation-based framework described in Section~\ref{sec:lensing}. Error bars correspond to the square root of the diagonal of the covariance matrix described in Section~\ref{sec:like}. The shaded regions around the best-fit curves denote 68\% confidence intervals. All model fits are computed by forward-modeling the predicted galaxy-matter correlation into $\kappa(\theta)$, applying an $L < 3000$ multipole cut and binning in the same annular bins as the ones used in the data. We attribute the worst fit in the case of the BGS sample to the rapid redshift evolution of the signal (as a function of angular scale), which is not accounted for in our fixed-snapshot approach.}
    \label{fig:tracer}
\end{figure}

\begin{table}
\centering
\begin{tabular}{lccc}
\hline
Param. & ${\rm Main LRGs}$ & ${\rm Extended LRGs}$ & ${\rm BGS}_{\log M > 10.5}$ \\
\hline
$\log M_{\rm cut}$ & $12.683^{+0.129}_{-0.063}$ & $12.491^{+0.07}_{-0.045}$ & $12.133^{+0.102}_{-0.125}$ \\ \vspace{0.2cm}
$\log M_1$ & $14.063^{+0.241}_{-0.395}$ & $14.196^{+0.144}_{-0.433}$ & $13.83^{+0.276}_{-0.67}$ \\ \vspace{0.2cm}
$\sigma_{\log M}$ & $0.133^{+0.179}_{-0.083}$ & $0.108^{+0.126}_{-0.06}$ & $0.107^{+0.194}_{-0.074}$ \\ \vspace{0.2cm}
$\alpha$ & $0.848^{+0.611}_{-0.362}$ & $0.642^{+0.725}_{-0.222}$ & $1.219^{+0.501}_{-0.77}$ \\ \vspace{0.2cm}
$\kappa$ & $1.245^{+0.924}_{-0.816}$ & $0.941^{+1.075}_{-0.737}$ & $1.069^{+0.992}_{-0.79}$ \\
\hline
$\bar{n} \times 1000$ & $0.745^{+0.223}_{-0.133}$ & $1.254^{+0.223}_{-0.198}$ & $3.608^{+1.615}_{-0.723}$ \\ \vspace{0.2cm}
$f_{\rm sat}$ & $0.08^{+0.104}_{-0.057}$ & $0.098^{+0.091}_{-0.078}$ & $0.108^{+0.195}_{-0.032}$ \\ \vspace{0.2cm}
$\log \bar{M}_{\rm h}$ & $13.179^{+0.029}_{-0.021}$ & $13.025^{+0.022}_{-0.016}$ & $13.022^{+0.091}_{-0.061}$ \\
\hline
\end{tabular}
\caption{Best-fit values and 68\% confidence intervals for the five HOD parameters and three derived parameters: comoving number density $\bar{n}$ (in $[{\rm Mpc}/h]^{-3}$), satellite fraction $f_{\rm sat}$, and mean halo mass $\langle M_{\rm halo} \rangle$ (in $M_\odot/h$). Results are shown for each of the three tracer samples: Main LRGs, Extended LRGs, and BGS. All mass units are in $M_\odot/h$. The masses correspond to the virial mass definition from \citet{1998ApJ...495...80B}. We budget around 7\% for the systematic bias on the mean halo mass, as described in the main text.} 
\label{tab:tracer}
\end{table}

In Fig.~\ref{fig:tracer}, we find that our halo model provides an excellent fit to the DR6 $\kappa$ signal for both Main and Extended LRGs. The BGS sample, which consists of lower-redshift and lower-bias galaxies, shows larger residuals between the data and the model prediction. This is not unexpected: BGS galaxies represent a more diverse population, including lower-mass and star-forming systems for which our simple five-parameter HOD may not be as accurate. In contrast, the LRG samples are dominated by red, massive galaxies, where the standard HOD form is known to perform well. The relatively poor fit for the BGS sample likely reflects two important factors: a) as the sample spans a wide range of low redshifts, $z = (0, 0.5)$, there is a substantial mixture of angular scales in the averaged $\kappa$ profiles; additionally, our theoretical model is evaluated at a fixed redshift, $z = 0.3$, and hence lacks the redshift evolution seen in the data profiles; b) the increasing importance of the two-halo term and the need to account for effects such as central/satellite assembly bias or more complex halo occupation dependencies, which are more pronounced for the diverse BGS objects.

{\bf Error on the mass estimate:} Nevertheless, since our primary goal is to infer the mean halo mass of these samples, the limitations of the model are acceptable. Systematic errors from HOD model mis-specification and unmodeled assembly bias and the stellar component are expected to affect the inferred halo mass at the $\sim$5\% level, which is comparable to other sources of uncertainty in the analysis. Since we take a $\sim$5\% uncertainty on modeling and a further (uncorrelated) $\sim$5\% uncertainty due to possible foreground biases, we shall quote a $\sim 5\sqrt{2} \approx 7\%$ systematic uncertainty per sample, noting that it dominates over the statistical uncertainty of the lensing measurement, but is still smaller than the uncertainty on the gas profiles from kSZ and therefore not the limiting factor in our analysis at present. 

One may also worry about the effect of baryonic feedback on CMB lensing, which can redistribute gas beyond the virial radius and lead to a mismatch between the true and modeled mass profiles. However, since we restrict our modeling to multipoles $L < 3000$, our analysis focuses on sufficiently large angular scales where baryonic effects are minimized and the full halo mass is still captured well. Baryon effects in CMB lensing have been studied in \cite{Chung:2019bsk} and shown to be negligible for most of the multipole $L$ range used in our analysis, though they may become sizeable around $L \approx 3000$ for the most extreme models. While smaller than the kSZ uncertainties in this analysis, they motivate a joint and self-consistent analysis of both kSZ and CMB lensing (for example, similar to what was performed in \cite{2025MNRAS.540..143M, Sunseri:2025hhj}), which is left to upcoming work. 
 

Table~\ref{tab:tracer} summarizes the best-fit Halo Occupation Distribution (HOD) parameters and derived quantities for the Main LRG, Extended LRG, and BGS samples. As expected, the Main LRGs occupy the most massive halos on average, followed by the Extended LRGs and then BGS, which reflects their target selection and redshift distributions. The satellite fraction and number density are highest for BGS, which is consistent with it being a low-redshift, high-density sample. On the other hand, they are lower for the Extended LRGs, and lowest for the Main LRGs. 

The five HOD parameters (threshold mass $M_{\rm cut}$, scatter $\sigma_{\log M}$, satellite mass scale $M_1$, slope $\alpha$, and satellite cutoff $\kappa$) all follow the trend of increasing with tracer mass and decreasing redshift. Main LRGs, being the most massive and sparsest sample, require higher occupation thresholds and steeper satellite scaling to match the clustering and number densities. Overall, the results across samples are internally consistent and reflect our expectations from both theory and observational selection.

\begin{figure}
    \centering
    \includegraphics[width=0.45\textwidth]{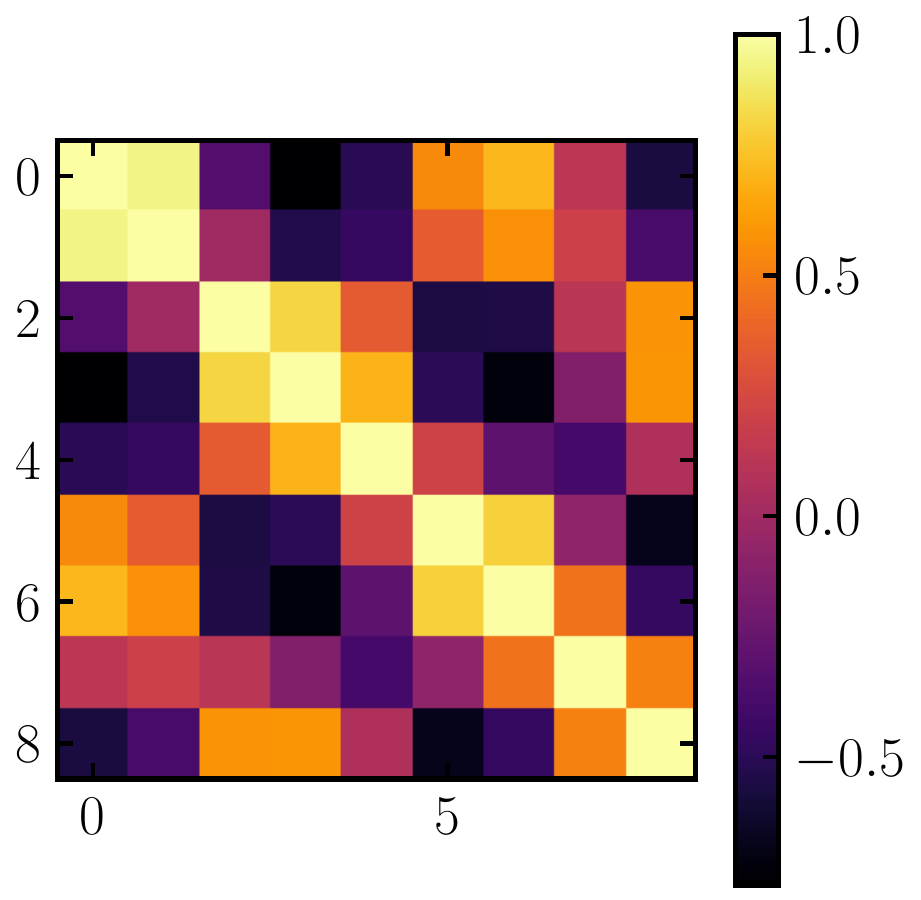}
    \caption{
    Correlation matrix of the lensing signal $\kappa(\theta)$ for the Extended sample, combining all four redshift bins. 
    The pattern is representative of all samples and redshift bins, so we show this case as a typical example. 
    We see a characteristic block structure with alternating positive and negative correlations, which arises due to the $\ell < 3000$ multipole cut applied to the lensing maps, as it induces correlations on angular scales of a few arcminutes.
    }
    \label{fig:corr}
\end{figure}

Fig.~\ref{fig:corr} shows the correlation matrix of the measured $\kappa(\theta)$ signal for the Extended sample, combining all four redshift bins. We normalize the covariance matrix to obtain the correlation coefficients between the angular bins. We note that the correlation matrix is qualitatively identical across the different samples. A clear block-like pattern emerges with alternating positive and negative correlated patches. This structure is characteristic of lensing measurements derived from filtered maps with an $\ell < 3000$ cut, which correlates structure on angular scales of order a few arcminutes, inducing coherent fluctuations across adjacent angular bins in the profile.

\subsection{Comparison with kSZ}

\begin{figure}[t]
    \centering
    \includegraphics[width=0.48\textwidth]{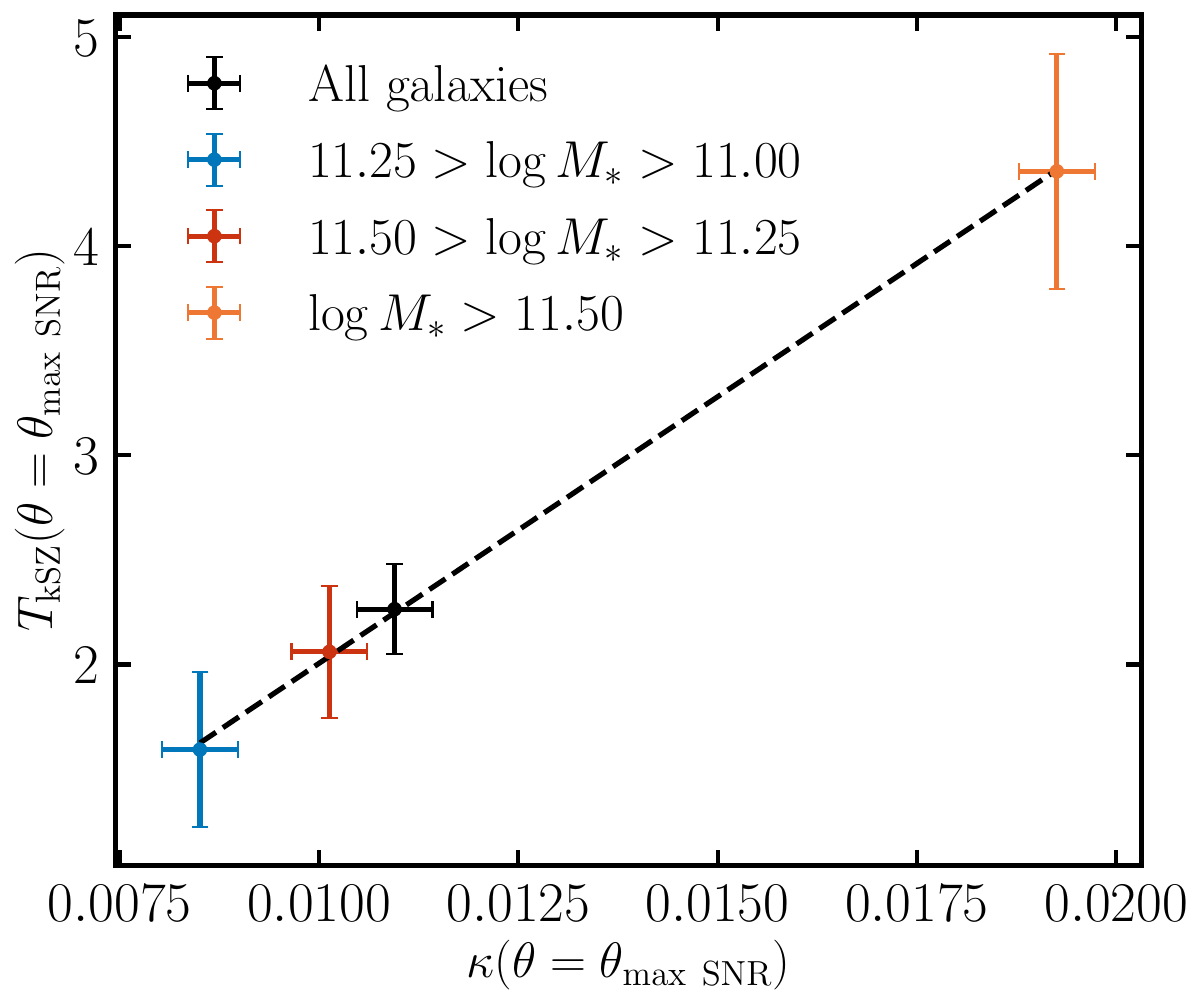}
    \caption{
        Scatter plot comparing the kSZ amplitude at the third radial bin (from \cite{2024arXiv240707152H}) to the $\kappa$ amplitude at the first radial bin (from this work) for the Main LRG sample.
        Each point represents a different stellar mass selection: All galaxies (black point), and the stellar mas bins $10^{11.0} < M_\ast < 10^{11.25}\,M_\odot$, $10^{11.25} < M_\ast < 10^{11.5}\,M_\odot$, and $M_\ast > 10^{11.5}\,M_\odot$. The dashed black line is fitted to the points with a simple linear model and serves to guide the eye. The approximately linear trend indicates that the kSZ and $\kappa$ signals scale proportionally with gas mass and total halo mass, respectively, supporting the interpretation that $M_{\mathrm{gas}} \propto M_{\mathrm{halo}}$.
    }
    \label{fig:ksz_mass}
\end{figure}

\begin{figure}[h]
    \centering
    \includegraphics[width=0.45\textwidth]{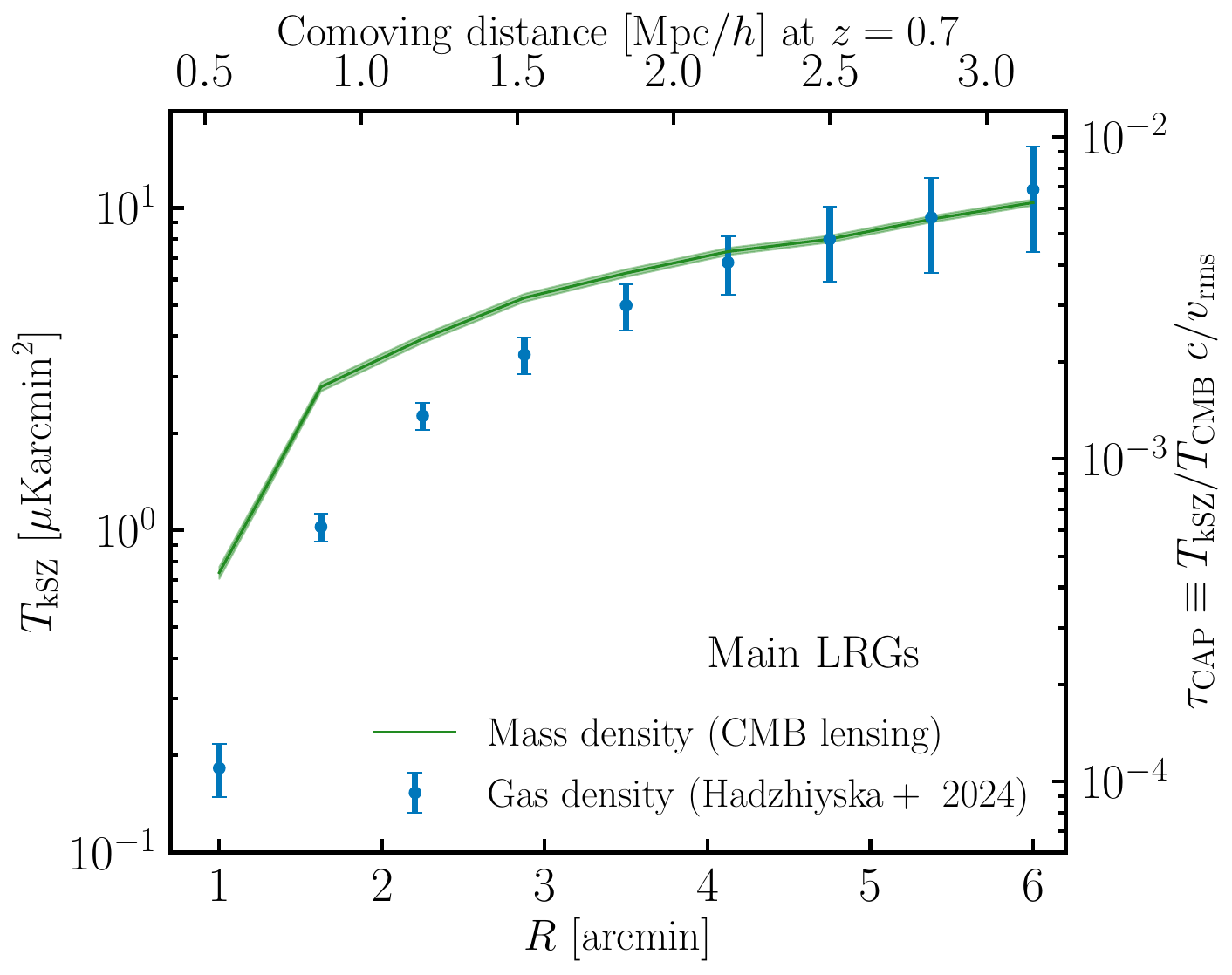}
    \includegraphics[width=0.45\textwidth]{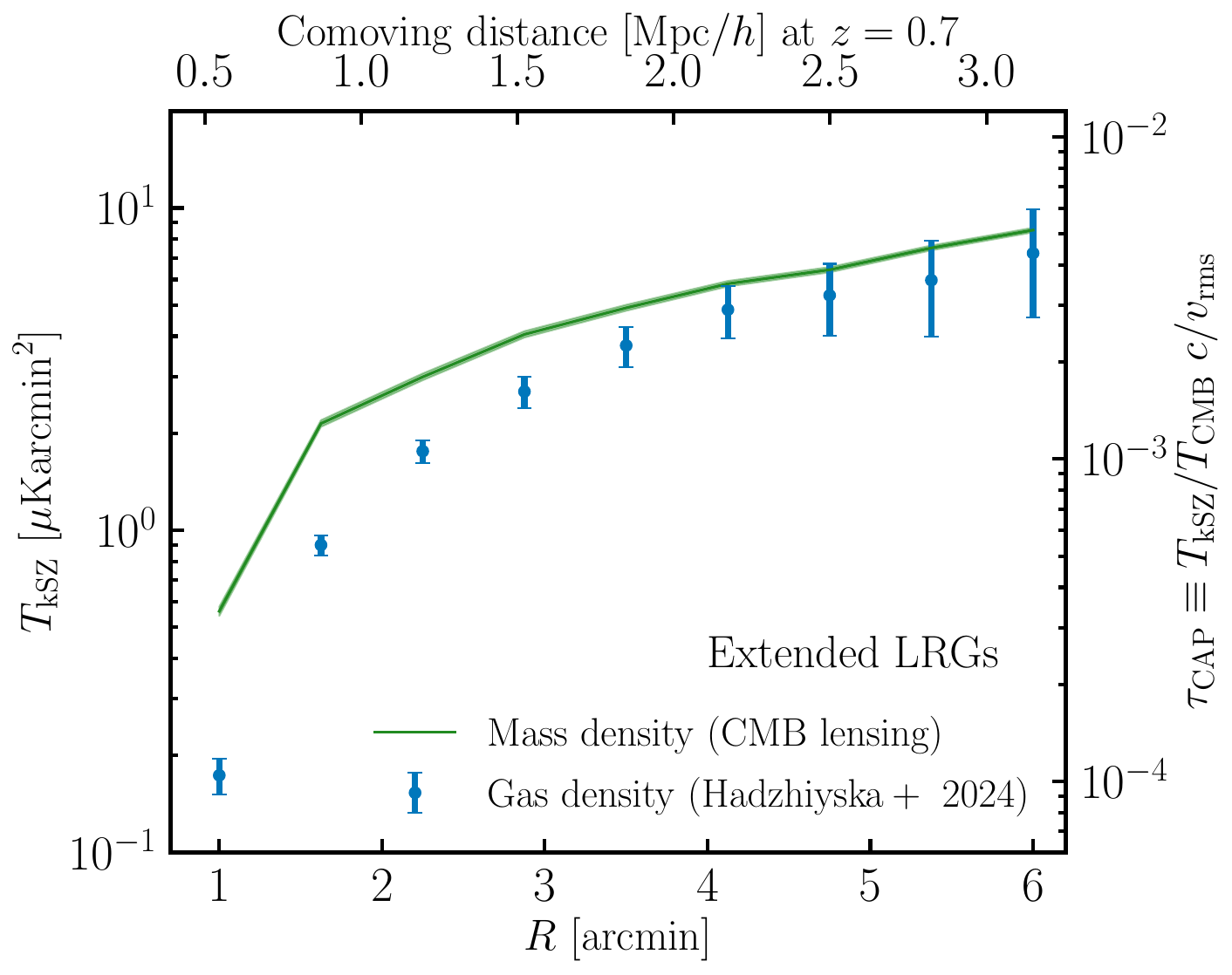}
    \caption{Comparison between the gas density profile measured from the kinematic Sunyaev-Zel’dovich (kSZ) effect \cite{2024arXiv240707152H} and the total matter (dark matter–dominated) profile inferred from our $\kappa(\theta)$ measurements, for the Main (top panel) and Extended (bottom panel) LRG samples. The $\kappa$-based profile has been translated into the same units as the kSZ profile, assuming that gas traces the total matter distribution and that there is no baryonic feedback. The profiles are plotted as a function of aperture radius (in comoving Mpc$/h$). The shaded regions around the $\kappa$-derived curves denote 68\% confidence intervals, and the error bars on the kSZ points indicate the error on the kSZ measurements. The agreement on large scales indicates that the baryon fraction is recovered sufficiently far from the center, indicating no evidence of large amounts of unbound gas.}
    \label{fig:kSZ_kappa}
\end{figure}

In this section, we compare the convergence profiles $\kappa(\theta)$ measured in this work against previous measurements of the kSZ profiles. To examine the connection between the kinetic Sunyaev-Zel'dovich (kSZ) effect and gravitational lensing convergence ($\kappa$), we compare the amplitudes of the two signals for the same galaxy sample in Fig.~\ref{fig:ksz_mass}. Specifically, we focus on the Main LRG sample and consider the kSZ amplitude at the third radial bin from \cite{2024arXiv240707152H}, and the $\kappa$ amplitude at the first radial bin from this work, corresponding to the highest SNR bins in our end-to-end simulation tests \footnote{To avoid potential bias from upward fluctuations, we do not select the bin with the highest SNR in the data, but instead use simulations. We also note that the first radial bin in the $\kappa$ measurements correlates best with the one-halo term amplitude, as it is sensitive to smaller scales.}. The four points in the plot correspond to different stellar mass selections: the full sample (All galaxies), and three stellar mass bins: $10^{11.0} < M_\ast/M_\odot < 10^{11.25}$, $10^{11.25} < M_\ast/M_\odot < 10^{11.5}$, and $M_\ast > 10^{11.5}\,M_\odot$. The dashed black line, fitted to the points with a linear model, serves to guide the eye.

We find that the points lie approximately on a straight line. This linear relationship provides strong empirical support for the idea that the kSZ signal traces the total gas mass within an aperture, while the $\kappa$ signal traces the total matter (halo) mass. The proportionality between the two signals is consistent with the expectation that $M_{\mathrm{gas}} \propto M_{\mathrm{halo}}$, further validating both observables as tracers of halo-scale baryon and matter content.

Next, we compare the radial profiles of the total matter and gas density. To this end, we take the best-fit HOD values for the Main and Extended LRG samples (see Table~\ref{tab:tracer}) and compute the galaxy-matter cross-power spectrum. We then convert it to a kSZ-like CAP measurement by adopting Eq.~\ref{eq:kSZ} but replacing the gas density profile, $\rho^{\rm 2D}_{\rm gas}(\theta)$, with the matter density profile, $\rho^{\rm 2D}_{m}(\theta)$, which is in turn obtained from:
\begin{equation}
    \rho^{\rm 2D}_{m}(\theta) = \int \frac{\ell\, d\ell}{2\pi} J_0(\ell \theta) C^{g m}_\ell,
\end{equation}
where $C^{g m}_\ell$ can be linked to the three-dimensional cross-power spectrum, $P^{g m}(k, z)$, analogously to Eq.~\ref{eq:cell}. 

Fig.~\ref{fig:kSZ_kappa} presents a direct comparison between the gas density profile around LRGs inferred from kSZ measurements and the total matter density profile derived from gravitational lensing. The two panels show results for the Main (top) and Extended (bottom) LRG samples. The lensing-based profiles (derived from $\kappa(\theta)$) are converted into the same units as the kSZ measurement as described above. 

What makes this comparison especially powerful is that, unlike earlier works interpreting the kSZ signal, we no longer need to marginalize over the halo mass or rescale the theoretical prediction. Our lensing measurement provides an independent and accurate mass calibration, enabling a largely parameter-free comparison. At large apertures (2–3 Mpc$/h$, i.e. a few virial radii), we find excellent agreement between the two profiles, supporting the expectation that on large scales, gas and dark matter trace each other well. Thus, the entire baryonic content is recovered sufficiently far from the group center. We note that for the Extended LRG sample, the gas appears to lie systematically lower than the matter, suggesting that a fraction of the gas surrounding these galaxies has been pushed out even further out and is unbound. Intuitively, this makes sense, as the mean mass of the Extended sample is slightly lower and thus the gravitational pull the gas feels towards the group center is slightly weaker.

However, at smaller apertures, we observe a significant discrepancy between the lensing-inferred and kSZ-inferred profiles at the level of $\sim 6.5\sigma$
, indicative of large baryonic feedback processes. These processes push gas beyond the virial radius, causing the gas profile to continue rising at larger scales, whereas the dark matter profile flattens out around $\sim$1.5 $R_{\rm vir}$. This behavior aligns with previous findings \cite{2024arXiv240707152H, 2024arXiv241203631H, RiedGuachalla:2025byu, Schaan21, 2021PhRvD.103f3514A} of large baryonic feedback and underscores the importance of feedback in redistributing baryons.

The only remaining source of systematic uncertainty comes from the velocity reconstruction used in the kSZ analysis, which affects the amplitude of the recovered gas profiles at the 10-15\% level (as quantified in \cite{2024arXiv240707152H, RiedGuachalla:2025byu}). Nonetheless, the comparison clearly demonstrates the complementarity of weak lensing and kSZ as probes of the matter and gas distribution, and marks an important step forward in constraining feedback models and cosmology robustly and self-consistently.

We note that the conversion from the measured $\kappa(\theta)$ profile to the corresponding CAP total matter density profile, $\rho_{\rm matter}(\theta)$, relies on an intermediate step involving simulations. However, this step is largely model-independent: it primarily serves to extend the range of the $\ell$-modes in the cross-spectrum $C_\ell^{\kappa g}$ -- which $\kappa(\theta)$ effectively probes, to smaller scales not directly accessible in the data. Since $C_\ell^{\kappa g}$ is a smooth and featureless function, this extrapolation could equally well be performed using a spline fit or similar functional extension without reference to a specific simulation. We verify in simulations that such approaches yield consistent results with our method, confirming that our inferred matter profiles are robust to the details of this modeling choice. 

\begin{figure}[h]
    \centering
    \includegraphics[width=0.45\textwidth]{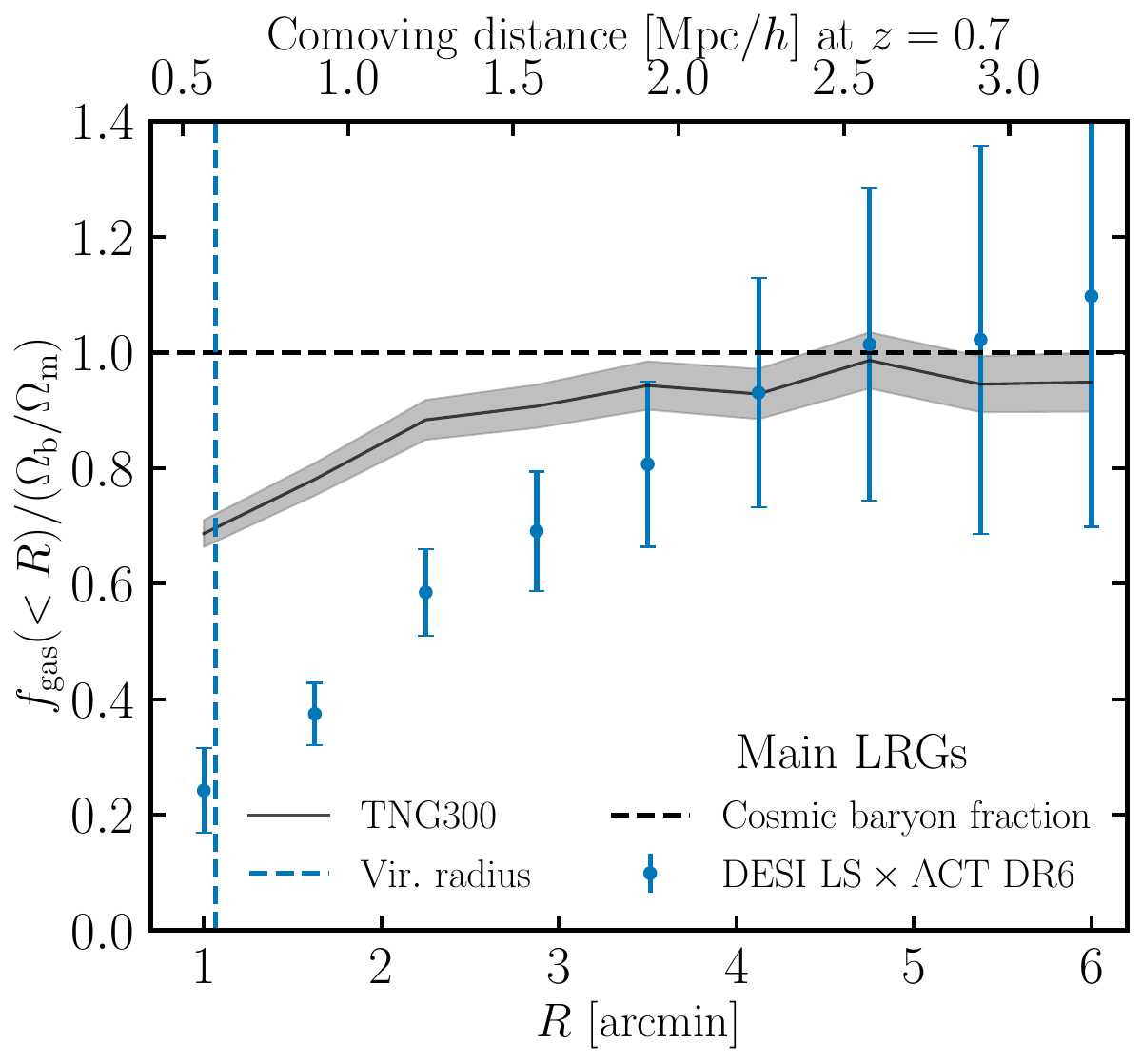}
    \includegraphics[width=0.45\textwidth]{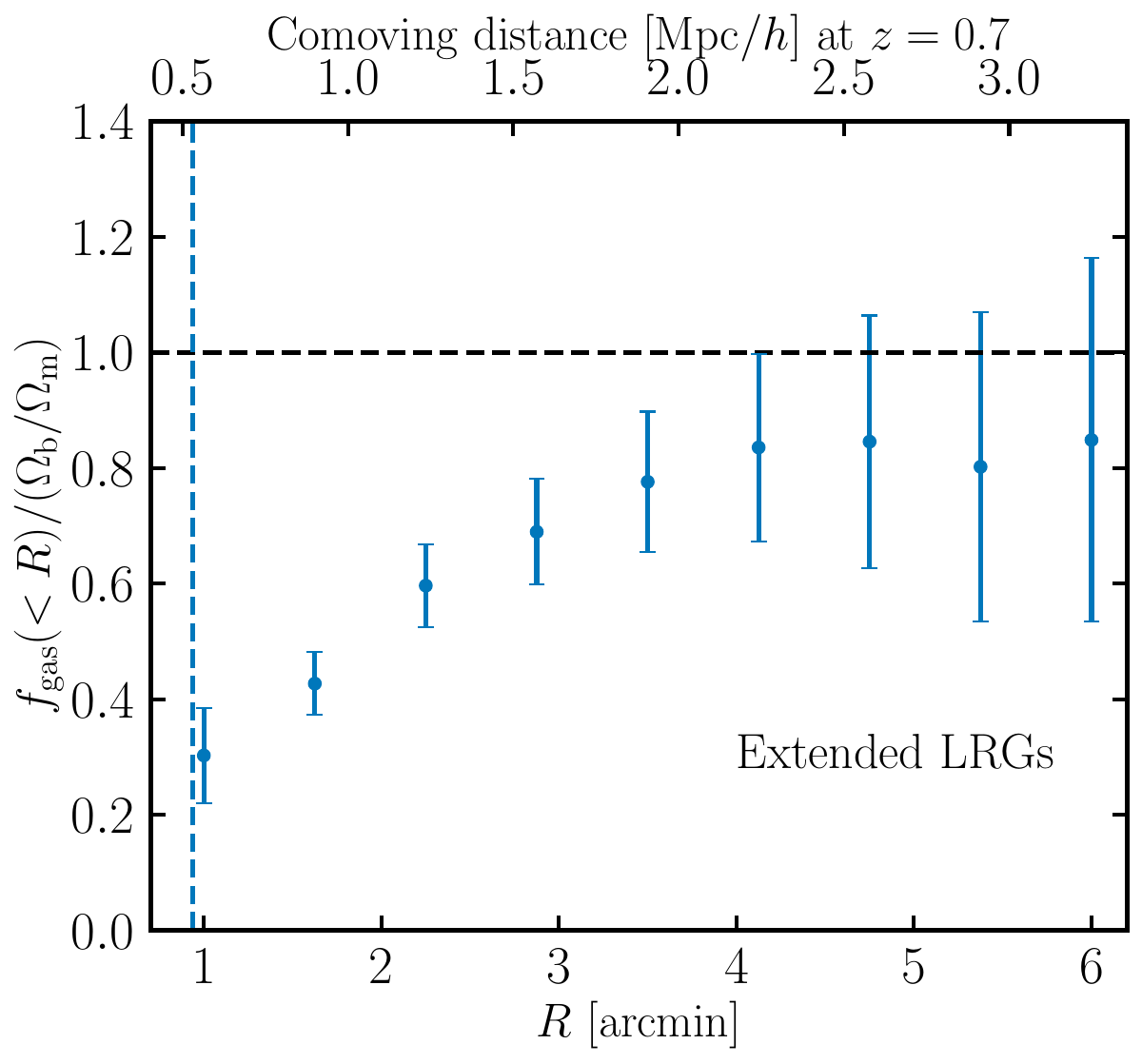}
    \caption{Cumulative baryon fraction, $f_{\rm gas}(<R) / (\Omega_b / \Omega_m)$, as a function of aperture radius for the Main (top) and Extended (bottom) LRG samples. This is computed by dividing the kSZ-inferred gas profile by the $\kappa$-based total matter profile, after correcting for the beam. A value of unity (horizontal dashed line) indicates the cosmic baryon fraction. Vertical dashed lines show the halo virial radii (0.6 and 0.53 ${\rm Mpc}/h$). The black shaded band shows results from TNG300 for a Main-like sample, using the same method. The simulation overpredicts the gas fraction by about 4$\sigma$, 
    reinforcing the case for stronger baryonic feedback in the real Universe.
    At the virial radius, our inferred gas fraction is in excellent agreement with the empirical relation derived by \citet{2024arXiv241116555P} using eROSITA data, which predicts $f_{\rm gas} (R_{200m})/(\Omega_{\rm b}/\Omega_{\rm m}) \approx 0.34$ for Main-LRG-like halos.
}
    \label{fig:fgas}
\end{figure}

Fig.~\ref{fig:fgas} shows the inferred baryon fraction $f_{\rm gas}(<R) / (\Omega_{\rm b} / \Omega_{\rm m})$ as a function of aperture radius for both the Main and Extended LRG samples, where $\Omega_{\rm b} / \Omega_{\rm m}$ is the cosmic baryon fraction. This ratio is computed by dividing the gas density profile derived from the kSZ signal \cite{2024arXiv240707152H} by the total matter profile inferred from our $\kappa(\theta)$ measurements, and correcting for the CMB beam suppression using a simulation-calibrated transfer function (note that this correction is about 3\%). Notably, this transfer function is only weakly sensitive to baryonic feedback effects, especially at large radii.

On large scales ($R \gtrsim 5$ arcmin), the Main sample recovers a baryon fraction consistent with unity, indicating that essentially all the expected baryons are accounted for in these regions. For the Extended sample, we observe a modest shortfall ($f_{\rm gas} / (\Omega_{\rm b} / \Omega_{\rm m}) < 1$), which could stem from statistical fluctuations, a slight mischaracterization of the correlation coefficient $r$ in the estimator due to the larger photometric noise of that sample, or the presence of gas that is no longer bound to the halo.

Most strikingly, at small radii ($R \lesssim 1$ arcmin), we find $f_{\rm gas}/(\Omega_{\rm b} / \Omega_{\rm m}) \sim 0.3$, indicating that a substantial fraction of baryons have been removed from the inner halo regions—strong evidence for the effects of feedback mechanisms such as AGN-driven outflows. This result provides direct observational support for baryon depletion in massive galaxy halos and highlights the complementarity of kSZ and $\kappa$ measurements in tracing baryon dynamics.

We compare our gas fraction at the virial radius with the empirical relation from \citet{2024arXiv241116555P}, based on eROSITA data for low-redshift ($z < 0.2$) systems, and find excellent agreement: $f_{\rm gas}^{\rm X-ray} (R_{200m})/(\Omega_{\rm b}/\Omega_{\rm m}) \approx 0.34$ compared with $f_{\rm gas}^{\rm kSZ} (R_{200m})/(\Omega_{\rm b}/\Omega_{\rm m}) \approx 0.3$ and well within the uncertainty. Their analysis, like ours, indicates that baryonic feedback must be significantly stronger in the real Universe than in most state-of-the-art hydrodynamical simulations, which they find to overpredict the gas content, particularly in the $10^{13.5}$–$10^{14.5},M_\odot$ mass range. While the redshift range differs from ours ($z \sim 0.7$), this consistency suggests that feedback capable of expelling gas from halos without overly suppressing star formation is a necessary ingredient across cosmic time (see also \cite{Lucie-Smith:2025hgj} for a related discussion). Similar conclusions were also recently obtained by \cite{Kovac:2025zqy} by using a ``baryonification'' prescription to relate $f_{\rm gas}$ measured with eROSITA data to kSZ profiles from BOSS galaxies. We leave a more direct comparison to X-ray data for future work.

To facilitate a robust comparison with simulations, we also compute the gas fraction in the TNG300 simulation for a Main-LRG-like sample, shown as a black shaded region. This is done by taking the ratio of the kSZ and dark matter profiles in the simulation, using the same beam-corrected method applied to the data. We find that the TNG300 gas fraction is consistently higher by $\gtrsim$4$\sigma$ 
than the observed one at all radii, only approaching unity at large distances. This underscores the fact that baryonic feedback in TNG300 appears to be weaker than in the real Universe. Because this method compares relative rather than absolute quantities, it is also less sensitive to uncertainties in normalization and mean halo mass, making it a more direct probe of feedback strength.

\begin{figure}[t]
    \centering
    \includegraphics[width=0.45\textwidth]{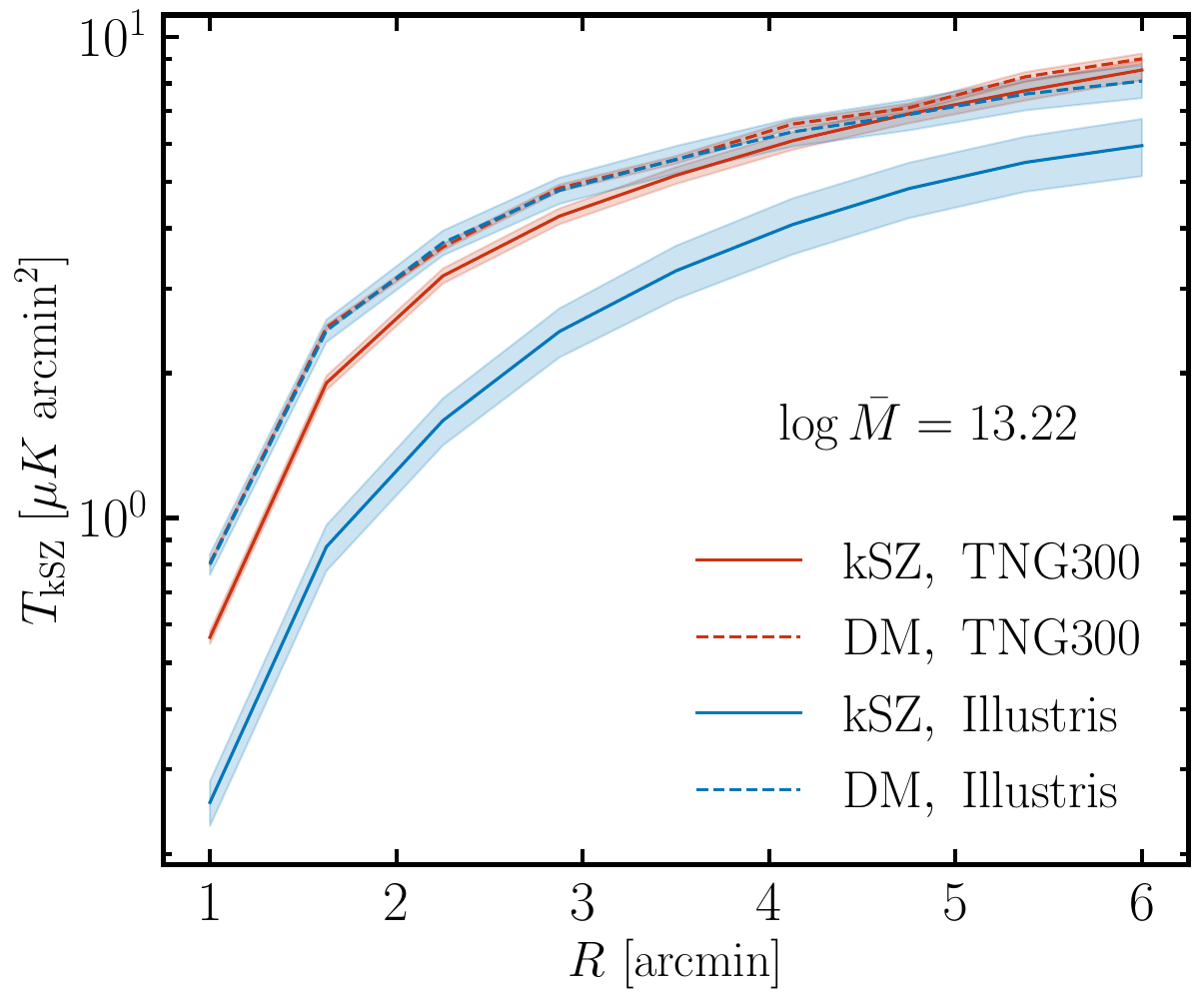}
    \caption{
        CAP profiles in kSZ units for gas (solid lines) and dark matter (dashed lines) from the Illustris and TNG300 simulations. Stellar-mass-selected samples are constructed such that the mean halo mass is $\log(\bar{M}_{\rm halo}/M_\odot\,h^{-1}) = 13.22$ in both cases, matching the Main LRG sample in the data (see Table~\ref{tab:tracer}). The gas profiles reflect the strength of baryonic feedback: Illustris (red) exhibits a significantly steeper profile and reduced amplitude at large apertures due to strong AGN feedback that pushes out, heats up, and unbinds gas from halos. TNG300 (blue) shows a shallower gas profile and no evidence for unbound gas. The dark matter profiles from both simulations are nearly identical, validating the halo matching and convergence of the N-body solvers. No artificial rescaling has been applied to the simulation outputs.
    }
    \label{fig:kSZ_sims}
\end{figure}

To better understand the effects of baryonic feedback on the gas distribution around LRG-like halos, we analyze the Compensated Aperture Photometry (CAP) profiles in kSZ units for two hydrodynamical simulations: Illustris and TNG300. For both simulations, we construct stellar-mass-selected galaxy samples using a number density cut such that the mean halo mass matches that of the observed sample, $\log_{10}(\bar{M}_{\rm halo}/M_\odot\,h^{-1}) = 13.22$. The effective number density cut that we apply to each is 0.9 and 1.1 $\times 10^{-4} \ [{\rm Mpc}/h]^{-3}$, respectively. This ensures a fair comparison with the data. We note that both of these simulations, and especially Illustris, lack the most massive galaxy groups due to their smaller volume, which artificially upweights the less massive galaxy groups when computing the average profiles.

In Fig.~\ref{fig:kSZ_sims}, we show the CAP profiles of gas (solid lines) and dark matter (dashed lines) for both simulations. As expected, baryonic feedback redistributes gas within halos, steepening the gas profiles relative to the dark matter. TNG300, which includes more moderate feedback prescriptions, shows only a modest steepening. Illustris, by contrast, exhibits much steeper gas profiles, indicating overly strong feedback that ejects gas out to large radii or beyond the halo's virial boundary. In addition, Illustris shows a suppressed gas profile amplitude at large apertures due to unbound gas escaping group-sized halos. This feature is not prominently seen in the data (except perhaps marginally in the Extended sample \cite{2024arXiv240707152H}), though the observed kSZ profile depends on the velocity–reconstruction cross-correlation coefficient $r$, which affects the amplitude.

Importantly, the dark matter profiles in both simulations agree remarkably well, demonstrating that the underlying mass distributions and stellar mass selections are consistent. These profiles are shown without any normalization or rescaling applied. Qualitatively, they agree extremely well with the data curves shown in Fig.~\ref{fig:kSZ_kappa}.

\subsection{Mass evolution}

\begin{figure}[h]
    \centering
    \includegraphics[width=0.45\textwidth]{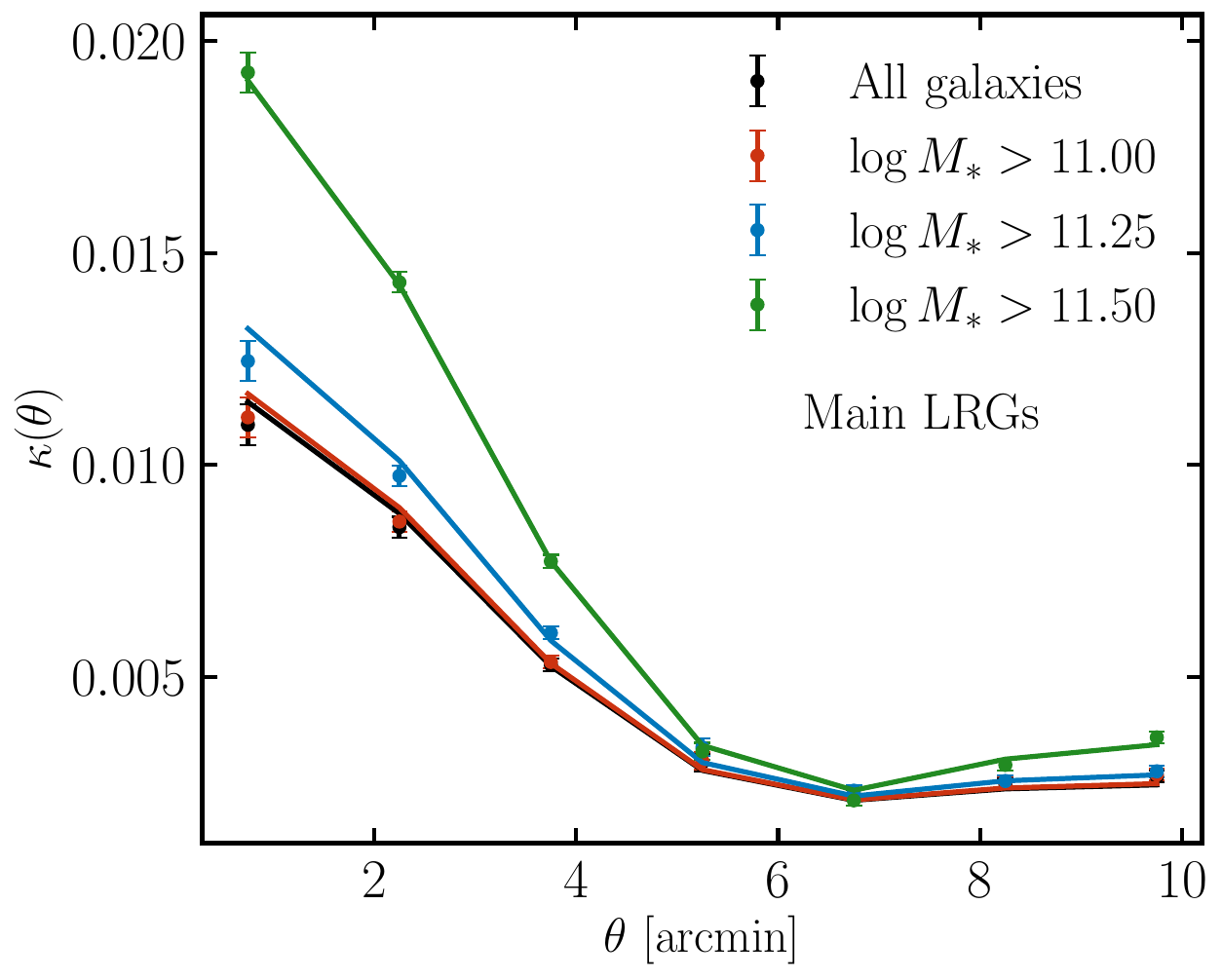}
    \includegraphics[width=0.45\textwidth]{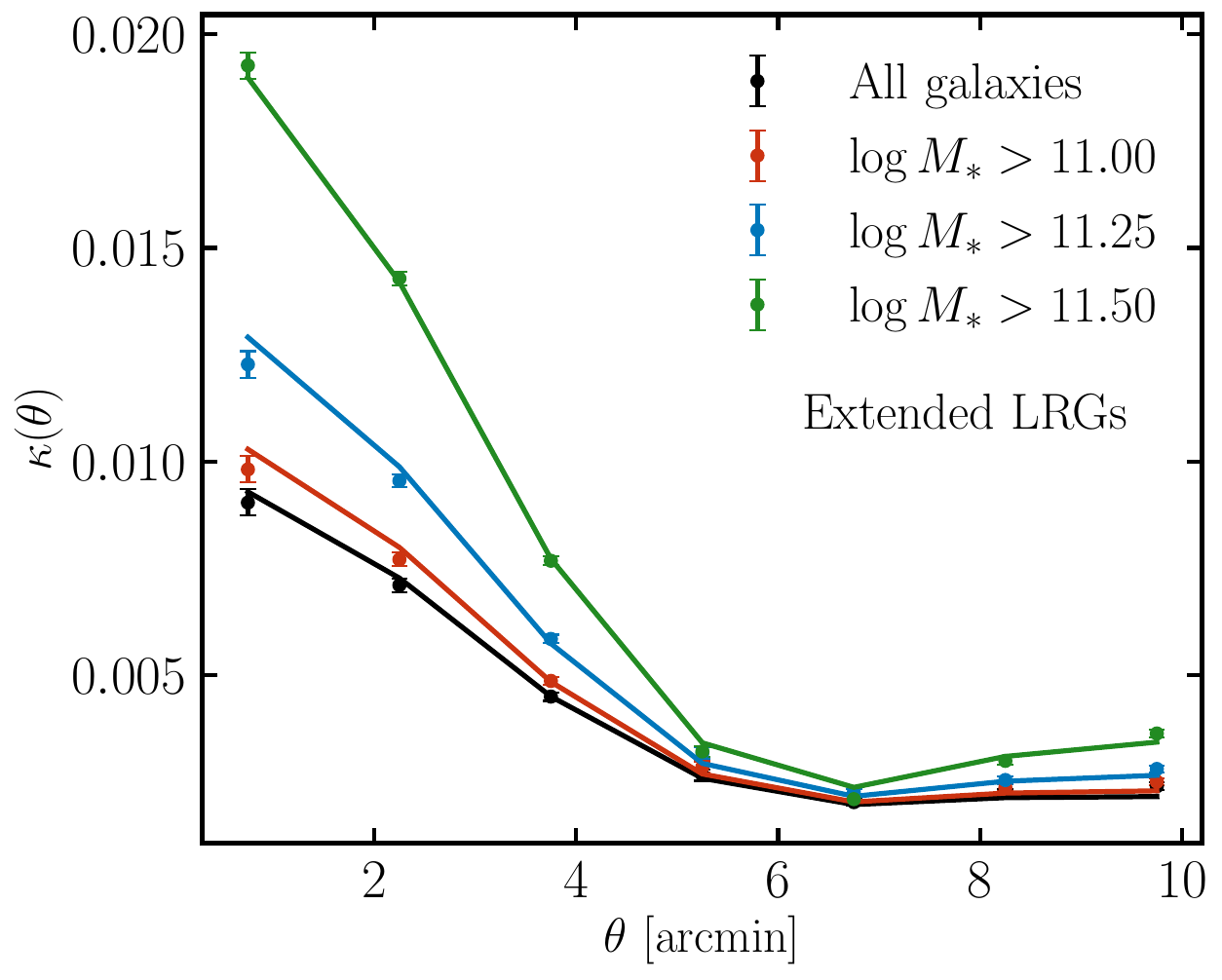}
    \caption{\textbf{Top:} Lensing signal $\kappa(\theta)$ for the Main sample, split into cumulative stellar mass bins: all galaxies (solid), $\log(M_\ast/M_\odot) > 11$ (dashed), $> 11.25$ (dash-dotted), and $> 11.5$ (dotted). The solid lines correspond to the best-fit model predictions, and the error bars are obtained from the diagonal of the covariance matrix.
    \textbf{Bottom:} Same, but for the Extended sample. 
    As expected, the signal amplitude increases with stellar mass threshold due to the corresponding increase in mean halo mass. 
    The Extended sample shows systematically lower amplitudes, reflecting its lower average halo mass.}
    \label{fig:mass_tracer}
\end{figure}

\begin{figure}[h]
    \centering
    \includegraphics[width=0.45\textwidth]{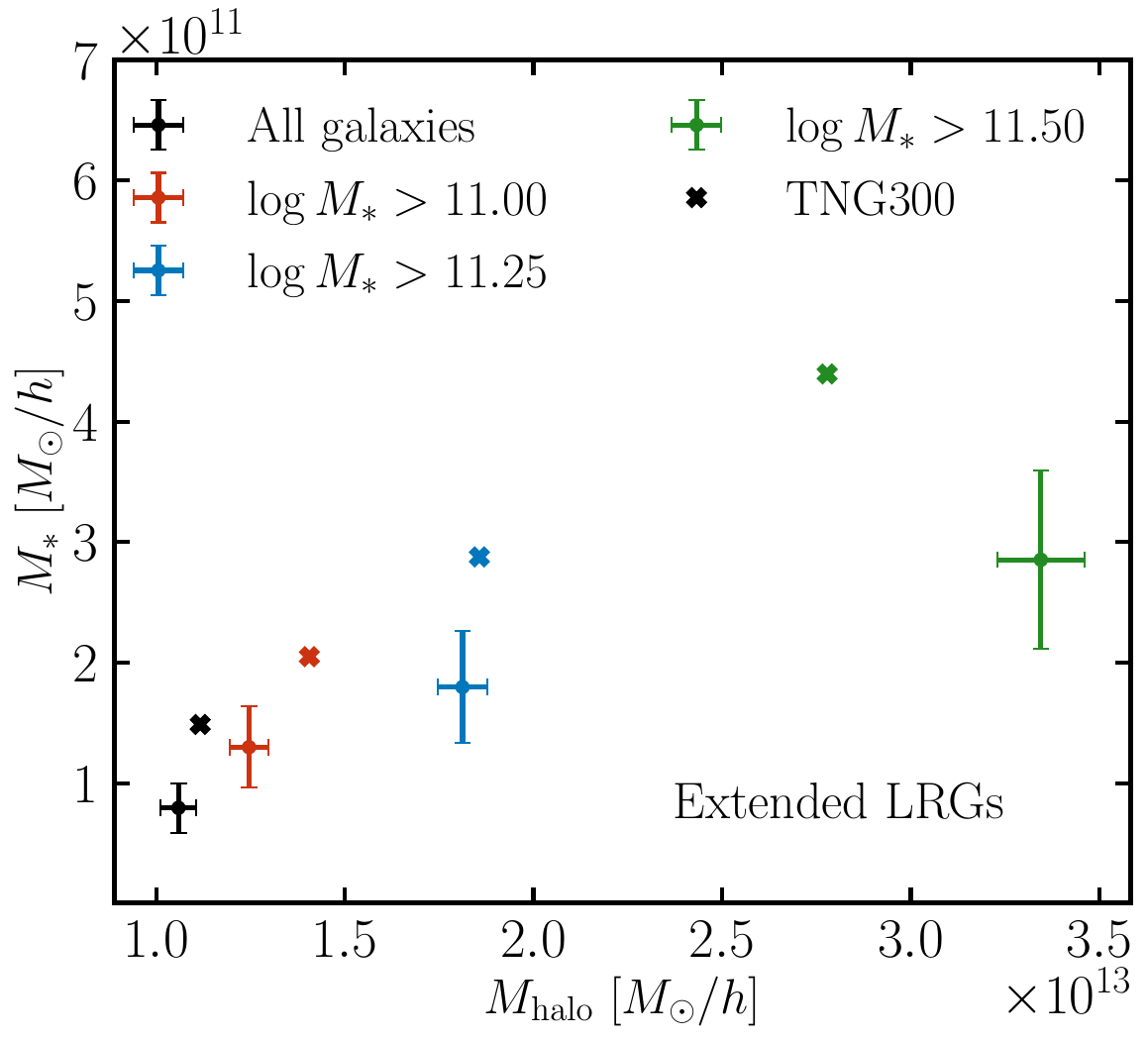}
    \caption{Stellar-halo mass relation for the Extended LRG samples from data and simulations. We show the mean stellar mass versus mean halo mass in four stellar mass-selected bins: all galaxies, $\log(M_\ast/M_\odot) > 11.0$, $> 11.25$, and $> 11.5$. Results are shown both for DESI data (circles) and the matched galaxies in the TNG300 simulation with added stellar mass scatter (crosses). The TNG300 results include a 0.3 dex Gaussian scatter in $\log M_\ast$ to emulate observational uncertainties. 
    The data points have error bars in the $x$ direction from the halo masses inferred in this work and error bars in the $y$ direction from the systematic bias of 0.1 dex on the stellar mass estimates reported in App. C of Ref.~\cite{2023AJ....165...58Z}.
    }
    \label{fig:shmr}
\end{figure}

In Fig.~\ref{fig:mass_tracer}, we show the measured convergence profiles $\kappa(\theta)$ for the Main and Extended samples, split into three cumulative stellar mass bins, along with the full sample. As expected, the amplitude of the signal increases with stellar mass threshold. This behavior reflects the well-established correlation between stellar mass and halo mass: more massive galaxies typically reside in more massive halos, which produce stronger lensing signals.

Since all samples lie at similar redshifts, the observed differences in signal amplitude are attributable to variations in halo mass rather than redshift evolution. Compared to the Main sample, the Extended sample exhibits systematically lower signal amplitudes, consistent with its lower average halo mass. The difference between the full sample and the $\log(M_\ast/M_\odot) > 11$ bin is also more pronounced for the Extended sample. This is because the stellar mass completeness limit for Extended is slightly lower, and the full sample includes a larger fraction of lower-mass galaxies, which reduces the mean halo mass and thus the overall signal amplitude.


Next, we study the stellar-to-halo mass ratio (SHMR) for the Main and Extended LRG samples when we split the galaxies into 4 bins. We first quote the mean halo masses per bin, and we comment on the SHMR, comparing it against the hydrodynamical simulation TNG300 \cite{2019ComAC...6....2N}.

We summarize the mean halo masses for the Main LRG and Extended LRG samples below, computed using our lensing-based inference method. The quoted uncertainties represent the $68\%$ confidence intervals, and all halo masses are reported in $\log(M_{\rm halo}/M_\odot\,h^{-1})$. 

\textbf{Main LRG:}
\begin{itemize}
    \item All galaxies: $13.179^{+0.029}_{-0.021}$
    \item $M_* > 10^{11.0} M_\odot$: $13.190^{+0.028}_{-0.020}$
    \item $M_* > 10^{11.25} M_\odot$: $13.282^{+0.026}_{-0.019}$
    \item $M_* > 10^{11.5} M_\odot$: $13.531^{+0.022}_{-0.019}$
\end{itemize}

\textbf{Extended LRG:}
\begin{itemize}
    \item All galaxies: $13.025^{+0.022}_{-0.016}$
    \item $M_* > 10^{11.0} M_\odot$: $13.096^{+0.020}_{-0.015}$
    \item $M_* > 10^{11.25} M_\odot$: $13.258^{+0.017}_{-0.014}$
    \item $M_* > 10^{11.5} M_\odot$: $13.524^{+0.015}_{-0.014}$
\end{itemize}
We observe a clear trend in both samples: the mean halo mass increases monotonically with the stellar mass threshold. This is consistent with expectations from galaxy-halo co-evolution models and validates that our lensing measurements robustly capture the stellar-to-halo mass relation across the LRG population.

We next examine the stellar-halo mass relation for the Extended LRG samples across four stellar mass bins in Fig.~\ref{fig:shmr} and compare it with simulations. We note that we add 0.3 dex of Gaussian scatter in $\log M_\ast$ to the simulated galaxies to mimic observational uncertainties in the mass determination of the DECaLS catalogs \cite{2021MNRAS.501.3309Z}. The stellar mass bins correspond to: (1) all galaxies, (2) $\log(M_\ast/M_\odot) > 11.0$, (3) $> 11.25$, and (4) $> 11.5$. For each bin, we show the mean stellar mass and mean halo mass inferred from the DESI DR6 CMB lensing measurements, and compare them against galaxies in the TNG300 simulation. For the Main LRG sample, the mean stellar masses across the four bins are:
$2.29 \times 10^{11} M_\odot$,
$2.36 \times 10^{11} M_\odot$,
$2.84 \times 10^{11} M_\odot$, and
$4.24 \times 10^{11} M_\odot$, respectively.
For the Extended LRG sample, the mean stellar masses are:
$1.18 \times 10^{11} M_\odot$,
$1.93 \times 10^{11} M_\odot$,
$2.67 \times 10^{11} M_\odot$, and
$4.24 \times 10^{11} M_\odot$, respectively.

Across all bins, we find that the mean halo masses inferred from the DESI data are in good agreement with those of the matched galaxy samples in the TNG300 simulation, validating our lensing-based mass estimates. The stellar mass trend is also well reproduced, with increasing halo mass corresponding to increasing stellar mass, as expected.

However, we observe that the TNG300 stellar masses are systematically higher than the DESI estimates in each bin. This offset is consistent with known findings in the literature: for instance, \cite{Pillepich:2017fcc} noted that TNG predicts slightly elevated stellar masses compared to data, which may point to insufficient quenching of star formation in massive galaxies in the simulation. Since our stellar mass estimates for DESI are derived from photometry-trained machine learning models \citep{2021MNRAS.501.3309Z}, some additional systematics may contribute to the offset, though the qualitative agreement remains good. We leave for future work the exploration of using more accurate stellar mass estimates (e.g., inferred from the full galaxy spectra as opposed to photometry).

\begin{figure}[h]
    \centering
    \includegraphics[width=0.5\textwidth]{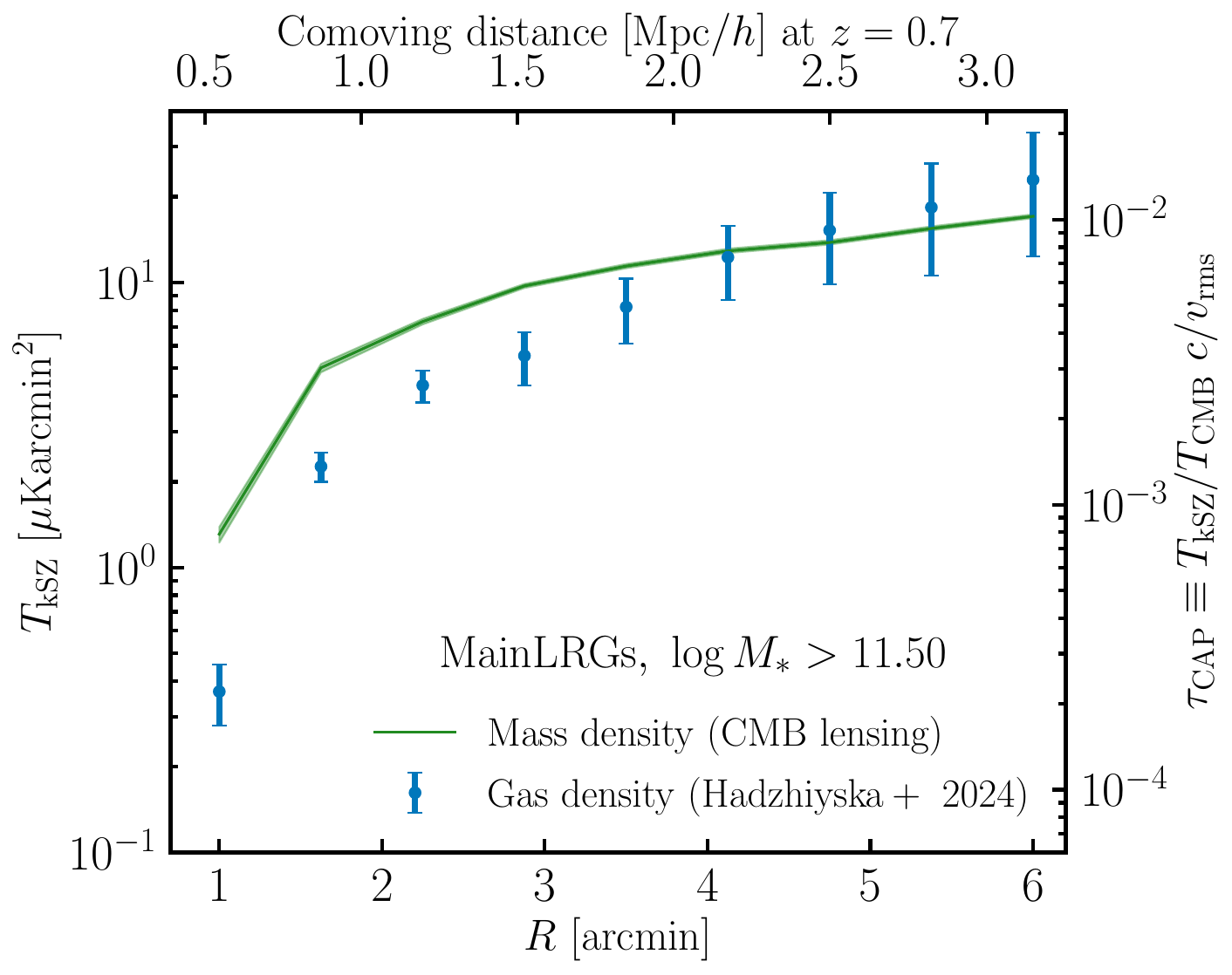}
    \includegraphics[width=0.42\textwidth]{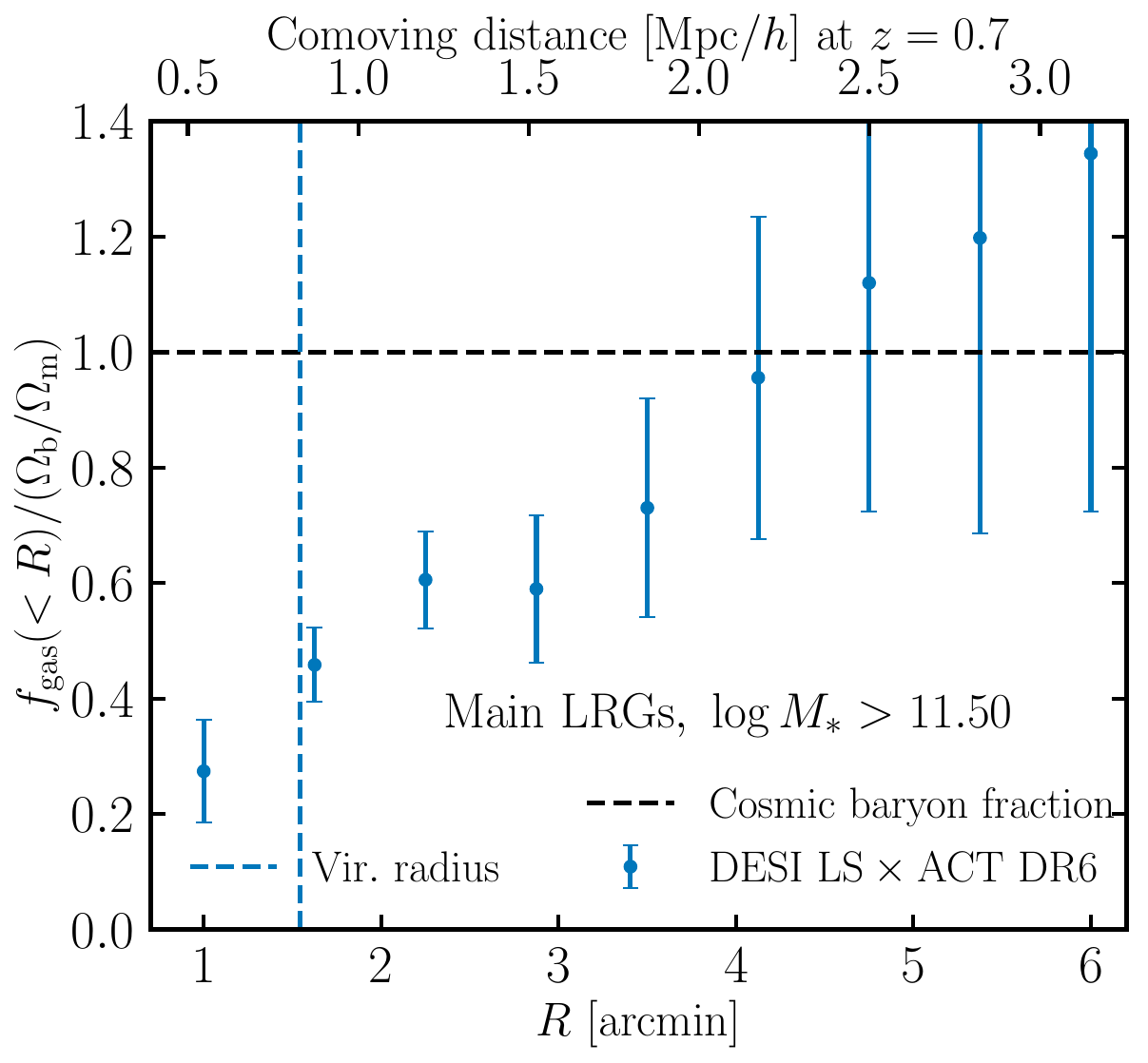}
    \caption{\textbf{Top panel:} Comparison between the gas density profile inferred from the kSZ signal \citep{2024arXiv240707152H} and the total matter profile inferred from our $\kappa(\theta)$ measurements for the highest stellar mass bin of the Main LRG sample ($\log_{10}(M_\ast/M_\odot) > 11.5$). The $\kappa$-derived profile is rescaled into kSZ units as outlined in Section~\ref{sec:method}. \textbf{Bottom panel:} Cumulative baryon fraction, $f_{\rm gas}(<R)/(\Omega_b/\Omega_m)$, computed from the ratio of the kSZ and $\kappa$ profiles and corrected for beam effects using a transfer function derived from simulations. The horizontal dashed line marks the cosmic baryon fraction, and the vertical dashed line denotes the halo virial radius for this subsample ($R_{\rm vir} = 0.83$ Mpc$/h$ comoving). At the virial radius, our inferred gas fraction matches the empirical relation from \citet{2024arXiv241116555P}, based on eROSITA observations very closely, which predicts $f_{\rm gas}(R_{200m})/(\Omega_{\rm b}/\Omega_{\rm m}) \approx 0.43$ for halos similar to those in the Main-LRG sample.}
    \label{fig:fgas_mass_main}
\end{figure}

In Fig.~\ref{fig:fgas_mass_main}, we focus on the most massive stellar mass bin of the Main sample ($\log_{10}(M_\ast/M_\odot) > 11.5$), for which a direct comparison with \citep{2024arXiv240707152H} is possible. The top panel shows the comparison between the gas profile inferred from the kSZ signal and the total matter profile derived from $\kappa(\theta)$ measurements. The bottom panel presents the corresponding cumulative gas fraction profile, $f_{\rm gas} (<R)/(\Omega_b/\Omega_m)$, corrected for the effect of the ACT beam using a transfer function calibrated on simulations. We restrict our detailed modeling to the highest bin, as stellar mass–selected subsamples with both lower and upper bounds lead to non-standard HOD shapes that are challenging to model robustly via the `vanilla' HOD formalism typically adopted for LRGs.

We find that, as with the full Main sample (cf. Fig.~\ref{fig:fgas}), the gas fraction asymptotes to unity at large radii, indicating that the baryons are fully accounted for beyond a few times the virial radius. Notably, the gas fraction in this high-mass bin is slightly elevated relative to the full sample, which is consistent with expectations: more massive halos have deeper gravitational potentials and are therefore less susceptible to baryon removal via feedback processes such as AGN activity. We also see indications that the baryon fraction reaches unit at smaller apertures compared with the full Main sample, which is consistent with the picture that less gas is expelled from high-mass halos. However, in this regime the error bars are also largest and thus follow-up with a larger cluster sample would be beneficial.

\subsection{Redshift evolution}

\begin{figure}[h]
    \centering
    \includegraphics[width=0.45\textwidth]{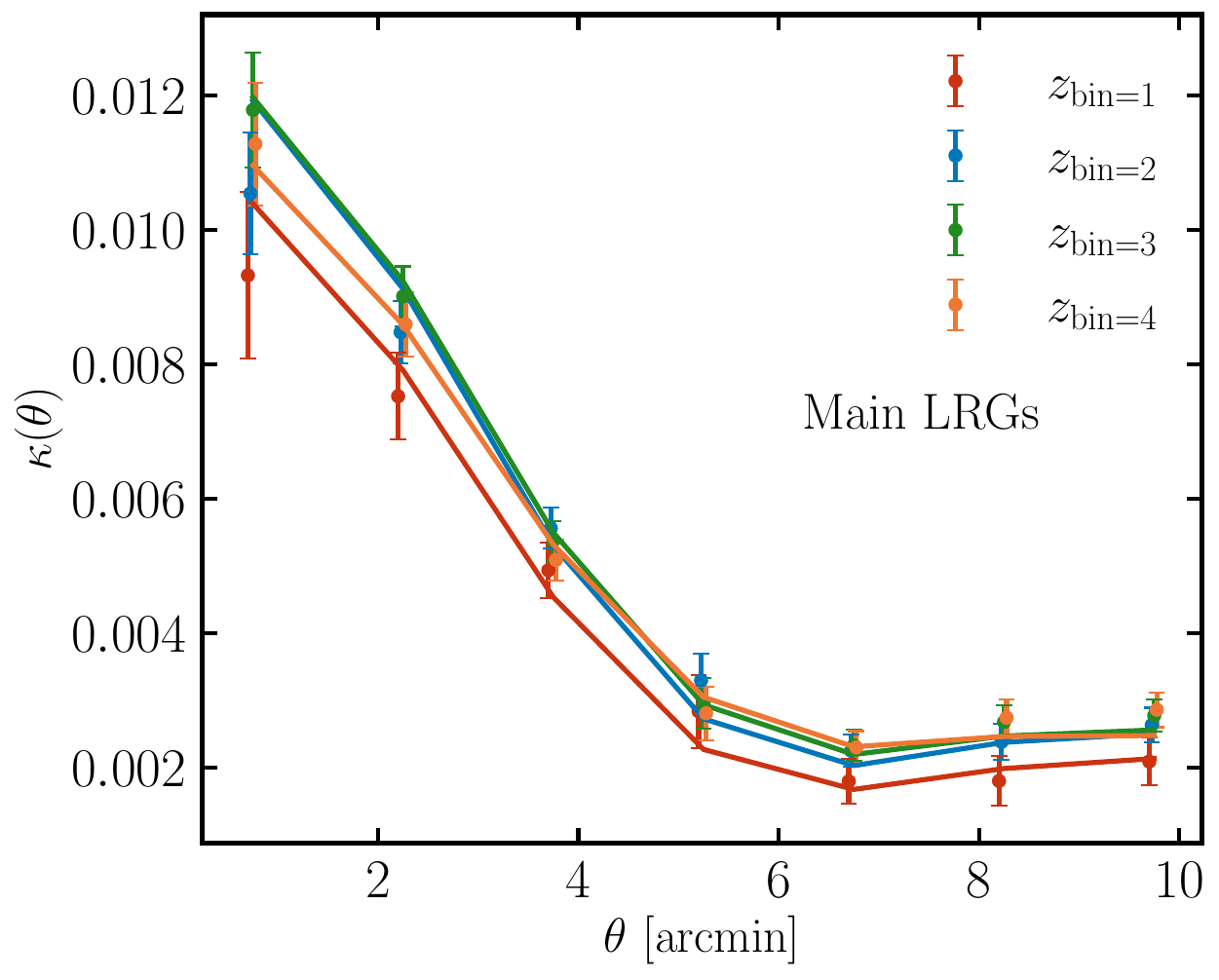}
    
    \includegraphics[width=0.45\textwidth]{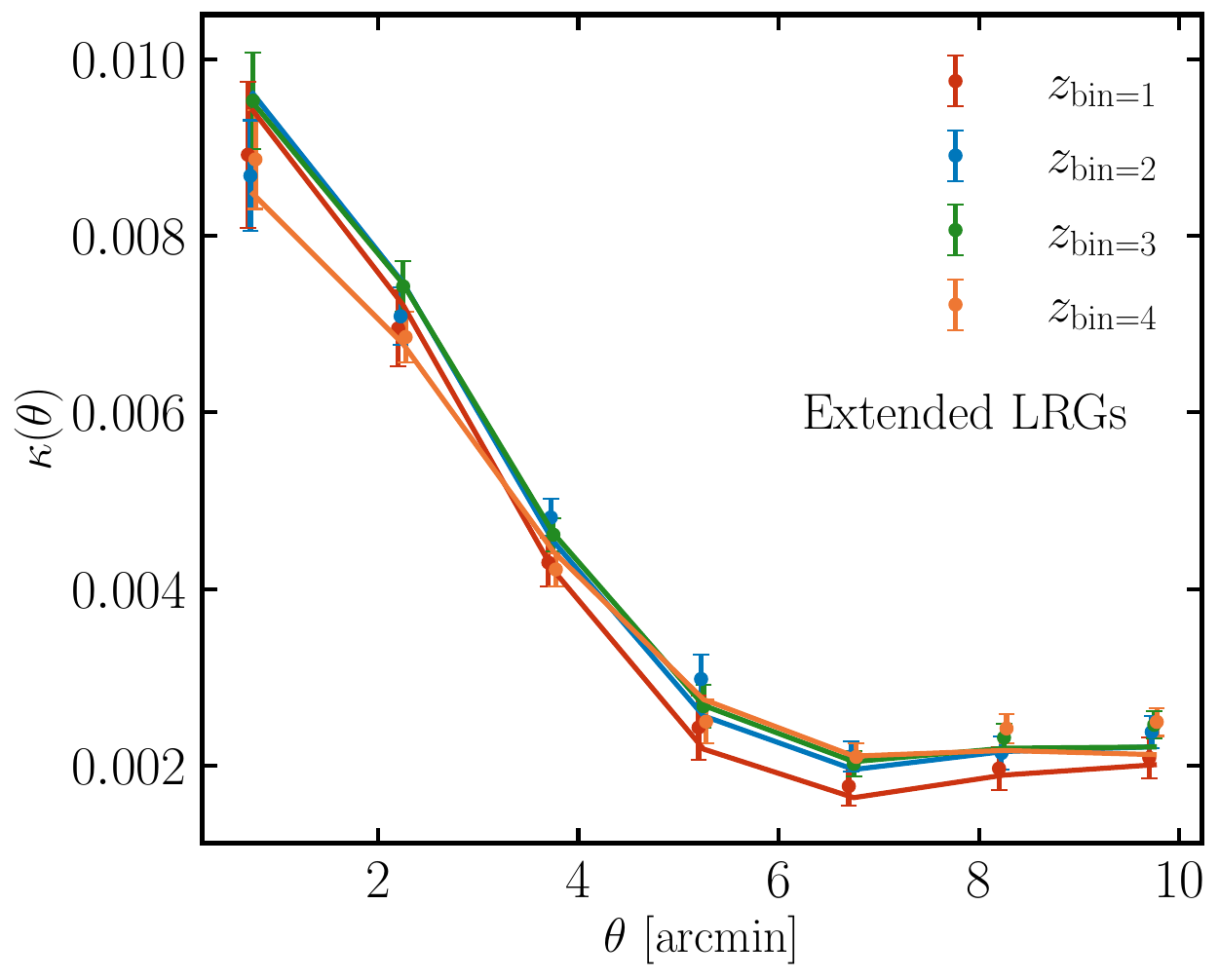}
    \caption{Redshift evolution of the lensing signal $\kappa(\theta)$ for the Main (top panel) and Extended (bottom panel) LRG samples, shown in four tomographic bins. We observe a mild increase in the signal amplitude with redshift, particularly near $\theta \sim 5 \ {\rm arcmin}$ where the lensing kernel is most sensitive. The curves correspond to the best-fit model predictions, and error bars are derived from the diagonal of the covariance matrices.}
    \label{fig:z_tracer}
\end{figure}

\begin{table}
\centering
\begin{tabular}{lcccc}
\hline
Param. & $z_1$ & $z_2$ & $z_3$ & $z_4$ \\
\hline
$\log M_{\rm cut}$ & $12.61^{+0.39}_{-0.15}$ & $12.63^{+0.23}_{-0.09}$ & $12.73^{+0.27}_{-0.1}$ & $12.63^{+0.32}_{-0.11}$ \\ \vspace{0.2cm}
$\log M_1$ & $13.91^{+0.33}_{-0.35}$ & $14.02^{+0.26}_{-0.41}$ & $13.98^{+0.29}_{-0.37}$ & $13.97^{+0.29}_{-0.37}$ \\ \vspace{0.2cm}
$\sigma_{\log M}$ & $0.3^{+0.29}_{-0.21}$ & $0.18^{+0.25}_{-0.13}$ & $0.21^{+0.23}_{-0.14}$ & $0.23^{+0.25}_{-0.15}$ \\ \vspace{0.2cm}
$\alpha$ & $1.0^{+0.56}_{-0.48}$ & $0.84^{+0.6}_{-0.39}$ & $0.96^{+0.57}_{-0.44}$ & $0.96^{+0.56}_{-0.45}$ \\ \vspace{0.2cm}
$\kappa$ & $1.3^{+0.92}_{-0.88}$ & $1.3^{+0.91}_{-0.9}$ & $1.33^{+0.89}_{-0.89}$ & $1.28^{+0.91}_{-0.85}$ \\
\hline
$\bar{n} \times 1000$ & $1.3^{+0.6}_{-0.41}$ & $1.09^{+0.37}_{-0.25}$ & $0.69^{+0.26}_{-0.19}$ & $0.91^{+0.34}_{-0.27}$ \\ \vspace{0.2cm}
$f_{\rm sat}$ & $0.08^{+0.09}_{-0.05}$ & $0.09^{+0.1}_{-0.06}$ & $0.08^{+0.09}_{-0.05}$ & $0.07^{+0.09}_{-0.05}$ \\ \vspace{0.2cm}
$\log \bar{M}_{\rm h}$ & $13.19^{+0.09}_{-0.06}$ & $13.22^{+0.05}_{-0.04}$ & $13.2^{+0.05}_{-0.03}$ & $13.12^{+0.06}_{-0.04}$ \\
\hline
\end{tabular}
\caption{Best-fit HOD and derived parameters for the Main LRG sample, shown across four redshift bins. The redshift bins correspond to: Bin 1: 0.4, 0.54, 0.713, 0.86,  $z_1 < z < 0.54$, Bin 2: $0.54 < z < 0.713$, Bin 3: $0.713 < z < 0.86$, Bin 4: $0.86 < z < 1.024$. All mass units are in $M_\odot/h$, and comoving number densities are in $[{\rm Mpc}/h]^{-3}$. The masses correspond to the virial mass definition from \citet{1998ApJ...495...80B}. We budget around 5\% for the systematic bias on the mean halo mass (see Section~\ref{sec:validation}).}
\label{tab:z_main}
\end{table}

Fig.~\ref{fig:z_tracer} displays the redshift evolution of the lensing convergence profile, $\kappa(\theta)$, for the Main and Extended LRG samples, split into four tomographic bins. The overall trend shows a mild increase in amplitude with redshift, particularly around $\theta \sim 5 \ {\rm arcmin}$. This monotonic behavior is consistent with the expectations from the lensing kernel, which peaks at higher values for more distant and more massive lenses. The observed redshift dependence complements the corresponding mass evolution summarized in Table \ref{tab:tracer}, confirming that the higher-redshift bins are, on average, associated with more massive halos. The agreement between data and model remains good across all bins, and no strong deviations from the fiducial HOD form are required to explain the redshift dependence.

In Table~\ref{tab:z_main}, we show the HOD fits and derived quantities for the Main LRG sample split into four redshift bins. Across the redshift range, the derived quantities (number density, satellite fraction, and mean halo mass) remain relatively stable, with no statistically significant trends. The stability in these parameters is expected, as the Main LRG selection targets a stable population of massive red galaxies across a relatively narrow redshift range. Minor fluctuations in the HOD parameters likely reflect small differences in completeness or selection efficiency, but the overall HOD structure remains consistent with a massive, primarily central-dominated population.

\begin{table}
\centering
\begin{tabular}{lcccc}
\hline
Param. & $z_1$ & $z_2$ & $z_3$ & $z_4$ \\
\hline
$\log M_{\rm cut}$ & $12.49^{+0.37}_{-0.1}$ & $12.44^{+0.17}_{-0.07}$ & $12.52^{+0.25}_{-0.08}$ & $12.38^{+0.33}_{-0.07}$ \\ \vspace{0.2cm}
$\log M_1$ & $13.96^{+0.29}_{-0.37}$ & $14.01^{+0.27}_{-0.49}$ & $14.05^{+0.24}_{-0.39}$ & $14.1^{+0.21}_{-0.39}$ \\ \vspace{0.2cm}
$\sigma_{\log M}$ & $0.24^{+0.31}_{-0.17}$ & $0.14^{+0.23}_{-0.1}$ & $0.17^{+0.25}_{-0.11}$ & $0.2^{+0.27}_{-0.13}$ \\ \vspace{0.2cm}
$\alpha$ & $0.92^{+0.58}_{-0.44}$ & $0.76^{+0.66}_{-0.34}$ & $0.89^{+0.61}_{-0.4}$ & $0.82^{+0.63}_{-0.35}$ \\ \vspace{0.2cm}
$\kappa$ & $1.28^{+0.94}_{-0.89}$ & $1.23^{+0.94}_{-0.85}$ & $1.23^{+0.94}_{-0.83}$ & $1.1^{+1.04}_{-0.79}$ \\
\hline
$\bar{n} \times 1000$ & $1.58^{+0.54}_{-0.43}$ & $1.66^{+0.54}_{-0.34}$ & $1.16^{+0.33}_{-0.28}$ & $1.72^{+0.35}_{-0.47}$ \\ \vspace{0.2cm}
$f_{\rm sat}$ & $0.07^{+0.1}_{-0.05}$ & $0.08^{+0.14}_{-0.06}$ & $0.06^{+0.09}_{-0.04}$ & $0.06^{+0.09}_{-0.04}$ \\ \vspace{0.2cm}
$\log \bar{M}_{\rm h}$ & $13.11^{+0.07}_{-0.04}$ & $13.09^{+0.04}_{-0.03}$ & $13.05^{+0.05}_{-0.03}$ & $12.92^{+0.06}_{-0.03}$ \\
\hline
\end{tabular}
\caption{Same as Table~\ref{tab:z_main}, but for the Extended LRG sample. The redshift bins are identical to those used in Table~\ref{tab:z_main}.}
\label{tab:z_extended}
\end{table}

Table~\ref{tab:z_extended} presents the redshift evolution of the HOD fits for the Extended LRG sample. We observe a mild decrease in the mean halo mass $\langle M_{\rm halo} \rangle$ with increasing redshift, consistent with the expectation that halos hosting galaxies of fixed stellar mass are less massive at earlier cosmic times. The satellite fraction and number density also exhibit modest variations across bins, reflecting changes in sample completeness and evolution in the underlying halo population. Nonetheless, the general HOD shape remains stable, supporting the robustness of our modeling framework across redshift.

\section{Summary and conclusions}
\label{sec:conclusions}

In this work, we measure the gravitational lensing convergence profile, $\kappa(\theta)$, around photometric galaxies in DESI using the CMB lensing maps from ACT DR6 and compare it with existing kSZ measurements. We focus on three DESI tracer samples: the Main Luminous Red Galaxies (LRGs), the Extended LRGs, and the Bright Galaxy Sample (BGS), and interpret the measurements through a simulation-based halo model, allowing us to infer key halo properties such as mean halo mass and satellite fraction. Our joint kSZ and CMB lensing analysis allows us to test whether we recover the full baryonic content predicted by cosmological models and hydrodynamical simulations. Moreover, because CMB lensing provides a direct estimate of halo mass, our results offer new insights into the stellar-to-halo mass relation (SHMR) and the mass evolution of DESI-selected galaxies. Unlike traditional galaxy-galaxy lensing (GGL), our approach is free from many of the noise and systematics issues that limit optical lensing, making it a powerful tool for future cosmological analyses.

We develop a simulation-based modeling framework, which we show to be fast, accurate and flexible. Overall, we find that our model provides an excellent fit to the $\kappa(\theta)$ measurements for both Main and Extended LRGs. The fit for the BGS sample is slightly worse likely due to the more complex halo occupation distribution (HOD) of the BGS, which includes a broader mix of galaxy types, including star-forming galaxies that are not as well described by the vanilla HOD model used here \citep[e.g.,][]{2021MNRAS.502.3599H,2023MNRAS.524.2524H}. We speculate that incorporating assembly bias and allowing for central and satellite galaxy biasing could further improve the fit, though this is beyond the scope of this work.

From the fits, we extract constraints on the five standard HOD parameters and three derived quantities (mean halo mass, number density, and satellite fraction). We find that 
the mean halo masses are highest for the Main sample, followed by the Extended and BGS samples, in line with expectations based on galaxy formation and evolution models. Satellite fractions and number densities follow the opposite trend, being highest for the BGS and lowest for the Main sample, again consistent with expectations.

A central result of this work is the comparison of our $\kappa(\theta)$-derived dark matter density profiles with gas density profiles inferred from the kSZ effect (from Ref.~\citep{2024arXiv240707152H}). By converting the $\kappa$ measurements into CAP density profiles in the same units as $T_{\rm kSZ}$, we perform a nearly assumption-free comparison between the gas and dark matter density. We find excellent agreement between the two profiles at large radii ($\sim 2$-$3~\mathrm{Mpc}/h$), consistent with the notion that gas and dark matter follow each other on large scales. However, at small apertures, the kSZ-inferred profiles lie below the lensing-based profiles, revealing the impact of baryonic feedback, which pushes gas out of the inner regions of halos. Both for the Main LRGs as well as the Extended LRGs, the baryon and matter curves are significantly discrepant. 
Note that this analysis is an improvement over Ref.~\citep{2024arXiv240707152H} since the amplitude no longer needs adjusting. Using the SNR definition from Ref.~\citep{2024arXiv240707152H}, this discrepancy between gas and dark matter translates to ${\rm SNR} = \sqrt{\Delta \chi^2} \approx 8.2$ (c.f. ${\rm SNR \approx 6}$ when fitting for a free amplitude parameter).

Additionally, we study the gas fractions inferred by the ratio of the kSZ and CMB lensing profiles. We find a $\gtrsim$4$\sigma$ discrepancy with the gas fractions of LRG-like galaxies in the hydrodynamical simulation TNG300 when comparing with our gas fraction measurements using kSZ and CMB lensing. Interestingly, the low gas fractions we find of about 0.3 for the LRG sample are in agreement with group-size gas fraction measurements using X-ray data. Upon studying the gas fraction of our highest-mass sample, we find that it is higher, as expected theoretically, and consistent with the gas fraction inferred from X-ray analyses of halos of similar mass. In accordance with previous findings, this suggests that large baryonic feedback is necessary to match the observed gas distribution. 

We explore the redshift evolution of the halo properties by performing the same analysis in four redshift bins for the Main and Extended LRG samples. We observe mild redshift trends in the halo masses, particularly for the Extended sample, where the inferred halo mass decreases at higher redshift. These trends are consistent with expectations from the evolving lensing kernel and halo mass function.

Finally, we compare the stellar-to-halo mass relation (SHMR) from the data against the hydrodynamical simulation TNG300. We find good agreement in the halo mass estimates across the four stellar mass bins (all galaxies, $\log (M_\ast/M_\odot) > 11, \ 11.25, \ 11.5$). However, TNG300 tends to overpredict the stellar masses compared to observations, hinting that its galaxies may not be sufficiently quenched. 
This finding aligns with previous research (e.g., \cite{Pillepich:2017fcc}) and highlights the value of combining CMB and galaxy survey data to investigate galaxy formation physics.

Our results demonstrate that joint analyses of $\kappa$ and kSZ can put tight constraints on the baryonic content and feedback processes by providing estimates of the halo mass. Future work leveraging high-resolution CMB lensing and temperature maps from upcoming experiments such as Simons Observatory and CMB-S4 
as well as galaxy survey data from DESI, {\it Euclid} and the Vera Rubin Observatory (LSST) will allow even more precise tests of baryonic and dark matter physics on small scales.

\acknowledgements

We thank Antón Baleato Lizancos, Esra Bulbul, Henry Liu, Mathew Madhavacheril, Frank Qu, Bernardita Ried Guachalla, Emmanuel Schaan, and Uro\v{s} Seljak for useful discussions. BH thanks the Miller Institute for supporting her postdoctoral research.
SF, GSF, NS, and RZ are supported by Lawrence Berkeley National Laboratory and the Director, Office of Science, Office of High Energy Physics of the U.S. Department of Energy under Contract No.\ DE-AC02-05CH11231.

\appendix
\section{2D Visualization of $\kappa$ and kSZ Stacks}

In this Appendix, we present 2D stacked images of the CMB lensing convergence $\kappa$ and the kinematic Sunyaev-Zel'dovich (kSZ) effect for the Main LRG sample to provide a visual comparison of the spatial distributions of dark matter and gas.

\begin{figure}
    \centering
    \includegraphics[width=0.48\textwidth]{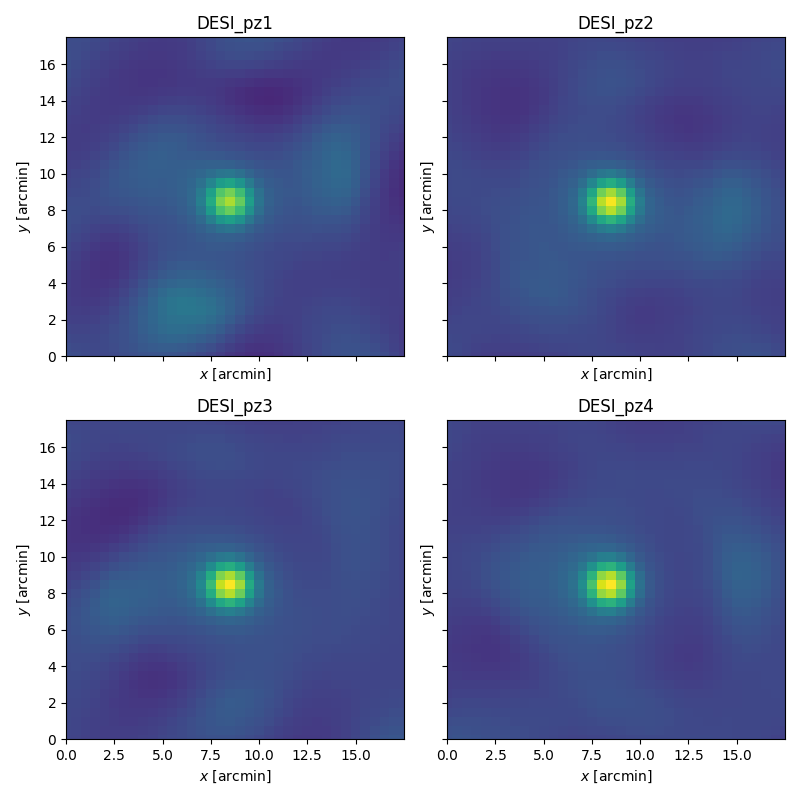}
    \includegraphics[width=0.48\textwidth]{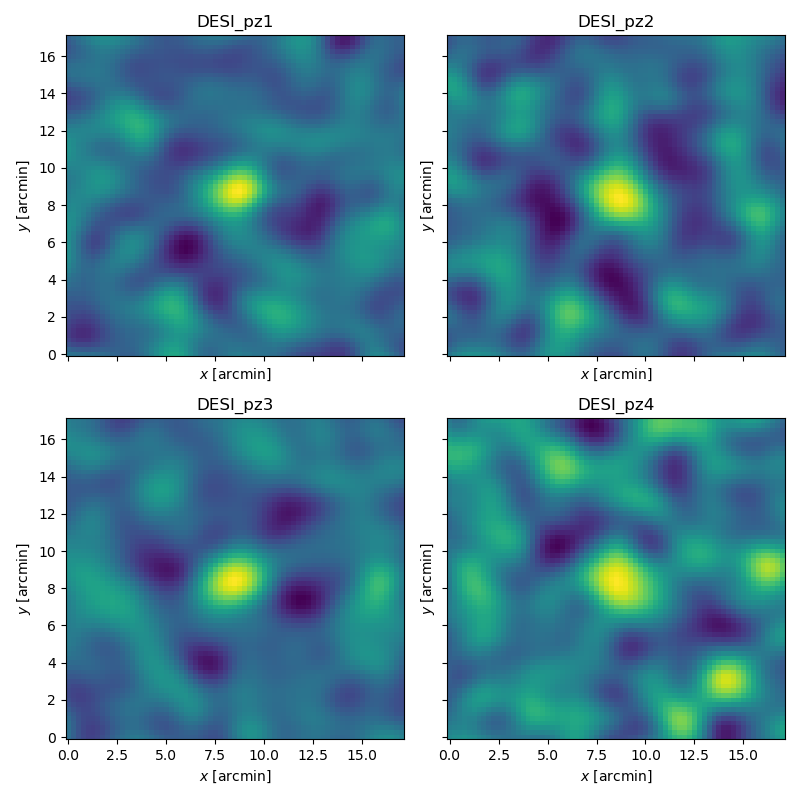}
    \caption{
    \textbf{Top:} 2D stacked images of ACT DR6 CMB lensing convergence $\kappa$ for the Main LRG sample. A transfer function correction, derived from AbacusSummit simulations, has been applied to approximately restore the shape of the small-scale structure removed by the $L > 3000$ cutoff in the ACT lensing pipeline, using halos of mass matching that of the Main sample, $M_{\rm vir} \sim 10^{13.2}~M_\odot/h$. This correction is used here for visualization only. 
    \textbf{Bottom:} Corresponding 2D stacks of the kSZ signal for the same galaxy sample, reproduced from \citet{2024arXiv240707152H}. Due to the faintness of the kSZ effect, the stacks are effectively high-pass filtered by the kSZ-to-CMB power spectrum ratio, suppressing large-scale modes. The broader appearance of the kSZ signal compared to $\kappa$ reflects the more extended nature of the gas, consistent with expectations from baryonic feedback processes.
    }
    \label{fig:2dstacks}
\end{figure}

The top panel of Fig.~\ref{fig:2dstacks} shows the 2D $\kappa$ stacks on ACT DR6 data for the Main LRG sample, arranged as $2 \times 2$ subfigures, each one corresponding to a different redshift bin. To account for the small-scale mode removal ($L < 3000$) applied in the ACT lensing quadratic estimator, we apply a correction derived from the AbacusSummit simulations. Specifically, we compute a transfer function by taking the ratio of simulated $\kappa$ maps with and without the $L > 3000$ cutoff (both convolved with the ACT 1.6' FWHM beam), matched to halos of mass $M_{\rm vir} \sim 10^{13.2}~M_\odot/h$ (i.e., the average halo mass for the Main sample). This transfer function is then applied multiplicatively to the observed $\kappa$ stacks to approximately restore large-scale power. The most visible smearing that we observe is thus due to the ACT beam. We check in simulations that this procedure works well; however, small residual effects may remain. We emphasize that these visualizations are intended for qualitative illustration only and are not used for any quantitative analysis.

The bottom panel shows the corresponding 2D kSZ stacks for the same sample, reproduced from \citet{2024arXiv240707152H}. The kSZ signal is subdominant relative to the total CMB anisotropies, so for visualization purposes only, we apply a high-pass filter governed by the ratio of the kSZ power spectrum to the total power. Inevitably, that suppresses some of the large-scale gas density signal in the 2D kSZ stacks. Nonetheless, it remains evident that the kSZ profiles appear more extended and diffuse compared to the $\kappa$ stacks.

This visual comparison highlights the core physical result of this work: baryonic feedback processes displace gas from the central regions of halos, leading to a more extended gas distribution relative to the underlying dark matter. The difference in spatial extent between the $\kappa$ and kSZ stacks is consistent with this interpretation.

\bibliography{apssamp}

\providecommand{\noopsort}[1]{}\providecommand{\singleletter}[1]{#1}%
\begin{thebibliography}{79}%
\makeatletter
\providecommand \@ifxundefined [1]{%
 \@ifx{#1\undefined}
}%
\providecommand \@ifnum [1]{%
 \ifnum #1\expandafter \@firstoftwo
 \else \expandafter \@secondoftwo
 \fi
}%
\providecommand \@ifx [1]{%
 \ifx #1\expandafter \@firstoftwo
 \else \expandafter \@secondoftwo
 \fi
}%
\providecommand \natexlab [1]{#1}%
\providecommand \enquote  [1]{``#1''}%
\providecommand \bibnamefont  [1]{#1}%
\providecommand \bibfnamefont [1]{#1}%
\providecommand \citenamefont [1]{#1}%
\providecommand \href@noop [0]{\@secondoftwo}%
\providecommand \href [0]{\begingroup \@sanitize@url \@href}%
\providecommand \@href[1]{\@@startlink{#1}\@@href}%
\providecommand \@@href[1]{\endgroup#1\@@endlink}%
\providecommand \@sanitize@url [0]{\catcode `\\12\catcode `\$12\catcode `\&12\catcode `\#12\catcode `\^12\catcode `\_12\catcode `\%12\relax}%
\providecommand \@@startlink[1]{}%
\providecommand \@@endlink[0]{}%
\providecommand \url  [0]{\begingroup\@sanitize@url \@url }%
\providecommand \@url [1]{\endgroup\@href {#1}{\urlprefix }}%
\providecommand \urlprefix  [0]{URL }%
\providecommand \Eprint [0]{\href }%
\providecommand \doibase [0]{https://doi.org/}%
\providecommand \selectlanguage [0]{\@gobble}%
\providecommand \bibinfo  [0]{\@secondoftwo}%
\providecommand \bibfield  [0]{\@secondoftwo}%
\providecommand \translation [1]{[#1]}%
\providecommand \BibitemOpen [0]{}%
\providecommand \bibitemStop [0]{}%
\providecommand \bibitemNoStop [0]{.\EOS\space}%
\providecommand \EOS [0]{\spacefactor3000\relax}%
\providecommand \BibitemShut  [1]{\csname bibitem#1\endcsname}%
\let\auto@bib@innerbib\@empty
\bibitem [{\citenamefont {{van Daalen}}\ \emph {et~al.}(2011)\citenamefont {{van Daalen}}, \citenamefont {{Schaye}}, \citenamefont {{Booth}},\ and\ \citenamefont {{Dalla Vecchia}}}]{2011MNRAS.415.3649V}%
  \BibitemOpen
  \bibfield  {author} {\bibinfo {author} {\bibfnamefont {M.~P.}\ \bibnamefont {{van Daalen}}}, \bibinfo {author} {\bibfnamefont {J.}~\bibnamefont {{Schaye}}}, \bibinfo {author} {\bibfnamefont {C.~M.}\ \bibnamefont {{Booth}}},\ and\ \bibinfo {author} {\bibfnamefont {C.}~\bibnamefont {{Dalla Vecchia}}},\ }\bibfield  {title} {\bibinfo {title} {{The effects of galaxy formation on the matter power spectrum: a challenge for precision cosmology}},\ }\href {https://doi.org/10.1111/j.1365-2966.2011.18981.x} {\bibfield  {journal} {\bibinfo  {journal} {\mnras}\ }\textbf {\bibinfo {volume} {415}},\ \bibinfo {pages} {3649} (\bibinfo {year} {2011})},\ \Eprint {https://arxiv.org/abs/1104.1174} {arXiv:1104.1174 [astro-ph.CO]} \BibitemShut {NoStop}%
\bibitem [{\citenamefont {{Chisari}}\ \emph {et~al.}(2019)\citenamefont {{Chisari}}, \citenamefont {{Mead}}, \citenamefont {{Joudaki}}, \citenamefont {{Ferreira}}, \citenamefont {{Schneider}}, \citenamefont {{Mohr}}, \citenamefont {{Tr{\"o}ster}}, \citenamefont {{Alonso}}, \citenamefont {{McCarthy}}, \citenamefont {{Martin-Alvarez}}, \citenamefont {{Devriendt}}, \citenamefont {{Slyz}},\ and\ \citenamefont {{van Daalen}}}]{2019OJAp....2E...4C}%
  \BibitemOpen
  \bibfield  {author} {\bibinfo {author} {\bibfnamefont {N.~E.}\ \bibnamefont {{Chisari}}}, \bibinfo {author} {\bibfnamefont {A.~J.}\ \bibnamefont {{Mead}}}, \bibinfo {author} {\bibfnamefont {S.}~\bibnamefont {{Joudaki}}}, \bibinfo {author} {\bibfnamefont {P.~G.}\ \bibnamefont {{Ferreira}}}, \bibinfo {author} {\bibfnamefont {A.}~\bibnamefont {{Schneider}}}, \bibinfo {author} {\bibfnamefont {J.}~\bibnamefont {{Mohr}}}, \bibinfo {author} {\bibfnamefont {T.}~\bibnamefont {{Tr{\"o}ster}}}, \bibinfo {author} {\bibfnamefont {D.}~\bibnamefont {{Alonso}}}, \bibinfo {author} {\bibfnamefont {I.~G.}\ \bibnamefont {{McCarthy}}}, \bibinfo {author} {\bibfnamefont {S.}~\bibnamefont {{Martin-Alvarez}}}, \bibinfo {author} {\bibfnamefont {J.}~\bibnamefont {{Devriendt}}}, \bibinfo {author} {\bibfnamefont {A.}~\bibnamefont {{Slyz}}},\ and\ \bibinfo {author} {\bibfnamefont {M.~P.}\ \bibnamefont {{van Daalen}}},\ }\bibfield  {title} {\bibinfo {title} {{Modelling baryonic feedback for survey cosmology}},\ }\href
  {https://doi.org/10.21105/astro.1905.06082} {\bibfield  {journal} {\bibinfo  {journal} {The Open Journal of Astrophysics}\ }\textbf {\bibinfo {volume} {2}},\ \bibinfo {eid} {4} (\bibinfo {year} {2019})},\ \Eprint {https://arxiv.org/abs/1905.06082} {arXiv:1905.06082 [astro-ph.CO]} \BibitemShut {NoStop}%
\bibitem [{\citenamefont {{Fukugita}}\ and\ \citenamefont {{Peebles}}(2004)}]{Fukugita_2004}%
  \BibitemOpen
  \bibfield  {author} {\bibinfo {author} {\bibfnamefont {M.}~\bibnamefont {{Fukugita}}}\ and\ \bibinfo {author} {\bibfnamefont {P.~J.~E.}\ \bibnamefont {{Peebles}}},\ }\bibfield  {title} {\bibinfo {title} {{The Cosmic Energy Inventory}},\ }\href {https://doi.org/10.1086/425155} {\bibfield  {journal} {\bibinfo  {journal} {\apj}\ }\textbf {\bibinfo {volume} {616}},\ \bibinfo {pages} {643} (\bibinfo {year} {2004})},\ \Eprint {https://arxiv.org/abs/astro-ph/0406095} {arXiv:astro-ph/0406095 [astro-ph]} \BibitemShut {NoStop}%
\bibitem [{\citenamefont {Cen}\ and\ \citenamefont {Ostriker}(2006)}]{Cen_2006}%
  \BibitemOpen
  \bibfield  {author} {\bibinfo {author} {\bibfnamefont {R.}~\bibnamefont {Cen}}\ and\ \bibinfo {author} {\bibfnamefont {J.~P.}\ \bibnamefont {Ostriker}},\ }\bibfield  {title} {\bibinfo {title} {Where are the baryons? {II}. feedback effects},\ }\href {https://doi.org/10.1086/506505} {\bibfield  {journal} {\bibinfo  {journal} {The Astrophysical Journal}\ }\textbf {\bibinfo {volume} {650}},\ \bibinfo {pages} {560} (\bibinfo {year} {2006})}\BibitemShut {NoStop}%
\bibitem [{\citenamefont {Schneider}\ \emph {et~al.}(2019)\citenamefont {Schneider}, \citenamefont {Teyssier}, \citenamefont {Stadel}, \citenamefont {Chisari}, \citenamefont {Le~Brun}, \citenamefont {Amara},\ and\ \citenamefont {Refregier}}]{Schneider:2018pfw}%
  \BibitemOpen
  \bibfield  {author} {\bibinfo {author} {\bibfnamefont {A.}~\bibnamefont {Schneider}}, \bibinfo {author} {\bibfnamefont {R.}~\bibnamefont {Teyssier}}, \bibinfo {author} {\bibfnamefont {J.}~\bibnamefont {Stadel}}, \bibinfo {author} {\bibfnamefont {N.~E.}\ \bibnamefont {Chisari}}, \bibinfo {author} {\bibfnamefont {A.~M.~C.}\ \bibnamefont {Le~Brun}}, \bibinfo {author} {\bibfnamefont {A.}~\bibnamefont {Amara}},\ and\ \bibinfo {author} {\bibfnamefont {A.}~\bibnamefont {Refregier}},\ }\bibfield  {title} {\bibinfo {title} {{Quantifying baryon effects on the matter power spectrum and the weak lensing shear correlation}},\ }\href {https://doi.org/10.1088/1475-7516/2019/03/020} {\bibfield  {journal} {\bibinfo  {journal} {JCAP}\ }\textbf {\bibinfo {volume} {03}}\bibfield  {number} {\bibinfo  {number} { (3)},\ \bibinfo {pages} {020}},\ }\Eprint {https://arxiv.org/abs/1810.08629} {arXiv:1810.08629 [astro-ph.CO]} \BibitemShut {NoStop}%
\bibitem [{\citenamefont {{Bigwood}}\ \emph {et~al.}(2024{\natexlab{a}})\citenamefont {{Bigwood}}, \citenamefont {{Amon}}, \citenamefont {{Schneider}}, \citenamefont {{Salcido}}, \citenamefont {{McCarthy}}, \citenamefont {{Preston}}, \citenamefont {{Sanchez}}, \citenamefont {{Sijacki}}, \citenamefont {{Schaan}}, \citenamefont {{Ferraro}}, \citenamefont {{Battaglia}}, \citenamefont {{Chen}}, \citenamefont {{Dodelson}}, \citenamefont {{Roodman}}, \citenamefont {{Pieres}}, \citenamefont {{Ferte}}, \citenamefont {{Alarcon}}, \citenamefont {{Drlica-Wagner}}, \citenamefont {{Choi}}, \citenamefont {{Navarro-Alsina}}, \citenamefont {{Campos}}, \citenamefont {{Ross}}, \citenamefont {{Carnero Rosell}}, \citenamefont {{Yin}}, \citenamefont {{Yanny}}, \citenamefont {{Sanchez}}, \citenamefont {{Chang}}, \citenamefont {{Davis}}, \citenamefont {{Doux}}, \citenamefont {{Gruen}}, \citenamefont {{Rykoff}}, \citenamefont {{Huff}}, \citenamefont {{Sheldon}}, \citenamefont {{Tarsitano}}, \citenamefont {{Andrade-Oliveira}},
  \citenamefont {{Bernstein}}, \citenamefont {{Giannini}}, \citenamefont {{Diehl}}, \citenamefont {{Huang}}, \citenamefont {{Harrison}}, \citenamefont {{Sevilla-Noarbe}}, \citenamefont {{Tutusaus}}, \citenamefont {{Elvin-Poole}}, \citenamefont {{McCullough}}, \citenamefont {{Zuntz}}, \citenamefont {{Blazek}}, \citenamefont {{DeRose}}, \citenamefont {{Cordero}}, \citenamefont {{Prat}}, \citenamefont {{Myles}}, \citenamefont {{Eckert}}, \citenamefont {{Bechtol}}, \citenamefont {{Herner}}, \citenamefont {{Secco}}, \citenamefont {{Gatti}}, \citenamefont {{Raveri}}, \citenamefont {{Carrasco Kind}}, \citenamefont {{Becker}}, \citenamefont {{Troxel}}, \citenamefont {{Jarvis}}, \citenamefont {{MacCrann}}, \citenamefont {{Friedrich}}, \citenamefont {{Alves}}, \citenamefont {{Leget}}, \citenamefont {{Chen}}, \citenamefont {{Rollins}}, \citenamefont {{Wechsler}}, \citenamefont {{Gruendl}}, \citenamefont {{Cawthon}}, \citenamefont {{Allam}}, \citenamefont {{Bridle}}, \citenamefont {{Pandey}}, \citenamefont {{Everett}},
  \citenamefont {{Shin}}, \citenamefont {{Hartley}}, \citenamefont {{Fang}}, \citenamefont {{Zhang}}, \citenamefont {{Aguena}}, \citenamefont {{Annis}}, \citenamefont {{Bacon}}, \citenamefont {{Bertin}}, \citenamefont {{Bocquet}}, \citenamefont {{Brooks}}, \citenamefont {{Carretero}}, \citenamefont {{Castander}}, \citenamefont {{da Costa}}, \citenamefont {{Pereira}}, \citenamefont {{De Vicente}}, \citenamefont {{Desai}}, \citenamefont {{Doel}}, \citenamefont {{Ferrero}}, \citenamefont {{Flaugher}}, \citenamefont {{Frieman}}, \citenamefont {{Garcia-Bellido}}, \citenamefont {{Gaztanaga}}, \citenamefont {{Gutierrez}}, \citenamefont {{Hinton}}, \citenamefont {{Hollowood}}, \citenamefont {{Honscheid}}, \citenamefont {{Huterer}}, \citenamefont {{James}}, \citenamefont {{Kuehn}}, \citenamefont {{Lahav}}, \citenamefont {{Lee}}, \citenamefont {{Marshall}}, \citenamefont {{Mena-Fernandez}}, \citenamefont {{Miquel}}, \citenamefont {{Muir}}, \citenamefont {{Paterno}}, \citenamefont {{Plazas Malagon}}, \citenamefont
  {{Porredon}}, \citenamefont {{Romer}}, \citenamefont {{Samuroff}}, \citenamefont {{Sanchez}}, \citenamefont {{Sanchez Cid}}, \citenamefont {{Smith}}, \citenamefont {{Soares-Santos}}, \citenamefont {{Suchyta}}, \citenamefont {{Swanson}}, \citenamefont {{Tarle}}, \citenamefont {{To}}, \citenamefont {{Weaverdyck}}, \citenamefont {{Weller}}, \citenamefont {{Wiseman}},\ and\ \citenamefont {{Yamamoto}}}]{2024arXiv240406098B}%
  \BibitemOpen
  \bibfield  {author} {\bibinfo {author} {\bibfnamefont {L.}~\bibnamefont {{Bigwood}}}, \bibinfo {author} {\bibfnamefont {A.}~\bibnamefont {{Amon}}}, \bibinfo {author} {\bibfnamefont {A.}~\bibnamefont {{Schneider}}}, \bibinfo {author} {\bibfnamefont {J.}~\bibnamefont {{Salcido}}}, \bibinfo {author} {\bibfnamefont {I.~G.}\ \bibnamefont {{McCarthy}}}, \bibinfo {author} {\bibfnamefont {C.}~\bibnamefont {{Preston}}}, \bibinfo {author} {\bibfnamefont {D.}~\bibnamefont {{Sanchez}}}, \bibinfo {author} {\bibfnamefont {D.}~\bibnamefont {{Sijacki}}}, \bibinfo {author} {\bibfnamefont {E.}~\bibnamefont {{Schaan}}}, \bibinfo {author} {\bibfnamefont {S.}~\bibnamefont {{Ferraro}}}, \bibinfo {author} {\bibfnamefont {N.}~\bibnamefont {{Battaglia}}}, \bibinfo {author} {\bibfnamefont {A.}~\bibnamefont {{Chen}}}, \bibinfo {author} {\bibfnamefont {S.}~\bibnamefont {{Dodelson}}}, \bibinfo {author} {\bibfnamefont {A.}~\bibnamefont {{Roodman}}}, \bibinfo {author} {\bibfnamefont {A.}~\bibnamefont {{Pieres}}}, \bibinfo {author}
  {\bibfnamefont {A.}~\bibnamefont {{Ferte}}}, \bibinfo {author} {\bibfnamefont {A.}~\bibnamefont {{Alarcon}}}, \bibinfo {author} {\bibfnamefont {A.}~\bibnamefont {{Drlica-Wagner}}}, \bibinfo {author} {\bibfnamefont {A.}~\bibnamefont {{Choi}}}, \bibinfo {author} {\bibfnamefont {A.}~\bibnamefont {{Navarro-Alsina}}}, \bibinfo {author} {\bibfnamefont {A.}~\bibnamefont {{Campos}}}, \bibinfo {author} {\bibfnamefont {A.~J.}\ \bibnamefont {{Ross}}}, \bibinfo {author} {\bibfnamefont {A.}~\bibnamefont {{Carnero Rosell}}}, \bibinfo {author} {\bibfnamefont {B.}~\bibnamefont {{Yin}}}, \bibinfo {author} {\bibfnamefont {B.}~\bibnamefont {{Yanny}}}, \bibinfo {author} {\bibfnamefont {C.}~\bibnamefont {{Sanchez}}}, \bibinfo {author} {\bibfnamefont {C.}~\bibnamefont {{Chang}}}, \bibinfo {author} {\bibfnamefont {C.}~\bibnamefont {{Davis}}}, \bibinfo {author} {\bibfnamefont {C.}~\bibnamefont {{Doux}}}, \bibinfo {author} {\bibfnamefont {D.}~\bibnamefont {{Gruen}}}, \bibinfo {author} {\bibfnamefont {E.~S.}\ \bibnamefont
  {{Rykoff}}}, \bibinfo {author} {\bibfnamefont {E.~M.}\ \bibnamefont {{Huff}}}, \bibinfo {author} {\bibfnamefont {E.}~\bibnamefont {{Sheldon}}}, \bibinfo {author} {\bibfnamefont {F.}~\bibnamefont {{Tarsitano}}}, \bibinfo {author} {\bibfnamefont {F.}~\bibnamefont {{Andrade-Oliveira}}}, \bibinfo {author} {\bibfnamefont {G.~M.}\ \bibnamefont {{Bernstein}}}, \bibinfo {author} {\bibfnamefont {G.}~\bibnamefont {{Giannini}}}, \bibinfo {author} {\bibfnamefont {H.~T.}\ \bibnamefont {{Diehl}}}, \bibinfo {author} {\bibfnamefont {H.}~\bibnamefont {{Huang}}}, \bibinfo {author} {\bibfnamefont {I.}~\bibnamefont {{Harrison}}}, \bibinfo {author} {\bibfnamefont {I.}~\bibnamefont {{Sevilla-Noarbe}}}, \bibinfo {author} {\bibfnamefont {I.}~\bibnamefont {{Tutusaus}}}, \bibinfo {author} {\bibfnamefont {J.}~\bibnamefont {{Elvin-Poole}}}, \bibinfo {author} {\bibfnamefont {J.}~\bibnamefont {{McCullough}}}, \bibinfo {author} {\bibfnamefont {J.}~\bibnamefont {{Zuntz}}}, \bibinfo {author} {\bibfnamefont {J.}~\bibnamefont {{Blazek}}},
  \bibinfo {author} {\bibfnamefont {J.}~\bibnamefont {{DeRose}}}, \bibinfo {author} {\bibfnamefont {J.}~\bibnamefont {{Cordero}}}, \bibinfo {author} {\bibfnamefont {J.}~\bibnamefont {{Prat}}}, \bibinfo {author} {\bibfnamefont {J.}~\bibnamefont {{Myles}}}, \bibinfo {author} {\bibfnamefont {K.}~\bibnamefont {{Eckert}}}, \bibinfo {author} {\bibfnamefont {K.}~\bibnamefont {{Bechtol}}}, \bibinfo {author} {\bibfnamefont {K.}~\bibnamefont {{Herner}}}, \bibinfo {author} {\bibfnamefont {L.~F.}\ \bibnamefont {{Secco}}}, \bibinfo {author} {\bibfnamefont {M.}~\bibnamefont {{Gatti}}}, \bibinfo {author} {\bibfnamefont {M.}~\bibnamefont {{Raveri}}}, \bibinfo {author} {\bibfnamefont {M.}~\bibnamefont {{Carrasco Kind}}}, \bibinfo {author} {\bibfnamefont {M.~R.}\ \bibnamefont {{Becker}}}, \bibinfo {author} {\bibfnamefont {M.~A.}\ \bibnamefont {{Troxel}}}, \bibinfo {author} {\bibfnamefont {M.}~\bibnamefont {{Jarvis}}}, \bibinfo {author} {\bibfnamefont {N.}~\bibnamefont {{MacCrann}}}, \bibinfo {author} {\bibfnamefont
  {O.}~\bibnamefont {{Friedrich}}}, \bibinfo {author} {\bibfnamefont {O.}~\bibnamefont {{Alves}}}, \bibinfo {author} {\bibfnamefont {P.~F.}\ \bibnamefont {{Leget}}}, \bibinfo {author} {\bibfnamefont {R.}~\bibnamefont {{Chen}}}, \bibinfo {author} {\bibfnamefont {R.~P.}\ \bibnamefont {{Rollins}}}, \bibinfo {author} {\bibfnamefont {R.~H.}\ \bibnamefont {{Wechsler}}}, \bibinfo {author} {\bibfnamefont {R.~A.}\ \bibnamefont {{Gruendl}}}, \bibinfo {author} {\bibfnamefont {R.}~\bibnamefont {{Cawthon}}}, \bibinfo {author} {\bibfnamefont {S.}~\bibnamefont {{Allam}}}, \bibinfo {author} {\bibfnamefont {S.~L.}\ \bibnamefont {{Bridle}}}, \bibinfo {author} {\bibfnamefont {S.}~\bibnamefont {{Pandey}}}, \bibinfo {author} {\bibfnamefont {S.}~\bibnamefont {{Everett}}}, \bibinfo {author} {\bibfnamefont {T.}~\bibnamefont {{Shin}}}, \bibinfo {author} {\bibfnamefont {W.~G.}\ \bibnamefont {{Hartley}}}, \bibinfo {author} {\bibfnamefont {X.}~\bibnamefont {{Fang}}}, \bibinfo {author} {\bibfnamefont {Y.}~\bibnamefont {{Zhang}}},
  \bibinfo {author} {\bibfnamefont {M.}~\bibnamefont {{Aguena}}}, \bibinfo {author} {\bibfnamefont {J.}~\bibnamefont {{Annis}}}, \bibinfo {author} {\bibfnamefont {D.}~\bibnamefont {{Bacon}}}, \bibinfo {author} {\bibfnamefont {E.}~\bibnamefont {{Bertin}}}, \bibinfo {author} {\bibfnamefont {S.}~\bibnamefont {{Bocquet}}}, \bibinfo {author} {\bibfnamefont {D.}~\bibnamefont {{Brooks}}}, \bibinfo {author} {\bibfnamefont {J.}~\bibnamefont {{Carretero}}}, \bibinfo {author} {\bibfnamefont {F.~J.}\ \bibnamefont {{Castander}}}, \bibinfo {author} {\bibfnamefont {L.~N.}\ \bibnamefont {{da Costa}}}, \bibinfo {author} {\bibfnamefont {M.~E.~S.}\ \bibnamefont {{Pereira}}}, \bibinfo {author} {\bibfnamefont {J.}~\bibnamefont {{De Vicente}}}, \bibinfo {author} {\bibfnamefont {S.}~\bibnamefont {{Desai}}}, \bibinfo {author} {\bibfnamefont {P.}~\bibnamefont {{Doel}}}, \bibinfo {author} {\bibfnamefont {I.}~\bibnamefont {{Ferrero}}}, \bibinfo {author} {\bibfnamefont {B.}~\bibnamefont {{Flaugher}}}, \bibinfo {author} {\bibfnamefont
  {J.}~\bibnamefont {{Frieman}}}, \bibinfo {author} {\bibfnamefont {J.}~\bibnamefont {{Garcia-Bellido}}}, \bibinfo {author} {\bibfnamefont {E.}~\bibnamefont {{Gaztanaga}}}, \bibinfo {author} {\bibfnamefont {G.}~\bibnamefont {{Gutierrez}}}, \bibinfo {author} {\bibfnamefont {S.~R.}\ \bibnamefont {{Hinton}}}, \bibinfo {author} {\bibfnamefont {D.~L.}\ \bibnamefont {{Hollowood}}}, \bibinfo {author} {\bibfnamefont {K.}~\bibnamefont {{Honscheid}}}, \bibinfo {author} {\bibfnamefont {D.}~\bibnamefont {{Huterer}}}, \bibinfo {author} {\bibfnamefont {D.~J.}\ \bibnamefont {{James}}}, \bibinfo {author} {\bibfnamefont {K.}~\bibnamefont {{Kuehn}}}, \bibinfo {author} {\bibfnamefont {O.}~\bibnamefont {{Lahav}}}, \bibinfo {author} {\bibfnamefont {S.}~\bibnamefont {{Lee}}}, \bibinfo {author} {\bibfnamefont {J.~L.}\ \bibnamefont {{Marshall}}}, \bibinfo {author} {\bibfnamefont {J.}~\bibnamefont {{Mena-Fernandez}}}, \bibinfo {author} {\bibfnamefont {R.}~\bibnamefont {{Miquel}}}, \bibinfo {author} {\bibfnamefont {J.}~\bibnamefont
  {{Muir}}}, \bibinfo {author} {\bibfnamefont {M.}~\bibnamefont {{Paterno}}}, \bibinfo {author} {\bibfnamefont {A.~A.}\ \bibnamefont {{Plazas Malagon}}}, \bibinfo {author} {\bibfnamefont {A.}~\bibnamefont {{Porredon}}}, \bibinfo {author} {\bibfnamefont {A.~K.}\ \bibnamefont {{Romer}}}, \bibinfo {author} {\bibfnamefont {S.}~\bibnamefont {{Samuroff}}}, \bibinfo {author} {\bibfnamefont {E.}~\bibnamefont {{Sanchez}}}, \bibinfo {author} {\bibfnamefont {D.}~\bibnamefont {{Sanchez Cid}}}, \bibinfo {author} {\bibfnamefont {M.}~\bibnamefont {{Smith}}}, \bibinfo {author} {\bibfnamefont {M.}~\bibnamefont {{Soares-Santos}}}, \bibinfo {author} {\bibfnamefont {E.}~\bibnamefont {{Suchyta}}}, \bibinfo {author} {\bibfnamefont {M.~E.~C.}\ \bibnamefont {{Swanson}}}, \bibinfo {author} {\bibfnamefont {G.}~\bibnamefont {{Tarle}}}, \bibinfo {author} {\bibfnamefont {C.}~\bibnamefont {{To}}}, \bibinfo {author} {\bibfnamefont {N.}~\bibnamefont {{Weaverdyck}}}, \bibinfo {author} {\bibfnamefont {J.}~\bibnamefont {{Weller}}}, \bibinfo
  {author} {\bibfnamefont {P.}~\bibnamefont {{Wiseman}}},\ and\ \bibinfo {author} {\bibfnamefont {M.}~\bibnamefont {{Yamamoto}}},\ }\bibfield  {title} {\bibinfo {title} {{Weak lensing combined with the kinetic Sunyaev Zel'dovich effect: A study of baryonic feedback}},\ }\href {https://doi.org/10.48550/arXiv.2404.06098} {\bibfield  {journal} {\bibinfo  {journal} {arXiv e-prints}\ ,\ \bibinfo {eid} {arXiv:2404.06098}} (\bibinfo {year} {2024}{\natexlab{a}})},\ \Eprint {https://arxiv.org/abs/2404.06098} {arXiv:2404.06098 [astro-ph.CO]} \BibitemShut {NoStop}%
\bibitem [{\citenamefont {{Schneider}}\ \emph {et~al.}(2019)\citenamefont {{Schneider}}, \citenamefont {{Teyssier}}, \citenamefont {{Stadel}}, \citenamefont {{Chisari}}, \citenamefont {{Le Brun}}, \citenamefont {{Amara}},\ and\ \citenamefont {{Refregier}}}]{2019JCAP...03..020S}%
  \BibitemOpen
  \bibfield  {author} {\bibinfo {author} {\bibfnamefont {A.}~\bibnamefont {{Schneider}}}, \bibinfo {author} {\bibfnamefont {R.}~\bibnamefont {{Teyssier}}}, \bibinfo {author} {\bibfnamefont {J.}~\bibnamefont {{Stadel}}}, \bibinfo {author} {\bibfnamefont {N.~E.}\ \bibnamefont {{Chisari}}}, \bibinfo {author} {\bibfnamefont {A.~M.~C.}\ \bibnamefont {{Le Brun}}}, \bibinfo {author} {\bibfnamefont {A.}~\bibnamefont {{Amara}}},\ and\ \bibinfo {author} {\bibfnamefont {A.}~\bibnamefont {{Refregier}}},\ }\bibfield  {title} {\bibinfo {title} {{Quantifying baryon effects on the matter power spectrum and the weak lensing shear correlation}},\ }\href {https://doi.org/10.1088/1475-7516/2019/03/020} {\bibfield  {journal} {\bibinfo  {journal} {\jcap}\ }\textbf {\bibinfo {volume} {2019}},\ \bibinfo {eid} {020} (\bibinfo {year} {2019})},\ \Eprint {https://arxiv.org/abs/1810.08629} {arXiv:1810.08629 [astro-ph.CO]} \BibitemShut {NoStop}%
\bibitem [{\citenamefont {{Amon}}\ and\ \citenamefont {{Efstathiou}}(2022)}]{2022MNRAS.516.5355A}%
  \BibitemOpen
  \bibfield  {author} {\bibinfo {author} {\bibfnamefont {A.}~\bibnamefont {{Amon}}}\ and\ \bibinfo {author} {\bibfnamefont {G.}~\bibnamefont {{Efstathiou}}},\ }\bibfield  {title} {\bibinfo {title} {{A non-linear solution to the S$_{8}$ tension?}},\ }\href {https://doi.org/10.1093/mnras/stac2429} {\bibfield  {journal} {\bibinfo  {journal} {\mnras}\ }\textbf {\bibinfo {volume} {516}},\ \bibinfo {pages} {5355} (\bibinfo {year} {2022})},\ \Eprint {https://arxiv.org/abs/2206.11794} {arXiv:2206.11794 [astro-ph.CO]} \BibitemShut {NoStop}%
\bibitem [{\citenamefont {{Elbers}}\ \emph {et~al.}(2025)\citenamefont {{Elbers}}, \citenamefont {{Frenk}}, \citenamefont {{Jenkins}}, \citenamefont {{Li}}, \citenamefont {{Helly}}, \citenamefont {{Kugel}}, \citenamefont {{Schaller}}, \citenamefont {{Schaye}}, \citenamefont {{Braspenning}}, \citenamefont {{Kwan}}, \citenamefont {{McCarthy}}, \citenamefont {{Salcido}}, \citenamefont {{van Daalen}}, \citenamefont {{Vandenbroucke}},\ and\ \citenamefont {{Pascoli}}}]{2025MNRAS.537.2160E}%
  \BibitemOpen
  \bibfield  {author} {\bibinfo {author} {\bibfnamefont {W.}~\bibnamefont {{Elbers}}}, \bibinfo {author} {\bibfnamefont {C.~S.}\ \bibnamefont {{Frenk}}}, \bibinfo {author} {\bibfnamefont {A.}~\bibnamefont {{Jenkins}}}, \bibinfo {author} {\bibfnamefont {B.}~\bibnamefont {{Li}}}, \bibinfo {author} {\bibfnamefont {J.~C.}\ \bibnamefont {{Helly}}}, \bibinfo {author} {\bibfnamefont {R.}~\bibnamefont {{Kugel}}}, \bibinfo {author} {\bibfnamefont {M.}~\bibnamefont {{Schaller}}}, \bibinfo {author} {\bibfnamefont {J.}~\bibnamefont {{Schaye}}}, \bibinfo {author} {\bibfnamefont {J.}~\bibnamefont {{Braspenning}}}, \bibinfo {author} {\bibfnamefont {J.}~\bibnamefont {{Kwan}}}, \bibinfo {author} {\bibfnamefont {I.~G.}\ \bibnamefont {{McCarthy}}}, \bibinfo {author} {\bibfnamefont {J.}~\bibnamefont {{Salcido}}}, \bibinfo {author} {\bibfnamefont {M.~P.}\ \bibnamefont {{van Daalen}}}, \bibinfo {author} {\bibfnamefont {B.}~\bibnamefont {{Vandenbroucke}}},\ and\ \bibinfo {author} {\bibfnamefont {S.}~\bibnamefont {{Pascoli}}},\
  }\bibfield  {title} {\bibinfo {title} {{The FLAMINGO project: the coupling between baryonic feedback and cosmology in light of the S$_{8}$ tension}},\ }\href {https://doi.org/10.1093/mnras/staf093} {\bibfield  {journal} {\bibinfo  {journal} {\mnras}\ }\textbf {\bibinfo {volume} {537}},\ \bibinfo {pages} {2160} (\bibinfo {year} {2025})},\ \Eprint {https://arxiv.org/abs/2403.12967} {arXiv:2403.12967 [astro-ph.CO]} \BibitemShut {NoStop}%
\bibitem [{\citenamefont {{Huang}}\ \emph {et~al.}(2019)\citenamefont {{Huang}}, \citenamefont {{Eifler}}, \citenamefont {{Mandelbaum}},\ and\ \citenamefont {{Dodelson}}}]{2019MNRAS.488.1652H}%
  \BibitemOpen
  \bibfield  {author} {\bibinfo {author} {\bibfnamefont {H.-J.}\ \bibnamefont {{Huang}}}, \bibinfo {author} {\bibfnamefont {T.}~\bibnamefont {{Eifler}}}, \bibinfo {author} {\bibfnamefont {R.}~\bibnamefont {{Mandelbaum}}},\ and\ \bibinfo {author} {\bibfnamefont {S.}~\bibnamefont {{Dodelson}}},\ }\bibfield  {title} {\bibinfo {title} {{Modelling baryonic physics in future weak lensing surveys}},\ }\href {https://doi.org/10.1093/mnras/stz1714} {\bibfield  {journal} {\bibinfo  {journal} {\mnras}\ }\textbf {\bibinfo {volume} {488}},\ \bibinfo {pages} {1652} (\bibinfo {year} {2019})},\ \Eprint {https://arxiv.org/abs/1809.01146} {arXiv:1809.01146 [astro-ph.CO]} \BibitemShut {NoStop}%
\bibitem [{\citenamefont {{Mroczkowski}}\ \emph {et~al.}(2019)\citenamefont {{Mroczkowski}}, \citenamefont {{Nagai}}, \citenamefont {{Basu}}, \citenamefont {{Chluba}}, \citenamefont {{Sayers}}, \citenamefont {{Adam}}, \citenamefont {{Churazov}}, \citenamefont {{Crites}}, \citenamefont {{Di Mascolo}}, \citenamefont {{Eckert}}, \citenamefont {{Macias-Perez}}, \citenamefont {{Mayet}}, \citenamefont {{Perotto}}, \citenamefont {{Pointecouteau}}, \citenamefont {{Romero}}, \citenamefont {{Ruppin}}, \citenamefont {{Scannapieco}},\ and\ \citenamefont {{ZuHone}}}]{2019SSRv..215...17M}%
  \BibitemOpen
  \bibfield  {author} {\bibinfo {author} {\bibfnamefont {T.}~\bibnamefont {{Mroczkowski}}}, \bibinfo {author} {\bibfnamefont {D.}~\bibnamefont {{Nagai}}}, \bibinfo {author} {\bibfnamefont {K.}~\bibnamefont {{Basu}}}, \bibinfo {author} {\bibfnamefont {J.}~\bibnamefont {{Chluba}}}, \bibinfo {author} {\bibfnamefont {J.}~\bibnamefont {{Sayers}}}, \bibinfo {author} {\bibfnamefont {R.}~\bibnamefont {{Adam}}}, \bibinfo {author} {\bibfnamefont {E.}~\bibnamefont {{Churazov}}}, \bibinfo {author} {\bibfnamefont {A.}~\bibnamefont {{Crites}}}, \bibinfo {author} {\bibfnamefont {L.}~\bibnamefont {{Di Mascolo}}}, \bibinfo {author} {\bibfnamefont {D.}~\bibnamefont {{Eckert}}}, \bibinfo {author} {\bibfnamefont {J.}~\bibnamefont {{Macias-Perez}}}, \bibinfo {author} {\bibfnamefont {F.}~\bibnamefont {{Mayet}}}, \bibinfo {author} {\bibfnamefont {L.}~\bibnamefont {{Perotto}}}, \bibinfo {author} {\bibfnamefont {E.}~\bibnamefont {{Pointecouteau}}}, \bibinfo {author} {\bibfnamefont {C.}~\bibnamefont {{Romero}}}, \bibinfo {author}
  {\bibfnamefont {F.}~\bibnamefont {{Ruppin}}}, \bibinfo {author} {\bibfnamefont {E.}~\bibnamefont {{Scannapieco}}},\ and\ \bibinfo {author} {\bibfnamefont {J.}~\bibnamefont {{ZuHone}}},\ }\bibfield  {title} {\bibinfo {title} {{Astrophysics with the Spatially and Spectrally Resolved Sunyaev-Zeldovich Effects. A Millimetre/Submillimetre Probe of the Warm and Hot Universe}},\ }\href {https://doi.org/10.1007/s11214-019-0581-2} {\bibfield  {journal} {\bibinfo  {journal} {\ssr}\ }\textbf {\bibinfo {volume} {215}},\ \bibinfo {eid} {17} (\bibinfo {year} {2019})},\ \Eprint {https://arxiv.org/abs/1811.02310} {arXiv:1811.02310 [astro-ph.CO]} \BibitemShut {NoStop}%
\bibitem [{\citenamefont {{Schaan}}\ \emph {et~al.}(2021{\natexlab{a}})\citenamefont {{Schaan}}, \citenamefont {{Ferraro}}, \citenamefont {{Amodeo}}, \citenamefont {{Battaglia}}, \citenamefont {{Aiola}}, \citenamefont {{Austermann}}, \citenamefont {{Beall}}, \citenamefont {{Bean}}, \citenamefont {{Becker}}, \citenamefont {{Bond}}, \citenamefont {{Calabrese}}, \citenamefont {{Calafut}}, \citenamefont {{Choi}}, \citenamefont {{Denison}}, \citenamefont {{Devlin}}, \citenamefont {{Duff}}, \citenamefont {{Duivenvoorden}}, \citenamefont {{Dunkley}}, \citenamefont {{D{\"u}nner}}, \citenamefont {{Gallardo}}, \citenamefont {{Guan}}, \citenamefont {{Han}}, \citenamefont {{Hill}}, \citenamefont {{Hilton}}, \citenamefont {{Hilton}}, \citenamefont {{Hlo{\v{z}}ek}}, \citenamefont {{Hubmayr}}, \citenamefont {{Huffenberger}}, \citenamefont {{Hughes}}, \citenamefont {{Koopman}}, \citenamefont {{MacInnis}}, \citenamefont {{McMahon}}, \citenamefont {{Madhavacheril}}, \citenamefont {{Moodley}}, \citenamefont {{Mroczkowski}},
  \citenamefont {{Naess}}, \citenamefont {{Nati}}, \citenamefont {{Newburgh}}, \citenamefont {{Niemack}}, \citenamefont {{Page}}, \citenamefont {{Partridge}}, \citenamefont {{Salatino}}, \citenamefont {{Sehgal}}, \citenamefont {{Schillaci}}, \citenamefont {{Sif{\'o}n}}, \citenamefont {{Smith}}, \citenamefont {{Spergel}}, \citenamefont {{Staggs}}, \citenamefont {{Storer}}, \citenamefont {{Trac}}, \citenamefont {{Ullom}}, \citenamefont {{Van Lanen}}, \citenamefont {{Vale}}, \citenamefont {{van Engelen}}, \citenamefont {{Maga{\~n}a}}, \citenamefont {{Vavagiakis}}, \citenamefont {{Wollack}}, \citenamefont {{Xu}},\ and\ \citenamefont {{Atacama Cosmology Telescope Collaboration}}}]{Schaan_2021}%
  \BibitemOpen
  \bibfield  {author} {\bibinfo {author} {\bibfnamefont {E.}~\bibnamefont {{Schaan}}}, \bibinfo {author} {\bibfnamefont {S.}~\bibnamefont {{Ferraro}}}, \bibinfo {author} {\bibfnamefont {S.}~\bibnamefont {{Amodeo}}}, \bibinfo {author} {\bibfnamefont {N.}~\bibnamefont {{Battaglia}}}, \bibinfo {author} {\bibfnamefont {S.}~\bibnamefont {{Aiola}}}, \bibinfo {author} {\bibfnamefont {J.~E.}\ \bibnamefont {{Austermann}}}, \bibinfo {author} {\bibfnamefont {J.~A.}\ \bibnamefont {{Beall}}}, \bibinfo {author} {\bibfnamefont {R.}~\bibnamefont {{Bean}}}, \bibinfo {author} {\bibfnamefont {D.~T.}\ \bibnamefont {{Becker}}}, \bibinfo {author} {\bibfnamefont {R.~J.}\ \bibnamefont {{Bond}}}, \bibinfo {author} {\bibfnamefont {E.}~\bibnamefont {{Calabrese}}}, \bibinfo {author} {\bibfnamefont {V.}~\bibnamefont {{Calafut}}}, \bibinfo {author} {\bibfnamefont {S.~K.}\ \bibnamefont {{Choi}}}, \bibinfo {author} {\bibfnamefont {E.~V.}\ \bibnamefont {{Denison}}}, \bibinfo {author} {\bibfnamefont {M.~J.}\ \bibnamefont {{Devlin}}}, \bibinfo
  {author} {\bibfnamefont {S.~M.}\ \bibnamefont {{Duff}}}, \bibinfo {author} {\bibfnamefont {A.~J.}\ \bibnamefont {{Duivenvoorden}}}, \bibinfo {author} {\bibfnamefont {J.}~\bibnamefont {{Dunkley}}}, \bibinfo {author} {\bibfnamefont {R.}~\bibnamefont {{D{\"u}nner}}}, \bibinfo {author} {\bibfnamefont {P.~A.}\ \bibnamefont {{Gallardo}}}, \bibinfo {author} {\bibfnamefont {Y.}~\bibnamefont {{Guan}}}, \bibinfo {author} {\bibfnamefont {D.}~\bibnamefont {{Han}}}, \bibinfo {author} {\bibfnamefont {J.~C.}\ \bibnamefont {{Hill}}}, \bibinfo {author} {\bibfnamefont {G.~C.}\ \bibnamefont {{Hilton}}}, \bibinfo {author} {\bibfnamefont {M.}~\bibnamefont {{Hilton}}}, \bibinfo {author} {\bibfnamefont {R.}~\bibnamefont {{Hlo{\v{z}}ek}}}, \bibinfo {author} {\bibfnamefont {J.}~\bibnamefont {{Hubmayr}}}, \bibinfo {author} {\bibfnamefont {K.~M.}\ \bibnamefont {{Huffenberger}}}, \bibinfo {author} {\bibfnamefont {J.~P.}\ \bibnamefont {{Hughes}}}, \bibinfo {author} {\bibfnamefont {B.~J.}\ \bibnamefont {{Koopman}}}, \bibinfo {author}
  {\bibfnamefont {A.}~\bibnamefont {{MacInnis}}}, \bibinfo {author} {\bibfnamefont {J.}~\bibnamefont {{McMahon}}}, \bibinfo {author} {\bibfnamefont {M.~S.}\ \bibnamefont {{Madhavacheril}}}, \bibinfo {author} {\bibfnamefont {K.}~\bibnamefont {{Moodley}}}, \bibinfo {author} {\bibfnamefont {T.}~\bibnamefont {{Mroczkowski}}}, \bibinfo {author} {\bibfnamefont {S.}~\bibnamefont {{Naess}}}, \bibinfo {author} {\bibfnamefont {F.}~\bibnamefont {{Nati}}}, \bibinfo {author} {\bibfnamefont {L.~B.}\ \bibnamefont {{Newburgh}}}, \bibinfo {author} {\bibfnamefont {M.~D.}\ \bibnamefont {{Niemack}}}, \bibinfo {author} {\bibfnamefont {L.~A.}\ \bibnamefont {{Page}}}, \bibinfo {author} {\bibfnamefont {B.}~\bibnamefont {{Partridge}}}, \bibinfo {author} {\bibfnamefont {M.}~\bibnamefont {{Salatino}}}, \bibinfo {author} {\bibfnamefont {N.}~\bibnamefont {{Sehgal}}}, \bibinfo {author} {\bibfnamefont {A.}~\bibnamefont {{Schillaci}}}, \bibinfo {author} {\bibfnamefont {C.}~\bibnamefont {{Sif{\'o}n}}}, \bibinfo {author} {\bibfnamefont
  {K.~M.}\ \bibnamefont {{Smith}}}, \bibinfo {author} {\bibfnamefont {D.~N.}\ \bibnamefont {{Spergel}}}, \bibinfo {author} {\bibfnamefont {S.}~\bibnamefont {{Staggs}}}, \bibinfo {author} {\bibfnamefont {E.~R.}\ \bibnamefont {{Storer}}}, \bibinfo {author} {\bibfnamefont {H.}~\bibnamefont {{Trac}}}, \bibinfo {author} {\bibfnamefont {J.~N.}\ \bibnamefont {{Ullom}}}, \bibinfo {author} {\bibfnamefont {J.}~\bibnamefont {{Van Lanen}}}, \bibinfo {author} {\bibfnamefont {L.~R.}\ \bibnamefont {{Vale}}}, \bibinfo {author} {\bibfnamefont {A.}~\bibnamefont {{van Engelen}}}, \bibinfo {author} {\bibfnamefont {M.~V.}\ \bibnamefont {{Maga{\~n}a}}}, \bibinfo {author} {\bibfnamefont {E.~M.}\ \bibnamefont {{Vavagiakis}}}, \bibinfo {author} {\bibfnamefont {E.~J.}\ \bibnamefont {{Wollack}}}, \bibinfo {author} {\bibfnamefont {Z.}~\bibnamefont {{Xu}}},\ and\ \bibinfo {author} {\bibnamefont {{Atacama Cosmology Telescope Collaboration}}},\ }\bibfield  {title} {\bibinfo {title} {{Atacama Cosmology Telescope: Combined kinematic and
  thermal Sunyaev-Zel'dovich measurements from BOSS CMASS and LOWZ halos}},\ }\href {https://doi.org/10.1103/PhysRevD.103.063513} {\bibfield  {journal} {\bibinfo  {journal} {\prd}\ }\textbf {\bibinfo {volume} {103}},\ \bibinfo {eid} {063513} (\bibinfo {year} {2021}{\natexlab{a}})},\ \Eprint {https://arxiv.org/abs/2009.05557} {arXiv:2009.05557 [astro-ph.CO]} \BibitemShut {NoStop}%
\bibitem [{\citenamefont {{Hadzhiyska}}\ \emph {et~al.}(2024{\natexlab{a}})\citenamefont {{Hadzhiyska}}, \citenamefont {{Ferraro}}, \citenamefont {{Ried Guachalla}}, \citenamefont {{Schaan}}, \citenamefont {{Aguilar}}, \citenamefont {{Battaglia}}, \citenamefont {{Bond}}, \citenamefont {{Brooks}}, \citenamefont {{Calabrese}}, \citenamefont {{Choi}}, \citenamefont {{Claybaugh}}, \citenamefont {{Coulton}}, \citenamefont {{Dawson}}, \citenamefont {{Devlin}}, \citenamefont {{Dey}}, \citenamefont {{Doel}}, \citenamefont {{Duivenvoorden}}, \citenamefont {{Dunkley}}, \citenamefont {{Farren}}, \citenamefont {{Font-Ribera}}, \citenamefont {{Forero-Romero}}, \citenamefont {{Gallardo}}, \citenamefont {{Gazta{\~n}aga}}, \citenamefont {{Gontcho Gontcho}}, \citenamefont {{Gralla}}, \citenamefont {{Le Guillou}}, \citenamefont {{Gutierrez}}, \citenamefont {{Guy}}, \citenamefont {{Hill}}, \citenamefont {{Hlo{\v{z}}ek}}, \citenamefont {{Honscheid}}, \citenamefont {{Juneau}}, \citenamefont {{Kisner}}, \citenamefont {{Kremin}},
  \citenamefont {{Landriau}}, \citenamefont {{Liu}}, \citenamefont {{Louis}}, \citenamefont {{MacCrann}}, \citenamefont {{de Macorra}}, \citenamefont {{Madhavacheril}}, \citenamefont {{Manera}}, \citenamefont {{Meisner}}, \citenamefont {{Miquel}}, \citenamefont {{Moodley}}, \citenamefont {{Moustakas}}, \citenamefont {{Mroczkowski}}, \citenamefont {{Naess}}, \citenamefont {{Newman}}, \citenamefont {{Niemack}}, \citenamefont {{Niz}}, \citenamefont {{Page}}, \citenamefont {{Palanque-Delabrouille}}, \citenamefont {{Partridge}}, \citenamefont {{Percival}}, \citenamefont {{Prada}}, \citenamefont {{Qu}}, \citenamefont {{Rossi}}, \citenamefont {{Sanchez}}, \citenamefont {{Schlegel}}, \citenamefont {{Schubnell}}, \citenamefont {{Sehgal}}, \citenamefont {{Seo}}, \citenamefont {{Sif{\'o}n}}, \citenamefont {{Spergel}}, \citenamefont {{Sprayberry}}, \citenamefont {{Staggs}}, \citenamefont {{Tarl{\'e}}}, \citenamefont {{Vargas}}, \citenamefont {{Vavagiakis}}, \citenamefont {{Weaver}}, \citenamefont {{Wollack}},
  \citenamefont {{Zhou}},\ and\ \citenamefont {{Zou}}}]{2024arXiv240707152H}%
  \BibitemOpen
  \bibfield  {author} {\bibinfo {author} {\bibfnamefont {B.}~\bibnamefont {{Hadzhiyska}}}, \bibinfo {author} {\bibfnamefont {S.}~\bibnamefont {{Ferraro}}}, \bibinfo {author} {\bibfnamefont {B.}~\bibnamefont {{Ried Guachalla}}}, \bibinfo {author} {\bibfnamefont {E.}~\bibnamefont {{Schaan}}}, \bibinfo {author} {\bibfnamefont {J.}~\bibnamefont {{Aguilar}}}, \bibinfo {author} {\bibfnamefont {N.}~\bibnamefont {{Battaglia}}}, \bibinfo {author} {\bibfnamefont {J.~R.}\ \bibnamefont {{Bond}}}, \bibinfo {author} {\bibfnamefont {D.}~\bibnamefont {{Brooks}}}, \bibinfo {author} {\bibfnamefont {E.}~\bibnamefont {{Calabrese}}}, \bibinfo {author} {\bibfnamefont {S.~K.}\ \bibnamefont {{Choi}}}, \bibinfo {author} {\bibfnamefont {T.}~\bibnamefont {{Claybaugh}}}, \bibinfo {author} {\bibfnamefont {W.~R.}\ \bibnamefont {{Coulton}}}, \bibinfo {author} {\bibfnamefont {K.}~\bibnamefont {{Dawson}}}, \bibinfo {author} {\bibfnamefont {M.}~\bibnamefont {{Devlin}}}, \bibinfo {author} {\bibfnamefont {B.}~\bibnamefont {{Dey}}}, \bibinfo
  {author} {\bibfnamefont {P.}~\bibnamefont {{Doel}}}, \bibinfo {author} {\bibfnamefont {A.~J.}\ \bibnamefont {{Duivenvoorden}}}, \bibinfo {author} {\bibfnamefont {J.}~\bibnamefont {{Dunkley}}}, \bibinfo {author} {\bibfnamefont {G.~S.}\ \bibnamefont {{Farren}}}, \bibinfo {author} {\bibfnamefont {A.}~\bibnamefont {{Font-Ribera}}}, \bibinfo {author} {\bibfnamefont {J.~E.}\ \bibnamefont {{Forero-Romero}}}, \bibinfo {author} {\bibfnamefont {P.~A.}\ \bibnamefont {{Gallardo}}}, \bibinfo {author} {\bibfnamefont {E.}~\bibnamefont {{Gazta{\~n}aga}}}, \bibinfo {author} {\bibfnamefont {S.}~\bibnamefont {{Gontcho Gontcho}}}, \bibinfo {author} {\bibfnamefont {M.}~\bibnamefont {{Gralla}}}, \bibinfo {author} {\bibfnamefont {L.}~\bibnamefont {{Le Guillou}}}, \bibinfo {author} {\bibfnamefont {G.}~\bibnamefont {{Gutierrez}}}, \bibinfo {author} {\bibfnamefont {J.}~\bibnamefont {{Guy}}}, \bibinfo {author} {\bibfnamefont {J.~C.}\ \bibnamefont {{Hill}}}, \bibinfo {author} {\bibfnamefont {R.}~\bibnamefont {{Hlo{\v{z}}ek}}},
  \bibinfo {author} {\bibfnamefont {K.}~\bibnamefont {{Honscheid}}}, \bibinfo {author} {\bibfnamefont {S.}~\bibnamefont {{Juneau}}}, \bibinfo {author} {\bibfnamefont {T.}~\bibnamefont {{Kisner}}}, \bibinfo {author} {\bibfnamefont {A.}~\bibnamefont {{Kremin}}}, \bibinfo {author} {\bibfnamefont {M.}~\bibnamefont {{Landriau}}}, \bibinfo {author} {\bibfnamefont {R.~H.}\ \bibnamefont {{Liu}}}, \bibinfo {author} {\bibfnamefont {T.}~\bibnamefont {{Louis}}}, \bibinfo {author} {\bibfnamefont {N.}~\bibnamefont {{MacCrann}}}, \bibinfo {author} {\bibfnamefont {A.}~\bibnamefont {{de Macorra}}}, \bibinfo {author} {\bibfnamefont {M.}~\bibnamefont {{Madhavacheril}}}, \bibinfo {author} {\bibfnamefont {M.}~\bibnamefont {{Manera}}}, \bibinfo {author} {\bibfnamefont {A.}~\bibnamefont {{Meisner}}}, \bibinfo {author} {\bibfnamefont {R.}~\bibnamefont {{Miquel}}}, \bibinfo {author} {\bibfnamefont {K.}~\bibnamefont {{Moodley}}}, \bibinfo {author} {\bibfnamefont {J.}~\bibnamefont {{Moustakas}}}, \bibinfo {author} {\bibfnamefont
  {T.}~\bibnamefont {{Mroczkowski}}}, \bibinfo {author} {\bibfnamefont {S.}~\bibnamefont {{Naess}}}, \bibinfo {author} {\bibfnamefont {J.}~\bibnamefont {{Newman}}}, \bibinfo {author} {\bibfnamefont {M.~D.}\ \bibnamefont {{Niemack}}}, \bibinfo {author} {\bibfnamefont {G.}~\bibnamefont {{Niz}}}, \bibinfo {author} {\bibfnamefont {L.}~\bibnamefont {{Page}}}, \bibinfo {author} {\bibfnamefont {N.}~\bibnamefont {{Palanque-Delabrouille}}}, \bibinfo {author} {\bibfnamefont {B.}~\bibnamefont {{Partridge}}}, \bibinfo {author} {\bibfnamefont {W.~J.}\ \bibnamefont {{Percival}}}, \bibinfo {author} {\bibfnamefont {F.}~\bibnamefont {{Prada}}}, \bibinfo {author} {\bibfnamefont {F.~J.}\ \bibnamefont {{Qu}}}, \bibinfo {author} {\bibfnamefont {G.}~\bibnamefont {{Rossi}}}, \bibinfo {author} {\bibfnamefont {E.}~\bibnamefont {{Sanchez}}}, \bibinfo {author} {\bibfnamefont {D.}~\bibnamefont {{Schlegel}}}, \bibinfo {author} {\bibfnamefont {M.}~\bibnamefont {{Schubnell}}}, \bibinfo {author} {\bibfnamefont {N.}~\bibnamefont {{Sehgal}}},
  \bibinfo {author} {\bibfnamefont {H.}~\bibnamefont {{Seo}}}, \bibinfo {author} {\bibfnamefont {C.}~\bibnamefont {{Sif{\'o}n}}}, \bibinfo {author} {\bibfnamefont {D.}~\bibnamefont {{Spergel}}}, \bibinfo {author} {\bibfnamefont {D.}~\bibnamefont {{Sprayberry}}}, \bibinfo {author} {\bibfnamefont {S.}~\bibnamefont {{Staggs}}}, \bibinfo {author} {\bibfnamefont {G.}~\bibnamefont {{Tarl{\'e}}}}, \bibinfo {author} {\bibfnamefont {C.}~\bibnamefont {{Vargas}}}, \bibinfo {author} {\bibfnamefont {E.~M.}\ \bibnamefont {{Vavagiakis}}}, \bibinfo {author} {\bibfnamefont {B.~A.}\ \bibnamefont {{Weaver}}}, \bibinfo {author} {\bibfnamefont {E.~J.}\ \bibnamefont {{Wollack}}}, \bibinfo {author} {\bibfnamefont {R.}~\bibnamefont {{Zhou}}},\ and\ \bibinfo {author} {\bibfnamefont {H.}~\bibnamefont {{Zou}}},\ }\bibfield  {title} {\bibinfo {title} {{Evidence for large baryonic feedback at low and intermediate redshifts from kinematic Sunyaev-Zel'dovich observations with ACT and DESI photometric galaxies}},\ }\href
  {https://doi.org/10.48550/arXiv.2407.07152} {\bibfield  {journal} {\bibinfo  {journal} {arXiv e-prints}\ ,\ \bibinfo {eid} {arXiv:2407.07152}} (\bibinfo {year} {2024}{\natexlab{a}})},\ \Eprint {https://arxiv.org/abs/2407.07152} {arXiv:2407.07152 [astro-ph.CO]} \BibitemShut {NoStop}%
\bibitem [{\citenamefont {{Hadzhiyska}}\ \emph {et~al.}(2024{\natexlab{b}})\citenamefont {{Hadzhiyska}}, \citenamefont {{Ferraro}},\ and\ \citenamefont {{Zhou}}}]{2024arXiv241203631H}%
  \BibitemOpen
  \bibfield  {author} {\bibinfo {author} {\bibfnamefont {B.}~\bibnamefont {{Hadzhiyska}}}, \bibinfo {author} {\bibfnamefont {S.}~\bibnamefont {{Ferraro}}},\ and\ \bibinfo {author} {\bibfnamefont {R.}~\bibnamefont {{Zhou}}},\ }\bibfield  {title} {\bibinfo {title} {{Tracing cosmic gas in filaments and halos: Low-redshift insights from the kinematic Sunyaev-Zel'dovich effect}},\ }\href {https://doi.org/10.48550/arXiv.2412.03631} {\bibfield  {journal} {\bibinfo  {journal} {arXiv e-prints}\ ,\ \bibinfo {eid} {arXiv:2412.03631}} (\bibinfo {year} {2024}{\natexlab{b}})},\ \Eprint {https://arxiv.org/abs/2412.03631} {arXiv:2412.03631 [astro-ph.CO]} \BibitemShut {NoStop}%
\bibitem [{\citenamefont {{Ried Guachalla}}\ \emph {et~al.}(2025)\citenamefont {{Ried Guachalla}}, \citenamefont {{Schaan}}, \citenamefont {{Hadzhiyska}}, \citenamefont {{Ferraro}}, \citenamefont {{Aguilar}}, \citenamefont {{Ahlen}}, \citenamefont {{Battaglia}}, \citenamefont {{Bianchi}}, \citenamefont {{Bond}}, \citenamefont {{Brooks}}, \citenamefont {{Claybaugh}}, \citenamefont {{Coulton}}, \citenamefont {{de la Macorra}}, \citenamefont {{Devlin}}, \citenamefont {{Dey}}, \citenamefont {{Doel}}, \citenamefont {{Dunkley}}, \citenamefont {{Fanning}}, \citenamefont {{Forero-Romero}}, \citenamefont {{Gaztanaga}}, \citenamefont {{Gontcho}}, \citenamefont {{Gutierrez}}, \citenamefont {{Guy}}, \citenamefont {{Hill}}, \citenamefont {{Honscheid}}, \citenamefont {{Juneau}}, \citenamefont {{Kisner}}, \citenamefont {{Kremin}}, \citenamefont {{Lambert}}, \citenamefont {{Landriau}}, \citenamefont {{Le Guillou}}, \citenamefont {{MacCrann}}, \citenamefont {{Manera}}, \citenamefont {{Meisner}}, \citenamefont {{Miquel}},
  \citenamefont {{Moodley}}, \citenamefont {{Moustakas}}, \citenamefont {{Mroczkowski}}, \citenamefont {{Myers}}, \citenamefont {{Niemack}}, \citenamefont {{Niz}}, \citenamefont {{Palanque-Delabrouille}}, \citenamefont {{Percival}}, \citenamefont {{P\textbackslash'erez-R\textbackslash`afols}}, \citenamefont {{Poppett}}, \citenamefont {{Prada}}, \citenamefont {{Qu}}, \citenamefont {{Rossi}}, \citenamefont {{Sanchez}}, \citenamefont {{Schlegel}}, \citenamefont {{Schubnell}}, \citenamefont {{Seo}}, \citenamefont {{Sif\textbackslash'on}}, \citenamefont {{Spergel}}, \citenamefont {{Sprayberry}}, \citenamefont {{Tarl\textbackslash'e}}, \citenamefont {{Vargas-Magana}}, \citenamefont {{Vavagiakis}}, \citenamefont {{Weaver}}, \citenamefont {{Wollack}},\ and\ \citenamefont {{Zarrouk}}}]{RiedGuachalla:2025byu}%
  \BibitemOpen
  \bibfield  {author} {\bibinfo {author} {\bibfnamefont {B.}~\bibnamefont {{Ried Guachalla}}}, \bibinfo {author} {\bibfnamefont {E.}~\bibnamefont {{Schaan}}}, \bibinfo {author} {\bibfnamefont {B.}~\bibnamefont {{Hadzhiyska}}}, \bibinfo {author} {\bibfnamefont {S.}~\bibnamefont {{Ferraro}}}, \bibinfo {author} {\bibfnamefont {J.~N.}\ \bibnamefont {{Aguilar}}}, \bibinfo {author} {\bibfnamefont {S.}~\bibnamefont {{Ahlen}}}, \bibinfo {author} {\bibfnamefont {N.}~\bibnamefont {{Battaglia}}}, \bibinfo {author} {\bibfnamefont {D.}~\bibnamefont {{Bianchi}}}, \bibinfo {author} {\bibfnamefont {R.}~\bibnamefont {{Bond}}}, \bibinfo {author} {\bibfnamefont {D.}~\bibnamefont {{Brooks}}}, \bibinfo {author} {\bibfnamefont {T.}~\bibnamefont {{Claybaugh}}}, \bibinfo {author} {\bibfnamefont {W.~R.}\ \bibnamefont {{Coulton}}}, \bibinfo {author} {\bibfnamefont {A.}~\bibnamefont {{de la Macorra}}}, \bibinfo {author} {\bibfnamefont {M.~J.}\ \bibnamefont {{Devlin}}}, \bibinfo {author} {\bibfnamefont {A.}~\bibnamefont {{Dey}}},
  \bibinfo {author} {\bibfnamefont {P.}~\bibnamefont {{Doel}}}, \bibinfo {author} {\bibfnamefont {J.}~\bibnamefont {{Dunkley}}}, \bibinfo {author} {\bibfnamefont {K.}~\bibnamefont {{Fanning}}}, \bibinfo {author} {\bibfnamefont {J.}~\bibnamefont {{Forero-Romero}}}, \bibinfo {author} {\bibfnamefont {E.}~\bibnamefont {{Gaztanaga}}}, \bibinfo {author} {\bibfnamefont {S.~G.~A.}\ \bibnamefont {{Gontcho}}}, \bibinfo {author} {\bibfnamefont {G.}~\bibnamefont {{Gutierrez}}}, \bibinfo {author} {\bibfnamefont {J.}~\bibnamefont {{Guy}}}, \bibinfo {author} {\bibfnamefont {J.~C.}\ \bibnamefont {{Hill}}}, \bibinfo {author} {\bibfnamefont {K.}~\bibnamefont {{Honscheid}}}, \bibinfo {author} {\bibfnamefont {S.}~\bibnamefont {{Juneau}}}, \bibinfo {author} {\bibfnamefont {T.}~\bibnamefont {{Kisner}}}, \bibinfo {author} {\bibfnamefont {A.}~\bibnamefont {{Kremin}}}, \bibinfo {author} {\bibfnamefont {A.}~\bibnamefont {{Lambert}}}, \bibinfo {author} {\bibfnamefont {M.}~\bibnamefont {{Landriau}}}, \bibinfo {author} {\bibfnamefont
  {L.}~\bibnamefont {{Le Guillou}}}, \bibinfo {author} {\bibfnamefont {N.}~\bibnamefont {{MacCrann}}}, \bibinfo {author} {\bibfnamefont {M.}~\bibnamefont {{Manera}}}, \bibinfo {author} {\bibfnamefont {A.}~\bibnamefont {{Meisner}}}, \bibinfo {author} {\bibfnamefont {R.}~\bibnamefont {{Miquel}}}, \bibinfo {author} {\bibfnamefont {K.}~\bibnamefont {{Moodley}}}, \bibinfo {author} {\bibfnamefont {J.}~\bibnamefont {{Moustakas}}}, \bibinfo {author} {\bibfnamefont {T.}~\bibnamefont {{Mroczkowski}}}, \bibinfo {author} {\bibfnamefont {A.~D.}\ \bibnamefont {{Myers}}}, \bibinfo {author} {\bibfnamefont {M.~D.}\ \bibnamefont {{Niemack}}}, \bibinfo {author} {\bibfnamefont {G.}~\bibnamefont {{Niz}}}, \bibinfo {author} {\bibfnamefont {N.}~\bibnamefont {{Palanque-Delabrouille}}}, \bibinfo {author} {\bibfnamefont {W.}~\bibnamefont {{Percival}}}, \bibinfo {author} {\bibfnamefont {I.}~\bibnamefont {{P\textbackslash'erez-R\textbackslash`afols}}}, \bibinfo {author} {\bibfnamefont {C.}~\bibnamefont {{Poppett}}}, \bibinfo {author}
  {\bibfnamefont {F.}~\bibnamefont {{Prada}}}, \bibinfo {author} {\bibfnamefont {F.~J.}\ \bibnamefont {{Qu}}}, \bibinfo {author} {\bibfnamefont {G.}~\bibnamefont {{Rossi}}}, \bibinfo {author} {\bibfnamefont {E.}~\bibnamefont {{Sanchez}}}, \bibinfo {author} {\bibfnamefont {D.}~\bibnamefont {{Schlegel}}}, \bibinfo {author} {\bibfnamefont {M.}~\bibnamefont {{Schubnell}}}, \bibinfo {author} {\bibfnamefont {H.-J.}\ \bibnamefont {{Seo}}}, \bibinfo {author} {\bibfnamefont {C.}~\bibnamefont {{Sif\textbackslash'on}}}, \bibinfo {author} {\bibfnamefont {D.~N.}\ \bibnamefont {{Spergel}}}, \bibinfo {author} {\bibfnamefont {D.}~\bibnamefont {{Sprayberry}}}, \bibinfo {author} {\bibfnamefont {G.}~\bibnamefont {{Tarl\textbackslash'e}}}, \bibinfo {author} {\bibfnamefont {M.}~\bibnamefont {{Vargas-Magana}}}, \bibinfo {author} {\bibfnamefont {E.~M.}\ \bibnamefont {{Vavagiakis}}}, \bibinfo {author} {\bibfnamefont {B.~A.}\ \bibnamefont {{Weaver}}}, \bibinfo {author} {\bibfnamefont {E.~J.}\ \bibnamefont {{Wollack}}},\ and\ \bibinfo
  {author} {\bibfnamefont {P.}~\bibnamefont {{Zarrouk}}},\ }\bibfield  {title} {\bibinfo {title} {{Backlighting extended gas halos around luminous red galaxies: kinematic Sunyaev-Zel'dovich effect from DESI Y1 x ACT}},\ }\href {https://doi.org/10.48550/arXiv.2503.19870} {\bibfield  {journal} {\bibinfo  {journal} {arXiv e-prints}\ ,\ \bibinfo {eid} {arXiv:2503.19870}} (\bibinfo {year} {2025})},\ \Eprint {https://arxiv.org/abs/2503.19870} {arXiv:2503.19870 [astro-ph.GA]} \BibitemShut {NoStop}%
\bibitem [{\citenamefont {{Moser}}\ \emph {et~al.}(2021)\citenamefont {{Moser}}, \citenamefont {{Amodeo}}, \citenamefont {{Battaglia}}, \citenamefont {{Alvarez}}, \citenamefont {{Ferraro}},\ and\ \citenamefont {{Schaan}}}]{2021ApJ...919....2M}%
  \BibitemOpen
  \bibfield  {author} {\bibinfo {author} {\bibfnamefont {E.}~\bibnamefont {{Moser}}}, \bibinfo {author} {\bibfnamefont {S.}~\bibnamefont {{Amodeo}}}, \bibinfo {author} {\bibfnamefont {N.}~\bibnamefont {{Battaglia}}}, \bibinfo {author} {\bibfnamefont {M.~A.}\ \bibnamefont {{Alvarez}}}, \bibinfo {author} {\bibfnamefont {S.}~\bibnamefont {{Ferraro}}},\ and\ \bibinfo {author} {\bibfnamefont {E.}~\bibnamefont {{Schaan}}},\ }\bibfield  {title} {\bibinfo {title} {{The Impacts of Modeling Choices on the Inference of Circumgalactic Medium Properties from Sunyaev-Zeldovich Observations}},\ }\href {https://doi.org/10.3847/1538-4357/ac0cea} {\bibfield  {journal} {\bibinfo  {journal} {\apj}\ }\textbf {\bibinfo {volume} {919}},\ \bibinfo {eid} {2} (\bibinfo {year} {2021})},\ \Eprint {https://arxiv.org/abs/2103.02469} {arXiv:2103.02469 [astro-ph.GA]} \BibitemShut {NoStop}%
\bibitem [{\citenamefont {{Amodeo}}\ \emph {et~al.}(2021)\citenamefont {{Amodeo}}, \citenamefont {{Battaglia}}, \citenamefont {{Schaan}}, \citenamefont {{Ferraro}}, \citenamefont {{Moser}}, \citenamefont {{Aiola}}, \citenamefont {{Austermann}}, \citenamefont {{Beall}}, \citenamefont {{Bean}}, \citenamefont {{Becker}}, \citenamefont {{Bond}}, \citenamefont {{Calabrese}}, \citenamefont {{Calafut}}, \citenamefont {{Choi}}, \citenamefont {{Denison}}, \citenamefont {{Devlin}}, \citenamefont {{Duff}}, \citenamefont {{Duivenvoorden}}, \citenamefont {{Dunkley}}, \citenamefont {{D{\"u}nner}}, \citenamefont {{Gallardo}}, \citenamefont {{Hall}}, \citenamefont {{Han}}, \citenamefont {{Hill}}, \citenamefont {{Hilton}}, \citenamefont {{Hilton}}, \citenamefont {{Hlo{\v{z}}ek}}, \citenamefont {{Hubmayr}}, \citenamefont {{Huffenberger}}, \citenamefont {{Hughes}}, \citenamefont {{Koopman}}, \citenamefont {{MacInnis}}, \citenamefont {{McMahon}}, \citenamefont {{Madhavacheril}}, \citenamefont {{Moodley}}, \citenamefont
  {{Mroczkowski}}, \citenamefont {{Naess}}, \citenamefont {{Nati}}, \citenamefont {{Newburgh}}, \citenamefont {{Niemack}}, \citenamefont {{Page}}, \citenamefont {{Partridge}}, \citenamefont {{Schillaci}}, \citenamefont {{Sehgal}}, \citenamefont {{Sif{\'o}n}}, \citenamefont {{Spergel}}, \citenamefont {{Staggs}}, \citenamefont {{Storer}}, \citenamefont {{Ullom}}, \citenamefont {{Vale}}, \citenamefont {{van Engelen}}, \citenamefont {{Van Lanen}}, \citenamefont {{Vavagiakis}}, \citenamefont {{Wollack}},\ and\ \citenamefont {{Xu}}}]{2021PhRvD.103f3514A}%
  \BibitemOpen
  \bibfield  {author} {\bibinfo {author} {\bibfnamefont {S.}~\bibnamefont {{Amodeo}}}, \bibinfo {author} {\bibfnamefont {N.}~\bibnamefont {{Battaglia}}}, \bibinfo {author} {\bibfnamefont {E.}~\bibnamefont {{Schaan}}}, \bibinfo {author} {\bibfnamefont {S.}~\bibnamefont {{Ferraro}}}, \bibinfo {author} {\bibfnamefont {E.}~\bibnamefont {{Moser}}}, \bibinfo {author} {\bibfnamefont {S.}~\bibnamefont {{Aiola}}}, \bibinfo {author} {\bibfnamefont {J.~E.}\ \bibnamefont {{Austermann}}}, \bibinfo {author} {\bibfnamefont {J.~A.}\ \bibnamefont {{Beall}}}, \bibinfo {author} {\bibfnamefont {R.}~\bibnamefont {{Bean}}}, \bibinfo {author} {\bibfnamefont {D.~T.}\ \bibnamefont {{Becker}}}, \bibinfo {author} {\bibfnamefont {R.~J.}\ \bibnamefont {{Bond}}}, \bibinfo {author} {\bibfnamefont {E.}~\bibnamefont {{Calabrese}}}, \bibinfo {author} {\bibfnamefont {V.}~\bibnamefont {{Calafut}}}, \bibinfo {author} {\bibfnamefont {S.~K.}\ \bibnamefont {{Choi}}}, \bibinfo {author} {\bibfnamefont {E.~V.}\ \bibnamefont {{Denison}}}, \bibinfo
  {author} {\bibfnamefont {M.}~\bibnamefont {{Devlin}}}, \bibinfo {author} {\bibfnamefont {S.~M.}\ \bibnamefont {{Duff}}}, \bibinfo {author} {\bibfnamefont {A.~J.}\ \bibnamefont {{Duivenvoorden}}}, \bibinfo {author} {\bibfnamefont {J.}~\bibnamefont {{Dunkley}}}, \bibinfo {author} {\bibfnamefont {R.}~\bibnamefont {{D{\"u}nner}}}, \bibinfo {author} {\bibfnamefont {P.~A.}\ \bibnamefont {{Gallardo}}}, \bibinfo {author} {\bibfnamefont {K.~R.}\ \bibnamefont {{Hall}}}, \bibinfo {author} {\bibfnamefont {D.}~\bibnamefont {{Han}}}, \bibinfo {author} {\bibfnamefont {J.~C.}\ \bibnamefont {{Hill}}}, \bibinfo {author} {\bibfnamefont {G.~C.}\ \bibnamefont {{Hilton}}}, \bibinfo {author} {\bibfnamefont {M.}~\bibnamefont {{Hilton}}}, \bibinfo {author} {\bibfnamefont {R.}~\bibnamefont {{Hlo{\v{z}}ek}}}, \bibinfo {author} {\bibfnamefont {J.}~\bibnamefont {{Hubmayr}}}, \bibinfo {author} {\bibfnamefont {K.~M.}\ \bibnamefont {{Huffenberger}}}, \bibinfo {author} {\bibfnamefont {J.~P.}\ \bibnamefont {{Hughes}}}, \bibinfo {author}
  {\bibfnamefont {B.~J.}\ \bibnamefont {{Koopman}}}, \bibinfo {author} {\bibfnamefont {A.}~\bibnamefont {{MacInnis}}}, \bibinfo {author} {\bibfnamefont {J.}~\bibnamefont {{McMahon}}}, \bibinfo {author} {\bibfnamefont {M.~S.}\ \bibnamefont {{Madhavacheril}}}, \bibinfo {author} {\bibfnamefont {K.}~\bibnamefont {{Moodley}}}, \bibinfo {author} {\bibfnamefont {T.}~\bibnamefont {{Mroczkowski}}}, \bibinfo {author} {\bibfnamefont {S.}~\bibnamefont {{Naess}}}, \bibinfo {author} {\bibfnamefont {F.}~\bibnamefont {{Nati}}}, \bibinfo {author} {\bibfnamefont {L.~B.}\ \bibnamefont {{Newburgh}}}, \bibinfo {author} {\bibfnamefont {M.~D.}\ \bibnamefont {{Niemack}}}, \bibinfo {author} {\bibfnamefont {L.~A.}\ \bibnamefont {{Page}}}, \bibinfo {author} {\bibfnamefont {B.}~\bibnamefont {{Partridge}}}, \bibinfo {author} {\bibfnamefont {A.}~\bibnamefont {{Schillaci}}}, \bibinfo {author} {\bibfnamefont {N.}~\bibnamefont {{Sehgal}}}, \bibinfo {author} {\bibfnamefont {C.}~\bibnamefont {{Sif{\'o}n}}}, \bibinfo {author} {\bibfnamefont
  {D.~N.}\ \bibnamefont {{Spergel}}}, \bibinfo {author} {\bibfnamefont {S.}~\bibnamefont {{Staggs}}}, \bibinfo {author} {\bibfnamefont {E.~R.}\ \bibnamefont {{Storer}}}, \bibinfo {author} {\bibfnamefont {J.~N.}\ \bibnamefont {{Ullom}}}, \bibinfo {author} {\bibfnamefont {L.~R.}\ \bibnamefont {{Vale}}}, \bibinfo {author} {\bibfnamefont {A.}~\bibnamefont {{van Engelen}}}, \bibinfo {author} {\bibfnamefont {J.}~\bibnamefont {{Van Lanen}}}, \bibinfo {author} {\bibfnamefont {E.~M.}\ \bibnamefont {{Vavagiakis}}}, \bibinfo {author} {\bibfnamefont {E.~J.}\ \bibnamefont {{Wollack}}},\ and\ \bibinfo {author} {\bibfnamefont {Z.}~\bibnamefont {{Xu}}},\ }\bibfield  {title} {\bibinfo {title} {{Atacama Cosmology Telescope: Modeling the gas thermodynamics in BOSS CMASS galaxies from kinematic and thermal Sunyaev-Zel'dovich measurements}},\ }\href {https://doi.org/10.1103/PhysRevD.103.063514} {\bibfield  {journal} {\bibinfo  {journal} {\prd}\ }\textbf {\bibinfo {volume} {103}},\ \bibinfo {eid} {063514} (\bibinfo {year}
  {2021})},\ \Eprint {https://arxiv.org/abs/2009.05558} {arXiv:2009.05558 [astro-ph.CO]} \BibitemShut {NoStop}%
\bibitem [{\citenamefont {{Baxter}}\ \emph {et~al.}(2015)\citenamefont {{Baxter}}, \citenamefont {{Keisler}}, \citenamefont {{Dodelson}}, \citenamefont {{Aird}}, \citenamefont {{Allen}}, \citenamefont {{Ashby}}, \citenamefont {{Bautz}}, \citenamefont {{Bayliss}}, \citenamefont {{Benson}}, \citenamefont {{Bleem}}, \citenamefont {{Bocquet}}, \citenamefont {{Brodwin}}, \citenamefont {{Carlstrom}}, \citenamefont {{Chang}}, \citenamefont {{Chiu}}, \citenamefont {{Cho}}, \citenamefont {{Clocchiatti}}, \citenamefont {{Crawford}}, \citenamefont {{Crites}}, \citenamefont {{Desai}}, \citenamefont {{Dietrich}}, \citenamefont {{de Haan}}, \citenamefont {{Dobbs}}, \citenamefont {{Foley}}, \citenamefont {{Forman}}, \citenamefont {{George}}, \citenamefont {{Gladders}}, \citenamefont {{Gonzalez}}, \citenamefont {{Halverson}}, \citenamefont {{Harrington}}, \citenamefont {{Hennig}}, \citenamefont {{Hoekstra}}, \citenamefont {{Holder}}, \citenamefont {{Holzapfel}}, \citenamefont {{Hou}}, \citenamefont {{Hrubes}}, \citenamefont
  {{Jones}}, \citenamefont {{Knox}}, \citenamefont {{Lee}}, \citenamefont {{Leitch}}, \citenamefont {{Liu}}, \citenamefont {{Lueker}}, \citenamefont {{Luong-Van}}, \citenamefont {{Mantz}}, \citenamefont {{Marrone}}, \citenamefont {{McDonald}}, \citenamefont {{McMahon}}, \citenamefont {{Meyer}}, \citenamefont {{Millea}}, \citenamefont {{Mocanu}}, \citenamefont {{Murray}}, \citenamefont {{Padin}}, \citenamefont {{Pryke}}, \citenamefont {{Reichardt}}, \citenamefont {{Rest}}, \citenamefont {{Ruhl}}, \citenamefont {{Saliwanchik}}, \citenamefont {{Saro}}, \citenamefont {{Sayre}}, \citenamefont {{Schaffer}}, \citenamefont {{Shirokoff}}, \citenamefont {{Song}}, \citenamefont {{Spieler}}, \citenamefont {{Stalder}}, \citenamefont {{Stanford}}, \citenamefont {{Staniszewski}}, \citenamefont {{Stark}}, \citenamefont {{Story}}, \citenamefont {{van Engelen}}, \citenamefont {{Vanderlinde}}, \citenamefont {{Vieira}}, \citenamefont {{Vikhlinin}}, \citenamefont {{Williamson}}, \citenamefont {{Zahn}},\ and\ \citenamefont
  {{Zenteno}}}]{2015ApJ...806..247B}%
  \BibitemOpen
  \bibfield  {author} {\bibinfo {author} {\bibfnamefont {E.~J.}\ \bibnamefont {{Baxter}}}, \bibinfo {author} {\bibfnamefont {R.}~\bibnamefont {{Keisler}}}, \bibinfo {author} {\bibfnamefont {S.}~\bibnamefont {{Dodelson}}}, \bibinfo {author} {\bibfnamefont {K.~A.}\ \bibnamefont {{Aird}}}, \bibinfo {author} {\bibfnamefont {S.~W.}\ \bibnamefont {{Allen}}}, \bibinfo {author} {\bibfnamefont {M.~L.~N.}\ \bibnamefont {{Ashby}}}, \bibinfo {author} {\bibfnamefont {M.}~\bibnamefont {{Bautz}}}, \bibinfo {author} {\bibfnamefont {M.}~\bibnamefont {{Bayliss}}}, \bibinfo {author} {\bibfnamefont {B.~A.}\ \bibnamefont {{Benson}}}, \bibinfo {author} {\bibfnamefont {L.~E.}\ \bibnamefont {{Bleem}}}, \bibinfo {author} {\bibfnamefont {S.}~\bibnamefont {{Bocquet}}}, \bibinfo {author} {\bibfnamefont {M.}~\bibnamefont {{Brodwin}}}, \bibinfo {author} {\bibfnamefont {J.~E.}\ \bibnamefont {{Carlstrom}}}, \bibinfo {author} {\bibfnamefont {C.~L.}\ \bibnamefont {{Chang}}}, \bibinfo {author} {\bibfnamefont {I.}~\bibnamefont {{Chiu}}},
  \bibinfo {author} {\bibfnamefont {H.~M.}\ \bibnamefont {{Cho}}}, \bibinfo {author} {\bibfnamefont {A.}~\bibnamefont {{Clocchiatti}}}, \bibinfo {author} {\bibfnamefont {T.~M.}\ \bibnamefont {{Crawford}}}, \bibinfo {author} {\bibfnamefont {A.~T.}\ \bibnamefont {{Crites}}}, \bibinfo {author} {\bibfnamefont {S.}~\bibnamefont {{Desai}}}, \bibinfo {author} {\bibfnamefont {J.~P.}\ \bibnamefont {{Dietrich}}}, \bibinfo {author} {\bibfnamefont {T.}~\bibnamefont {{de Haan}}}, \bibinfo {author} {\bibfnamefont {M.~A.}\ \bibnamefont {{Dobbs}}}, \bibinfo {author} {\bibfnamefont {R.~J.}\ \bibnamefont {{Foley}}}, \bibinfo {author} {\bibfnamefont {W.~R.}\ \bibnamefont {{Forman}}}, \bibinfo {author} {\bibfnamefont {E.~M.}\ \bibnamefont {{George}}}, \bibinfo {author} {\bibfnamefont {M.~D.}\ \bibnamefont {{Gladders}}}, \bibinfo {author} {\bibfnamefont {A.~H.}\ \bibnamefont {{Gonzalez}}}, \bibinfo {author} {\bibfnamefont {N.~W.}\ \bibnamefont {{Halverson}}}, \bibinfo {author} {\bibfnamefont {N.~L.}\ \bibnamefont {{Harrington}}},
  \bibinfo {author} {\bibfnamefont {C.}~\bibnamefont {{Hennig}}}, \bibinfo {author} {\bibfnamefont {H.}~\bibnamefont {{Hoekstra}}}, \bibinfo {author} {\bibfnamefont {G.~P.}\ \bibnamefont {{Holder}}}, \bibinfo {author} {\bibfnamefont {W.~L.}\ \bibnamefont {{Holzapfel}}}, \bibinfo {author} {\bibfnamefont {Z.}~\bibnamefont {{Hou}}}, \bibinfo {author} {\bibfnamefont {J.~D.}\ \bibnamefont {{Hrubes}}}, \bibinfo {author} {\bibfnamefont {C.}~\bibnamefont {{Jones}}}, \bibinfo {author} {\bibfnamefont {L.}~\bibnamefont {{Knox}}}, \bibinfo {author} {\bibfnamefont {A.~T.}\ \bibnamefont {{Lee}}}, \bibinfo {author} {\bibfnamefont {E.~M.}\ \bibnamefont {{Leitch}}}, \bibinfo {author} {\bibfnamefont {J.}~\bibnamefont {{Liu}}}, \bibinfo {author} {\bibfnamefont {M.}~\bibnamefont {{Lueker}}}, \bibinfo {author} {\bibfnamefont {D.}~\bibnamefont {{Luong-Van}}}, \bibinfo {author} {\bibfnamefont {A.}~\bibnamefont {{Mantz}}}, \bibinfo {author} {\bibfnamefont {D.~P.}\ \bibnamefont {{Marrone}}}, \bibinfo {author} {\bibfnamefont
  {M.}~\bibnamefont {{McDonald}}}, \bibinfo {author} {\bibfnamefont {J.~J.}\ \bibnamefont {{McMahon}}}, \bibinfo {author} {\bibfnamefont {S.~S.}\ \bibnamefont {{Meyer}}}, \bibinfo {author} {\bibfnamefont {M.}~\bibnamefont {{Millea}}}, \bibinfo {author} {\bibfnamefont {L.~M.}\ \bibnamefont {{Mocanu}}}, \bibinfo {author} {\bibfnamefont {S.~S.}\ \bibnamefont {{Murray}}}, \bibinfo {author} {\bibfnamefont {S.}~\bibnamefont {{Padin}}}, \bibinfo {author} {\bibfnamefont {C.}~\bibnamefont {{Pryke}}}, \bibinfo {author} {\bibfnamefont {C.~L.}\ \bibnamefont {{Reichardt}}}, \bibinfo {author} {\bibfnamefont {A.}~\bibnamefont {{Rest}}}, \bibinfo {author} {\bibfnamefont {J.~E.}\ \bibnamefont {{Ruhl}}}, \bibinfo {author} {\bibfnamefont {B.~R.}\ \bibnamefont {{Saliwanchik}}}, \bibinfo {author} {\bibfnamefont {A.}~\bibnamefont {{Saro}}}, \bibinfo {author} {\bibfnamefont {J.~T.}\ \bibnamefont {{Sayre}}}, \bibinfo {author} {\bibfnamefont {K.~K.}\ \bibnamefont {{Schaffer}}}, \bibinfo {author} {\bibfnamefont {E.}~\bibnamefont
  {{Shirokoff}}}, \bibinfo {author} {\bibfnamefont {J.}~\bibnamefont {{Song}}}, \bibinfo {author} {\bibfnamefont {H.~G.}\ \bibnamefont {{Spieler}}}, \bibinfo {author} {\bibfnamefont {B.}~\bibnamefont {{Stalder}}}, \bibinfo {author} {\bibfnamefont {S.~A.}\ \bibnamefont {{Stanford}}}, \bibinfo {author} {\bibfnamefont {Z.}~\bibnamefont {{Staniszewski}}}, \bibinfo {author} {\bibfnamefont {A.~A.}\ \bibnamefont {{Stark}}}, \bibinfo {author} {\bibfnamefont {K.~T.}\ \bibnamefont {{Story}}}, \bibinfo {author} {\bibfnamefont {A.}~\bibnamefont {{van Engelen}}}, \bibinfo {author} {\bibfnamefont {K.}~\bibnamefont {{Vanderlinde}}}, \bibinfo {author} {\bibfnamefont {J.~D.}\ \bibnamefont {{Vieira}}}, \bibinfo {author} {\bibfnamefont {A.}~\bibnamefont {{Vikhlinin}}}, \bibinfo {author} {\bibfnamefont {R.}~\bibnamefont {{Williamson}}}, \bibinfo {author} {\bibfnamefont {O.}~\bibnamefont {{Zahn}}},\ and\ \bibinfo {author} {\bibfnamefont {A.}~\bibnamefont {{Zenteno}}},\ }\bibfield  {title} {\bibinfo {title} {{A Measurement of
  Gravitational Lensing of the Cosmic Microwave Background by Galaxy Clusters Using Data from the South Pole Telescope}},\ }\href {https://doi.org/10.1088/0004-637X/806/2/247} {\bibfield  {journal} {\bibinfo  {journal} {\apj}\ }\textbf {\bibinfo {volume} {806}},\ \bibinfo {eid} {247} (\bibinfo {year} {2015})},\ \Eprint {https://arxiv.org/abs/1412.7521} {arXiv:1412.7521 [astro-ph.CO]} \BibitemShut {NoStop}%
\bibitem [{\citenamefont {{Madhavacheril}}\ \emph {et~al.}(2020)\citenamefont {{Madhavacheril}}, \citenamefont {{Sif{\'o}n}}, \citenamefont {{Battaglia}}, \citenamefont {{Aiola}}, \citenamefont {{Amodeo}}, \citenamefont {{Austermann}}, \citenamefont {{Beall}}, \citenamefont {{Becker}}, \citenamefont {{Bond}}, \citenamefont {{Calabrese}}, \citenamefont {{Choi}}, \citenamefont {{Denison}}, \citenamefont {{Devlin}}, \citenamefont {{Dicker}}, \citenamefont {{Duff}}, \citenamefont {{Duivenvoorden}}, \citenamefont {{Dunkley}}, \citenamefont {{D{\"u}nner}}, \citenamefont {{Ferraro}}, \citenamefont {{Gallardo}}, \citenamefont {{Guan}}, \citenamefont {{Han}}, \citenamefont {{Hill}}, \citenamefont {{Hilton}}, \citenamefont {{Hilton}}, \citenamefont {{Hubmayr}}, \citenamefont {{Huffenberger}}, \citenamefont {{Hughes}}, \citenamefont {{Koopman}}, \citenamefont {{Kosowsky}}, \citenamefont {{Van Lanen}}, \citenamefont {{Lee}}, \citenamefont {{Louis}}, \citenamefont {{MacInnis}}, \citenamefont {{McMahon}}, \citenamefont
  {{Moodley}}, \citenamefont {{Naess}}, \citenamefont {{Namikawa}}, \citenamefont {{Nati}}, \citenamefont {{Newburgh}}, \citenamefont {{Niemack}}, \citenamefont {{Page}}, \citenamefont {{Partridge}}, \citenamefont {{Qu}}, \citenamefont {{Robertson}}, \citenamefont {{Salatino}}, \citenamefont {{Schaan}}, \citenamefont {{Schillaci}}, \citenamefont {{Schmitt}}, \citenamefont {{Sehgal}}, \citenamefont {{Sherwin}}, \citenamefont {{Simon}}, \citenamefont {{Spergel}}, \citenamefont {{Staggs}}, \citenamefont {{Storer}}, \citenamefont {{Ullom}}, \citenamefont {{Vale}}, \citenamefont {{van Engelen}}, \citenamefont {{Vavagiakis}}, \citenamefont {{Wollack}},\ and\ \citenamefont {{Xu}}}]{2020ApJ...903L..13M}%
  \BibitemOpen
  \bibfield  {author} {\bibinfo {author} {\bibfnamefont {M.~S.}\ \bibnamefont {{Madhavacheril}}}, \bibinfo {author} {\bibfnamefont {C.}~\bibnamefont {{Sif{\'o}n}}}, \bibinfo {author} {\bibfnamefont {N.}~\bibnamefont {{Battaglia}}}, \bibinfo {author} {\bibfnamefont {S.}~\bibnamefont {{Aiola}}}, \bibinfo {author} {\bibfnamefont {S.}~\bibnamefont {{Amodeo}}}, \bibinfo {author} {\bibfnamefont {J.~E.}\ \bibnamefont {{Austermann}}}, \bibinfo {author} {\bibfnamefont {J.~A.}\ \bibnamefont {{Beall}}}, \bibinfo {author} {\bibfnamefont {D.~T.}\ \bibnamefont {{Becker}}}, \bibinfo {author} {\bibfnamefont {J.~R.}\ \bibnamefont {{Bond}}}, \bibinfo {author} {\bibfnamefont {E.}~\bibnamefont {{Calabrese}}}, \bibinfo {author} {\bibfnamefont {S.~K.}\ \bibnamefont {{Choi}}}, \bibinfo {author} {\bibfnamefont {E.~V.}\ \bibnamefont {{Denison}}}, \bibinfo {author} {\bibfnamefont {M.~J.}\ \bibnamefont {{Devlin}}}, \bibinfo {author} {\bibfnamefont {S.~R.}\ \bibnamefont {{Dicker}}}, \bibinfo {author} {\bibfnamefont {S.~M.}\ \bibnamefont
  {{Duff}}}, \bibinfo {author} {\bibfnamefont {A.~J.}\ \bibnamefont {{Duivenvoorden}}}, \bibinfo {author} {\bibfnamefont {J.}~\bibnamefont {{Dunkley}}}, \bibinfo {author} {\bibfnamefont {R.}~\bibnamefont {{D{\"u}nner}}}, \bibinfo {author} {\bibfnamefont {S.}~\bibnamefont {{Ferraro}}}, \bibinfo {author} {\bibfnamefont {P.~A.}\ \bibnamefont {{Gallardo}}}, \bibinfo {author} {\bibfnamefont {Y.}~\bibnamefont {{Guan}}}, \bibinfo {author} {\bibfnamefont {D.}~\bibnamefont {{Han}}}, \bibinfo {author} {\bibfnamefont {J.~C.}\ \bibnamefont {{Hill}}}, \bibinfo {author} {\bibfnamefont {G.~C.}\ \bibnamefont {{Hilton}}}, \bibinfo {author} {\bibfnamefont {M.}~\bibnamefont {{Hilton}}}, \bibinfo {author} {\bibfnamefont {J.}~\bibnamefont {{Hubmayr}}}, \bibinfo {author} {\bibfnamefont {K.~M.}\ \bibnamefont {{Huffenberger}}}, \bibinfo {author} {\bibfnamefont {J.~P.}\ \bibnamefont {{Hughes}}}, \bibinfo {author} {\bibfnamefont {B.~J.}\ \bibnamefont {{Koopman}}}, \bibinfo {author} {\bibfnamefont {A.}~\bibnamefont {{Kosowsky}}},
  \bibinfo {author} {\bibfnamefont {J.}~\bibnamefont {{Van Lanen}}}, \bibinfo {author} {\bibfnamefont {E.}~\bibnamefont {{Lee}}}, \bibinfo {author} {\bibfnamefont {T.}~\bibnamefont {{Louis}}}, \bibinfo {author} {\bibfnamefont {A.}~\bibnamefont {{MacInnis}}}, \bibinfo {author} {\bibfnamefont {J.}~\bibnamefont {{McMahon}}}, \bibinfo {author} {\bibfnamefont {K.}~\bibnamefont {{Moodley}}}, \bibinfo {author} {\bibfnamefont {S.}~\bibnamefont {{Naess}}}, \bibinfo {author} {\bibfnamefont {T.}~\bibnamefont {{Namikawa}}}, \bibinfo {author} {\bibfnamefont {F.}~\bibnamefont {{Nati}}}, \bibinfo {author} {\bibfnamefont {L.}~\bibnamefont {{Newburgh}}}, \bibinfo {author} {\bibfnamefont {M.~D.}\ \bibnamefont {{Niemack}}}, \bibinfo {author} {\bibfnamefont {L.~A.}\ \bibnamefont {{Page}}}, \bibinfo {author} {\bibfnamefont {B.}~\bibnamefont {{Partridge}}}, \bibinfo {author} {\bibfnamefont {F.~J.}\ \bibnamefont {{Qu}}}, \bibinfo {author} {\bibfnamefont {N.~C.}\ \bibnamefont {{Robertson}}}, \bibinfo {author} {\bibfnamefont
  {M.}~\bibnamefont {{Salatino}}}, \bibinfo {author} {\bibfnamefont {E.}~\bibnamefont {{Schaan}}}, \bibinfo {author} {\bibfnamefont {A.}~\bibnamefont {{Schillaci}}}, \bibinfo {author} {\bibfnamefont {B.~L.}\ \bibnamefont {{Schmitt}}}, \bibinfo {author} {\bibfnamefont {N.}~\bibnamefont {{Sehgal}}}, \bibinfo {author} {\bibfnamefont {B.~D.}\ \bibnamefont {{Sherwin}}}, \bibinfo {author} {\bibfnamefont {S.~M.}\ \bibnamefont {{Simon}}}, \bibinfo {author} {\bibfnamefont {D.~N.}\ \bibnamefont {{Spergel}}}, \bibinfo {author} {\bibfnamefont {S.}~\bibnamefont {{Staggs}}}, \bibinfo {author} {\bibfnamefont {E.~R.}\ \bibnamefont {{Storer}}}, \bibinfo {author} {\bibfnamefont {J.~N.}\ \bibnamefont {{Ullom}}}, \bibinfo {author} {\bibfnamefont {L.~R.}\ \bibnamefont {{Vale}}}, \bibinfo {author} {\bibfnamefont {A.}~\bibnamefont {{van Engelen}}}, \bibinfo {author} {\bibfnamefont {E.~M.}\ \bibnamefont {{Vavagiakis}}}, \bibinfo {author} {\bibfnamefont {E.~J.}\ \bibnamefont {{Wollack}}},\ and\ \bibinfo {author} {\bibfnamefont
  {Z.}~\bibnamefont {{Xu}}},\ }\bibfield  {title} {\bibinfo {title} {{The Atacama Cosmology Telescope: Weighing Distant Clusters with the Most Ancient Light}},\ }\href {https://doi.org/10.3847/2041-8213/abbccb} {\bibfield  {journal} {\bibinfo  {journal} {\apjl}\ }\textbf {\bibinfo {volume} {903}},\ \bibinfo {eid} {L13} (\bibinfo {year} {2020})},\ \Eprint {https://arxiv.org/abs/2009.07772} {arXiv:2009.07772 [astro-ph.CO]} \BibitemShut {NoStop}%
\bibitem [{\citenamefont {{Sunseri}}\ \emph {et~al.}(2025)\citenamefont {{Sunseri}}, \citenamefont {{Amon}}, \citenamefont {{Dunkley}}, \citenamefont {{Battaglia}}, \citenamefont {{Ferraro}}, \citenamefont {{Hadzhiyska}}, \citenamefont {{Ried Guachalla}},\ and\ \citenamefont {{Schaan}}}]{Sunseri:2025hhj}%
  \BibitemOpen
  \bibfield  {author} {\bibinfo {author} {\bibfnamefont {J.}~\bibnamefont {{Sunseri}}}, \bibinfo {author} {\bibfnamefont {A.}~\bibnamefont {{Amon}}}, \bibinfo {author} {\bibfnamefont {J.}~\bibnamefont {{Dunkley}}}, \bibinfo {author} {\bibfnamefont {N.}~\bibnamefont {{Battaglia}}}, \bibinfo {author} {\bibfnamefont {S.}~\bibnamefont {{Ferraro}}}, \bibinfo {author} {\bibfnamefont {B.}~\bibnamefont {{Hadzhiyska}}}, \bibinfo {author} {\bibfnamefont {B.}~\bibnamefont {{Ried Guachalla}}},\ and\ \bibinfo {author} {\bibfnamefont {E.}~\bibnamefont {{Schaan}}},\ }\bibfield  {title} {\bibinfo {title} {{Disentangling the Halo: Joint Model for Measurements of the Kinetic Sunyaev-Zeldovich Effect and Galaxy-Galaxy Lensing}},\ }\href {https://doi.org/10.48550/arXiv.2505.20413} {\bibfield  {journal} {\bibinfo  {journal} {arXiv e-prints}\ ,\ \bibinfo {eid} {arXiv:2505.20413}} (\bibinfo {year} {2025})},\ \Eprint {https://arxiv.org/abs/2505.20413} {arXiv:2505.20413 [astro-ph.CO]} \BibitemShut {NoStop}%
\bibitem [{\citenamefont {{McCarthy}}\ \emph {et~al.}(2025)\citenamefont {{McCarthy}}, \citenamefont {{Amon}}, \citenamefont {{Schaye}}, \citenamefont {{Schaan}}, \citenamefont {{Angulo}}, \citenamefont {{Salcido}}, \citenamefont {{Schaller}}, \citenamefont {{Bigwood}}, \citenamefont {{Elbers}}, \citenamefont {{Kugel}}, \citenamefont {{Helly}}, \citenamefont {{Forouhar Moreno}}, \citenamefont {{Frenk}}, \citenamefont {{McGibbon}}, \citenamefont {{Ondaro-Mallea}},\ and\ \citenamefont {{van Daalen}}}]{2025MNRAS.540..143M}%
  \BibitemOpen
  \bibfield  {author} {\bibinfo {author} {\bibfnamefont {I.~G.}\ \bibnamefont {{McCarthy}}}, \bibinfo {author} {\bibfnamefont {A.}~\bibnamefont {{Amon}}}, \bibinfo {author} {\bibfnamefont {J.}~\bibnamefont {{Schaye}}}, \bibinfo {author} {\bibfnamefont {E.}~\bibnamefont {{Schaan}}}, \bibinfo {author} {\bibfnamefont {R.~E.}\ \bibnamefont {{Angulo}}}, \bibinfo {author} {\bibfnamefont {J.}~\bibnamefont {{Salcido}}}, \bibinfo {author} {\bibfnamefont {M.}~\bibnamefont {{Schaller}}}, \bibinfo {author} {\bibfnamefont {L.}~\bibnamefont {{Bigwood}}}, \bibinfo {author} {\bibfnamefont {W.}~\bibnamefont {{Elbers}}}, \bibinfo {author} {\bibfnamefont {R.}~\bibnamefont {{Kugel}}}, \bibinfo {author} {\bibfnamefont {J.~C.}\ \bibnamefont {{Helly}}}, \bibinfo {author} {\bibfnamefont {V.~J.}\ \bibnamefont {{Forouhar Moreno}}}, \bibinfo {author} {\bibfnamefont {C.~S.}\ \bibnamefont {{Frenk}}}, \bibinfo {author} {\bibfnamefont {R.~J.}\ \bibnamefont {{McGibbon}}}, \bibinfo {author} {\bibfnamefont {L.}~\bibnamefont
  {{Ondaro-Mallea}}},\ and\ \bibinfo {author} {\bibfnamefont {M.~P.}\ \bibnamefont {{van Daalen}}},\ }\bibfield  {title} {\bibinfo {title} {{FLAMINGO: combining kinetic SZ effect and galaxy{\textendash}galaxy lensing measurements to gauge the impact of feedback on large-scale structure}},\ }\href {https://doi.org/10.1093/mnras/staf731} {\bibfield  {journal} {\bibinfo  {journal} {\mnras}\ }\textbf {\bibinfo {volume} {540}},\ \bibinfo {pages} {143} (\bibinfo {year} {2025})},\ \Eprint {https://arxiv.org/abs/2410.19905} {arXiv:2410.19905 [astro-ph.CO]} \BibitemShut {NoStop}%
\bibitem [{\citenamefont {{Bigwood}}\ \emph {et~al.}(2024{\natexlab{b}})\citenamefont {{Bigwood}}, \citenamefont {{Amon}}, \citenamefont {{Schneider}}, \citenamefont {{Salcido}}, \citenamefont {{McCarthy}}, \citenamefont {{Preston}}, \citenamefont {{Sanchez}}, \citenamefont {{Sijacki}}, \citenamefont {{Schaan}}, \citenamefont {{Ferraro}}, \citenamefont {{Battaglia}}, \citenamefont {{Chen}}, \citenamefont {{Dodelson}}, \citenamefont {{Roodman}}, \citenamefont {{Pieres}}, \citenamefont {{Fert{\'e}}}, \citenamefont {{Alarcon}}, \citenamefont {{Drlica-Wagner}}, \citenamefont {{Choi}}, \citenamefont {{Navarro-Alsina}}, \citenamefont {{Campos}}, \citenamefont {{Ross}}, \citenamefont {{Carnero Rosell}}, \citenamefont {{Yin}}, \citenamefont {{Yanny}}, \citenamefont {{S{\'a}nchez}}, \citenamefont {{Chang}}, \citenamefont {{Davis}}, \citenamefont {{Doux}}, \citenamefont {{Gruen}}, \citenamefont {{Rykoff}}, \citenamefont {{Huff}}, \citenamefont {{Sheldon}}, \citenamefont {{Tarsitano}}, \citenamefont
  {{Andrade-Oliveira}}, \citenamefont {{Bernstein}}, \citenamefont {{Giannini}}, \citenamefont {{Diehl}}, \citenamefont {{Huang}}, \citenamefont {{Harrison}}, \citenamefont {{Sevilla-Noarbe}}, \citenamefont {{Tutusaus}}, \citenamefont {{Elvin-Poole}}, \citenamefont {{McCullough}}, \citenamefont {{Zuntz}}, \citenamefont {{Blazek}}, \citenamefont {{DeRose}}, \citenamefont {{Cordero}}, \citenamefont {{Prat}}, \citenamefont {{Myles}}, \citenamefont {{Eckert}}, \citenamefont {{Bechtol}}, \citenamefont {{Herner}}, \citenamefont {{Secco}}, \citenamefont {{Gatti}}, \citenamefont {{Raveri}}, \citenamefont {{Kind}}, \citenamefont {{Becker}}, \citenamefont {{Troxel}}, \citenamefont {{Jarvis}}, \citenamefont {{MacCrann}}, \citenamefont {{Friedrich}}, \citenamefont {{Alves}}, \citenamefont {{Leget}}, \citenamefont {{Chen}}, \citenamefont {{Rollins}}, \citenamefont {{Wechsler}}, \citenamefont {{Gruendl}}, \citenamefont {{Cawthon}}, \citenamefont {{Allam}}, \citenamefont {{Bridle}}, \citenamefont {{Pandey}}, \citenamefont
  {{Everett}}, \citenamefont {{Shin}}, \citenamefont {{Hartley}}, \citenamefont {{Fang}}, \citenamefont {{Zhang}}, \citenamefont {{Aguena}}, \citenamefont {{Annis}}, \citenamefont {{Bacon}}, \citenamefont {{Bertin}}, \citenamefont {{Bocquet}}, \citenamefont {{Brooks}}, \citenamefont {{Carretero}}, \citenamefont {{Castander}}, \citenamefont {{da Costa}}, \citenamefont {{Pereira}}, \citenamefont {{De Vicente}}, \citenamefont {{Desai}}, \citenamefont {{Doel}}, \citenamefont {{Ferrero}}, \citenamefont {{Flaugher}}, \citenamefont {{Frieman}}, \citenamefont {{Garc{\'\i}a-Bellido}}, \citenamefont {{Gaztanaga}}, \citenamefont {{Gutierrez}}, \citenamefont {{Hinton}}, \citenamefont {{Hollowood}}, \citenamefont {{Honscheid}}, \citenamefont {{Huterer}}, \citenamefont {{James}}, \citenamefont {{Kuehn}}, \citenamefont {{Lahav}}, \citenamefont {{Lee}}, \citenamefont {{Marshall}}, \citenamefont {{Mena-Fern{\'a}ndez}}, \citenamefont {{Miquel}}, \citenamefont {{Muir}}, \citenamefont {{Paterno}}, \citenamefont {{Plazas
  Malag{\'o}n}}, \citenamefont {{Porredon}}, \citenamefont {{Romer}}, \citenamefont {{Samuroff}}, \citenamefont {{Sanchez}}, \citenamefont {{Sanchez Cid}}, \citenamefont {{Smith}}, \citenamefont {{Soares-Santos}}, \citenamefont {{Suchyta}}, \citenamefont {{Swanson}}, \citenamefont {{Tarle}}, \citenamefont {{To}}, \citenamefont {{Weaverdyck}}, \citenamefont {{Weller}}, \citenamefont {{Wiseman}},\ and\ \citenamefont {{Yamamoto}}}]{2024MNRAS.534..655B}%
  \BibitemOpen
  \bibfield  {author} {\bibinfo {author} {\bibfnamefont {L.}~\bibnamefont {{Bigwood}}}, \bibinfo {author} {\bibfnamefont {A.}~\bibnamefont {{Amon}}}, \bibinfo {author} {\bibfnamefont {A.}~\bibnamefont {{Schneider}}}, \bibinfo {author} {\bibfnamefont {J.}~\bibnamefont {{Salcido}}}, \bibinfo {author} {\bibfnamefont {I.~G.}\ \bibnamefont {{McCarthy}}}, \bibinfo {author} {\bibfnamefont {C.}~\bibnamefont {{Preston}}}, \bibinfo {author} {\bibfnamefont {D.}~\bibnamefont {{Sanchez}}}, \bibinfo {author} {\bibfnamefont {D.}~\bibnamefont {{Sijacki}}}, \bibinfo {author} {\bibfnamefont {E.}~\bibnamefont {{Schaan}}}, \bibinfo {author} {\bibfnamefont {S.}~\bibnamefont {{Ferraro}}}, \bibinfo {author} {\bibfnamefont {N.}~\bibnamefont {{Battaglia}}}, \bibinfo {author} {\bibfnamefont {A.}~\bibnamefont {{Chen}}}, \bibinfo {author} {\bibfnamefont {S.}~\bibnamefont {{Dodelson}}}, \bibinfo {author} {\bibfnamefont {A.}~\bibnamefont {{Roodman}}}, \bibinfo {author} {\bibfnamefont {A.}~\bibnamefont {{Pieres}}}, \bibinfo {author}
  {\bibfnamefont {A.}~\bibnamefont {{Fert{\'e}}}}, \bibinfo {author} {\bibfnamefont {A.}~\bibnamefont {{Alarcon}}}, \bibinfo {author} {\bibfnamefont {A.}~\bibnamefont {{Drlica-Wagner}}}, \bibinfo {author} {\bibfnamefont {A.}~\bibnamefont {{Choi}}}, \bibinfo {author} {\bibfnamefont {A.}~\bibnamefont {{Navarro-Alsina}}}, \bibinfo {author} {\bibfnamefont {A.}~\bibnamefont {{Campos}}}, \bibinfo {author} {\bibfnamefont {A.~J.}\ \bibnamefont {{Ross}}}, \bibinfo {author} {\bibfnamefont {A.}~\bibnamefont {{Carnero Rosell}}}, \bibinfo {author} {\bibfnamefont {B.}~\bibnamefont {{Yin}}}, \bibinfo {author} {\bibfnamefont {B.}~\bibnamefont {{Yanny}}}, \bibinfo {author} {\bibfnamefont {C.}~\bibnamefont {{S{\'a}nchez}}}, \bibinfo {author} {\bibfnamefont {C.}~\bibnamefont {{Chang}}}, \bibinfo {author} {\bibfnamefont {C.}~\bibnamefont {{Davis}}}, \bibinfo {author} {\bibfnamefont {C.}~\bibnamefont {{Doux}}}, \bibinfo {author} {\bibfnamefont {D.}~\bibnamefont {{Gruen}}}, \bibinfo {author} {\bibfnamefont {E.~S.}\ \bibnamefont
  {{Rykoff}}}, \bibinfo {author} {\bibfnamefont {E.~M.}\ \bibnamefont {{Huff}}}, \bibinfo {author} {\bibfnamefont {E.}~\bibnamefont {{Sheldon}}}, \bibinfo {author} {\bibfnamefont {F.}~\bibnamefont {{Tarsitano}}}, \bibinfo {author} {\bibfnamefont {F.}~\bibnamefont {{Andrade-Oliveira}}}, \bibinfo {author} {\bibfnamefont {G.~M.}\ \bibnamefont {{Bernstein}}}, \bibinfo {author} {\bibfnamefont {G.}~\bibnamefont {{Giannini}}}, \bibinfo {author} {\bibfnamefont {H.~T.}\ \bibnamefont {{Diehl}}}, \bibinfo {author} {\bibfnamefont {H.}~\bibnamefont {{Huang}}}, \bibinfo {author} {\bibfnamefont {I.}~\bibnamefont {{Harrison}}}, \bibinfo {author} {\bibfnamefont {I.}~\bibnamefont {{Sevilla-Noarbe}}}, \bibinfo {author} {\bibfnamefont {I.}~\bibnamefont {{Tutusaus}}}, \bibinfo {author} {\bibfnamefont {J.}~\bibnamefont {{Elvin-Poole}}}, \bibinfo {author} {\bibfnamefont {J.}~\bibnamefont {{McCullough}}}, \bibinfo {author} {\bibfnamefont {J.}~\bibnamefont {{Zuntz}}}, \bibinfo {author} {\bibfnamefont {J.}~\bibnamefont {{Blazek}}},
  \bibinfo {author} {\bibfnamefont {J.}~\bibnamefont {{DeRose}}}, \bibinfo {author} {\bibfnamefont {J.}~\bibnamefont {{Cordero}}}, \bibinfo {author} {\bibfnamefont {J.}~\bibnamefont {{Prat}}}, \bibinfo {author} {\bibfnamefont {J.}~\bibnamefont {{Myles}}}, \bibinfo {author} {\bibfnamefont {K.}~\bibnamefont {{Eckert}}}, \bibinfo {author} {\bibfnamefont {K.}~\bibnamefont {{Bechtol}}}, \bibinfo {author} {\bibfnamefont {K.}~\bibnamefont {{Herner}}}, \bibinfo {author} {\bibfnamefont {L.~F.}\ \bibnamefont {{Secco}}}, \bibinfo {author} {\bibfnamefont {M.}~\bibnamefont {{Gatti}}}, \bibinfo {author} {\bibfnamefont {M.}~\bibnamefont {{Raveri}}}, \bibinfo {author} {\bibfnamefont {M.~C.}\ \bibnamefont {{Kind}}}, \bibinfo {author} {\bibfnamefont {M.~R.}\ \bibnamefont {{Becker}}}, \bibinfo {author} {\bibfnamefont {M.~A.}\ \bibnamefont {{Troxel}}}, \bibinfo {author} {\bibfnamefont {M.}~\bibnamefont {{Jarvis}}}, \bibinfo {author} {\bibfnamefont {N.}~\bibnamefont {{MacCrann}}}, \bibinfo {author} {\bibfnamefont
  {O.}~\bibnamefont {{Friedrich}}}, \bibinfo {author} {\bibfnamefont {O.}~\bibnamefont {{Alves}}}, \bibinfo {author} {\bibfnamefont {P.~F.}\ \bibnamefont {{Leget}}}, \bibinfo {author} {\bibfnamefont {R.}~\bibnamefont {{Chen}}}, \bibinfo {author} {\bibfnamefont {R.~P.}\ \bibnamefont {{Rollins}}}, \bibinfo {author} {\bibfnamefont {R.~H.}\ \bibnamefont {{Wechsler}}}, \bibinfo {author} {\bibfnamefont {R.~A.}\ \bibnamefont {{Gruendl}}}, \bibinfo {author} {\bibfnamefont {R.}~\bibnamefont {{Cawthon}}}, \bibinfo {author} {\bibfnamefont {S.}~\bibnamefont {{Allam}}}, \bibinfo {author} {\bibfnamefont {S.~L.}\ \bibnamefont {{Bridle}}}, \bibinfo {author} {\bibfnamefont {S.}~\bibnamefont {{Pandey}}}, \bibinfo {author} {\bibfnamefont {S.}~\bibnamefont {{Everett}}}, \bibinfo {author} {\bibfnamefont {T.}~\bibnamefont {{Shin}}}, \bibinfo {author} {\bibfnamefont {W.~G.}\ \bibnamefont {{Hartley}}}, \bibinfo {author} {\bibfnamefont {X.}~\bibnamefont {{Fang}}}, \bibinfo {author} {\bibfnamefont {Y.}~\bibnamefont {{Zhang}}},
  \bibinfo {author} {\bibfnamefont {M.}~\bibnamefont {{Aguena}}}, \bibinfo {author} {\bibfnamefont {J.}~\bibnamefont {{Annis}}}, \bibinfo {author} {\bibfnamefont {D.}~\bibnamefont {{Bacon}}}, \bibinfo {author} {\bibfnamefont {E.}~\bibnamefont {{Bertin}}}, \bibinfo {author} {\bibfnamefont {S.}~\bibnamefont {{Bocquet}}}, \bibinfo {author} {\bibfnamefont {D.}~\bibnamefont {{Brooks}}}, \bibinfo {author} {\bibfnamefont {J.}~\bibnamefont {{Carretero}}}, \bibinfo {author} {\bibfnamefont {F.~J.}\ \bibnamefont {{Castander}}}, \bibinfo {author} {\bibfnamefont {L.~N.}\ \bibnamefont {{da Costa}}}, \bibinfo {author} {\bibfnamefont {M.~E.~S.}\ \bibnamefont {{Pereira}}}, \bibinfo {author} {\bibfnamefont {J.}~\bibnamefont {{De Vicente}}}, \bibinfo {author} {\bibfnamefont {S.}~\bibnamefont {{Desai}}}, \bibinfo {author} {\bibfnamefont {P.}~\bibnamefont {{Doel}}}, \bibinfo {author} {\bibfnamefont {I.}~\bibnamefont {{Ferrero}}}, \bibinfo {author} {\bibfnamefont {B.}~\bibnamefont {{Flaugher}}}, \bibinfo {author} {\bibfnamefont
  {J.}~\bibnamefont {{Frieman}}}, \bibinfo {author} {\bibfnamefont {J.}~\bibnamefont {{Garc{\'\i}a-Bellido}}}, \bibinfo {author} {\bibfnamefont {E.}~\bibnamefont {{Gaztanaga}}}, \bibinfo {author} {\bibfnamefont {G.}~\bibnamefont {{Gutierrez}}}, \bibinfo {author} {\bibfnamefont {S.~R.}\ \bibnamefont {{Hinton}}}, \bibinfo {author} {\bibfnamefont {D.~L.}\ \bibnamefont {{Hollowood}}}, \bibinfo {author} {\bibfnamefont {K.}~\bibnamefont {{Honscheid}}}, \bibinfo {author} {\bibfnamefont {D.}~\bibnamefont {{Huterer}}}, \bibinfo {author} {\bibfnamefont {D.~J.}\ \bibnamefont {{James}}}, \bibinfo {author} {\bibfnamefont {K.}~\bibnamefont {{Kuehn}}}, \bibinfo {author} {\bibfnamefont {O.}~\bibnamefont {{Lahav}}}, \bibinfo {author} {\bibfnamefont {S.}~\bibnamefont {{Lee}}}, \bibinfo {author} {\bibfnamefont {J.~L.}\ \bibnamefont {{Marshall}}}, \bibinfo {author} {\bibfnamefont {J.}~\bibnamefont {{Mena-Fern{\'a}ndez}}}, \bibinfo {author} {\bibfnamefont {R.}~\bibnamefont {{Miquel}}}, \bibinfo {author} {\bibfnamefont
  {J.}~\bibnamefont {{Muir}}}, \bibinfo {author} {\bibfnamefont {M.}~\bibnamefont {{Paterno}}}, \bibinfo {author} {\bibfnamefont {A.~A.}\ \bibnamefont {{Plazas Malag{\'o}n}}}, \bibinfo {author} {\bibfnamefont {A.}~\bibnamefont {{Porredon}}}, \bibinfo {author} {\bibfnamefont {A.~K.}\ \bibnamefont {{Romer}}}, \bibinfo {author} {\bibfnamefont {S.}~\bibnamefont {{Samuroff}}}, \bibinfo {author} {\bibfnamefont {E.}~\bibnamefont {{Sanchez}}}, \bibinfo {author} {\bibfnamefont {D.}~\bibnamefont {{Sanchez Cid}}}, \bibinfo {author} {\bibfnamefont {M.}~\bibnamefont {{Smith}}}, \bibinfo {author} {\bibfnamefont {M.}~\bibnamefont {{Soares-Santos}}}, \bibinfo {author} {\bibfnamefont {E.}~\bibnamefont {{Suchyta}}}, \bibinfo {author} {\bibfnamefont {M.~E.~C.}\ \bibnamefont {{Swanson}}}, \bibinfo {author} {\bibfnamefont {G.}~\bibnamefont {{Tarle}}}, \bibinfo {author} {\bibfnamefont {C.}~\bibnamefont {{To}}}, \bibinfo {author} {\bibfnamefont {N.}~\bibnamefont {{Weaverdyck}}}, \bibinfo {author} {\bibfnamefont {J.}~\bibnamefont
  {{Weller}}}, \bibinfo {author} {\bibfnamefont {P.}~\bibnamefont {{Wiseman}}},\ and\ \bibinfo {author} {\bibfnamefont {M.}~\bibnamefont {{Yamamoto}}},\ }\bibfield  {title} {\bibinfo {title} {{Weak lensing combined with the kinetic Sunyaev-Zel'dovich effect: a study of baryonic feedback}},\ }\href {https://doi.org/10.1093/mnras/stae2100} {\bibfield  {journal} {\bibinfo  {journal} {\mnras}\ }\textbf {\bibinfo {volume} {534}},\ \bibinfo {pages} {655} (\bibinfo {year} {2024}{\natexlab{b}})},\ \Eprint {https://arxiv.org/abs/2404.06098} {arXiv:2404.06098 [astro-ph.CO]} \BibitemShut {NoStop}%
\bibitem [{\citenamefont {{DESI Collaboration}}\ \emph {et~al.}(2022)\citenamefont {{DESI Collaboration}}, \citenamefont {{Abareshi}}, \citenamefont {{Aguilar}}, \citenamefont {{Ahlen}}, \citenamefont {{Alam}}, \citenamefont {{Alexander}}, \citenamefont {{Alfarsy}}, \citenamefont {{Allen}}, \citenamefont {{Allende Prieto}}, \citenamefont {{Alves}}, \citenamefont {{Ameel}}, \citenamefont {{Armengaud}}, \citenamefont {{Asorey}}, \citenamefont {{Aviles}}, \citenamefont {{Bailey}}, \citenamefont {{Balaguera-Antol{\'\i}nez}}, \citenamefont {{Ballester}}, \citenamefont {{Baltay}}, \citenamefont {{Bault}}, \citenamefont {{Beltran}}, \citenamefont {{Benavides}}, \citenamefont {{BenZvi}}, \citenamefont {{Berti}}, \citenamefont {{Besuner}}, \citenamefont {{Beutler}}, \citenamefont {{Bianchi}}, \citenamefont {{Blake}}, \citenamefont {{Blanc}}, \citenamefont {{Blum}}, \citenamefont {{Bolton}}, \citenamefont {{Bose}}, \citenamefont {{Bramall}}, \citenamefont {{Brieden}}, \citenamefont {{Brodzeller}}, \citenamefont
  {{Brooks}}, \citenamefont {{Brownewell}}, \citenamefont {{Buckley-Geer}}, \citenamefont {{Cahn}}, \citenamefont {{Cai}}, \citenamefont {{Canning}}, \citenamefont {{Capasso}}, \citenamefont {{Carnero Rosell}}, \citenamefont {{Carton}}, \citenamefont {{Casas}}, \citenamefont {{Castander}}, \citenamefont {{Cervantes-Cota}}, \citenamefont {{Chabanier}}, \citenamefont {{Chaussidon}}, \citenamefont {{Chuang}}, \citenamefont {{Circosta}}, \citenamefont {{Cole}}, \citenamefont {{Cooper}}, \citenamefont {{da Costa}}, \citenamefont {{Cousinou}}, \citenamefont {{Cuceu}}, \citenamefont {{Davis}}, \citenamefont {{Dawson}}, \citenamefont {{de la Cruz-Noriega}}, \citenamefont {{de la Macorra}}, \citenamefont {{de Mattia}}, \citenamefont {{Della Costa}}, \citenamefont {{Demmer}}, \citenamefont {{Derwent}}, \citenamefont {{Dey}}, \citenamefont {{Dey}}, \citenamefont {{Dhungana}}, \citenamefont {{Ding}}, \citenamefont {{Dobson}}, \citenamefont {{Doel}}, \citenamefont {{Donald-McCann}}, \citenamefont {{Donaldson}},
  \citenamefont {{Douglass}}, \citenamefont {{Duan}}, \citenamefont {{Dunlop}}, \citenamefont {{Edelstein}}, \citenamefont {{Eftekharzadeh}}, \citenamefont {{Eisenstein}}, \citenamefont {{Enriquez-Vargas}}, \citenamefont {{Escoffier}}, \citenamefont {{Evatt}}, \citenamefont {{Fagrelius}}, \citenamefont {{Fan}}, \citenamefont {{Fanning}}, \citenamefont {{Fawcett}}, \citenamefont {{Ferraro}}, \citenamefont {{Ereza}}, \citenamefont {{Flaugher}}, \citenamefont {{Font-Ribera}}, \citenamefont {{Forero-Romero}}, \citenamefont {{Frenk}}, \citenamefont {{Fromenteau}}, \citenamefont {{G{\"a}nsicke}}, \citenamefont {{Garcia-Quintero}}, \citenamefont {{Garrison}}, \citenamefont {{Gazta{\~n}aga}}, \citenamefont {{Gerardi}}, \citenamefont {{Gil-Mar{\'\i}n}}, \citenamefont {{Gontcho a Gontcho}}, \citenamefont {{Gonzalez-Morales}}, \citenamefont {{Gonzalez-de-Rivera}}, \citenamefont {{Gonzalez-Perez}}, \citenamefont {{Gordon}}, \citenamefont {{Graur}}, \citenamefont {{Green}}, \citenamefont {{Grove}}, \citenamefont
  {{Gruen}}, \citenamefont {{Gutierrez}}, \citenamefont {{Guy}}, \citenamefont {{Hahn}}, \citenamefont {{Harris}}, \citenamefont {{Herrera}}, \citenamefont {{Herrera-Alcantar}}, \citenamefont {{Honscheid}}, \citenamefont {{Howlett}}, \citenamefont {{Huterer}}, \citenamefont {{Ir{\v{s}}i{\v{c}}}}, \citenamefont {{Ishak}}, \citenamefont {{Jelinsky}}, \citenamefont {{Jiang}}, \citenamefont {{Jimenez}}, \citenamefont {{Jing}}, \citenamefont {{Joyce}}, \citenamefont {{Jullo}}, \citenamefont {{Juneau}}, \citenamefont {{Kara{\c{c}}ayl{\i}}}, \citenamefont {{Karamanis}}, \citenamefont {{Karcher}}, \citenamefont {{Karim}}, \citenamefont {{Kehoe}}, \citenamefont {{Kent}}, \citenamefont {{Kirkby}}, \citenamefont {{Kisner}}, \citenamefont {{Kitaura}}, \citenamefont {{Koposov}}, \citenamefont {{Kov{\'a}cs}}, \citenamefont {{Kremin}}, \citenamefont {{Krolewski}}, \citenamefont {{L'Huillier}}, \citenamefont {{Lahav}}, \citenamefont {{Lambert}}, \citenamefont {{Lamman}}, \citenamefont {{Lan}}, \citenamefont {{Landriau}},
  \citenamefont {{Lane}}, \citenamefont {{Lang}}, \citenamefont {{Lange}}, \citenamefont {{Lasker}}, \citenamefont {{Le Guillou}}, \citenamefont {{Leauthaud}}, \citenamefont {{Le Van Suu}}, \citenamefont {{Levi}}, \citenamefont {{Li}}, \citenamefont {{Magneville}}, \citenamefont {{Manera}}, \citenamefont {{Manser}}, \citenamefont {{Marshall}}, \citenamefont {{Martini}}, \citenamefont {{McCollam}}, \citenamefont {{McDonald}}, \citenamefont {{Meisner}}, \citenamefont {{Mena-Fern{\'a}ndez}}, \citenamefont {{Meneses-Rizo}}, \citenamefont {{Mezcua}}, \citenamefont {{Miller}}, \citenamefont {{Miquel}}, \citenamefont {{Montero-Camacho}}, \citenamefont {{Moon}}, \citenamefont {{Moustakas}}, \citenamefont {{Mueller}}, \citenamefont {{Mu{\~n}oz-Guti{\'e}rrez}}, \citenamefont {{Myers}}, \citenamefont {{Nadathur}}, \citenamefont {{Najita}}, \citenamefont {{Napolitano}}, \citenamefont {{Neilsen}}, \citenamefont {{Newman}}, \citenamefont {{Nie}}, \citenamefont {{Ning}}, \citenamefont {{Niz}}, \citenamefont {{Norberg}},
  \citenamefont {{Noriega}}, \citenamefont {{O'Brien}}, \citenamefont {{Obuljen}}, \citenamefont {{Palanque-Delabrouille}}, \citenamefont {{Palmese}}, \citenamefont {{Zhiwei}}, \citenamefont {{Pappalardo}}, \citenamefont {{PENG}}, \citenamefont {{Percival}}, \citenamefont {{Perruchot}}, \citenamefont {{Pogge}}, \citenamefont {{Poppett}}, \citenamefont {{Porredon}}, \citenamefont {{Prada}}, \citenamefont {{Prochaska}}, \citenamefont {{Pucha}}, \citenamefont {{P{\'e}rez-Fern{\'a}ndez}}, \citenamefont {{P{\'e}rez-R{\`a}fols}}, \citenamefont {{Rabinowitz}}, \citenamefont {{Raichoor}}, \citenamefont {{Ramirez-Solano}}, \citenamefont {{Ram{\'\i}rez-P{\'e}rez}}, \citenamefont {{Ravoux}}, \citenamefont {{Reil}}, \citenamefont {{Rezaie}}, \citenamefont {{Rocher}}, \citenamefont {{Rockosi}}, \citenamefont {{Roe}}, \citenamefont {{Roodman}}, \citenamefont {{Ross}}, \citenamefont {{Rossi}}, \citenamefont {{Ruggeri}}, \citenamefont {{Ruhlmann-Kleider}}, \citenamefont {{Sabiu}}, \citenamefont {{Safonova}}, \citenamefont
  {{Said}}, \citenamefont {{Saintonge}}, \citenamefont {{Salas Catonga}}, \citenamefont {{Samushia}}, \citenamefont {{Sanchez}}, \citenamefont {{Saulder}}, \citenamefont {{Schaan}}, \citenamefont {{Schlafly}}, \citenamefont {{Schlegel}}, \citenamefont {{Schmoll}}, \citenamefont {{Scholte}}, \citenamefont {{Schubnell}}, \citenamefont {{Secroun}}, \citenamefont {{Seo}}, \citenamefont {{Serrano}}, \citenamefont {{Sharples}}, \citenamefont {{Sholl}}, \citenamefont {{Silber}}, \citenamefont {{Silva}}, \citenamefont {{Sirk}}, \citenamefont {{Siudek}}, \citenamefont {{Smith}}, \citenamefont {{Sprayberry}}, \citenamefont {{Staten}}, \citenamefont {{Stupak}}, \citenamefont {{Tan}}, \citenamefont {{Tarl{\'e}}}, \citenamefont {{Tie}}, \citenamefont {{Tojeiro}}, \citenamefont {{Ure{\~n}a-L{\'o}pez}}, \citenamefont {{Valdes}}, \citenamefont {{Valenzuela}}, \citenamefont {{Valluri}}, \citenamefont {{Vargas-Maga{\~n}a}}, \citenamefont {{Verde}}, \citenamefont {{Walther}}, \citenamefont {{Wang}}, \citenamefont {{Wang}},
  \citenamefont {{Weaver}}, \citenamefont {{Weaverdyck}}, \citenamefont {{Wechsler}}, \citenamefont {{Wilson}}, \citenamefont {{Yang}}, \citenamefont {{Yu}}, \citenamefont {{Yuan}}, \citenamefont {{Y{\`e}che}}, \citenamefont {{Zhang}}, \citenamefont {{Zhang}}, \citenamefont {{Zhao}}, \citenamefont {{Zhou}}, \citenamefont {{Zhou}}, \citenamefont {{Zou}}, \citenamefont {{Zou}}, \citenamefont {{Zou}}, \citenamefont {{Zu}},\ and\ \citenamefont {{DESI Collaboration}}}]{2022AJ....164..207D}%
  \BibitemOpen
  \bibfield  {author} {\bibinfo {author} {\bibnamefont {{DESI Collaboration}}}, \bibinfo {author} {\bibfnamefont {B.}~\bibnamefont {{Abareshi}}}, \bibinfo {author} {\bibfnamefont {J.}~\bibnamefont {{Aguilar}}}, \bibinfo {author} {\bibfnamefont {S.}~\bibnamefont {{Ahlen}}}, \bibinfo {author} {\bibfnamefont {S.}~\bibnamefont {{Alam}}}, \bibinfo {author} {\bibfnamefont {D.~M.}\ \bibnamefont {{Alexander}}}, \bibinfo {author} {\bibfnamefont {R.}~\bibnamefont {{Alfarsy}}}, \bibinfo {author} {\bibfnamefont {L.}~\bibnamefont {{Allen}}}, \bibinfo {author} {\bibfnamefont {C.}~\bibnamefont {{Allende Prieto}}}, \bibinfo {author} {\bibfnamefont {O.}~\bibnamefont {{Alves}}}, \bibinfo {author} {\bibfnamefont {J.}~\bibnamefont {{Ameel}}}, \bibinfo {author} {\bibfnamefont {E.}~\bibnamefont {{Armengaud}}}, \bibinfo {author} {\bibfnamefont {J.}~\bibnamefont {{Asorey}}}, \bibinfo {author} {\bibfnamefont {A.}~\bibnamefont {{Aviles}}}, \bibinfo {author} {\bibfnamefont {S.}~\bibnamefont {{Bailey}}}, \bibinfo {author} {\bibfnamefont
  {A.}~\bibnamefont {{Balaguera-Antol{\'\i}nez}}}, \bibinfo {author} {\bibfnamefont {O.}~\bibnamefont {{Ballester}}}, \bibinfo {author} {\bibfnamefont {C.}~\bibnamefont {{Baltay}}}, \bibinfo {author} {\bibfnamefont {A.}~\bibnamefont {{Bault}}}, \bibinfo {author} {\bibfnamefont {S.~F.}\ \bibnamefont {{Beltran}}}, \bibinfo {author} {\bibfnamefont {B.}~\bibnamefont {{Benavides}}}, \bibinfo {author} {\bibfnamefont {S.}~\bibnamefont {{BenZvi}}}, \bibinfo {author} {\bibfnamefont {A.}~\bibnamefont {{Berti}}}, \bibinfo {author} {\bibfnamefont {R.}~\bibnamefont {{Besuner}}}, \bibinfo {author} {\bibfnamefont {F.}~\bibnamefont {{Beutler}}}, \bibinfo {author} {\bibfnamefont {D.}~\bibnamefont {{Bianchi}}}, \bibinfo {author} {\bibfnamefont {C.}~\bibnamefont {{Blake}}}, \bibinfo {author} {\bibfnamefont {P.}~\bibnamefont {{Blanc}}}, \bibinfo {author} {\bibfnamefont {R.}~\bibnamefont {{Blum}}}, \bibinfo {author} {\bibfnamefont {A.}~\bibnamefont {{Bolton}}}, \bibinfo {author} {\bibfnamefont {S.}~\bibnamefont {{Bose}}},
  \bibinfo {author} {\bibfnamefont {D.}~\bibnamefont {{Bramall}}}, \bibinfo {author} {\bibfnamefont {S.}~\bibnamefont {{Brieden}}}, \bibinfo {author} {\bibfnamefont {A.}~\bibnamefont {{Brodzeller}}}, \bibinfo {author} {\bibfnamefont {D.}~\bibnamefont {{Brooks}}}, \bibinfo {author} {\bibfnamefont {C.}~\bibnamefont {{Brownewell}}}, \bibinfo {author} {\bibfnamefont {E.}~\bibnamefont {{Buckley-Geer}}}, \bibinfo {author} {\bibfnamefont {R.~N.}\ \bibnamefont {{Cahn}}}, \bibinfo {author} {\bibfnamefont {Z.}~\bibnamefont {{Cai}}}, \bibinfo {author} {\bibfnamefont {R.}~\bibnamefont {{Canning}}}, \bibinfo {author} {\bibfnamefont {R.}~\bibnamefont {{Capasso}}}, \bibinfo {author} {\bibfnamefont {A.}~\bibnamefont {{Carnero Rosell}}}, \bibinfo {author} {\bibfnamefont {P.}~\bibnamefont {{Carton}}}, \bibinfo {author} {\bibfnamefont {R.}~\bibnamefont {{Casas}}}, \bibinfo {author} {\bibfnamefont {F.~J.}\ \bibnamefont {{Castander}}}, \bibinfo {author} {\bibfnamefont {J.~L.}\ \bibnamefont {{Cervantes-Cota}}}, \bibinfo {author}
  {\bibfnamefont {S.}~\bibnamefont {{Chabanier}}}, \bibinfo {author} {\bibfnamefont {E.}~\bibnamefont {{Chaussidon}}}, \bibinfo {author} {\bibfnamefont {C.}~\bibnamefont {{Chuang}}}, \bibinfo {author} {\bibfnamefont {C.}~\bibnamefont {{Circosta}}}, \bibinfo {author} {\bibfnamefont {S.}~\bibnamefont {{Cole}}}, \bibinfo {author} {\bibfnamefont {A.~P.}\ \bibnamefont {{Cooper}}}, \bibinfo {author} {\bibfnamefont {L.}~\bibnamefont {{da Costa}}}, \bibinfo {author} {\bibfnamefont {M.~C.}\ \bibnamefont {{Cousinou}}}, \bibinfo {author} {\bibfnamefont {A.}~\bibnamefont {{Cuceu}}}, \bibinfo {author} {\bibfnamefont {T.~M.}\ \bibnamefont {{Davis}}}, \bibinfo {author} {\bibfnamefont {K.}~\bibnamefont {{Dawson}}}, \bibinfo {author} {\bibfnamefont {R.}~\bibnamefont {{de la Cruz-Noriega}}}, \bibinfo {author} {\bibfnamefont {A.}~\bibnamefont {{de la Macorra}}}, \bibinfo {author} {\bibfnamefont {A.}~\bibnamefont {{de Mattia}}}, \bibinfo {author} {\bibfnamefont {J.}~\bibnamefont {{Della Costa}}}, \bibinfo {author} {\bibfnamefont
  {P.}~\bibnamefont {{Demmer}}}, \bibinfo {author} {\bibfnamefont {M.}~\bibnamefont {{Derwent}}}, \bibinfo {author} {\bibfnamefont {A.}~\bibnamefont {{Dey}}}, \bibinfo {author} {\bibfnamefont {B.}~\bibnamefont {{Dey}}}, \bibinfo {author} {\bibfnamefont {G.}~\bibnamefont {{Dhungana}}}, \bibinfo {author} {\bibfnamefont {Z.}~\bibnamefont {{Ding}}}, \bibinfo {author} {\bibfnamefont {C.}~\bibnamefont {{Dobson}}}, \bibinfo {author} {\bibfnamefont {P.}~\bibnamefont {{Doel}}}, \bibinfo {author} {\bibfnamefont {J.}~\bibnamefont {{Donald-McCann}}}, \bibinfo {author} {\bibfnamefont {J.}~\bibnamefont {{Donaldson}}}, \bibinfo {author} {\bibfnamefont {K.}~\bibnamefont {{Douglass}}}, \bibinfo {author} {\bibfnamefont {Y.}~\bibnamefont {{Duan}}}, \bibinfo {author} {\bibfnamefont {P.}~\bibnamefont {{Dunlop}}}, \bibinfo {author} {\bibfnamefont {J.}~\bibnamefont {{Edelstein}}}, \bibinfo {author} {\bibfnamefont {S.}~\bibnamefont {{Eftekharzadeh}}}, \bibinfo {author} {\bibfnamefont {D.~J.}\ \bibnamefont {{Eisenstein}}}, \bibinfo
  {author} {\bibfnamefont {M.}~\bibnamefont {{Enriquez-Vargas}}}, \bibinfo {author} {\bibfnamefont {S.}~\bibnamefont {{Escoffier}}}, \bibinfo {author} {\bibfnamefont {M.}~\bibnamefont {{Evatt}}}, \bibinfo {author} {\bibfnamefont {P.}~\bibnamefont {{Fagrelius}}}, \bibinfo {author} {\bibfnamefont {X.}~\bibnamefont {{Fan}}}, \bibinfo {author} {\bibfnamefont {K.}~\bibnamefont {{Fanning}}}, \bibinfo {author} {\bibfnamefont {V.~A.}\ \bibnamefont {{Fawcett}}}, \bibinfo {author} {\bibfnamefont {S.}~\bibnamefont {{Ferraro}}}, \bibinfo {author} {\bibfnamefont {J.}~\bibnamefont {{Ereza}}}, \bibinfo {author} {\bibfnamefont {B.}~\bibnamefont {{Flaugher}}}, \bibinfo {author} {\bibfnamefont {A.}~\bibnamefont {{Font-Ribera}}}, \bibinfo {author} {\bibfnamefont {J.~E.}\ \bibnamefont {{Forero-Romero}}}, \bibinfo {author} {\bibfnamefont {C.~S.}\ \bibnamefont {{Frenk}}}, \bibinfo {author} {\bibfnamefont {S.}~\bibnamefont {{Fromenteau}}}, \bibinfo {author} {\bibfnamefont {B.~T.}\ \bibnamefont {{G{\"a}nsicke}}}, \bibinfo {author}
  {\bibfnamefont {C.}~\bibnamefont {{Garcia-Quintero}}}, \bibinfo {author} {\bibfnamefont {L.}~\bibnamefont {{Garrison}}}, \bibinfo {author} {\bibfnamefont {E.}~\bibnamefont {{Gazta{\~n}aga}}}, \bibinfo {author} {\bibfnamefont {F.}~\bibnamefont {{Gerardi}}}, \bibinfo {author} {\bibfnamefont {H.}~\bibnamefont {{Gil-Mar{\'\i}n}}}, \bibinfo {author} {\bibfnamefont {S.}~\bibnamefont {{Gontcho a Gontcho}}}, \bibinfo {author} {\bibfnamefont {A.~X.}\ \bibnamefont {{Gonzalez-Morales}}}, \bibinfo {author} {\bibfnamefont {G.}~\bibnamefont {{Gonzalez-de-Rivera}}}, \bibinfo {author} {\bibfnamefont {V.}~\bibnamefont {{Gonzalez-Perez}}}, \bibinfo {author} {\bibfnamefont {C.}~\bibnamefont {{Gordon}}}, \bibinfo {author} {\bibfnamefont {O.}~\bibnamefont {{Graur}}}, \bibinfo {author} {\bibfnamefont {D.}~\bibnamefont {{Green}}}, \bibinfo {author} {\bibfnamefont {C.}~\bibnamefont {{Grove}}}, \bibinfo {author} {\bibfnamefont {D.}~\bibnamefont {{Gruen}}}, \bibinfo {author} {\bibfnamefont {G.}~\bibnamefont {{Gutierrez}}}, \bibinfo
  {author} {\bibfnamefont {J.}~\bibnamefont {{Guy}}}, \bibinfo {author} {\bibfnamefont {C.}~\bibnamefont {{Hahn}}}, \bibinfo {author} {\bibfnamefont {S.}~\bibnamefont {{Harris}}}, \bibinfo {author} {\bibfnamefont {D.}~\bibnamefont {{Herrera}}}, \bibinfo {author} {\bibfnamefont {H.~K.}\ \bibnamefont {{Herrera-Alcantar}}}, \bibinfo {author} {\bibfnamefont {K.}~\bibnamefont {{Honscheid}}}, \bibinfo {author} {\bibfnamefont {C.}~\bibnamefont {{Howlett}}}, \bibinfo {author} {\bibfnamefont {D.}~\bibnamefont {{Huterer}}}, \bibinfo {author} {\bibfnamefont {V.}~\bibnamefont {{Ir{\v{s}}i{\v{c}}}}}, \bibinfo {author} {\bibfnamefont {M.}~\bibnamefont {{Ishak}}}, \bibinfo {author} {\bibfnamefont {P.}~\bibnamefont {{Jelinsky}}}, \bibinfo {author} {\bibfnamefont {L.}~\bibnamefont {{Jiang}}}, \bibinfo {author} {\bibfnamefont {J.}~\bibnamefont {{Jimenez}}}, \bibinfo {author} {\bibfnamefont {Y.~P.}\ \bibnamefont {{Jing}}}, \bibinfo {author} {\bibfnamefont {R.}~\bibnamefont {{Joyce}}}, \bibinfo {author} {\bibfnamefont
  {E.}~\bibnamefont {{Jullo}}}, \bibinfo {author} {\bibfnamefont {S.}~\bibnamefont {{Juneau}}}, \bibinfo {author} {\bibfnamefont {N.~G.}\ \bibnamefont {{Kara{\c{c}}ayl{\i}}}}, \bibinfo {author} {\bibfnamefont {M.}~\bibnamefont {{Karamanis}}}, \bibinfo {author} {\bibfnamefont {A.}~\bibnamefont {{Karcher}}}, \bibinfo {author} {\bibfnamefont {T.}~\bibnamefont {{Karim}}}, \bibinfo {author} {\bibfnamefont {R.}~\bibnamefont {{Kehoe}}}, \bibinfo {author} {\bibfnamefont {S.}~\bibnamefont {{Kent}}}, \bibinfo {author} {\bibfnamefont {D.}~\bibnamefont {{Kirkby}}}, \bibinfo {author} {\bibfnamefont {T.}~\bibnamefont {{Kisner}}}, \bibinfo {author} {\bibfnamefont {F.}~\bibnamefont {{Kitaura}}}, \bibinfo {author} {\bibfnamefont {S.~E.}\ \bibnamefont {{Koposov}}}, \bibinfo {author} {\bibfnamefont {A.}~\bibnamefont {{Kov{\'a}cs}}}, \bibinfo {author} {\bibfnamefont {A.}~\bibnamefont {{Kremin}}}, \bibinfo {author} {\bibfnamefont {A.}~\bibnamefont {{Krolewski}}}, \bibinfo {author} {\bibfnamefont {B.}~\bibnamefont {{L'Huillier}}},
  \bibinfo {author} {\bibfnamefont {O.}~\bibnamefont {{Lahav}}}, \bibinfo {author} {\bibfnamefont {A.}~\bibnamefont {{Lambert}}}, \bibinfo {author} {\bibfnamefont {C.}~\bibnamefont {{Lamman}}}, \bibinfo {author} {\bibfnamefont {T.-W.}\ \bibnamefont {{Lan}}}, \bibinfo {author} {\bibfnamefont {M.}~\bibnamefont {{Landriau}}}, \bibinfo {author} {\bibfnamefont {S.}~\bibnamefont {{Lane}}}, \bibinfo {author} {\bibfnamefont {D.}~\bibnamefont {{Lang}}}, \bibinfo {author} {\bibfnamefont {J.~U.}\ \bibnamefont {{Lange}}}, \bibinfo {author} {\bibfnamefont {J.}~\bibnamefont {{Lasker}}}, \bibinfo {author} {\bibfnamefont {L.}~\bibnamefont {{Le Guillou}}}, \bibinfo {author} {\bibfnamefont {A.}~\bibnamefont {{Leauthaud}}}, \bibinfo {author} {\bibfnamefont {A.}~\bibnamefont {{Le Van Suu}}}, \bibinfo {author} {\bibfnamefont {M.~E.}\ \bibnamefont {{Levi}}}, \bibinfo {author} {\bibfnamefont {T.~S.}\ \bibnamefont {{Li}}}, \bibinfo {author} {\bibfnamefont {C.}~\bibnamefont {{Magneville}}}, \bibinfo {author} {\bibfnamefont
  {M.}~\bibnamefont {{Manera}}}, \bibinfo {author} {\bibfnamefont {C.~J.}\ \bibnamefont {{Manser}}}, \bibinfo {author} {\bibfnamefont {B.}~\bibnamefont {{Marshall}}}, \bibinfo {author} {\bibfnamefont {P.}~\bibnamefont {{Martini}}}, \bibinfo {author} {\bibfnamefont {W.}~\bibnamefont {{McCollam}}}, \bibinfo {author} {\bibfnamefont {P.}~\bibnamefont {{McDonald}}}, \bibinfo {author} {\bibfnamefont {A.~M.}\ \bibnamefont {{Meisner}}}, \bibinfo {author} {\bibfnamefont {J.}~\bibnamefont {{Mena-Fern{\'a}ndez}}}, \bibinfo {author} {\bibfnamefont {J.}~\bibnamefont {{Meneses-Rizo}}}, \bibinfo {author} {\bibfnamefont {M.}~\bibnamefont {{Mezcua}}}, \bibinfo {author} {\bibfnamefont {T.}~\bibnamefont {{Miller}}}, \bibinfo {author} {\bibfnamefont {R.}~\bibnamefont {{Miquel}}}, \bibinfo {author} {\bibfnamefont {P.}~\bibnamefont {{Montero-Camacho}}}, \bibinfo {author} {\bibfnamefont {J.}~\bibnamefont {{Moon}}}, \bibinfo {author} {\bibfnamefont {J.}~\bibnamefont {{Moustakas}}}, \bibinfo {author} {\bibfnamefont {E.}~\bibnamefont
  {{Mueller}}}, \bibinfo {author} {\bibfnamefont {A.}~\bibnamefont {{Mu{\~n}oz-Guti{\'e}rrez}}}, \bibinfo {author} {\bibfnamefont {A.~D.}\ \bibnamefont {{Myers}}}, \bibinfo {author} {\bibfnamefont {S.}~\bibnamefont {{Nadathur}}}, \bibinfo {author} {\bibfnamefont {J.}~\bibnamefont {{Najita}}}, \bibinfo {author} {\bibfnamefont {L.}~\bibnamefont {{Napolitano}}}, \bibinfo {author} {\bibfnamefont {E.}~\bibnamefont {{Neilsen}}}, \bibinfo {author} {\bibfnamefont {J.~A.}\ \bibnamefont {{Newman}}}, \bibinfo {author} {\bibfnamefont {J.~D.}\ \bibnamefont {{Nie}}}, \bibinfo {author} {\bibfnamefont {Y.}~\bibnamefont {{Ning}}}, \bibinfo {author} {\bibfnamefont {G.}~\bibnamefont {{Niz}}}, \bibinfo {author} {\bibfnamefont {P.}~\bibnamefont {{Norberg}}}, \bibinfo {author} {\bibfnamefont {H.~E.}\ \bibnamefont {{Noriega}}}, \bibinfo {author} {\bibfnamefont {T.}~\bibnamefont {{O'Brien}}}, \bibinfo {author} {\bibfnamefont {A.}~\bibnamefont {{Obuljen}}}, \bibinfo {author} {\bibfnamefont {N.}~\bibnamefont
  {{Palanque-Delabrouille}}}, \bibinfo {author} {\bibfnamefont {A.}~\bibnamefont {{Palmese}}}, \bibinfo {author} {\bibfnamefont {P.}~\bibnamefont {{Zhiwei}}}, \bibinfo {author} {\bibfnamefont {D.}~\bibnamefont {{Pappalardo}}}, \bibinfo {author} {\bibfnamefont {X.}~\bibnamefont {{PENG}}}, \bibinfo {author} {\bibfnamefont {W.~J.}\ \bibnamefont {{Percival}}}, \bibinfo {author} {\bibfnamefont {S.}~\bibnamefont {{Perruchot}}}, \bibinfo {author} {\bibfnamefont {R.}~\bibnamefont {{Pogge}}}, \bibinfo {author} {\bibfnamefont {C.}~\bibnamefont {{Poppett}}}, \bibinfo {author} {\bibfnamefont {A.}~\bibnamefont {{Porredon}}}, \bibinfo {author} {\bibfnamefont {F.}~\bibnamefont {{Prada}}}, \bibinfo {author} {\bibfnamefont {J.}~\bibnamefont {{Prochaska}}}, \bibinfo {author} {\bibfnamefont {R.}~\bibnamefont {{Pucha}}}, \bibinfo {author} {\bibfnamefont {A.}~\bibnamefont {{P{\'e}rez-Fern{\'a}ndez}}}, \bibinfo {author} {\bibfnamefont {I.}~\bibnamefont {{P{\'e}rez-R{\`a}fols}}}, \bibinfo {author} {\bibfnamefont {D.}~\bibnamefont
  {{Rabinowitz}}}, \bibinfo {author} {\bibfnamefont {A.}~\bibnamefont {{Raichoor}}}, \bibinfo {author} {\bibfnamefont {S.}~\bibnamefont {{Ramirez-Solano}}}, \bibinfo {author} {\bibfnamefont {C.}~\bibnamefont {{Ram{\'\i}rez-P{\'e}rez}}}, \bibinfo {author} {\bibfnamefont {C.}~\bibnamefont {{Ravoux}}}, \bibinfo {author} {\bibfnamefont {K.}~\bibnamefont {{Reil}}}, \bibinfo {author} {\bibfnamefont {M.}~\bibnamefont {{Rezaie}}}, \bibinfo {author} {\bibfnamefont {A.}~\bibnamefont {{Rocher}}}, \bibinfo {author} {\bibfnamefont {C.}~\bibnamefont {{Rockosi}}}, \bibinfo {author} {\bibfnamefont {N.~A.}\ \bibnamefont {{Roe}}}, \bibinfo {author} {\bibfnamefont {A.}~\bibnamefont {{Roodman}}}, \bibinfo {author} {\bibfnamefont {A.~J.}\ \bibnamefont {{Ross}}}, \bibinfo {author} {\bibfnamefont {G.}~\bibnamefont {{Rossi}}}, \bibinfo {author} {\bibfnamefont {R.}~\bibnamefont {{Ruggeri}}}, \bibinfo {author} {\bibfnamefont {V.}~\bibnamefont {{Ruhlmann-Kleider}}}, \bibinfo {author} {\bibfnamefont {C.~G.}\ \bibnamefont {{Sabiu}}},
  \bibinfo {author} {\bibfnamefont {S.}~\bibnamefont {{Safonova}}}, \bibinfo {author} {\bibfnamefont {K.}~\bibnamefont {{Said}}}, \bibinfo {author} {\bibfnamefont {A.}~\bibnamefont {{Saintonge}}}, \bibinfo {author} {\bibfnamefont {J.}~\bibnamefont {{Salas Catonga}}}, \bibinfo {author} {\bibfnamefont {L.}~\bibnamefont {{Samushia}}}, \bibinfo {author} {\bibfnamefont {E.}~\bibnamefont {{Sanchez}}}, \bibinfo {author} {\bibfnamefont {C.}~\bibnamefont {{Saulder}}}, \bibinfo {author} {\bibfnamefont {E.}~\bibnamefont {{Schaan}}}, \bibinfo {author} {\bibfnamefont {E.}~\bibnamefont {{Schlafly}}}, \bibinfo {author} {\bibfnamefont {D.}~\bibnamefont {{Schlegel}}}, \bibinfo {author} {\bibfnamefont {J.}~\bibnamefont {{Schmoll}}}, \bibinfo {author} {\bibfnamefont {D.}~\bibnamefont {{Scholte}}}, \bibinfo {author} {\bibfnamefont {M.}~\bibnamefont {{Schubnell}}}, \bibinfo {author} {\bibfnamefont {A.}~\bibnamefont {{Secroun}}}, \bibinfo {author} {\bibfnamefont {H.}~\bibnamefont {{Seo}}}, \bibinfo {author} {\bibfnamefont
  {S.}~\bibnamefont {{Serrano}}}, \bibinfo {author} {\bibfnamefont {R.~M.}\ \bibnamefont {{Sharples}}}, \bibinfo {author} {\bibfnamefont {M.~J.}\ \bibnamefont {{Sholl}}}, \bibinfo {author} {\bibfnamefont {J.~H.}\ \bibnamefont {{Silber}}}, \bibinfo {author} {\bibfnamefont {D.~R.}\ \bibnamefont {{Silva}}}, \bibinfo {author} {\bibfnamefont {M.}~\bibnamefont {{Sirk}}}, \bibinfo {author} {\bibfnamefont {M.}~\bibnamefont {{Siudek}}}, \bibinfo {author} {\bibfnamefont {A.}~\bibnamefont {{Smith}}}, \bibinfo {author} {\bibfnamefont {D.}~\bibnamefont {{Sprayberry}}}, \bibinfo {author} {\bibfnamefont {R.}~\bibnamefont {{Staten}}}, \bibinfo {author} {\bibfnamefont {B.}~\bibnamefont {{Stupak}}}, \bibinfo {author} {\bibfnamefont {T.}~\bibnamefont {{Tan}}}, \bibinfo {author} {\bibfnamefont {G.}~\bibnamefont {{Tarl{\'e}}}}, \bibinfo {author} {\bibfnamefont {S.~S.}\ \bibnamefont {{Tie}}}, \bibinfo {author} {\bibfnamefont {R.}~\bibnamefont {{Tojeiro}}}, \bibinfo {author} {\bibfnamefont {L.~A.}\ \bibnamefont
  {{Ure{\~n}a-L{\'o}pez}}}, \bibinfo {author} {\bibfnamefont {F.}~\bibnamefont {{Valdes}}}, \bibinfo {author} {\bibfnamefont {O.}~\bibnamefont {{Valenzuela}}}, \bibinfo {author} {\bibfnamefont {M.}~\bibnamefont {{Valluri}}}, \bibinfo {author} {\bibfnamefont {M.}~\bibnamefont {{Vargas-Maga{\~n}a}}}, \bibinfo {author} {\bibfnamefont {L.}~\bibnamefont {{Verde}}}, \bibinfo {author} {\bibfnamefont {M.}~\bibnamefont {{Walther}}}, \bibinfo {author} {\bibfnamefont {B.}~\bibnamefont {{Wang}}}, \bibinfo {author} {\bibfnamefont {M.~S.}\ \bibnamefont {{Wang}}}, \bibinfo {author} {\bibfnamefont {B.~A.}\ \bibnamefont {{Weaver}}}, \bibinfo {author} {\bibfnamefont {C.}~\bibnamefont {{Weaverdyck}}}, \bibinfo {author} {\bibfnamefont {R.}~\bibnamefont {{Wechsler}}}, \bibinfo {author} {\bibfnamefont {M.~J.}\ \bibnamefont {{Wilson}}}, \bibinfo {author} {\bibfnamefont {J.}~\bibnamefont {{Yang}}}, \bibinfo {author} {\bibfnamefont {Y.}~\bibnamefont {{Yu}}}, \bibinfo {author} {\bibfnamefont {S.}~\bibnamefont {{Yuan}}}, \bibinfo
  {author} {\bibfnamefont {C.}~\bibnamefont {{Y{\`e}che}}}, \bibinfo {author} {\bibfnamefont {H.}~\bibnamefont {{Zhang}}}, \bibinfo {author} {\bibfnamefont {K.}~\bibnamefont {{Zhang}}}, \bibinfo {author} {\bibfnamefont {C.}~\bibnamefont {{Zhao}}}, \bibinfo {author} {\bibfnamefont {R.}~\bibnamefont {{Zhou}}}, \bibinfo {author} {\bibfnamefont {Z.}~\bibnamefont {{Zhou}}}, \bibinfo {author} {\bibfnamefont {H.}~\bibnamefont {{Zou}}}, \bibinfo {author} {\bibfnamefont {J.}~\bibnamefont {{Zou}}}, \bibinfo {author} {\bibfnamefont {S.}~\bibnamefont {{Zou}}}, \bibinfo {author} {\bibfnamefont {Y.}~\bibnamefont {{Zu}}},\ and\ \bibinfo {author} {\bibnamefont {{DESI Collaboration}}},\ }\bibfield  {title} {\bibinfo {title} {{Overview of the Instrumentation for the Dark Energy Spectroscopic Instrument}},\ }\href {https://doi.org/10.3847/1538-3881/ac882b} {\bibfield  {journal} {\bibinfo  {journal} {\aj}\ }\textbf {\bibinfo {volume} {164}},\ \bibinfo {eid} {207} (\bibinfo {year} {2022})},\ \Eprint
  {https://arxiv.org/abs/2205.10939} {arXiv:2205.10939 [astro-ph.IM]} \BibitemShut {NoStop}%
\bibitem [{\citenamefont {{DESI Collaboration}}\ \emph {et~al.}(2016{\natexlab{a}})\citenamefont {{DESI Collaboration}}, \citenamefont {{Aghamousa}}, \citenamefont {{Aguilar}}, \citenamefont {{Ahlen}}, \citenamefont {{Alam}}, \citenamefont {{Allen}}, \citenamefont {{Allende Prieto}}, \citenamefont {{Annis}}, \citenamefont {{Bailey}}, \citenamefont {{Balland}}, \citenamefont {{Ballester}}, \citenamefont {{Baltay}}, \citenamefont {{Beaufore}}, \citenamefont {{Bebek}}, \citenamefont {{Beers}}, \citenamefont {{Bell}}, \citenamefont {{Bernal}}, \citenamefont {{Besuner}}, \citenamefont {{Beutler}}, \citenamefont {{Blake}}, \citenamefont {{Bleuler}}, \citenamefont {{Blomqvist}}, \citenamefont {{Blum}}, \citenamefont {{Bolton}}, \citenamefont {{Briceno}}, \citenamefont {{Brooks}}, \citenamefont {{Brownstein}}, \citenamefont {{Buckley-Geer}}, \citenamefont {{Burden}}, \citenamefont {{Burtin}}, \citenamefont {{Busca}}, \citenamefont {{Cahn}}, \citenamefont {{Cai}}, \citenamefont {{Cardiel-Sas}}, \citenamefont
  {{Carlberg}}, \citenamefont {{Carton}}, \citenamefont {{Casas}}, \citenamefont {{Castander}}, \citenamefont {{Cervantes-Cota}}, \citenamefont {{Claybaugh}}, \citenamefont {{Close}}, \citenamefont {{Coker}}, \citenamefont {{Cole}}, \citenamefont {{Comparat}}, \citenamefont {{Cooper}}, \citenamefont {{Cousinou}}, \citenamefont {{Crocce}}, \citenamefont {{Cuby}}, \citenamefont {{Cunningham}}, \citenamefont {{Davis}}, \citenamefont {{Dawson}}, \citenamefont {{de la Macorra}}, \citenamefont {{De Vicente}}, \citenamefont {{Delubac}}, \citenamefont {{Derwent}}, \citenamefont {{Dey}}, \citenamefont {{Dhungana}}, \citenamefont {{Ding}}, \citenamefont {{Doel}}, \citenamefont {{Duan}}, \citenamefont {{Ealet}}, \citenamefont {{Edelstein}}, \citenamefont {{Eftekharzadeh}}, \citenamefont {{Eisenstein}}, \citenamefont {{Elliott}}, \citenamefont {{Escoffier}}, \citenamefont {{Evatt}}, \citenamefont {{Fagrelius}}, \citenamefont {{Fan}}, \citenamefont {{Fanning}}, \citenamefont {{Farahi}}, \citenamefont {{Farihi}},
  \citenamefont {{Favole}}, \citenamefont {{Feng}}, \citenamefont {{Fernandez}}, \citenamefont {{Findlay}}, \citenamefont {{Finkbeiner}}, \citenamefont {{Fitzpatrick}}, \citenamefont {{Flaugher}}, \citenamefont {{Flender}}, \citenamefont {{Font-Ribera}}, \citenamefont {{Forero-Romero}}, \citenamefont {{Fosalba}}, \citenamefont {{Frenk}}, \citenamefont {{Fumagalli}}, \citenamefont {{Gaensicke}}, \citenamefont {{Gallo}}, \citenamefont {{Garcia-Bellido}}, \citenamefont {{Gaztanaga}}, \citenamefont {{Pietro Gentile Fusillo}}, \citenamefont {{Gerard}}, \citenamefont {{Gershkovich}}, \citenamefont {{Giannantonio}}, \citenamefont {{Gillet}}, \citenamefont {{Gonzalez-de-Rivera}}, \citenamefont {{Gonzalez-Perez}}, \citenamefont {{Gott}}, \citenamefont {{Graur}}, \citenamefont {{Gutierrez}}, \citenamefont {{Guy}}, \citenamefont {{Habib}}, \citenamefont {{Heetderks}}, \citenamefont {{Heetderks}}, \citenamefont {{Heitmann}}, \citenamefont {{Hellwing}}, \citenamefont {{Herrera}}, \citenamefont {{Ho}}, \citenamefont
  {{Holland}}, \citenamefont {{Honscheid}}, \citenamefont {{Huff}}, \citenamefont {{Hutchinson}}, \citenamefont {{Huterer}}, \citenamefont {{Hwang}}, \citenamefont {{Illa Laguna}}, \citenamefont {{Ishikawa}}, \citenamefont {{Jacobs}}, \citenamefont {{Jeffrey}}, \citenamefont {{Jelinsky}}, \citenamefont {{Jennings}}, \citenamefont {{Jiang}}, \citenamefont {{Jimenez}}, \citenamefont {{Johnson}}, \citenamefont {{Joyce}}, \citenamefont {{Jullo}}, \citenamefont {{Juneau}}, \citenamefont {{Kama}}, \citenamefont {{Karcher}}, \citenamefont {{Karkar}}, \citenamefont {{Kehoe}}, \citenamefont {{Kennamer}}, \citenamefont {{Kent}}, \citenamefont {{Kilbinger}}, \citenamefont {{Kim}}, \citenamefont {{Kirkby}}, \citenamefont {{Kisner}}, \citenamefont {{Kitanidis}}, \citenamefont {{Kneib}}, \citenamefont {{Koposov}}, \citenamefont {{Kovacs}}, \citenamefont {{Koyama}}, \citenamefont {{Kremin}}, \citenamefont {{Kron}}, \citenamefont {{Kronig}}, \citenamefont {{Kueter-Young}}, \citenamefont {{Lacey}}, \citenamefont {{Lafever}},
  \citenamefont {{Lahav}}, \citenamefont {{Lambert}}, \citenamefont {{Lampton}}, \citenamefont {{Landriau}}, \citenamefont {{Lang}}, \citenamefont {{Lauer}}, \citenamefont {{Le Goff}}, \citenamefont {{Le Guillou}}, \citenamefont {{Le Van Suu}}, \citenamefont {{Lee}}, \citenamefont {{Lee}}, \citenamefont {{Leitner}}, \citenamefont {{Lesser}}, \citenamefont {{Levi}}, \citenamefont {{L'Huillier}}, \citenamefont {{Li}}, \citenamefont {{Liang}}, \citenamefont {{Lin}}, \citenamefont {{Linder}}, \citenamefont {{Loebman}}, \citenamefont {{Luki{\'c}}}, \citenamefont {{Ma}}, \citenamefont {{MacCrann}}, \citenamefont {{Magneville}}, \citenamefont {{Makarem}}, \citenamefont {{Manera}}, \citenamefont {{Manser}}, \citenamefont {{Marshall}}, \citenamefont {{Martini}}, \citenamefont {{Massey}}, \citenamefont {{Matheson}}, \citenamefont {{McCauley}}, \citenamefont {{McDonald}}, \citenamefont {{McGreer}}, \citenamefont {{Meisner}}, \citenamefont {{Metcalfe}}, \citenamefont {{Miller}}, \citenamefont {{Miquel}}, \citenamefont
  {{Moustakas}}, \citenamefont {{Myers}}, \citenamefont {{Naik}}, \citenamefont {{Newman}}, \citenamefont {{Nichol}}, \citenamefont {{Nicola}}, \citenamefont {{Nicolati da Costa}}, \citenamefont {{Nie}}, \citenamefont {{Niz}}, \citenamefont {{Norberg}}, \citenamefont {{Nord}}, \citenamefont {{Norman}}, \citenamefont {{Nugent}}, \citenamefont {{O'Brien}}, \citenamefont {{Oh}}, \citenamefont {{Olsen}}, \citenamefont {{Padilla}}, \citenamefont {{Padmanabhan}}, \citenamefont {{Padmanabhan}}, \citenamefont {{Palanque-Delabrouille}}, \citenamefont {{Palmese}}, \citenamefont {{Pappalardo}}, \citenamefont {{P{\^a}ris}}, \citenamefont {{Park}}, \citenamefont {{Patej}}, \citenamefont {{Peacock}}, \citenamefont {{Peiris}}, \citenamefont {{Peng}}, \citenamefont {{Percival}}, \citenamefont {{Perruchot}}, \citenamefont {{Pieri}}, \citenamefont {{Pogge}}, \citenamefont {{Pollack}}, \citenamefont {{Poppett}}, \citenamefont {{Prada}}, \citenamefont {{Prakash}}, \citenamefont {{Probst}}, \citenamefont {{Rabinowitz}},
  \citenamefont {{Raichoor}}, \citenamefont {{Ree}}, \citenamefont {{Refregier}}, \citenamefont {{Regal}}, \citenamefont {{Reid}}, \citenamefont {{Reil}}, \citenamefont {{Rezaie}}, \citenamefont {{Rockosi}}, \citenamefont {{Roe}}, \citenamefont {{Ronayette}}, \citenamefont {{Roodman}}, \citenamefont {{Ross}}, \citenamefont {{Ross}}, \citenamefont {{Rossi}}, \citenamefont {{Rozo}}, \citenamefont {{Ruhlmann-Kleider}}, \citenamefont {{Rykoff}}, \citenamefont {{Sabiu}}, \citenamefont {{Samushia}}, \citenamefont {{Sanchez}}, \citenamefont {{Sanchez}}, \citenamefont {{Schlegel}}, \citenamefont {{Schneider}}, \citenamefont {{Schubnell}}, \citenamefont {{Secroun}}, \citenamefont {{Seljak}}, \citenamefont {{Seo}}, \citenamefont {{Serrano}}, \citenamefont {{Shafieloo}}, \citenamefont {{Shan}}, \citenamefont {{Sharples}}, \citenamefont {{Sholl}}, \citenamefont {{Shourt}}, \citenamefont {{Silber}}, \citenamefont {{Silva}}, \citenamefont {{Sirk}}, \citenamefont {{Slosar}}, \citenamefont {{Smith}}, \citenamefont {{Smoot}},
  \citenamefont {{Som}}, \citenamefont {{Song}}, \citenamefont {{Sprayberry}}, \citenamefont {{Staten}}, \citenamefont {{Stefanik}}, \citenamefont {{Tarle}}, \citenamefont {{Sien Tie}}, \citenamefont {{Tinker}}, \citenamefont {{Tojeiro}}, \citenamefont {{Valdes}}, \citenamefont {{Valenzuela}}, \citenamefont {{Valluri}}, \citenamefont {{Vargas-Magana}}, \citenamefont {{Verde}}, \citenamefont {{Walker}}, \citenamefont {{Wang}}, \citenamefont {{Wang}}, \citenamefont {{Weaver}}, \citenamefont {{Weaverdyck}}, \citenamefont {{Wechsler}}, \citenamefont {{Weinberg}}, \citenamefont {{White}}, \citenamefont {{Yang}}, \citenamefont {{Yeche}}, \citenamefont {{Zhang}}, \citenamefont {{Zhao}}, \citenamefont {{Zheng}}, \citenamefont {{Zhou}}, \citenamefont {{Zhou}}, \citenamefont {{Zhu}}, \citenamefont {{Zou}},\ and\ \citenamefont {{Zu}}}]{2016arXiv161100037D}%
  \BibitemOpen
  \bibfield  {author} {\bibinfo {author} {\bibnamefont {{DESI Collaboration}}}, \bibinfo {author} {\bibfnamefont {A.}~\bibnamefont {{Aghamousa}}}, \bibinfo {author} {\bibfnamefont {J.}~\bibnamefont {{Aguilar}}}, \bibinfo {author} {\bibfnamefont {S.}~\bibnamefont {{Ahlen}}}, \bibinfo {author} {\bibfnamefont {S.}~\bibnamefont {{Alam}}}, \bibinfo {author} {\bibfnamefont {L.~E.}\ \bibnamefont {{Allen}}}, \bibinfo {author} {\bibfnamefont {C.}~\bibnamefont {{Allende Prieto}}}, \bibinfo {author} {\bibfnamefont {J.}~\bibnamefont {{Annis}}}, \bibinfo {author} {\bibfnamefont {S.}~\bibnamefont {{Bailey}}}, \bibinfo {author} {\bibfnamefont {C.}~\bibnamefont {{Balland}}}, \bibinfo {author} {\bibfnamefont {O.}~\bibnamefont {{Ballester}}}, \bibinfo {author} {\bibfnamefont {C.}~\bibnamefont {{Baltay}}}, \bibinfo {author} {\bibfnamefont {L.}~\bibnamefont {{Beaufore}}}, \bibinfo {author} {\bibfnamefont {C.}~\bibnamefont {{Bebek}}}, \bibinfo {author} {\bibfnamefont {T.~C.}\ \bibnamefont {{Beers}}}, \bibinfo {author}
  {\bibfnamefont {E.~F.}\ \bibnamefont {{Bell}}}, \bibinfo {author} {\bibfnamefont {J.~L.}\ \bibnamefont {{Bernal}}}, \bibinfo {author} {\bibfnamefont {R.}~\bibnamefont {{Besuner}}}, \bibinfo {author} {\bibfnamefont {F.}~\bibnamefont {{Beutler}}}, \bibinfo {author} {\bibfnamefont {C.}~\bibnamefont {{Blake}}}, \bibinfo {author} {\bibfnamefont {H.}~\bibnamefont {{Bleuler}}}, \bibinfo {author} {\bibfnamefont {M.}~\bibnamefont {{Blomqvist}}}, \bibinfo {author} {\bibfnamefont {R.}~\bibnamefont {{Blum}}}, \bibinfo {author} {\bibfnamefont {A.~S.}\ \bibnamefont {{Bolton}}}, \bibinfo {author} {\bibfnamefont {C.}~\bibnamefont {{Briceno}}}, \bibinfo {author} {\bibfnamefont {D.}~\bibnamefont {{Brooks}}}, \bibinfo {author} {\bibfnamefont {J.~R.}\ \bibnamefont {{Brownstein}}}, \bibinfo {author} {\bibfnamefont {E.}~\bibnamefont {{Buckley-Geer}}}, \bibinfo {author} {\bibfnamefont {A.}~\bibnamefont {{Burden}}}, \bibinfo {author} {\bibfnamefont {E.}~\bibnamefont {{Burtin}}}, \bibinfo {author} {\bibfnamefont {N.~G.}\
  \bibnamefont {{Busca}}}, \bibinfo {author} {\bibfnamefont {R.~N.}\ \bibnamefont {{Cahn}}}, \bibinfo {author} {\bibfnamefont {Y.-C.}\ \bibnamefont {{Cai}}}, \bibinfo {author} {\bibfnamefont {L.}~\bibnamefont {{Cardiel-Sas}}}, \bibinfo {author} {\bibfnamefont {R.~G.}\ \bibnamefont {{Carlberg}}}, \bibinfo {author} {\bibfnamefont {P.-H.}\ \bibnamefont {{Carton}}}, \bibinfo {author} {\bibfnamefont {R.}~\bibnamefont {{Casas}}}, \bibinfo {author} {\bibfnamefont {F.~J.}\ \bibnamefont {{Castander}}}, \bibinfo {author} {\bibfnamefont {J.~L.}\ \bibnamefont {{Cervantes-Cota}}}, \bibinfo {author} {\bibfnamefont {T.~M.}\ \bibnamefont {{Claybaugh}}}, \bibinfo {author} {\bibfnamefont {M.}~\bibnamefont {{Close}}}, \bibinfo {author} {\bibfnamefont {C.~T.}\ \bibnamefont {{Coker}}}, \bibinfo {author} {\bibfnamefont {S.}~\bibnamefont {{Cole}}}, \bibinfo {author} {\bibfnamefont {J.}~\bibnamefont {{Comparat}}}, \bibinfo {author} {\bibfnamefont {A.~P.}\ \bibnamefont {{Cooper}}}, \bibinfo {author} {\bibfnamefont {M.~C.}\
  \bibnamefont {{Cousinou}}}, \bibinfo {author} {\bibfnamefont {M.}~\bibnamefont {{Crocce}}}, \bibinfo {author} {\bibfnamefont {J.-G.}\ \bibnamefont {{Cuby}}}, \bibinfo {author} {\bibfnamefont {D.~P.}\ \bibnamefont {{Cunningham}}}, \bibinfo {author} {\bibfnamefont {T.~M.}\ \bibnamefont {{Davis}}}, \bibinfo {author} {\bibfnamefont {K.~S.}\ \bibnamefont {{Dawson}}}, \bibinfo {author} {\bibfnamefont {A.}~\bibnamefont {{de la Macorra}}}, \bibinfo {author} {\bibfnamefont {J.}~\bibnamefont {{De Vicente}}}, \bibinfo {author} {\bibfnamefont {T.}~\bibnamefont {{Delubac}}}, \bibinfo {author} {\bibfnamefont {M.}~\bibnamefont {{Derwent}}}, \bibinfo {author} {\bibfnamefont {A.}~\bibnamefont {{Dey}}}, \bibinfo {author} {\bibfnamefont {G.}~\bibnamefont {{Dhungana}}}, \bibinfo {author} {\bibfnamefont {Z.}~\bibnamefont {{Ding}}}, \bibinfo {author} {\bibfnamefont {P.}~\bibnamefont {{Doel}}}, \bibinfo {author} {\bibfnamefont {Y.~T.}\ \bibnamefont {{Duan}}}, \bibinfo {author} {\bibfnamefont {A.}~\bibnamefont {{Ealet}}}, \bibinfo
  {author} {\bibfnamefont {J.}~\bibnamefont {{Edelstein}}}, \bibinfo {author} {\bibfnamefont {S.}~\bibnamefont {{Eftekharzadeh}}}, \bibinfo {author} {\bibfnamefont {D.~J.}\ \bibnamefont {{Eisenstein}}}, \bibinfo {author} {\bibfnamefont {A.}~\bibnamefont {{Elliott}}}, \bibinfo {author} {\bibfnamefont {S.}~\bibnamefont {{Escoffier}}}, \bibinfo {author} {\bibfnamefont {M.}~\bibnamefont {{Evatt}}}, \bibinfo {author} {\bibfnamefont {P.}~\bibnamefont {{Fagrelius}}}, \bibinfo {author} {\bibfnamefont {X.}~\bibnamefont {{Fan}}}, \bibinfo {author} {\bibfnamefont {K.}~\bibnamefont {{Fanning}}}, \bibinfo {author} {\bibfnamefont {A.}~\bibnamefont {{Farahi}}}, \bibinfo {author} {\bibfnamefont {J.}~\bibnamefont {{Farihi}}}, \bibinfo {author} {\bibfnamefont {G.}~\bibnamefont {{Favole}}}, \bibinfo {author} {\bibfnamefont {Y.}~\bibnamefont {{Feng}}}, \bibinfo {author} {\bibfnamefont {E.}~\bibnamefont {{Fernandez}}}, \bibinfo {author} {\bibfnamefont {J.~R.}\ \bibnamefont {{Findlay}}}, \bibinfo {author} {\bibfnamefont {D.~P.}\
  \bibnamefont {{Finkbeiner}}}, \bibinfo {author} {\bibfnamefont {M.~J.}\ \bibnamefont {{Fitzpatrick}}}, \bibinfo {author} {\bibfnamefont {B.}~\bibnamefont {{Flaugher}}}, \bibinfo {author} {\bibfnamefont {S.}~\bibnamefont {{Flender}}}, \bibinfo {author} {\bibfnamefont {A.}~\bibnamefont {{Font-Ribera}}}, \bibinfo {author} {\bibfnamefont {J.~E.}\ \bibnamefont {{Forero-Romero}}}, \bibinfo {author} {\bibfnamefont {P.}~\bibnamefont {{Fosalba}}}, \bibinfo {author} {\bibfnamefont {C.~S.}\ \bibnamefont {{Frenk}}}, \bibinfo {author} {\bibfnamefont {M.}~\bibnamefont {{Fumagalli}}}, \bibinfo {author} {\bibfnamefont {B.~T.}\ \bibnamefont {{Gaensicke}}}, \bibinfo {author} {\bibfnamefont {G.}~\bibnamefont {{Gallo}}}, \bibinfo {author} {\bibfnamefont {J.}~\bibnamefont {{Garcia-Bellido}}}, \bibinfo {author} {\bibfnamefont {E.}~\bibnamefont {{Gaztanaga}}}, \bibinfo {author} {\bibfnamefont {N.}~\bibnamefont {{Pietro Gentile Fusillo}}}, \bibinfo {author} {\bibfnamefont {T.}~\bibnamefont {{Gerard}}}, \bibinfo {author}
  {\bibfnamefont {I.}~\bibnamefont {{Gershkovich}}}, \bibinfo {author} {\bibfnamefont {T.}~\bibnamefont {{Giannantonio}}}, \bibinfo {author} {\bibfnamefont {D.}~\bibnamefont {{Gillet}}}, \bibinfo {author} {\bibfnamefont {G.}~\bibnamefont {{Gonzalez-de-Rivera}}}, \bibinfo {author} {\bibfnamefont {V.}~\bibnamefont {{Gonzalez-Perez}}}, \bibinfo {author} {\bibfnamefont {S.}~\bibnamefont {{Gott}}}, \bibinfo {author} {\bibfnamefont {O.}~\bibnamefont {{Graur}}}, \bibinfo {author} {\bibfnamefont {G.}~\bibnamefont {{Gutierrez}}}, \bibinfo {author} {\bibfnamefont {J.}~\bibnamefont {{Guy}}}, \bibinfo {author} {\bibfnamefont {S.}~\bibnamefont {{Habib}}}, \bibinfo {author} {\bibfnamefont {H.}~\bibnamefont {{Heetderks}}}, \bibinfo {author} {\bibfnamefont {I.}~\bibnamefont {{Heetderks}}}, \bibinfo {author} {\bibfnamefont {K.}~\bibnamefont {{Heitmann}}}, \bibinfo {author} {\bibfnamefont {W.~A.}\ \bibnamefont {{Hellwing}}}, \bibinfo {author} {\bibfnamefont {D.~A.}\ \bibnamefont {{Herrera}}}, \bibinfo {author} {\bibfnamefont
  {S.}~\bibnamefont {{Ho}}}, \bibinfo {author} {\bibfnamefont {S.}~\bibnamefont {{Holland}}}, \bibinfo {author} {\bibfnamefont {K.}~\bibnamefont {{Honscheid}}}, \bibinfo {author} {\bibfnamefont {E.}~\bibnamefont {{Huff}}}, \bibinfo {author} {\bibfnamefont {T.~A.}\ \bibnamefont {{Hutchinson}}}, \bibinfo {author} {\bibfnamefont {D.}~\bibnamefont {{Huterer}}}, \bibinfo {author} {\bibfnamefont {H.~S.}\ \bibnamefont {{Hwang}}}, \bibinfo {author} {\bibfnamefont {J.~M.}\ \bibnamefont {{Illa Laguna}}}, \bibinfo {author} {\bibfnamefont {Y.}~\bibnamefont {{Ishikawa}}}, \bibinfo {author} {\bibfnamefont {D.}~\bibnamefont {{Jacobs}}}, \bibinfo {author} {\bibfnamefont {N.}~\bibnamefont {{Jeffrey}}}, \bibinfo {author} {\bibfnamefont {P.}~\bibnamefont {{Jelinsky}}}, \bibinfo {author} {\bibfnamefont {E.}~\bibnamefont {{Jennings}}}, \bibinfo {author} {\bibfnamefont {L.}~\bibnamefont {{Jiang}}}, \bibinfo {author} {\bibfnamefont {J.}~\bibnamefont {{Jimenez}}}, \bibinfo {author} {\bibfnamefont {J.}~\bibnamefont {{Johnson}}},
  \bibinfo {author} {\bibfnamefont {R.}~\bibnamefont {{Joyce}}}, \bibinfo {author} {\bibfnamefont {E.}~\bibnamefont {{Jullo}}}, \bibinfo {author} {\bibfnamefont {S.}~\bibnamefont {{Juneau}}}, \bibinfo {author} {\bibfnamefont {S.}~\bibnamefont {{Kama}}}, \bibinfo {author} {\bibfnamefont {A.}~\bibnamefont {{Karcher}}}, \bibinfo {author} {\bibfnamefont {S.}~\bibnamefont {{Karkar}}}, \bibinfo {author} {\bibfnamefont {R.}~\bibnamefont {{Kehoe}}}, \bibinfo {author} {\bibfnamefont {N.}~\bibnamefont {{Kennamer}}}, \bibinfo {author} {\bibfnamefont {S.}~\bibnamefont {{Kent}}}, \bibinfo {author} {\bibfnamefont {M.}~\bibnamefont {{Kilbinger}}}, \bibinfo {author} {\bibfnamefont {A.~G.}\ \bibnamefont {{Kim}}}, \bibinfo {author} {\bibfnamefont {D.}~\bibnamefont {{Kirkby}}}, \bibinfo {author} {\bibfnamefont {T.}~\bibnamefont {{Kisner}}}, \bibinfo {author} {\bibfnamefont {E.}~\bibnamefont {{Kitanidis}}}, \bibinfo {author} {\bibfnamefont {J.-P.}\ \bibnamefont {{Kneib}}}, \bibinfo {author} {\bibfnamefont {S.}~\bibnamefont
  {{Koposov}}}, \bibinfo {author} {\bibfnamefont {E.}~\bibnamefont {{Kovacs}}}, \bibinfo {author} {\bibfnamefont {K.}~\bibnamefont {{Koyama}}}, \bibinfo {author} {\bibfnamefont {A.}~\bibnamefont {{Kremin}}}, \bibinfo {author} {\bibfnamefont {R.}~\bibnamefont {{Kron}}}, \bibinfo {author} {\bibfnamefont {L.}~\bibnamefont {{Kronig}}}, \bibinfo {author} {\bibfnamefont {A.}~\bibnamefont {{Kueter-Young}}}, \bibinfo {author} {\bibfnamefont {C.~G.}\ \bibnamefont {{Lacey}}}, \bibinfo {author} {\bibfnamefont {R.}~\bibnamefont {{Lafever}}}, \bibinfo {author} {\bibfnamefont {O.}~\bibnamefont {{Lahav}}}, \bibinfo {author} {\bibfnamefont {A.}~\bibnamefont {{Lambert}}}, \bibinfo {author} {\bibfnamefont {M.}~\bibnamefont {{Lampton}}}, \bibinfo {author} {\bibfnamefont {M.}~\bibnamefont {{Landriau}}}, \bibinfo {author} {\bibfnamefont {D.}~\bibnamefont {{Lang}}}, \bibinfo {author} {\bibfnamefont {T.~R.}\ \bibnamefont {{Lauer}}}, \bibinfo {author} {\bibfnamefont {J.-M.}\ \bibnamefont {{Le Goff}}}, \bibinfo {author}
  {\bibfnamefont {L.}~\bibnamefont {{Le Guillou}}}, \bibinfo {author} {\bibfnamefont {A.}~\bibnamefont {{Le Van Suu}}}, \bibinfo {author} {\bibfnamefont {J.~H.}\ \bibnamefont {{Lee}}}, \bibinfo {author} {\bibfnamefont {S.-J.}\ \bibnamefont {{Lee}}}, \bibinfo {author} {\bibfnamefont {D.}~\bibnamefont {{Leitner}}}, \bibinfo {author} {\bibfnamefont {M.}~\bibnamefont {{Lesser}}}, \bibinfo {author} {\bibfnamefont {M.~E.}\ \bibnamefont {{Levi}}}, \bibinfo {author} {\bibfnamefont {B.}~\bibnamefont {{L'Huillier}}}, \bibinfo {author} {\bibfnamefont {B.}~\bibnamefont {{Li}}}, \bibinfo {author} {\bibfnamefont {M.}~\bibnamefont {{Liang}}}, \bibinfo {author} {\bibfnamefont {H.}~\bibnamefont {{Lin}}}, \bibinfo {author} {\bibfnamefont {E.}~\bibnamefont {{Linder}}}, \bibinfo {author} {\bibfnamefont {S.~R.}\ \bibnamefont {{Loebman}}}, \bibinfo {author} {\bibfnamefont {Z.}~\bibnamefont {{Luki{\'c}}}}, \bibinfo {author} {\bibfnamefont {J.}~\bibnamefont {{Ma}}}, \bibinfo {author} {\bibfnamefont {N.}~\bibnamefont {{MacCrann}}},
  \bibinfo {author} {\bibfnamefont {C.}~\bibnamefont {{Magneville}}}, \bibinfo {author} {\bibfnamefont {L.}~\bibnamefont {{Makarem}}}, \bibinfo {author} {\bibfnamefont {M.}~\bibnamefont {{Manera}}}, \bibinfo {author} {\bibfnamefont {C.~J.}\ \bibnamefont {{Manser}}}, \bibinfo {author} {\bibfnamefont {R.}~\bibnamefont {{Marshall}}}, \bibinfo {author} {\bibfnamefont {P.}~\bibnamefont {{Martini}}}, \bibinfo {author} {\bibfnamefont {R.}~\bibnamefont {{Massey}}}, \bibinfo {author} {\bibfnamefont {T.}~\bibnamefont {{Matheson}}}, \bibinfo {author} {\bibfnamefont {J.}~\bibnamefont {{McCauley}}}, \bibinfo {author} {\bibfnamefont {P.}~\bibnamefont {{McDonald}}}, \bibinfo {author} {\bibfnamefont {I.~D.}\ \bibnamefont {{McGreer}}}, \bibinfo {author} {\bibfnamefont {A.}~\bibnamefont {{Meisner}}}, \bibinfo {author} {\bibfnamefont {N.}~\bibnamefont {{Metcalfe}}}, \bibinfo {author} {\bibfnamefont {T.~N.}\ \bibnamefont {{Miller}}}, \bibinfo {author} {\bibfnamefont {R.}~\bibnamefont {{Miquel}}}, \bibinfo {author} {\bibfnamefont
  {J.}~\bibnamefont {{Moustakas}}}, \bibinfo {author} {\bibfnamefont {A.}~\bibnamefont {{Myers}}}, \bibinfo {author} {\bibfnamefont {M.}~\bibnamefont {{Naik}}}, \bibinfo {author} {\bibfnamefont {J.~A.}\ \bibnamefont {{Newman}}}, \bibinfo {author} {\bibfnamefont {R.~C.}\ \bibnamefont {{Nichol}}}, \bibinfo {author} {\bibfnamefont {A.}~\bibnamefont {{Nicola}}}, \bibinfo {author} {\bibfnamefont {L.}~\bibnamefont {{Nicolati da Costa}}}, \bibinfo {author} {\bibfnamefont {J.}~\bibnamefont {{Nie}}}, \bibinfo {author} {\bibfnamefont {G.}~\bibnamefont {{Niz}}}, \bibinfo {author} {\bibfnamefont {P.}~\bibnamefont {{Norberg}}}, \bibinfo {author} {\bibfnamefont {B.}~\bibnamefont {{Nord}}}, \bibinfo {author} {\bibfnamefont {D.}~\bibnamefont {{Norman}}}, \bibinfo {author} {\bibfnamefont {P.}~\bibnamefont {{Nugent}}}, \bibinfo {author} {\bibfnamefont {T.}~\bibnamefont {{O'Brien}}}, \bibinfo {author} {\bibfnamefont {M.}~\bibnamefont {{Oh}}}, \bibinfo {author} {\bibfnamefont {K.~A.~G.}\ \bibnamefont {{Olsen}}}, \bibinfo
  {author} {\bibfnamefont {C.}~\bibnamefont {{Padilla}}}, \bibinfo {author} {\bibfnamefont {H.}~\bibnamefont {{Padmanabhan}}}, \bibinfo {author} {\bibfnamefont {N.}~\bibnamefont {{Padmanabhan}}}, \bibinfo {author} {\bibfnamefont {N.}~\bibnamefont {{Palanque-Delabrouille}}}, \bibinfo {author} {\bibfnamefont {A.}~\bibnamefont {{Palmese}}}, \bibinfo {author} {\bibfnamefont {D.}~\bibnamefont {{Pappalardo}}}, \bibinfo {author} {\bibfnamefont {I.}~\bibnamefont {{P{\^a}ris}}}, \bibinfo {author} {\bibfnamefont {C.}~\bibnamefont {{Park}}}, \bibinfo {author} {\bibfnamefont {A.}~\bibnamefont {{Patej}}}, \bibinfo {author} {\bibfnamefont {J.~A.}\ \bibnamefont {{Peacock}}}, \bibinfo {author} {\bibfnamefont {H.~V.}\ \bibnamefont {{Peiris}}}, \bibinfo {author} {\bibfnamefont {X.}~\bibnamefont {{Peng}}}, \bibinfo {author} {\bibfnamefont {W.~J.}\ \bibnamefont {{Percival}}}, \bibinfo {author} {\bibfnamefont {S.}~\bibnamefont {{Perruchot}}}, \bibinfo {author} {\bibfnamefont {M.~M.}\ \bibnamefont {{Pieri}}}, \bibinfo {author}
  {\bibfnamefont {R.}~\bibnamefont {{Pogge}}}, \bibinfo {author} {\bibfnamefont {J.~E.}\ \bibnamefont {{Pollack}}}, \bibinfo {author} {\bibfnamefont {C.}~\bibnamefont {{Poppett}}}, \bibinfo {author} {\bibfnamefont {F.}~\bibnamefont {{Prada}}}, \bibinfo {author} {\bibfnamefont {A.}~\bibnamefont {{Prakash}}}, \bibinfo {author} {\bibfnamefont {R.~G.}\ \bibnamefont {{Probst}}}, \bibinfo {author} {\bibfnamefont {D.}~\bibnamefont {{Rabinowitz}}}, \bibinfo {author} {\bibfnamefont {A.}~\bibnamefont {{Raichoor}}}, \bibinfo {author} {\bibfnamefont {C.~H.}\ \bibnamefont {{Ree}}}, \bibinfo {author} {\bibfnamefont {A.}~\bibnamefont {{Refregier}}}, \bibinfo {author} {\bibfnamefont {X.}~\bibnamefont {{Regal}}}, \bibinfo {author} {\bibfnamefont {B.}~\bibnamefont {{Reid}}}, \bibinfo {author} {\bibfnamefont {K.}~\bibnamefont {{Reil}}}, \bibinfo {author} {\bibfnamefont {M.}~\bibnamefont {{Rezaie}}}, \bibinfo {author} {\bibfnamefont {C.~M.}\ \bibnamefont {{Rockosi}}}, \bibinfo {author} {\bibfnamefont {N.}~\bibnamefont {{Roe}}},
  \bibinfo {author} {\bibfnamefont {S.}~\bibnamefont {{Ronayette}}}, \bibinfo {author} {\bibfnamefont {A.}~\bibnamefont {{Roodman}}}, \bibinfo {author} {\bibfnamefont {A.~J.}\ \bibnamefont {{Ross}}}, \bibinfo {author} {\bibfnamefont {N.~P.}\ \bibnamefont {{Ross}}}, \bibinfo {author} {\bibfnamefont {G.}~\bibnamefont {{Rossi}}}, \bibinfo {author} {\bibfnamefont {E.}~\bibnamefont {{Rozo}}}, \bibinfo {author} {\bibfnamefont {V.}~\bibnamefont {{Ruhlmann-Kleider}}}, \bibinfo {author} {\bibfnamefont {E.~S.}\ \bibnamefont {{Rykoff}}}, \bibinfo {author} {\bibfnamefont {C.}~\bibnamefont {{Sabiu}}}, \bibinfo {author} {\bibfnamefont {L.}~\bibnamefont {{Samushia}}}, \bibinfo {author} {\bibfnamefont {E.}~\bibnamefont {{Sanchez}}}, \bibinfo {author} {\bibfnamefont {J.}~\bibnamefont {{Sanchez}}}, \bibinfo {author} {\bibfnamefont {D.~J.}\ \bibnamefont {{Schlegel}}}, \bibinfo {author} {\bibfnamefont {M.}~\bibnamefont {{Schneider}}}, \bibinfo {author} {\bibfnamefont {M.}~\bibnamefont {{Schubnell}}}, \bibinfo {author}
  {\bibfnamefont {A.}~\bibnamefont {{Secroun}}}, \bibinfo {author} {\bibfnamefont {U.}~\bibnamefont {{Seljak}}}, \bibinfo {author} {\bibfnamefont {H.-J.}\ \bibnamefont {{Seo}}}, \bibinfo {author} {\bibfnamefont {S.}~\bibnamefont {{Serrano}}}, \bibinfo {author} {\bibfnamefont {A.}~\bibnamefont {{Shafieloo}}}, \bibinfo {author} {\bibfnamefont {H.}~\bibnamefont {{Shan}}}, \bibinfo {author} {\bibfnamefont {R.}~\bibnamefont {{Sharples}}}, \bibinfo {author} {\bibfnamefont {M.~J.}\ \bibnamefont {{Sholl}}}, \bibinfo {author} {\bibfnamefont {W.~V.}\ \bibnamefont {{Shourt}}}, \bibinfo {author} {\bibfnamefont {J.~H.}\ \bibnamefont {{Silber}}}, \bibinfo {author} {\bibfnamefont {D.~R.}\ \bibnamefont {{Silva}}}, \bibinfo {author} {\bibfnamefont {M.~M.}\ \bibnamefont {{Sirk}}}, \bibinfo {author} {\bibfnamefont {A.}~\bibnamefont {{Slosar}}}, \bibinfo {author} {\bibfnamefont {A.}~\bibnamefont {{Smith}}}, \bibinfo {author} {\bibfnamefont {G.~F.}\ \bibnamefont {{Smoot}}}, \bibinfo {author} {\bibfnamefont {D.}~\bibnamefont
  {{Som}}}, \bibinfo {author} {\bibfnamefont {Y.-S.}\ \bibnamefont {{Song}}}, \bibinfo {author} {\bibfnamefont {D.}~\bibnamefont {{Sprayberry}}}, \bibinfo {author} {\bibfnamefont {R.}~\bibnamefont {{Staten}}}, \bibinfo {author} {\bibfnamefont {A.}~\bibnamefont {{Stefanik}}}, \bibinfo {author} {\bibfnamefont {G.}~\bibnamefont {{Tarle}}}, \bibinfo {author} {\bibfnamefont {S.}~\bibnamefont {{Sien Tie}}}, \bibinfo {author} {\bibfnamefont {J.~L.}\ \bibnamefont {{Tinker}}}, \bibinfo {author} {\bibfnamefont {R.}~\bibnamefont {{Tojeiro}}}, \bibinfo {author} {\bibfnamefont {F.}~\bibnamefont {{Valdes}}}, \bibinfo {author} {\bibfnamefont {O.}~\bibnamefont {{Valenzuela}}}, \bibinfo {author} {\bibfnamefont {M.}~\bibnamefont {{Valluri}}}, \bibinfo {author} {\bibfnamefont {M.}~\bibnamefont {{Vargas-Magana}}}, \bibinfo {author} {\bibfnamefont {L.}~\bibnamefont {{Verde}}}, \bibinfo {author} {\bibfnamefont {A.~R.}\ \bibnamefont {{Walker}}}, \bibinfo {author} {\bibfnamefont {J.}~\bibnamefont {{Wang}}}, \bibinfo {author}
  {\bibfnamefont {Y.}~\bibnamefont {{Wang}}}, \bibinfo {author} {\bibfnamefont {B.~A.}\ \bibnamefont {{Weaver}}}, \bibinfo {author} {\bibfnamefont {C.}~\bibnamefont {{Weaverdyck}}}, \bibinfo {author} {\bibfnamefont {R.~H.}\ \bibnamefont {{Wechsler}}}, \bibinfo {author} {\bibfnamefont {D.~H.}\ \bibnamefont {{Weinberg}}}, \bibinfo {author} {\bibfnamefont {M.}~\bibnamefont {{White}}}, \bibinfo {author} {\bibfnamefont {Q.}~\bibnamefont {{Yang}}}, \bibinfo {author} {\bibfnamefont {C.}~\bibnamefont {{Yeche}}}, \bibinfo {author} {\bibfnamefont {T.}~\bibnamefont {{Zhang}}}, \bibinfo {author} {\bibfnamefont {G.-B.}\ \bibnamefont {{Zhao}}}, \bibinfo {author} {\bibfnamefont {Y.}~\bibnamefont {{Zheng}}}, \bibinfo {author} {\bibfnamefont {X.}~\bibnamefont {{Zhou}}}, \bibinfo {author} {\bibfnamefont {Z.}~\bibnamefont {{Zhou}}}, \bibinfo {author} {\bibfnamefont {Y.}~\bibnamefont {{Zhu}}}, \bibinfo {author} {\bibfnamefont {H.}~\bibnamefont {{Zou}}},\ and\ \bibinfo {author} {\bibfnamefont {Y.}~\bibnamefont {{Zu}}},\
  }\bibfield  {title} {\bibinfo {title} {{The DESI Experiment Part II: Instrument Design}},\ }\href@noop {} {\bibfield  {journal} {\bibinfo  {journal} {arXiv e-prints}\ ,\ \bibinfo {eid} {arXiv:1611.00037}} (\bibinfo {year} {2016}{\natexlab{a}})},\ \Eprint {https://arxiv.org/abs/1611.00037} {arXiv:1611.00037 [astro-ph.IM]} \BibitemShut {NoStop}%
\bibitem [{\citenamefont {{Silber}}\ \emph {et~al.}(2023)\citenamefont {{Silber}}, \citenamefont {{Fagrelius}}, \citenamefont {{Fanning}}, \citenamefont {{Schubnell}}, \citenamefont {{Aguilar}}, \citenamefont {{Ahlen}}, \citenamefont {{Ameel}}, \citenamefont {{Ballester}}, \citenamefont {{Baltay}}, \citenamefont {{Bebek}}, \citenamefont {{Benton Beard}}, \citenamefont {{Besuner}}, \citenamefont {{Cardiel-Sas}}, \citenamefont {{Casas}}, \citenamefont {{Castander}}, \citenamefont {{Claybaugh}}, \citenamefont {{Dobson}}, \citenamefont {{Duan}}, \citenamefont {{Dunlop}}, \citenamefont {{Edelstein}}, \citenamefont {{Emmet}}, \citenamefont {{Elliott}}, \citenamefont {{Evatt}}, \citenamefont {{Gershkovich}}, \citenamefont {{Guy}}, \citenamefont {{Harris}}, \citenamefont {{Heetderks}}, \citenamefont {{Heetderks}}, \citenamefont {{Honscheid}}, \citenamefont {{Illa}}, \citenamefont {{Jelinsky}}, \citenamefont {{Jelinsky}}, \citenamefont {{Jimenez}}, \citenamefont {{Karcher}}, \citenamefont {{Kent}}, \citenamefont
  {{Kirkby}}, \citenamefont {{Kneib}}, \citenamefont {{Lambert}}, \citenamefont {{Lampton}}, \citenamefont {{Leitner}}, \citenamefont {{Levi}}, \citenamefont {{McCauley}}, \citenamefont {{Meisner}}, \citenamefont {{Miller}}, \citenamefont {{Miquel}}, \citenamefont {{Mundet}}, \citenamefont {{Poppett}}, \citenamefont {{Rabinowitz}}, \citenamefont {{Reil}}, \citenamefont {{Roman}}, \citenamefont {{Schlegel}}, \citenamefont {{Serrano}}, \citenamefont {{Van Shourt}}, \citenamefont {{Sprayberry}}, \citenamefont {{Tarl{\'e}}}, \citenamefont {{Tie}}, \citenamefont {{Weaverdyck}}, \citenamefont {{Zhang}}, \citenamefont {{Azzaro}}, \citenamefont {{Bailey}}, \citenamefont {{Becerril}}, \citenamefont {{Blackwell}}, \citenamefont {{Bouri}}, \citenamefont {{Brooks}}, \citenamefont {{Buckley-Geer}}, \citenamefont {{Castro}}, \citenamefont {{Derwent}}, \citenamefont {{Dey}}, \citenamefont {{Dhungana}}, \citenamefont {{Doel}}, \citenamefont {{Eisenstein}}, \citenamefont {{Fahim}}, \citenamefont {{Garcia-Bellido}},
  \citenamefont {{Gazta{\~n}aga}}, \citenamefont {{A Gontcho}}, \citenamefont {{Gutierrez}}, \citenamefont {{H{\"o}rler}}, \citenamefont {{Kehoe}}, \citenamefont {{Kisner}}, \citenamefont {{Kremin}}, \citenamefont {{Kronig}}, \citenamefont {{Landriau}}, \citenamefont {{Le Guillou}}, \citenamefont {{Martini}}, \citenamefont {{Moustakas}}, \citenamefont {{Palanque-Delabrouille}}, \citenamefont {{Peng}}, \citenamefont {{Percival}}, \citenamefont {{Prada}}, \citenamefont {{Allende Prieto}}, \citenamefont {{de Rivera}}, \citenamefont {{Sanchez}}, \citenamefont {{Sanchez}}, \citenamefont {{Sharples}}, \citenamefont {{Soares-Santos}}, \citenamefont {{Schlafly}}, \citenamefont {{Weaver}}, \citenamefont {{Zhou}}, \citenamefont {{Zhu}}, \citenamefont {{Zou}},\ and\ \citenamefont {{DESI Collaboration}}}]{2023AJ....165....9S}%
  \BibitemOpen
  \bibfield  {author} {\bibinfo {author} {\bibfnamefont {J.~H.}\ \bibnamefont {{Silber}}}, \bibinfo {author} {\bibfnamefont {P.}~\bibnamefont {{Fagrelius}}}, \bibinfo {author} {\bibfnamefont {K.}~\bibnamefont {{Fanning}}}, \bibinfo {author} {\bibfnamefont {M.}~\bibnamefont {{Schubnell}}}, \bibinfo {author} {\bibfnamefont {J.~N.}\ \bibnamefont {{Aguilar}}}, \bibinfo {author} {\bibfnamefont {S.}~\bibnamefont {{Ahlen}}}, \bibinfo {author} {\bibfnamefont {J.}~\bibnamefont {{Ameel}}}, \bibinfo {author} {\bibfnamefont {O.}~\bibnamefont {{Ballester}}}, \bibinfo {author} {\bibfnamefont {C.}~\bibnamefont {{Baltay}}}, \bibinfo {author} {\bibfnamefont {C.}~\bibnamefont {{Bebek}}}, \bibinfo {author} {\bibfnamefont {D.}~\bibnamefont {{Benton Beard}}}, \bibinfo {author} {\bibfnamefont {R.}~\bibnamefont {{Besuner}}}, \bibinfo {author} {\bibfnamefont {L.}~\bibnamefont {{Cardiel-Sas}}}, \bibinfo {author} {\bibfnamefont {R.}~\bibnamefont {{Casas}}}, \bibinfo {author} {\bibfnamefont {F.~J.}\ \bibnamefont {{Castander}}},
  \bibinfo {author} {\bibfnamefont {T.}~\bibnamefont {{Claybaugh}}}, \bibinfo {author} {\bibfnamefont {C.}~\bibnamefont {{Dobson}}}, \bibinfo {author} {\bibfnamefont {Y.}~\bibnamefont {{Duan}}}, \bibinfo {author} {\bibfnamefont {P.}~\bibnamefont {{Dunlop}}}, \bibinfo {author} {\bibfnamefont {J.}~\bibnamefont {{Edelstein}}}, \bibinfo {author} {\bibfnamefont {W.~T.}\ \bibnamefont {{Emmet}}}, \bibinfo {author} {\bibfnamefont {A.}~\bibnamefont {{Elliott}}}, \bibinfo {author} {\bibfnamefont {M.}~\bibnamefont {{Evatt}}}, \bibinfo {author} {\bibfnamefont {I.}~\bibnamefont {{Gershkovich}}}, \bibinfo {author} {\bibfnamefont {J.}~\bibnamefont {{Guy}}}, \bibinfo {author} {\bibfnamefont {S.}~\bibnamefont {{Harris}}}, \bibinfo {author} {\bibfnamefont {H.}~\bibnamefont {{Heetderks}}}, \bibinfo {author} {\bibfnamefont {I.}~\bibnamefont {{Heetderks}}}, \bibinfo {author} {\bibfnamefont {K.}~\bibnamefont {{Honscheid}}}, \bibinfo {author} {\bibfnamefont {J.~M.}\ \bibnamefont {{Illa}}}, \bibinfo {author} {\bibfnamefont
  {P.}~\bibnamefont {{Jelinsky}}}, \bibinfo {author} {\bibfnamefont {S.~R.}\ \bibnamefont {{Jelinsky}}}, \bibinfo {author} {\bibfnamefont {J.}~\bibnamefont {{Jimenez}}}, \bibinfo {author} {\bibfnamefont {A.}~\bibnamefont {{Karcher}}}, \bibinfo {author} {\bibfnamefont {S.}~\bibnamefont {{Kent}}}, \bibinfo {author} {\bibfnamefont {D.}~\bibnamefont {{Kirkby}}}, \bibinfo {author} {\bibfnamefont {J.-P.}\ \bibnamefont {{Kneib}}}, \bibinfo {author} {\bibfnamefont {A.}~\bibnamefont {{Lambert}}}, \bibinfo {author} {\bibfnamefont {M.}~\bibnamefont {{Lampton}}}, \bibinfo {author} {\bibfnamefont {D.}~\bibnamefont {{Leitner}}}, \bibinfo {author} {\bibfnamefont {M.}~\bibnamefont {{Levi}}}, \bibinfo {author} {\bibfnamefont {J.}~\bibnamefont {{McCauley}}}, \bibinfo {author} {\bibfnamefont {A.}~\bibnamefont {{Meisner}}}, \bibinfo {author} {\bibfnamefont {T.~N.}\ \bibnamefont {{Miller}}}, \bibinfo {author} {\bibfnamefont {R.}~\bibnamefont {{Miquel}}}, \bibinfo {author} {\bibfnamefont {J.}~\bibnamefont {{Mundet}}}, \bibinfo
  {author} {\bibfnamefont {C.}~\bibnamefont {{Poppett}}}, \bibinfo {author} {\bibfnamefont {D.}~\bibnamefont {{Rabinowitz}}}, \bibinfo {author} {\bibfnamefont {K.}~\bibnamefont {{Reil}}}, \bibinfo {author} {\bibfnamefont {D.}~\bibnamefont {{Roman}}}, \bibinfo {author} {\bibfnamefont {D.}~\bibnamefont {{Schlegel}}}, \bibinfo {author} {\bibfnamefont {S.}~\bibnamefont {{Serrano}}}, \bibinfo {author} {\bibfnamefont {W.}~\bibnamefont {{Van Shourt}}}, \bibinfo {author} {\bibfnamefont {D.}~\bibnamefont {{Sprayberry}}}, \bibinfo {author} {\bibfnamefont {G.}~\bibnamefont {{Tarl{\'e}}}}, \bibinfo {author} {\bibfnamefont {S.~S.}\ \bibnamefont {{Tie}}}, \bibinfo {author} {\bibfnamefont {C.}~\bibnamefont {{Weaverdyck}}}, \bibinfo {author} {\bibfnamefont {K.}~\bibnamefont {{Zhang}}}, \bibinfo {author} {\bibfnamefont {M.}~\bibnamefont {{Azzaro}}}, \bibinfo {author} {\bibfnamefont {S.}~\bibnamefont {{Bailey}}}, \bibinfo {author} {\bibfnamefont {S.}~\bibnamefont {{Becerril}}}, \bibinfo {author} {\bibfnamefont
  {T.}~\bibnamefont {{Blackwell}}}, \bibinfo {author} {\bibfnamefont {M.}~\bibnamefont {{Bouri}}}, \bibinfo {author} {\bibfnamefont {D.}~\bibnamefont {{Brooks}}}, \bibinfo {author} {\bibfnamefont {E.}~\bibnamefont {{Buckley-Geer}}}, \bibinfo {author} {\bibfnamefont {J.~P.}\ \bibnamefont {{Castro}}}, \bibinfo {author} {\bibfnamefont {M.}~\bibnamefont {{Derwent}}}, \bibinfo {author} {\bibfnamefont {A.}~\bibnamefont {{Dey}}}, \bibinfo {author} {\bibfnamefont {G.}~\bibnamefont {{Dhungana}}}, \bibinfo {author} {\bibfnamefont {P.}~\bibnamefont {{Doel}}}, \bibinfo {author} {\bibfnamefont {D.~J.}\ \bibnamefont {{Eisenstein}}}, \bibinfo {author} {\bibfnamefont {N.}~\bibnamefont {{Fahim}}}, \bibinfo {author} {\bibfnamefont {J.}~\bibnamefont {{Garcia-Bellido}}}, \bibinfo {author} {\bibfnamefont {E.}~\bibnamefont {{Gazta{\~n}aga}}}, \bibinfo {author} {\bibfnamefont {S.~G.}\ \bibnamefont {{A Gontcho}}}, \bibinfo {author} {\bibfnamefont {G.}~\bibnamefont {{Gutierrez}}}, \bibinfo {author} {\bibfnamefont {P.}~\bibnamefont
  {{H{\"o}rler}}}, \bibinfo {author} {\bibfnamefont {R.}~\bibnamefont {{Kehoe}}}, \bibinfo {author} {\bibfnamefont {T.}~\bibnamefont {{Kisner}}}, \bibinfo {author} {\bibfnamefont {A.}~\bibnamefont {{Kremin}}}, \bibinfo {author} {\bibfnamefont {L.}~\bibnamefont {{Kronig}}}, \bibinfo {author} {\bibfnamefont {M.}~\bibnamefont {{Landriau}}}, \bibinfo {author} {\bibfnamefont {L.}~\bibnamefont {{Le Guillou}}}, \bibinfo {author} {\bibfnamefont {P.}~\bibnamefont {{Martini}}}, \bibinfo {author} {\bibfnamefont {J.}~\bibnamefont {{Moustakas}}}, \bibinfo {author} {\bibfnamefont {N.}~\bibnamefont {{Palanque-Delabrouille}}}, \bibinfo {author} {\bibfnamefont {X.}~\bibnamefont {{Peng}}}, \bibinfo {author} {\bibfnamefont {W.}~\bibnamefont {{Percival}}}, \bibinfo {author} {\bibfnamefont {F.}~\bibnamefont {{Prada}}}, \bibinfo {author} {\bibfnamefont {C.}~\bibnamefont {{Allende Prieto}}}, \bibinfo {author} {\bibfnamefont {G.~G.}\ \bibnamefont {{de Rivera}}}, \bibinfo {author} {\bibfnamefont {E.}~\bibnamefont {{Sanchez}}},
  \bibinfo {author} {\bibfnamefont {J.}~\bibnamefont {{Sanchez}}}, \bibinfo {author} {\bibfnamefont {R.}~\bibnamefont {{Sharples}}}, \bibinfo {author} {\bibfnamefont {M.}~\bibnamefont {{Soares-Santos}}}, \bibinfo {author} {\bibfnamefont {E.}~\bibnamefont {{Schlafly}}}, \bibinfo {author} {\bibfnamefont {B.~A.}\ \bibnamefont {{Weaver}}}, \bibinfo {author} {\bibfnamefont {Z.}~\bibnamefont {{Zhou}}}, \bibinfo {author} {\bibfnamefont {Y.}~\bibnamefont {{Zhu}}}, \bibinfo {author} {\bibfnamefont {H.}~\bibnamefont {{Zou}}},\ and\ \bibinfo {author} {\bibnamefont {{DESI Collaboration}}},\ }\bibfield  {title} {\bibinfo {title} {{The Robotic Multiobject Focal Plane System of the Dark Energy Spectroscopic Instrument (DESI)}},\ }\href {https://doi.org/10.3847/1538-3881/ac9ab1} {\bibfield  {journal} {\bibinfo  {journal} {\aj}\ }\textbf {\bibinfo {volume} {165}},\ \bibinfo {eid} {9} (\bibinfo {year} {2023})},\ \Eprint {https://arxiv.org/abs/2205.09014} {arXiv:2205.09014 [astro-ph.IM]} \BibitemShut {NoStop}%
\bibitem [{\citenamefont {{Miller}}\ \emph {et~al.}(2023)\citenamefont {{Miller}}, \citenamefont {{Doel}}, \citenamefont {{Gutierrez}}, \citenamefont {{Besuner}}, \citenamefont {{Brooks}}, \citenamefont {{Gallo}}, \citenamefont {{Heetderks}}, \citenamefont {{Jelinsky}}, \citenamefont {{Kent}}, \citenamefont {{Lampton}}, \citenamefont {{Levi}}, \citenamefont {{Liang}}, \citenamefont {{Meisner}}, \citenamefont {{Sholl}}, \citenamefont {{Silber}}, \citenamefont {{Sprayberry}}, \citenamefont {{Aguilar}}, \citenamefont {{de la Macorra}}, \citenamefont {{Eisenstein}}, \citenamefont {{Fanning}}, \citenamefont {{Font-Ribera}}, \citenamefont {{Gaztanaga}}, \citenamefont {{Gontcho}}, \citenamefont {{Honscheid}}, \citenamefont {{Jimenez}}, \citenamefont {{Joyce}}, \citenamefont {{Kehoe}}, \citenamefont {{Kisner}}, \citenamefont {{Kremin}}, \citenamefont {{Landriau}}, \citenamefont {{Le Guillou}}, \citenamefont {{Magneville}}, \citenamefont {{Martini}}, \citenamefont {{Miquel}}, \citenamefont {{Moustakas}}, \citenamefont
  {{Nie}}, \citenamefont {{Percival}}, \citenamefont {{Poppett}}, \citenamefont {{Prada}}, \citenamefont {{Rossi}}, \citenamefont {{Schlegel}}, \citenamefont {{Schubnell}}, \citenamefont {{Seo}}, \citenamefont {{Sharples}}, \citenamefont {{Tarle}}, \citenamefont {{Vargas-Magana}},\ and\ \citenamefont {{Zhou}}}]{2023arXiv230606310M}%
  \BibitemOpen
  \bibfield  {author} {\bibinfo {author} {\bibfnamefont {T.~N.}\ \bibnamefont {{Miller}}}, \bibinfo {author} {\bibfnamefont {P.}~\bibnamefont {{Doel}}}, \bibinfo {author} {\bibfnamefont {G.}~\bibnamefont {{Gutierrez}}}, \bibinfo {author} {\bibfnamefont {R.}~\bibnamefont {{Besuner}}}, \bibinfo {author} {\bibfnamefont {D.}~\bibnamefont {{Brooks}}}, \bibinfo {author} {\bibfnamefont {G.}~\bibnamefont {{Gallo}}}, \bibinfo {author} {\bibfnamefont {H.}~\bibnamefont {{Heetderks}}}, \bibinfo {author} {\bibfnamefont {P.}~\bibnamefont {{Jelinsky}}}, \bibinfo {author} {\bibfnamefont {S.~M.}\ \bibnamefont {{Kent}}}, \bibinfo {author} {\bibfnamefont {M.}~\bibnamefont {{Lampton}}}, \bibinfo {author} {\bibfnamefont {M.}~\bibnamefont {{Levi}}}, \bibinfo {author} {\bibfnamefont {M.}~\bibnamefont {{Liang}}}, \bibinfo {author} {\bibfnamefont {A.}~\bibnamefont {{Meisner}}}, \bibinfo {author} {\bibfnamefont {M.~J.}\ \bibnamefont {{Sholl}}}, \bibinfo {author} {\bibfnamefont {J.~H.}\ \bibnamefont {{Silber}}}, \bibinfo {author}
  {\bibfnamefont {D.}~\bibnamefont {{Sprayberry}}}, \bibinfo {author} {\bibfnamefont {J.~N.}\ \bibnamefont {{Aguilar}}}, \bibinfo {author} {\bibfnamefont {A.}~\bibnamefont {{de la Macorra}}}, \bibinfo {author} {\bibfnamefont {D.}~\bibnamefont {{Eisenstein}}}, \bibinfo {author} {\bibfnamefont {K.}~\bibnamefont {{Fanning}}}, \bibinfo {author} {\bibfnamefont {A.}~\bibnamefont {{Font-Ribera}}}, \bibinfo {author} {\bibfnamefont {E.}~\bibnamefont {{Gaztanaga}}}, \bibinfo {author} {\bibfnamefont {S.~G.~A.}\ \bibnamefont {{Gontcho}}}, \bibinfo {author} {\bibfnamefont {K.}~\bibnamefont {{Honscheid}}}, \bibinfo {author} {\bibfnamefont {J.}~\bibnamefont {{Jimenez}}}, \bibinfo {author} {\bibfnamefont {D.}~\bibnamefont {{Joyce}}}, \bibinfo {author} {\bibfnamefont {R.}~\bibnamefont {{Kehoe}}}, \bibinfo {author} {\bibfnamefont {T.}~\bibnamefont {{Kisner}}}, \bibinfo {author} {\bibfnamefont {A.}~\bibnamefont {{Kremin}}}, \bibinfo {author} {\bibfnamefont {M.}~\bibnamefont {{Landriau}}}, \bibinfo {author} {\bibfnamefont
  {L.}~\bibnamefont {{Le Guillou}}}, \bibinfo {author} {\bibfnamefont {C.}~\bibnamefont {{Magneville}}}, \bibinfo {author} {\bibfnamefont {P.}~\bibnamefont {{Martini}}}, \bibinfo {author} {\bibfnamefont {R.}~\bibnamefont {{Miquel}}}, \bibinfo {author} {\bibfnamefont {J.}~\bibnamefont {{Moustakas}}}, \bibinfo {author} {\bibfnamefont {J.}~\bibnamefont {{Nie}}}, \bibinfo {author} {\bibfnamefont {W.}~\bibnamefont {{Percival}}}, \bibinfo {author} {\bibfnamefont {C.}~\bibnamefont {{Poppett}}}, \bibinfo {author} {\bibfnamefont {F.}~\bibnamefont {{Prada}}}, \bibinfo {author} {\bibfnamefont {G.}~\bibnamefont {{Rossi}}}, \bibinfo {author} {\bibfnamefont {D.}~\bibnamefont {{Schlegel}}}, \bibinfo {author} {\bibfnamefont {M.}~\bibnamefont {{Schubnell}}}, \bibinfo {author} {\bibfnamefont {H.-J.}\ \bibnamefont {{Seo}}}, \bibinfo {author} {\bibfnamefont {R.}~\bibnamefont {{Sharples}}}, \bibinfo {author} {\bibfnamefont {G.}~\bibnamefont {{Tarle}}}, \bibinfo {author} {\bibfnamefont {M.}~\bibnamefont {{Vargas-Magana}}},\ and\
  \bibinfo {author} {\bibfnamefont {Z.}~\bibnamefont {{Zhou}}},\ }\bibfield  {title} {\bibinfo {title} {{The Optical Corrector for the Dark Energy Spectroscopic Instrument}},\ }\href {https://doi.org/10.48550/arXiv.2306.06310} {\bibfield  {journal} {\bibinfo  {journal} {arXiv e-prints}\ ,\ \bibinfo {eid} {arXiv:2306.06310}} (\bibinfo {year} {2023})},\ \Eprint {https://arxiv.org/abs/2306.06310} {arXiv:2306.06310 [astro-ph.IM]} \BibitemShut {NoStop}%
\bibitem [{\citenamefont {{Levi}}\ \emph {et~al.}(2013)\citenamefont {{Levi}}, \citenamefont {{Bebek}}, \citenamefont {{Beers}}, \citenamefont {{Blum}}, \citenamefont {{Cahn}}, \citenamefont {{Eisenstein}}, \citenamefont {{Flaugher}}, \citenamefont {{Honscheid}}, \citenamefont {{Kron}}, \citenamefont {{Lahav}}, \citenamefont {{McDonald}}, \citenamefont {{Roe}}, \citenamefont {{Schlegel}},\ and\ \citenamefont {{representing the DESI collaboration}}}]{2013arXiv1308.0847L}%
  \BibitemOpen
  \bibfield  {author} {\bibinfo {author} {\bibfnamefont {M.}~\bibnamefont {{Levi}}}, \bibinfo {author} {\bibfnamefont {C.}~\bibnamefont {{Bebek}}}, \bibinfo {author} {\bibfnamefont {T.}~\bibnamefont {{Beers}}}, \bibinfo {author} {\bibfnamefont {R.}~\bibnamefont {{Blum}}}, \bibinfo {author} {\bibfnamefont {R.}~\bibnamefont {{Cahn}}}, \bibinfo {author} {\bibfnamefont {D.}~\bibnamefont {{Eisenstein}}}, \bibinfo {author} {\bibfnamefont {B.}~\bibnamefont {{Flaugher}}}, \bibinfo {author} {\bibfnamefont {K.}~\bibnamefont {{Honscheid}}}, \bibinfo {author} {\bibfnamefont {R.}~\bibnamefont {{Kron}}}, \bibinfo {author} {\bibfnamefont {O.}~\bibnamefont {{Lahav}}}, \bibinfo {author} {\bibfnamefont {P.}~\bibnamefont {{McDonald}}}, \bibinfo {author} {\bibfnamefont {N.}~\bibnamefont {{Roe}}}, \bibinfo {author} {\bibfnamefont {D.}~\bibnamefont {{Schlegel}}},\ and\ \bibinfo {author} {\bibnamefont {{representing the DESI collaboration}}},\ }\bibfield  {title} {\bibinfo {title} {{The DESI Experiment, a whitepaper for Snowmass
  2013}},\ }\href@noop {} {\bibfield  {journal} {\bibinfo  {journal} {arXiv e-prints}\ ,\ \bibinfo {eid} {arXiv:1308.0847}} (\bibinfo {year} {2013})},\ \Eprint {https://arxiv.org/abs/1308.0847} {arXiv:1308.0847 [astro-ph.CO]} \BibitemShut {NoStop}%
\bibitem [{\citenamefont {{DESI Collaboration}}\ \emph {et~al.}(2016{\natexlab{b}})\citenamefont {{DESI Collaboration}}, \citenamefont {{Aghamousa}}, \citenamefont {{Aguilar}}, \citenamefont {{Ahlen}}, \citenamefont {{Alam}}, \citenamefont {{Allen}}, \citenamefont {{Allende Prieto}}, \citenamefont {{Annis}}, \citenamefont {{Bailey}}, \citenamefont {{Balland}}, \citenamefont {{Ballester}}, \citenamefont {{Baltay}}, \citenamefont {{Beaufore}}, \citenamefont {{Bebek}}, \citenamefont {{Beers}}, \citenamefont {{Bell}}, \citenamefont {{Bernal}}, \citenamefont {{Besuner}}, \citenamefont {{Beutler}}, \citenamefont {{Blake}}, \citenamefont {{Bleuler}}, \citenamefont {{Blomqvist}}, \citenamefont {{Blum}}, \citenamefont {{Bolton}}, \citenamefont {{Briceno}}, \citenamefont {{Brooks}}, \citenamefont {{Brownstein}}, \citenamefont {{Buckley-Geer}}, \citenamefont {{Burden}}, \citenamefont {{Burtin}}, \citenamefont {{Busca}}, \citenamefont {{Cahn}}, \citenamefont {{Cai}}, \citenamefont {{Cardiel-Sas}}, \citenamefont
  {{Carlberg}}, \citenamefont {{Carton}}, \citenamefont {{Casas}}, \citenamefont {{Castander}}, \citenamefont {{Cervantes-Cota}}, \citenamefont {{Claybaugh}}, \citenamefont {{Close}}, \citenamefont {{Coker}}, \citenamefont {{Cole}}, \citenamefont {{Comparat}}, \citenamefont {{Cooper}}, \citenamefont {{Cousinou}}, \citenamefont {{Crocce}}, \citenamefont {{Cuby}}, \citenamefont {{Cunningham}}, \citenamefont {{Davis}}, \citenamefont {{Dawson}}, \citenamefont {{de la Macorra}}, \citenamefont {{De Vicente}}, \citenamefont {{Delubac}}, \citenamefont {{Derwent}}, \citenamefont {{Dey}}, \citenamefont {{Dhungana}}, \citenamefont {{Ding}}, \citenamefont {{Doel}}, \citenamefont {{Duan}}, \citenamefont {{Ealet}}, \citenamefont {{Edelstein}}, \citenamefont {{Eftekharzadeh}}, \citenamefont {{Eisenstein}}, \citenamefont {{Elliott}}, \citenamefont {{Escoffier}}, \citenamefont {{Evatt}}, \citenamefont {{Fagrelius}}, \citenamefont {{Fan}}, \citenamefont {{Fanning}}, \citenamefont {{Farahi}}, \citenamefont {{Farihi}},
  \citenamefont {{Favole}}, \citenamefont {{Feng}}, \citenamefont {{Fernandez}}, \citenamefont {{Findlay}}, \citenamefont {{Finkbeiner}}, \citenamefont {{Fitzpatrick}}, \citenamefont {{Flaugher}}, \citenamefont {{Flender}}, \citenamefont {{Font-Ribera}}, \citenamefont {{Forero-Romero}}, \citenamefont {{Fosalba}}, \citenamefont {{Frenk}}, \citenamefont {{Fumagalli}}, \citenamefont {{Gaensicke}}, \citenamefont {{Gallo}}, \citenamefont {{Garcia-Bellido}}, \citenamefont {{Gaztanaga}}, \citenamefont {{Pietro Gentile Fusillo}}, \citenamefont {{Gerard}}, \citenamefont {{Gershkovich}}, \citenamefont {{Giannantonio}}, \citenamefont {{Gillet}}, \citenamefont {{Gonzalez-de-Rivera}}, \citenamefont {{Gonzalez-Perez}}, \citenamefont {{Gott}}, \citenamefont {{Graur}}, \citenamefont {{Gutierrez}}, \citenamefont {{Guy}}, \citenamefont {{Habib}}, \citenamefont {{Heetderks}}, \citenamefont {{Heetderks}}, \citenamefont {{Heitmann}}, \citenamefont {{Hellwing}}, \citenamefont {{Herrera}}, \citenamefont {{Ho}}, \citenamefont
  {{Holland}}, \citenamefont {{Honscheid}}, \citenamefont {{Huff}}, \citenamefont {{Hutchinson}}, \citenamefont {{Huterer}}, \citenamefont {{Hwang}}, \citenamefont {{Illa Laguna}}, \citenamefont {{Ishikawa}}, \citenamefont {{Jacobs}}, \citenamefont {{Jeffrey}}, \citenamefont {{Jelinsky}}, \citenamefont {{Jennings}}, \citenamefont {{Jiang}}, \citenamefont {{Jimenez}}, \citenamefont {{Johnson}}, \citenamefont {{Joyce}}, \citenamefont {{Jullo}}, \citenamefont {{Juneau}}, \citenamefont {{Kama}}, \citenamefont {{Karcher}}, \citenamefont {{Karkar}}, \citenamefont {{Kehoe}}, \citenamefont {{Kennamer}}, \citenamefont {{Kent}}, \citenamefont {{Kilbinger}}, \citenamefont {{Kim}}, \citenamefont {{Kirkby}}, \citenamefont {{Kisner}}, \citenamefont {{Kitanidis}}, \citenamefont {{Kneib}}, \citenamefont {{Koposov}}, \citenamefont {{Kovacs}}, \citenamefont {{Koyama}}, \citenamefont {{Kremin}}, \citenamefont {{Kron}}, \citenamefont {{Kronig}}, \citenamefont {{Kueter-Young}}, \citenamefont {{Lacey}}, \citenamefont {{Lafever}},
  \citenamefont {{Lahav}}, \citenamefont {{Lambert}}, \citenamefont {{Lampton}}, \citenamefont {{Landriau}}, \citenamefont {{Lang}}, \citenamefont {{Lauer}}, \citenamefont {{Le Goff}}, \citenamefont {{Le Guillou}}, \citenamefont {{Le Van Suu}}, \citenamefont {{Lee}}, \citenamefont {{Lee}}, \citenamefont {{Leitner}}, \citenamefont {{Lesser}}, \citenamefont {{Levi}}, \citenamefont {{L'Huillier}}, \citenamefont {{Li}}, \citenamefont {{Liang}}, \citenamefont {{Lin}}, \citenamefont {{Linder}}, \citenamefont {{Loebman}}, \citenamefont {{Luki{\'c}}}, \citenamefont {{Ma}}, \citenamefont {{MacCrann}}, \citenamefont {{Magneville}}, \citenamefont {{Makarem}}, \citenamefont {{Manera}}, \citenamefont {{Manser}}, \citenamefont {{Marshall}}, \citenamefont {{Martini}}, \citenamefont {{Massey}}, \citenamefont {{Matheson}}, \citenamefont {{McCauley}}, \citenamefont {{McDonald}}, \citenamefont {{McGreer}}, \citenamefont {{Meisner}}, \citenamefont {{Metcalfe}}, \citenamefont {{Miller}}, \citenamefont {{Miquel}}, \citenamefont
  {{Moustakas}}, \citenamefont {{Myers}}, \citenamefont {{Naik}}, \citenamefont {{Newman}}, \citenamefont {{Nichol}}, \citenamefont {{Nicola}}, \citenamefont {{Nicolati da Costa}}, \citenamefont {{Nie}}, \citenamefont {{Niz}}, \citenamefont {{Norberg}}, \citenamefont {{Nord}}, \citenamefont {{Norman}}, \citenamefont {{Nugent}}, \citenamefont {{O'Brien}}, \citenamefont {{Oh}}, \citenamefont {{Olsen}}, \citenamefont {{Padilla}}, \citenamefont {{Padmanabhan}}, \citenamefont {{Padmanabhan}}, \citenamefont {{Palanque-Delabrouille}}, \citenamefont {{Palmese}}, \citenamefont {{Pappalardo}}, \citenamefont {{P{\^a}ris}}, \citenamefont {{Park}}, \citenamefont {{Patej}}, \citenamefont {{Peacock}}, \citenamefont {{Peiris}}, \citenamefont {{Peng}}, \citenamefont {{Percival}}, \citenamefont {{Perruchot}}, \citenamefont {{Pieri}}, \citenamefont {{Pogge}}, \citenamefont {{Pollack}}, \citenamefont {{Poppett}}, \citenamefont {{Prada}}, \citenamefont {{Prakash}}, \citenamefont {{Probst}}, \citenamefont {{Rabinowitz}},
  \citenamefont {{Raichoor}}, \citenamefont {{Ree}}, \citenamefont {{Refregier}}, \citenamefont {{Regal}}, \citenamefont {{Reid}}, \citenamefont {{Reil}}, \citenamefont {{Rezaie}}, \citenamefont {{Rockosi}}, \citenamefont {{Roe}}, \citenamefont {{Ronayette}}, \citenamefont {{Roodman}}, \citenamefont {{Ross}}, \citenamefont {{Ross}}, \citenamefont {{Rossi}}, \citenamefont {{Rozo}}, \citenamefont {{Ruhlmann-Kleider}}, \citenamefont {{Rykoff}}, \citenamefont {{Sabiu}}, \citenamefont {{Samushia}}, \citenamefont {{Sanchez}}, \citenamefont {{Sanchez}}, \citenamefont {{Schlegel}}, \citenamefont {{Schneider}}, \citenamefont {{Schubnell}}, \citenamefont {{Secroun}}, \citenamefont {{Seljak}}, \citenamefont {{Seo}}, \citenamefont {{Serrano}}, \citenamefont {{Shafieloo}}, \citenamefont {{Shan}}, \citenamefont {{Sharples}}, \citenamefont {{Sholl}}, \citenamefont {{Shourt}}, \citenamefont {{Silber}}, \citenamefont {{Silva}}, \citenamefont {{Sirk}}, \citenamefont {{Slosar}}, \citenamefont {{Smith}}, \citenamefont {{Smoot}},
  \citenamefont {{Som}}, \citenamefont {{Song}}, \citenamefont {{Sprayberry}}, \citenamefont {{Staten}}, \citenamefont {{Stefanik}}, \citenamefont {{Tarle}}, \citenamefont {{Sien Tie}}, \citenamefont {{Tinker}}, \citenamefont {{Tojeiro}}, \citenamefont {{Valdes}}, \citenamefont {{Valenzuela}}, \citenamefont {{Valluri}}, \citenamefont {{Vargas-Magana}}, \citenamefont {{Verde}}, \citenamefont {{Walker}}, \citenamefont {{Wang}}, \citenamefont {{Wang}}, \citenamefont {{Weaver}}, \citenamefont {{Weaverdyck}}, \citenamefont {{Wechsler}}, \citenamefont {{Weinberg}}, \citenamefont {{White}}, \citenamefont {{Yang}}, \citenamefont {{Yeche}}, \citenamefont {{Zhang}}, \citenamefont {{Zhao}}, \citenamefont {{Zheng}}, \citenamefont {{Zhou}}, \citenamefont {{Zhou}}, \citenamefont {{Zhu}}, \citenamefont {{Zou}},\ and\ \citenamefont {{Zu}}}]{2016arXiv161100036D}%
  \BibitemOpen
  \bibfield  {author} {\bibinfo {author} {\bibnamefont {{DESI Collaboration}}}, \bibinfo {author} {\bibfnamefont {A.}~\bibnamefont {{Aghamousa}}}, \bibinfo {author} {\bibfnamefont {J.}~\bibnamefont {{Aguilar}}}, \bibinfo {author} {\bibfnamefont {S.}~\bibnamefont {{Ahlen}}}, \bibinfo {author} {\bibfnamefont {S.}~\bibnamefont {{Alam}}}, \bibinfo {author} {\bibfnamefont {L.~E.}\ \bibnamefont {{Allen}}}, \bibinfo {author} {\bibfnamefont {C.}~\bibnamefont {{Allende Prieto}}}, \bibinfo {author} {\bibfnamefont {J.}~\bibnamefont {{Annis}}}, \bibinfo {author} {\bibfnamefont {S.}~\bibnamefont {{Bailey}}}, \bibinfo {author} {\bibfnamefont {C.}~\bibnamefont {{Balland}}}, \bibinfo {author} {\bibfnamefont {O.}~\bibnamefont {{Ballester}}}, \bibinfo {author} {\bibfnamefont {C.}~\bibnamefont {{Baltay}}}, \bibinfo {author} {\bibfnamefont {L.}~\bibnamefont {{Beaufore}}}, \bibinfo {author} {\bibfnamefont {C.}~\bibnamefont {{Bebek}}}, \bibinfo {author} {\bibfnamefont {T.~C.}\ \bibnamefont {{Beers}}}, \bibinfo {author}
  {\bibfnamefont {E.~F.}\ \bibnamefont {{Bell}}}, \bibinfo {author} {\bibfnamefont {J.~L.}\ \bibnamefont {{Bernal}}}, \bibinfo {author} {\bibfnamefont {R.}~\bibnamefont {{Besuner}}}, \bibinfo {author} {\bibfnamefont {F.}~\bibnamefont {{Beutler}}}, \bibinfo {author} {\bibfnamefont {C.}~\bibnamefont {{Blake}}}, \bibinfo {author} {\bibfnamefont {H.}~\bibnamefont {{Bleuler}}}, \bibinfo {author} {\bibfnamefont {M.}~\bibnamefont {{Blomqvist}}}, \bibinfo {author} {\bibfnamefont {R.}~\bibnamefont {{Blum}}}, \bibinfo {author} {\bibfnamefont {A.~S.}\ \bibnamefont {{Bolton}}}, \bibinfo {author} {\bibfnamefont {C.}~\bibnamefont {{Briceno}}}, \bibinfo {author} {\bibfnamefont {D.}~\bibnamefont {{Brooks}}}, \bibinfo {author} {\bibfnamefont {J.~R.}\ \bibnamefont {{Brownstein}}}, \bibinfo {author} {\bibfnamefont {E.}~\bibnamefont {{Buckley-Geer}}}, \bibinfo {author} {\bibfnamefont {A.}~\bibnamefont {{Burden}}}, \bibinfo {author} {\bibfnamefont {E.}~\bibnamefont {{Burtin}}}, \bibinfo {author} {\bibfnamefont {N.~G.}\
  \bibnamefont {{Busca}}}, \bibinfo {author} {\bibfnamefont {R.~N.}\ \bibnamefont {{Cahn}}}, \bibinfo {author} {\bibfnamefont {Y.-C.}\ \bibnamefont {{Cai}}}, \bibinfo {author} {\bibfnamefont {L.}~\bibnamefont {{Cardiel-Sas}}}, \bibinfo {author} {\bibfnamefont {R.~G.}\ \bibnamefont {{Carlberg}}}, \bibinfo {author} {\bibfnamefont {P.-H.}\ \bibnamefont {{Carton}}}, \bibinfo {author} {\bibfnamefont {R.}~\bibnamefont {{Casas}}}, \bibinfo {author} {\bibfnamefont {F.~J.}\ \bibnamefont {{Castander}}}, \bibinfo {author} {\bibfnamefont {J.~L.}\ \bibnamefont {{Cervantes-Cota}}}, \bibinfo {author} {\bibfnamefont {T.~M.}\ \bibnamefont {{Claybaugh}}}, \bibinfo {author} {\bibfnamefont {M.}~\bibnamefont {{Close}}}, \bibinfo {author} {\bibfnamefont {C.~T.}\ \bibnamefont {{Coker}}}, \bibinfo {author} {\bibfnamefont {S.}~\bibnamefont {{Cole}}}, \bibinfo {author} {\bibfnamefont {J.}~\bibnamefont {{Comparat}}}, \bibinfo {author} {\bibfnamefont {A.~P.}\ \bibnamefont {{Cooper}}}, \bibinfo {author} {\bibfnamefont {M.~C.}\
  \bibnamefont {{Cousinou}}}, \bibinfo {author} {\bibfnamefont {M.}~\bibnamefont {{Crocce}}}, \bibinfo {author} {\bibfnamefont {J.-G.}\ \bibnamefont {{Cuby}}}, \bibinfo {author} {\bibfnamefont {D.~P.}\ \bibnamefont {{Cunningham}}}, \bibinfo {author} {\bibfnamefont {T.~M.}\ \bibnamefont {{Davis}}}, \bibinfo {author} {\bibfnamefont {K.~S.}\ \bibnamefont {{Dawson}}}, \bibinfo {author} {\bibfnamefont {A.}~\bibnamefont {{de la Macorra}}}, \bibinfo {author} {\bibfnamefont {J.}~\bibnamefont {{De Vicente}}}, \bibinfo {author} {\bibfnamefont {T.}~\bibnamefont {{Delubac}}}, \bibinfo {author} {\bibfnamefont {M.}~\bibnamefont {{Derwent}}}, \bibinfo {author} {\bibfnamefont {A.}~\bibnamefont {{Dey}}}, \bibinfo {author} {\bibfnamefont {G.}~\bibnamefont {{Dhungana}}}, \bibinfo {author} {\bibfnamefont {Z.}~\bibnamefont {{Ding}}}, \bibinfo {author} {\bibfnamefont {P.}~\bibnamefont {{Doel}}}, \bibinfo {author} {\bibfnamefont {Y.~T.}\ \bibnamefont {{Duan}}}, \bibinfo {author} {\bibfnamefont {A.}~\bibnamefont {{Ealet}}}, \bibinfo
  {author} {\bibfnamefont {J.}~\bibnamefont {{Edelstein}}}, \bibinfo {author} {\bibfnamefont {S.}~\bibnamefont {{Eftekharzadeh}}}, \bibinfo {author} {\bibfnamefont {D.~J.}\ \bibnamefont {{Eisenstein}}}, \bibinfo {author} {\bibfnamefont {A.}~\bibnamefont {{Elliott}}}, \bibinfo {author} {\bibfnamefont {S.}~\bibnamefont {{Escoffier}}}, \bibinfo {author} {\bibfnamefont {M.}~\bibnamefont {{Evatt}}}, \bibinfo {author} {\bibfnamefont {P.}~\bibnamefont {{Fagrelius}}}, \bibinfo {author} {\bibfnamefont {X.}~\bibnamefont {{Fan}}}, \bibinfo {author} {\bibfnamefont {K.}~\bibnamefont {{Fanning}}}, \bibinfo {author} {\bibfnamefont {A.}~\bibnamefont {{Farahi}}}, \bibinfo {author} {\bibfnamefont {J.}~\bibnamefont {{Farihi}}}, \bibinfo {author} {\bibfnamefont {G.}~\bibnamefont {{Favole}}}, \bibinfo {author} {\bibfnamefont {Y.}~\bibnamefont {{Feng}}}, \bibinfo {author} {\bibfnamefont {E.}~\bibnamefont {{Fernandez}}}, \bibinfo {author} {\bibfnamefont {J.~R.}\ \bibnamefont {{Findlay}}}, \bibinfo {author} {\bibfnamefont {D.~P.}\
  \bibnamefont {{Finkbeiner}}}, \bibinfo {author} {\bibfnamefont {M.~J.}\ \bibnamefont {{Fitzpatrick}}}, \bibinfo {author} {\bibfnamefont {B.}~\bibnamefont {{Flaugher}}}, \bibinfo {author} {\bibfnamefont {S.}~\bibnamefont {{Flender}}}, \bibinfo {author} {\bibfnamefont {A.}~\bibnamefont {{Font-Ribera}}}, \bibinfo {author} {\bibfnamefont {J.~E.}\ \bibnamefont {{Forero-Romero}}}, \bibinfo {author} {\bibfnamefont {P.}~\bibnamefont {{Fosalba}}}, \bibinfo {author} {\bibfnamefont {C.~S.}\ \bibnamefont {{Frenk}}}, \bibinfo {author} {\bibfnamefont {M.}~\bibnamefont {{Fumagalli}}}, \bibinfo {author} {\bibfnamefont {B.~T.}\ \bibnamefont {{Gaensicke}}}, \bibinfo {author} {\bibfnamefont {G.}~\bibnamefont {{Gallo}}}, \bibinfo {author} {\bibfnamefont {J.}~\bibnamefont {{Garcia-Bellido}}}, \bibinfo {author} {\bibfnamefont {E.}~\bibnamefont {{Gaztanaga}}}, \bibinfo {author} {\bibfnamefont {N.}~\bibnamefont {{Pietro Gentile Fusillo}}}, \bibinfo {author} {\bibfnamefont {T.}~\bibnamefont {{Gerard}}}, \bibinfo {author}
  {\bibfnamefont {I.}~\bibnamefont {{Gershkovich}}}, \bibinfo {author} {\bibfnamefont {T.}~\bibnamefont {{Giannantonio}}}, \bibinfo {author} {\bibfnamefont {D.}~\bibnamefont {{Gillet}}}, \bibinfo {author} {\bibfnamefont {G.}~\bibnamefont {{Gonzalez-de-Rivera}}}, \bibinfo {author} {\bibfnamefont {V.}~\bibnamefont {{Gonzalez-Perez}}}, \bibinfo {author} {\bibfnamefont {S.}~\bibnamefont {{Gott}}}, \bibinfo {author} {\bibfnamefont {O.}~\bibnamefont {{Graur}}}, \bibinfo {author} {\bibfnamefont {G.}~\bibnamefont {{Gutierrez}}}, \bibinfo {author} {\bibfnamefont {J.}~\bibnamefont {{Guy}}}, \bibinfo {author} {\bibfnamefont {S.}~\bibnamefont {{Habib}}}, \bibinfo {author} {\bibfnamefont {H.}~\bibnamefont {{Heetderks}}}, \bibinfo {author} {\bibfnamefont {I.}~\bibnamefont {{Heetderks}}}, \bibinfo {author} {\bibfnamefont {K.}~\bibnamefont {{Heitmann}}}, \bibinfo {author} {\bibfnamefont {W.~A.}\ \bibnamefont {{Hellwing}}}, \bibinfo {author} {\bibfnamefont {D.~A.}\ \bibnamefont {{Herrera}}}, \bibinfo {author} {\bibfnamefont
  {S.}~\bibnamefont {{Ho}}}, \bibinfo {author} {\bibfnamefont {S.}~\bibnamefont {{Holland}}}, \bibinfo {author} {\bibfnamefont {K.}~\bibnamefont {{Honscheid}}}, \bibinfo {author} {\bibfnamefont {E.}~\bibnamefont {{Huff}}}, \bibinfo {author} {\bibfnamefont {T.~A.}\ \bibnamefont {{Hutchinson}}}, \bibinfo {author} {\bibfnamefont {D.}~\bibnamefont {{Huterer}}}, \bibinfo {author} {\bibfnamefont {H.~S.}\ \bibnamefont {{Hwang}}}, \bibinfo {author} {\bibfnamefont {J.~M.}\ \bibnamefont {{Illa Laguna}}}, \bibinfo {author} {\bibfnamefont {Y.}~\bibnamefont {{Ishikawa}}}, \bibinfo {author} {\bibfnamefont {D.}~\bibnamefont {{Jacobs}}}, \bibinfo {author} {\bibfnamefont {N.}~\bibnamefont {{Jeffrey}}}, \bibinfo {author} {\bibfnamefont {P.}~\bibnamefont {{Jelinsky}}}, \bibinfo {author} {\bibfnamefont {E.}~\bibnamefont {{Jennings}}}, \bibinfo {author} {\bibfnamefont {L.}~\bibnamefont {{Jiang}}}, \bibinfo {author} {\bibfnamefont {J.}~\bibnamefont {{Jimenez}}}, \bibinfo {author} {\bibfnamefont {J.}~\bibnamefont {{Johnson}}},
  \bibinfo {author} {\bibfnamefont {R.}~\bibnamefont {{Joyce}}}, \bibinfo {author} {\bibfnamefont {E.}~\bibnamefont {{Jullo}}}, \bibinfo {author} {\bibfnamefont {S.}~\bibnamefont {{Juneau}}}, \bibinfo {author} {\bibfnamefont {S.}~\bibnamefont {{Kama}}}, \bibinfo {author} {\bibfnamefont {A.}~\bibnamefont {{Karcher}}}, \bibinfo {author} {\bibfnamefont {S.}~\bibnamefont {{Karkar}}}, \bibinfo {author} {\bibfnamefont {R.}~\bibnamefont {{Kehoe}}}, \bibinfo {author} {\bibfnamefont {N.}~\bibnamefont {{Kennamer}}}, \bibinfo {author} {\bibfnamefont {S.}~\bibnamefont {{Kent}}}, \bibinfo {author} {\bibfnamefont {M.}~\bibnamefont {{Kilbinger}}}, \bibinfo {author} {\bibfnamefont {A.~G.}\ \bibnamefont {{Kim}}}, \bibinfo {author} {\bibfnamefont {D.}~\bibnamefont {{Kirkby}}}, \bibinfo {author} {\bibfnamefont {T.}~\bibnamefont {{Kisner}}}, \bibinfo {author} {\bibfnamefont {E.}~\bibnamefont {{Kitanidis}}}, \bibinfo {author} {\bibfnamefont {J.-P.}\ \bibnamefont {{Kneib}}}, \bibinfo {author} {\bibfnamefont {S.}~\bibnamefont
  {{Koposov}}}, \bibinfo {author} {\bibfnamefont {E.}~\bibnamefont {{Kovacs}}}, \bibinfo {author} {\bibfnamefont {K.}~\bibnamefont {{Koyama}}}, \bibinfo {author} {\bibfnamefont {A.}~\bibnamefont {{Kremin}}}, \bibinfo {author} {\bibfnamefont {R.}~\bibnamefont {{Kron}}}, \bibinfo {author} {\bibfnamefont {L.}~\bibnamefont {{Kronig}}}, \bibinfo {author} {\bibfnamefont {A.}~\bibnamefont {{Kueter-Young}}}, \bibinfo {author} {\bibfnamefont {C.~G.}\ \bibnamefont {{Lacey}}}, \bibinfo {author} {\bibfnamefont {R.}~\bibnamefont {{Lafever}}}, \bibinfo {author} {\bibfnamefont {O.}~\bibnamefont {{Lahav}}}, \bibinfo {author} {\bibfnamefont {A.}~\bibnamefont {{Lambert}}}, \bibinfo {author} {\bibfnamefont {M.}~\bibnamefont {{Lampton}}}, \bibinfo {author} {\bibfnamefont {M.}~\bibnamefont {{Landriau}}}, \bibinfo {author} {\bibfnamefont {D.}~\bibnamefont {{Lang}}}, \bibinfo {author} {\bibfnamefont {T.~R.}\ \bibnamefont {{Lauer}}}, \bibinfo {author} {\bibfnamefont {J.-M.}\ \bibnamefont {{Le Goff}}}, \bibinfo {author}
  {\bibfnamefont {L.}~\bibnamefont {{Le Guillou}}}, \bibinfo {author} {\bibfnamefont {A.}~\bibnamefont {{Le Van Suu}}}, \bibinfo {author} {\bibfnamefont {J.~H.}\ \bibnamefont {{Lee}}}, \bibinfo {author} {\bibfnamefont {S.-J.}\ \bibnamefont {{Lee}}}, \bibinfo {author} {\bibfnamefont {D.}~\bibnamefont {{Leitner}}}, \bibinfo {author} {\bibfnamefont {M.}~\bibnamefont {{Lesser}}}, \bibinfo {author} {\bibfnamefont {M.~E.}\ \bibnamefont {{Levi}}}, \bibinfo {author} {\bibfnamefont {B.}~\bibnamefont {{L'Huillier}}}, \bibinfo {author} {\bibfnamefont {B.}~\bibnamefont {{Li}}}, \bibinfo {author} {\bibfnamefont {M.}~\bibnamefont {{Liang}}}, \bibinfo {author} {\bibfnamefont {H.}~\bibnamefont {{Lin}}}, \bibinfo {author} {\bibfnamefont {E.}~\bibnamefont {{Linder}}}, \bibinfo {author} {\bibfnamefont {S.~R.}\ \bibnamefont {{Loebman}}}, \bibinfo {author} {\bibfnamefont {Z.}~\bibnamefont {{Luki{\'c}}}}, \bibinfo {author} {\bibfnamefont {J.}~\bibnamefont {{Ma}}}, \bibinfo {author} {\bibfnamefont {N.}~\bibnamefont {{MacCrann}}},
  \bibinfo {author} {\bibfnamefont {C.}~\bibnamefont {{Magneville}}}, \bibinfo {author} {\bibfnamefont {L.}~\bibnamefont {{Makarem}}}, \bibinfo {author} {\bibfnamefont {M.}~\bibnamefont {{Manera}}}, \bibinfo {author} {\bibfnamefont {C.~J.}\ \bibnamefont {{Manser}}}, \bibinfo {author} {\bibfnamefont {R.}~\bibnamefont {{Marshall}}}, \bibinfo {author} {\bibfnamefont {P.}~\bibnamefont {{Martini}}}, \bibinfo {author} {\bibfnamefont {R.}~\bibnamefont {{Massey}}}, \bibinfo {author} {\bibfnamefont {T.}~\bibnamefont {{Matheson}}}, \bibinfo {author} {\bibfnamefont {J.}~\bibnamefont {{McCauley}}}, \bibinfo {author} {\bibfnamefont {P.}~\bibnamefont {{McDonald}}}, \bibinfo {author} {\bibfnamefont {I.~D.}\ \bibnamefont {{McGreer}}}, \bibinfo {author} {\bibfnamefont {A.}~\bibnamefont {{Meisner}}}, \bibinfo {author} {\bibfnamefont {N.}~\bibnamefont {{Metcalfe}}}, \bibinfo {author} {\bibfnamefont {T.~N.}\ \bibnamefont {{Miller}}}, \bibinfo {author} {\bibfnamefont {R.}~\bibnamefont {{Miquel}}}, \bibinfo {author} {\bibfnamefont
  {J.}~\bibnamefont {{Moustakas}}}, \bibinfo {author} {\bibfnamefont {A.}~\bibnamefont {{Myers}}}, \bibinfo {author} {\bibfnamefont {M.}~\bibnamefont {{Naik}}}, \bibinfo {author} {\bibfnamefont {J.~A.}\ \bibnamefont {{Newman}}}, \bibinfo {author} {\bibfnamefont {R.~C.}\ \bibnamefont {{Nichol}}}, \bibinfo {author} {\bibfnamefont {A.}~\bibnamefont {{Nicola}}}, \bibinfo {author} {\bibfnamefont {L.}~\bibnamefont {{Nicolati da Costa}}}, \bibinfo {author} {\bibfnamefont {J.}~\bibnamefont {{Nie}}}, \bibinfo {author} {\bibfnamefont {G.}~\bibnamefont {{Niz}}}, \bibinfo {author} {\bibfnamefont {P.}~\bibnamefont {{Norberg}}}, \bibinfo {author} {\bibfnamefont {B.}~\bibnamefont {{Nord}}}, \bibinfo {author} {\bibfnamefont {D.}~\bibnamefont {{Norman}}}, \bibinfo {author} {\bibfnamefont {P.}~\bibnamefont {{Nugent}}}, \bibinfo {author} {\bibfnamefont {T.}~\bibnamefont {{O'Brien}}}, \bibinfo {author} {\bibfnamefont {M.}~\bibnamefont {{Oh}}}, \bibinfo {author} {\bibfnamefont {K.~A.~G.}\ \bibnamefont {{Olsen}}}, \bibinfo
  {author} {\bibfnamefont {C.}~\bibnamefont {{Padilla}}}, \bibinfo {author} {\bibfnamefont {H.}~\bibnamefont {{Padmanabhan}}}, \bibinfo {author} {\bibfnamefont {N.}~\bibnamefont {{Padmanabhan}}}, \bibinfo {author} {\bibfnamefont {N.}~\bibnamefont {{Palanque-Delabrouille}}}, \bibinfo {author} {\bibfnamefont {A.}~\bibnamefont {{Palmese}}}, \bibinfo {author} {\bibfnamefont {D.}~\bibnamefont {{Pappalardo}}}, \bibinfo {author} {\bibfnamefont {I.}~\bibnamefont {{P{\^a}ris}}}, \bibinfo {author} {\bibfnamefont {C.}~\bibnamefont {{Park}}}, \bibinfo {author} {\bibfnamefont {A.}~\bibnamefont {{Patej}}}, \bibinfo {author} {\bibfnamefont {J.~A.}\ \bibnamefont {{Peacock}}}, \bibinfo {author} {\bibfnamefont {H.~V.}\ \bibnamefont {{Peiris}}}, \bibinfo {author} {\bibfnamefont {X.}~\bibnamefont {{Peng}}}, \bibinfo {author} {\bibfnamefont {W.~J.}\ \bibnamefont {{Percival}}}, \bibinfo {author} {\bibfnamefont {S.}~\bibnamefont {{Perruchot}}}, \bibinfo {author} {\bibfnamefont {M.~M.}\ \bibnamefont {{Pieri}}}, \bibinfo {author}
  {\bibfnamefont {R.}~\bibnamefont {{Pogge}}}, \bibinfo {author} {\bibfnamefont {J.~E.}\ \bibnamefont {{Pollack}}}, \bibinfo {author} {\bibfnamefont {C.}~\bibnamefont {{Poppett}}}, \bibinfo {author} {\bibfnamefont {F.}~\bibnamefont {{Prada}}}, \bibinfo {author} {\bibfnamefont {A.}~\bibnamefont {{Prakash}}}, \bibinfo {author} {\bibfnamefont {R.~G.}\ \bibnamefont {{Probst}}}, \bibinfo {author} {\bibfnamefont {D.}~\bibnamefont {{Rabinowitz}}}, \bibinfo {author} {\bibfnamefont {A.}~\bibnamefont {{Raichoor}}}, \bibinfo {author} {\bibfnamefont {C.~H.}\ \bibnamefont {{Ree}}}, \bibinfo {author} {\bibfnamefont {A.}~\bibnamefont {{Refregier}}}, \bibinfo {author} {\bibfnamefont {X.}~\bibnamefont {{Regal}}}, \bibinfo {author} {\bibfnamefont {B.}~\bibnamefont {{Reid}}}, \bibinfo {author} {\bibfnamefont {K.}~\bibnamefont {{Reil}}}, \bibinfo {author} {\bibfnamefont {M.}~\bibnamefont {{Rezaie}}}, \bibinfo {author} {\bibfnamefont {C.~M.}\ \bibnamefont {{Rockosi}}}, \bibinfo {author} {\bibfnamefont {N.}~\bibnamefont {{Roe}}},
  \bibinfo {author} {\bibfnamefont {S.}~\bibnamefont {{Ronayette}}}, \bibinfo {author} {\bibfnamefont {A.}~\bibnamefont {{Roodman}}}, \bibinfo {author} {\bibfnamefont {A.~J.}\ \bibnamefont {{Ross}}}, \bibinfo {author} {\bibfnamefont {N.~P.}\ \bibnamefont {{Ross}}}, \bibinfo {author} {\bibfnamefont {G.}~\bibnamefont {{Rossi}}}, \bibinfo {author} {\bibfnamefont {E.}~\bibnamefont {{Rozo}}}, \bibinfo {author} {\bibfnamefont {V.}~\bibnamefont {{Ruhlmann-Kleider}}}, \bibinfo {author} {\bibfnamefont {E.~S.}\ \bibnamefont {{Rykoff}}}, \bibinfo {author} {\bibfnamefont {C.}~\bibnamefont {{Sabiu}}}, \bibinfo {author} {\bibfnamefont {L.}~\bibnamefont {{Samushia}}}, \bibinfo {author} {\bibfnamefont {E.}~\bibnamefont {{Sanchez}}}, \bibinfo {author} {\bibfnamefont {J.}~\bibnamefont {{Sanchez}}}, \bibinfo {author} {\bibfnamefont {D.~J.}\ \bibnamefont {{Schlegel}}}, \bibinfo {author} {\bibfnamefont {M.}~\bibnamefont {{Schneider}}}, \bibinfo {author} {\bibfnamefont {M.}~\bibnamefont {{Schubnell}}}, \bibinfo {author}
  {\bibfnamefont {A.}~\bibnamefont {{Secroun}}}, \bibinfo {author} {\bibfnamefont {U.}~\bibnamefont {{Seljak}}}, \bibinfo {author} {\bibfnamefont {H.-J.}\ \bibnamefont {{Seo}}}, \bibinfo {author} {\bibfnamefont {S.}~\bibnamefont {{Serrano}}}, \bibinfo {author} {\bibfnamefont {A.}~\bibnamefont {{Shafieloo}}}, \bibinfo {author} {\bibfnamefont {H.}~\bibnamefont {{Shan}}}, \bibinfo {author} {\bibfnamefont {R.}~\bibnamefont {{Sharples}}}, \bibinfo {author} {\bibfnamefont {M.~J.}\ \bibnamefont {{Sholl}}}, \bibinfo {author} {\bibfnamefont {W.~V.}\ \bibnamefont {{Shourt}}}, \bibinfo {author} {\bibfnamefont {J.~H.}\ \bibnamefont {{Silber}}}, \bibinfo {author} {\bibfnamefont {D.~R.}\ \bibnamefont {{Silva}}}, \bibinfo {author} {\bibfnamefont {M.~M.}\ \bibnamefont {{Sirk}}}, \bibinfo {author} {\bibfnamefont {A.}~\bibnamefont {{Slosar}}}, \bibinfo {author} {\bibfnamefont {A.}~\bibnamefont {{Smith}}}, \bibinfo {author} {\bibfnamefont {G.~F.}\ \bibnamefont {{Smoot}}}, \bibinfo {author} {\bibfnamefont {D.}~\bibnamefont
  {{Som}}}, \bibinfo {author} {\bibfnamefont {Y.-S.}\ \bibnamefont {{Song}}}, \bibinfo {author} {\bibfnamefont {D.}~\bibnamefont {{Sprayberry}}}, \bibinfo {author} {\bibfnamefont {R.}~\bibnamefont {{Staten}}}, \bibinfo {author} {\bibfnamefont {A.}~\bibnamefont {{Stefanik}}}, \bibinfo {author} {\bibfnamefont {G.}~\bibnamefont {{Tarle}}}, \bibinfo {author} {\bibfnamefont {S.}~\bibnamefont {{Sien Tie}}}, \bibinfo {author} {\bibfnamefont {J.~L.}\ \bibnamefont {{Tinker}}}, \bibinfo {author} {\bibfnamefont {R.}~\bibnamefont {{Tojeiro}}}, \bibinfo {author} {\bibfnamefont {F.}~\bibnamefont {{Valdes}}}, \bibinfo {author} {\bibfnamefont {O.}~\bibnamefont {{Valenzuela}}}, \bibinfo {author} {\bibfnamefont {M.}~\bibnamefont {{Valluri}}}, \bibinfo {author} {\bibfnamefont {M.}~\bibnamefont {{Vargas-Magana}}}, \bibinfo {author} {\bibfnamefont {L.}~\bibnamefont {{Verde}}}, \bibinfo {author} {\bibfnamefont {A.~R.}\ \bibnamefont {{Walker}}}, \bibinfo {author} {\bibfnamefont {J.}~\bibnamefont {{Wang}}}, \bibinfo {author}
  {\bibfnamefont {Y.}~\bibnamefont {{Wang}}}, \bibinfo {author} {\bibfnamefont {B.~A.}\ \bibnamefont {{Weaver}}}, \bibinfo {author} {\bibfnamefont {C.}~\bibnamefont {{Weaverdyck}}}, \bibinfo {author} {\bibfnamefont {R.~H.}\ \bibnamefont {{Wechsler}}}, \bibinfo {author} {\bibfnamefont {D.~H.}\ \bibnamefont {{Weinberg}}}, \bibinfo {author} {\bibfnamefont {M.}~\bibnamefont {{White}}}, \bibinfo {author} {\bibfnamefont {Q.}~\bibnamefont {{Yang}}}, \bibinfo {author} {\bibfnamefont {C.}~\bibnamefont {{Yeche}}}, \bibinfo {author} {\bibfnamefont {T.}~\bibnamefont {{Zhang}}}, \bibinfo {author} {\bibfnamefont {G.-B.}\ \bibnamefont {{Zhao}}}, \bibinfo {author} {\bibfnamefont {Y.}~\bibnamefont {{Zheng}}}, \bibinfo {author} {\bibfnamefont {X.}~\bibnamefont {{Zhou}}}, \bibinfo {author} {\bibfnamefont {Z.}~\bibnamefont {{Zhou}}}, \bibinfo {author} {\bibfnamefont {Y.}~\bibnamefont {{Zhu}}}, \bibinfo {author} {\bibfnamefont {H.}~\bibnamefont {{Zou}}},\ and\ \bibinfo {author} {\bibfnamefont {Y.}~\bibnamefont {{Zu}}},\
  }\bibfield  {title} {\bibinfo {title} {{The DESI Experiment Part I: Science,Targeting, and Survey Design}},\ }\href {https://doi.org/10.48550/arXiv.1611.00036} {\bibfield  {journal} {\bibinfo  {journal} {arXiv e-prints}\ ,\ \bibinfo {eid} {arXiv:1611.00036}} (\bibinfo {year} {2016}{\natexlab{b}})},\ \Eprint {https://arxiv.org/abs/1611.00036} {arXiv:1611.00036 [astro-ph.IM]} \BibitemShut {NoStop}%
\bibitem [{\citenamefont {{Zhou}}\ \emph {et~al.}(2023{\natexlab{a}})\citenamefont {{Zhou}}, \citenamefont {{Dey}}, \citenamefont {{Newman}}, \citenamefont {{Eisenstein}}, \citenamefont {{Dawson}}, \citenamefont {{Bailey}}, \citenamefont {{Berti}}, \citenamefont {{Guy}}, \citenamefont {{Lan}}, \citenamefont {{Zou}}, \citenamefont {{Aguilar}}, \citenamefont {{Ahlen}}, \citenamefont {{Alam}}, \citenamefont {{Brooks}}, \citenamefont {{de la Macorra}}, \citenamefont {{Dey}}, \citenamefont {{Dhungana}}, \citenamefont {{Fanning}}, \citenamefont {{Font-Ribera}}, \citenamefont {{Gontcho}}, \citenamefont {{Honscheid}}, \citenamefont {{Ishak}}, \citenamefont {{Kisner}}, \citenamefont {{Kov{\'a}cs}}, \citenamefont {{Kremin}}, \citenamefont {{Landriau}}, \citenamefont {{Levi}}, \citenamefont {{Magneville}}, \citenamefont {{Manera}}, \citenamefont {{Martini}}, \citenamefont {{Meisner}}, \citenamefont {{Miquel}}, \citenamefont {{Moustakas}}, \citenamefont {{Myers}}, \citenamefont {{Nie}}, \citenamefont
  {{Palanque-Delabrouille}}, \citenamefont {{Percival}}, \citenamefont {{Poppett}}, \citenamefont {{Prada}}, \citenamefont {{Raichoor}}, \citenamefont {{Ross}}, \citenamefont {{Schlafly}}, \citenamefont {{Schlegel}}, \citenamefont {{Schubnell}}, \citenamefont {{Tarl{\'e}}}, \citenamefont {{Weaver}}, \citenamefont {{Wechsler}}, \citenamefont {{Y{\'e}che}},\ and\ \citenamefont {{Zhou}}}]{2023AJ....165...58Z}%
  \BibitemOpen
  \bibfield  {author} {\bibinfo {author} {\bibfnamefont {R.}~\bibnamefont {{Zhou}}}, \bibinfo {author} {\bibfnamefont {B.}~\bibnamefont {{Dey}}}, \bibinfo {author} {\bibfnamefont {J.~A.}\ \bibnamefont {{Newman}}}, \bibinfo {author} {\bibfnamefont {D.~J.}\ \bibnamefont {{Eisenstein}}}, \bibinfo {author} {\bibfnamefont {K.}~\bibnamefont {{Dawson}}}, \bibinfo {author} {\bibfnamefont {S.}~\bibnamefont {{Bailey}}}, \bibinfo {author} {\bibfnamefont {A.}~\bibnamefont {{Berti}}}, \bibinfo {author} {\bibfnamefont {J.}~\bibnamefont {{Guy}}}, \bibinfo {author} {\bibfnamefont {T.-W.}\ \bibnamefont {{Lan}}}, \bibinfo {author} {\bibfnamefont {H.}~\bibnamefont {{Zou}}}, \bibinfo {author} {\bibfnamefont {J.}~\bibnamefont {{Aguilar}}}, \bibinfo {author} {\bibfnamefont {S.}~\bibnamefont {{Ahlen}}}, \bibinfo {author} {\bibfnamefont {S.}~\bibnamefont {{Alam}}}, \bibinfo {author} {\bibfnamefont {D.}~\bibnamefont {{Brooks}}}, \bibinfo {author} {\bibfnamefont {A.}~\bibnamefont {{de la Macorra}}}, \bibinfo {author} {\bibfnamefont
  {A.}~\bibnamefont {{Dey}}}, \bibinfo {author} {\bibfnamefont {G.}~\bibnamefont {{Dhungana}}}, \bibinfo {author} {\bibfnamefont {K.}~\bibnamefont {{Fanning}}}, \bibinfo {author} {\bibfnamefont {A.}~\bibnamefont {{Font-Ribera}}}, \bibinfo {author} {\bibfnamefont {S.~G.~A.}\ \bibnamefont {{Gontcho}}}, \bibinfo {author} {\bibfnamefont {K.}~\bibnamefont {{Honscheid}}}, \bibinfo {author} {\bibfnamefont {M.}~\bibnamefont {{Ishak}}}, \bibinfo {author} {\bibfnamefont {T.}~\bibnamefont {{Kisner}}}, \bibinfo {author} {\bibfnamefont {A.}~\bibnamefont {{Kov{\'a}cs}}}, \bibinfo {author} {\bibfnamefont {A.}~\bibnamefont {{Kremin}}}, \bibinfo {author} {\bibfnamefont {M.}~\bibnamefont {{Landriau}}}, \bibinfo {author} {\bibfnamefont {M.~E.}\ \bibnamefont {{Levi}}}, \bibinfo {author} {\bibfnamefont {C.}~\bibnamefont {{Magneville}}}, \bibinfo {author} {\bibfnamefont {M.}~\bibnamefont {{Manera}}}, \bibinfo {author} {\bibfnamefont {P.}~\bibnamefont {{Martini}}}, \bibinfo {author} {\bibfnamefont {A.~M.}\ \bibnamefont
  {{Meisner}}}, \bibinfo {author} {\bibfnamefont {R.}~\bibnamefont {{Miquel}}}, \bibinfo {author} {\bibfnamefont {J.}~\bibnamefont {{Moustakas}}}, \bibinfo {author} {\bibfnamefont {A.~D.}\ \bibnamefont {{Myers}}}, \bibinfo {author} {\bibfnamefont {J.}~\bibnamefont {{Nie}}}, \bibinfo {author} {\bibfnamefont {N.}~\bibnamefont {{Palanque-Delabrouille}}}, \bibinfo {author} {\bibfnamefont {W.~J.}\ \bibnamefont {{Percival}}}, \bibinfo {author} {\bibfnamefont {C.}~\bibnamefont {{Poppett}}}, \bibinfo {author} {\bibfnamefont {F.}~\bibnamefont {{Prada}}}, \bibinfo {author} {\bibfnamefont {A.}~\bibnamefont {{Raichoor}}}, \bibinfo {author} {\bibfnamefont {A.~J.}\ \bibnamefont {{Ross}}}, \bibinfo {author} {\bibfnamefont {E.}~\bibnamefont {{Schlafly}}}, \bibinfo {author} {\bibfnamefont {D.}~\bibnamefont {{Schlegel}}}, \bibinfo {author} {\bibfnamefont {M.}~\bibnamefont {{Schubnell}}}, \bibinfo {author} {\bibfnamefont {G.}~\bibnamefont {{Tarl{\'e}}}}, \bibinfo {author} {\bibfnamefont {B.~A.}\ \bibnamefont {{Weaver}}},
  \bibinfo {author} {\bibfnamefont {R.~H.}\ \bibnamefont {{Wechsler}}}, \bibinfo {author} {\bibfnamefont {C.}~\bibnamefont {{Y{\'e}che}}},\ and\ \bibinfo {author} {\bibfnamefont {Z.}~\bibnamefont {{Zhou}}},\ }\bibfield  {title} {\bibinfo {title} {{Target Selection and Validation of DESI Luminous Red Galaxies}},\ }\href {https://doi.org/10.3847/1538-3881/aca5fb} {\bibfield  {journal} {\bibinfo  {journal} {\aj}\ }\textbf {\bibinfo {volume} {165}},\ \bibinfo {eid} {58} (\bibinfo {year} {2023}{\natexlab{a}})},\ \Eprint {https://arxiv.org/abs/2208.08515} {arXiv:2208.08515 [astro-ph.CO]} \BibitemShut {NoStop}%
\bibitem [{\citenamefont {{Zhou}}\ \emph {et~al.}(2023{\natexlab{b}})\citenamefont {{Zhou}}, \citenamefont {{Ferraro}}, \citenamefont {{White}}, \citenamefont {{DeRose}}, \citenamefont {{Sailer}}, \citenamefont {{Aguilar}}, \citenamefont {{Ahlen}}, \citenamefont {{Bailey}}, \citenamefont {{Brooks}}, \citenamefont {{Claybaugh}}, \citenamefont {{Dawson}}, \citenamefont {{de la Macorra}}, \citenamefont {{Dey}}, \citenamefont {{Doel}}, \citenamefont {{Font-Ribera}}, \citenamefont {{Forero-Romero}}, \citenamefont {{Gontcho A Gontcho}}, \citenamefont {{Guy}}, \citenamefont {{Kremin}}, \citenamefont {{Lambert}}, \citenamefont {{Le Guillou}}, \citenamefont {{Levi}}, \citenamefont {{Magneville}}, \citenamefont {{Manera}}, \citenamefont {{Meisner}}, \citenamefont {{Miquel}}, \citenamefont {{Moustakas}}, \citenamefont {{Myers}}, \citenamefont {{Newman}}, \citenamefont {{Nie}}, \citenamefont {{Percival}}, \citenamefont {{Rezaie}}, \citenamefont {{Rossi}}, \citenamefont {{Sanchez}}, \citenamefont {{Schlegel}},
  \citenamefont {{Schubnell}}, \citenamefont {{Seo}}, \citenamefont {{Tarl{\'e}}},\ and\ \citenamefont {{Zhou}}}]{Zhou:2023gji}%
  \BibitemOpen
  \bibfield  {author} {\bibinfo {author} {\bibfnamefont {R.}~\bibnamefont {{Zhou}}}, \bibinfo {author} {\bibfnamefont {S.}~\bibnamefont {{Ferraro}}}, \bibinfo {author} {\bibfnamefont {M.}~\bibnamefont {{White}}}, \bibinfo {author} {\bibfnamefont {J.}~\bibnamefont {{DeRose}}}, \bibinfo {author} {\bibfnamefont {N.}~\bibnamefont {{Sailer}}}, \bibinfo {author} {\bibfnamefont {J.}~\bibnamefont {{Aguilar}}}, \bibinfo {author} {\bibfnamefont {S.}~\bibnamefont {{Ahlen}}}, \bibinfo {author} {\bibfnamefont {S.}~\bibnamefont {{Bailey}}}, \bibinfo {author} {\bibfnamefont {D.}~\bibnamefont {{Brooks}}}, \bibinfo {author} {\bibfnamefont {T.}~\bibnamefont {{Claybaugh}}}, \bibinfo {author} {\bibfnamefont {K.}~\bibnamefont {{Dawson}}}, \bibinfo {author} {\bibfnamefont {A.}~\bibnamefont {{de la Macorra}}}, \bibinfo {author} {\bibfnamefont {B.}~\bibnamefont {{Dey}}}, \bibinfo {author} {\bibfnamefont {P.}~\bibnamefont {{Doel}}}, \bibinfo {author} {\bibfnamefont {A.}~\bibnamefont {{Font-Ribera}}}, \bibinfo {author} {\bibfnamefont
  {J.~E.}\ \bibnamefont {{Forero-Romero}}}, \bibinfo {author} {\bibfnamefont {S.}~\bibnamefont {{Gontcho A Gontcho}}}, \bibinfo {author} {\bibfnamefont {J.}~\bibnamefont {{Guy}}}, \bibinfo {author} {\bibfnamefont {A.}~\bibnamefont {{Kremin}}}, \bibinfo {author} {\bibfnamefont {A.}~\bibnamefont {{Lambert}}}, \bibinfo {author} {\bibfnamefont {L.}~\bibnamefont {{Le Guillou}}}, \bibinfo {author} {\bibfnamefont {M.}~\bibnamefont {{Levi}}}, \bibinfo {author} {\bibfnamefont {C.}~\bibnamefont {{Magneville}}}, \bibinfo {author} {\bibfnamefont {M.}~\bibnamefont {{Manera}}}, \bibinfo {author} {\bibfnamefont {A.}~\bibnamefont {{Meisner}}}, \bibinfo {author} {\bibfnamefont {R.}~\bibnamefont {{Miquel}}}, \bibinfo {author} {\bibfnamefont {J.}~\bibnamefont {{Moustakas}}}, \bibinfo {author} {\bibfnamefont {A.~D.}\ \bibnamefont {{Myers}}}, \bibinfo {author} {\bibfnamefont {J.~A.}\ \bibnamefont {{Newman}}}, \bibinfo {author} {\bibfnamefont {J.}~\bibnamefont {{Nie}}}, \bibinfo {author} {\bibfnamefont {W.}~\bibnamefont
  {{Percival}}}, \bibinfo {author} {\bibfnamefont {M.}~\bibnamefont {{Rezaie}}}, \bibinfo {author} {\bibfnamefont {G.}~\bibnamefont {{Rossi}}}, \bibinfo {author} {\bibfnamefont {E.}~\bibnamefont {{Sanchez}}}, \bibinfo {author} {\bibfnamefont {D.}~\bibnamefont {{Schlegel}}}, \bibinfo {author} {\bibfnamefont {M.}~\bibnamefont {{Schubnell}}}, \bibinfo {author} {\bibfnamefont {H.-J.}\ \bibnamefont {{Seo}}}, \bibinfo {author} {\bibfnamefont {G.}~\bibnamefont {{Tarl{\'e}}}},\ and\ \bibinfo {author} {\bibfnamefont {Z.}~\bibnamefont {{Zhou}}},\ }\bibfield  {title} {\bibinfo {title} {{DESI luminous red galaxy samples for cross-correlations}},\ }\href {https://doi.org/10.1088/1475-7516/2023/11/097} {\bibfield  {journal} {\bibinfo  {journal} {\jcap}\ }\textbf {\bibinfo {volume} {2023}},\ \bibinfo {eid} {097} (\bibinfo {year} {2023}{\natexlab{b}})},\ \Eprint {https://arxiv.org/abs/2309.06443} {arXiv:2309.06443 [astro-ph.CO]} \BibitemShut {NoStop}%
\bibitem [{\citenamefont {Hahn}\ \emph {et~al.}(2023)\citenamefont {Hahn} \emph {et~al.}}]{Hahn:2022dnf}%
  \BibitemOpen
  \bibfield  {author} {\bibinfo {author} {\bibfnamefont {C.}~\bibnamefont {Hahn}} \emph {et~al.},\ }\bibfield  {title} {\bibinfo {title} {{The DESI Bright Galaxy Survey: Final Target Selection, Design, and Validation}},\ }\href {https://doi.org/10.3847/1538-3881/accff8} {\bibfield  {journal} {\bibinfo  {journal} {Astron. J.}\ }\textbf {\bibinfo {volume} {165}},\ \bibinfo {pages} {253} (\bibinfo {year} {2023})},\ \Eprint {https://arxiv.org/abs/2208.08512} {arXiv:2208.08512 [astro-ph.CO]} \BibitemShut {NoStop}%
\bibitem [{\citenamefont {Chen}\ \emph {et~al.}(2024)\citenamefont {Chen} \emph {et~al.}}]{Chen:2024vvk}%
  \BibitemOpen
  \bibfield  {author} {\bibinfo {author} {\bibfnamefont {S.}~\bibnamefont {Chen}} \emph {et~al.},\ }\bibfield  {title} {\bibinfo {title} {{Analysis of DESI\texttimes{}DES using the Lagrangian effective theory of LSS}},\ }\href {https://doi.org/10.1103/PhysRevD.110.103518} {\bibfield  {journal} {\bibinfo  {journal} {Phys. Rev. D}\ }\textbf {\bibinfo {volume} {110}},\ \bibinfo {pages} {103518} (\bibinfo {year} {2024})},\ \Eprint {https://arxiv.org/abs/2407.04795} {arXiv:2407.04795 [astro-ph.CO]} \BibitemShut {NoStop}%
\bibitem [{\citenamefont {Madhavacheril}\ \emph {et~al.}(2024)\citenamefont {Madhavacheril} \emph {et~al.}}]{ACT:2023kun}%
  \BibitemOpen
  \bibfield  {author} {\bibinfo {author} {\bibfnamefont {M.~S.}\ \bibnamefont {Madhavacheril}} \emph {et~al.} (\bibinfo {collaboration} {ACT}),\ }\bibfield  {title} {\bibinfo {title} {{The Atacama Cosmology Telescope: DR6 Gravitational Lensing Map and Cosmological Parameters}},\ }\href {https://doi.org/10.3847/1538-4357/acff5f} {\bibfield  {journal} {\bibinfo  {journal} {Astrophys. J.}\ }\textbf {\bibinfo {volume} {962}},\ \bibinfo {pages} {113} (\bibinfo {year} {2024})},\ \Eprint {https://arxiv.org/abs/2304.05203} {arXiv:2304.05203 [astro-ph.CO]} \BibitemShut {NoStop}%
\bibitem [{\citenamefont {Qu}\ \emph {et~al.}(2024)\citenamefont {Qu} \emph {et~al.}}]{ACT:2023dou}%
  \BibitemOpen
  \bibfield  {author} {\bibinfo {author} {\bibfnamefont {F.~J.}\ \bibnamefont {Qu}} \emph {et~al.} (\bibinfo {collaboration} {ACT}),\ }\bibfield  {title} {\bibinfo {title} {{The Atacama Cosmology Telescope: A Measurement of the DR6 CMB Lensing Power Spectrum and Its Implications for Structure Growth}},\ }\href {https://doi.org/10.3847/1538-4357/acfe06} {\bibfield  {journal} {\bibinfo  {journal} {Astrophys. J.}\ }\textbf {\bibinfo {volume} {962}},\ \bibinfo {pages} {112} (\bibinfo {year} {2024})},\ \Eprint {https://arxiv.org/abs/2304.05202} {arXiv:2304.05202 [astro-ph.CO]} \BibitemShut {NoStop}%
\bibitem [{\citenamefont {MacCrann}\ \emph {et~al.}(2024)\citenamefont {MacCrann} \emph {et~al.}}]{ACT:2023ubw}%
  \BibitemOpen
  \bibfield  {author} {\bibinfo {author} {\bibfnamefont {N.}~\bibnamefont {MacCrann}} \emph {et~al.} (\bibinfo {collaboration} {ACT}),\ }\bibfield  {title} {\bibinfo {title} {{The Atacama Cosmology Telescope: Mitigating the Impact of Extragalactic Foregrounds for the DR6 Cosmic Microwave Background Lensing Analysis}},\ }\href {https://doi.org/10.3847/1538-4357/ad2610} {\bibfield  {journal} {\bibinfo  {journal} {Astrophys. J.}\ }\textbf {\bibinfo {volume} {966}},\ \bibinfo {pages} {138} (\bibinfo {year} {2024})},\ \Eprint {https://arxiv.org/abs/2304.05196} {arXiv:2304.05196 [astro-ph.CO]} \BibitemShut {NoStop}%
\bibitem [{\citenamefont {Farren}\ \emph {et~al.}(2024)\citenamefont {Farren} \emph {et~al.}}]{ACT:2023oei}%
  \BibitemOpen
  \bibfield  {author} {\bibinfo {author} {\bibfnamefont {G.~S.}\ \bibnamefont {Farren}} \emph {et~al.} (\bibinfo {collaboration} {ACT}),\ }\bibfield  {title} {\bibinfo {title} {{The Atacama Cosmology Telescope: Cosmology from Cross-correlations of unWISE Galaxies and ACT DR6 CMB Lensing}},\ }\href {https://doi.org/10.3847/1538-4357/ad31a5} {\bibfield  {journal} {\bibinfo  {journal} {Astrophys. J.}\ }\textbf {\bibinfo {volume} {966}},\ \bibinfo {pages} {157} (\bibinfo {year} {2024})},\ \Eprint {https://arxiv.org/abs/2309.05659} {arXiv:2309.05659 [astro-ph.CO]} \BibitemShut {NoStop}%
\bibitem [{\citenamefont {Sailer}\ \emph {et~al.}(2020)\citenamefont {Sailer}, \citenamefont {Schaan},\ and\ \citenamefont {Ferraro}}]{Sailer:2020lal}%
  \BibitemOpen
  \bibfield  {author} {\bibinfo {author} {\bibfnamefont {N.}~\bibnamefont {Sailer}}, \bibinfo {author} {\bibfnamefont {E.}~\bibnamefont {Schaan}},\ and\ \bibinfo {author} {\bibfnamefont {S.}~\bibnamefont {Ferraro}},\ }\bibfield  {title} {\bibinfo {title} {{Lower bias, lower noise CMB lensing with foreground-hardened estimators}},\ }\href {https://doi.org/10.1103/PhysRevD.102.063517} {\bibfield  {journal} {\bibinfo  {journal} {Phys. Rev. D}\ }\textbf {\bibinfo {volume} {102}},\ \bibinfo {pages} {063517} (\bibinfo {year} {2020})},\ \Eprint {https://arxiv.org/abs/2007.04325} {arXiv:2007.04325 [astro-ph.CO]} \BibitemShut {NoStop}%
\bibitem [{Note1()}]{Note1}%
  \BibitemOpen
  \bibinfo {note} {\protect \url {https://github.com/EmmanuelSchaan/ThumbStack}}\BibitemShut {NoStop}%
\bibitem [{\citenamefont {{White}}(2015)}]{2015MNRAS.450.3822W}%
  \BibitemOpen
  \bibfield  {author} {\bibinfo {author} {\bibfnamefont {M.}~\bibnamefont {{White}}},\ }\bibfield  {title} {\bibinfo {title} {{Reconstruction within the Zeldovich approximation}},\ }\href {https://doi.org/10.1093/mnras/stv842} {\bibfield  {journal} {\bibinfo  {journal} {\mnras}\ }\textbf {\bibinfo {volume} {450}},\ \bibinfo {pages} {3822} (\bibinfo {year} {2015})},\ \Eprint {https://arxiv.org/abs/1504.03677} {arXiv:1504.03677 [astro-ph.CO]} \BibitemShut {NoStop}%
\bibitem [{\citenamefont {{Hadzhiyska}}\ \emph {et~al.}(2024{\natexlab{c}})\citenamefont {{Hadzhiyska}}, \citenamefont {{Ferraro}}, \citenamefont {{Ried Guachalla}},\ and\ \citenamefont {{Schaan}}}]{2024PhRvD.109j3534H}%
  \BibitemOpen
  \bibfield  {author} {\bibinfo {author} {\bibfnamefont {B.}~\bibnamefont {{Hadzhiyska}}}, \bibinfo {author} {\bibfnamefont {S.}~\bibnamefont {{Ferraro}}}, \bibinfo {author} {\bibfnamefont {B.}~\bibnamefont {{Ried Guachalla}}},\ and\ \bibinfo {author} {\bibfnamefont {E.}~\bibnamefont {{Schaan}}},\ }\bibfield  {title} {\bibinfo {title} {{Velocity reconstruction in the era of DESI and Rubin/LSST. II. Realistic samples on the light cone}},\ }\href {https://doi.org/10.1103/PhysRevD.109.103534} {\bibfield  {journal} {\bibinfo  {journal} {\prd}\ }\textbf {\bibinfo {volume} {109}},\ \bibinfo {eid} {103534} (\bibinfo {year} {2024}{\natexlab{c}})},\ \Eprint {https://arxiv.org/abs/2312.12434} {arXiv:2312.12434 [astro-ph.CO]} \BibitemShut {NoStop}%
\bibitem [{\citenamefont {{Ried Guachalla}}\ \emph {et~al.}(2024)\citenamefont {{Ried Guachalla}}, \citenamefont {{Schaan}}, \citenamefont {{Hadzhiyska}},\ and\ \citenamefont {{Ferraro}}}]{2024PhRvD.109j3533R}%
  \BibitemOpen
  \bibfield  {author} {\bibinfo {author} {\bibfnamefont {B.}~\bibnamefont {{Ried Guachalla}}}, \bibinfo {author} {\bibfnamefont {E.}~\bibnamefont {{Schaan}}}, \bibinfo {author} {\bibfnamefont {B.}~\bibnamefont {{Hadzhiyska}}},\ and\ \bibinfo {author} {\bibfnamefont {S.}~\bibnamefont {{Ferraro}}},\ }\bibfield  {title} {\bibinfo {title} {{Velocity reconstruction in the era of DESI and Rubin/LSST. I. Exploring spectroscopic, photometric, and hybrid samples}},\ }\href {https://doi.org/10.1103/PhysRevD.109.103533} {\bibfield  {journal} {\bibinfo  {journal} {\prd}\ }\textbf {\bibinfo {volume} {109}},\ \bibinfo {eid} {103533} (\bibinfo {year} {2024})},\ \Eprint {https://arxiv.org/abs/2312.12435} {arXiv:2312.12435 [astro-ph.CO]} \BibitemShut {NoStop}%
\bibitem [{\citenamefont {{Maksimova}}\ \emph {et~al.}(2021)\citenamefont {{Maksimova}}, \citenamefont {{Garrison}}, \citenamefont {{Eisenstein}}, \citenamefont {{Hadzhiyska}}, \citenamefont {{Bose}},\ and\ \citenamefont {{Satterthwaite}}}]{2021MNRAS.508.4017M}%
  \BibitemOpen
  \bibfield  {author} {\bibinfo {author} {\bibfnamefont {N.~A.}\ \bibnamefont {{Maksimova}}}, \bibinfo {author} {\bibfnamefont {L.~H.}\ \bibnamefont {{Garrison}}}, \bibinfo {author} {\bibfnamefont {D.~J.}\ \bibnamefont {{Eisenstein}}}, \bibinfo {author} {\bibfnamefont {B.}~\bibnamefont {{Hadzhiyska}}}, \bibinfo {author} {\bibfnamefont {S.}~\bibnamefont {{Bose}}},\ and\ \bibinfo {author} {\bibfnamefont {T.~P.}\ \bibnamefont {{Satterthwaite}}},\ }\bibfield  {title} {\bibinfo {title} {{ABACUSSUMMIT: a massive set of high-accuracy, high-resolution N-body simulations}},\ }\href {https://doi.org/10.1093/mnras/stab2484} {\bibfield  {journal} {\bibinfo  {journal} {\mnras}\ }\textbf {\bibinfo {volume} {508}},\ \bibinfo {pages} {4017} (\bibinfo {year} {2021})},\ \Eprint {https://arxiv.org/abs/2110.11398} {arXiv:2110.11398 [astro-ph.CO]} \BibitemShut {NoStop}%
\bibitem [{\citenamefont {{Hadzhiyska}}\ \emph {et~al.}(2022{\natexlab{a}})\citenamefont {{Hadzhiyska}}, \citenamefont {{Garrison}}, \citenamefont {{Eisenstein}},\ and\ \citenamefont {{Bose}}}]{2022MNRAS.509.2194H}%
  \BibitemOpen
  \bibfield  {author} {\bibinfo {author} {\bibfnamefont {B.}~\bibnamefont {{Hadzhiyska}}}, \bibinfo {author} {\bibfnamefont {L.~H.}\ \bibnamefont {{Garrison}}}, \bibinfo {author} {\bibfnamefont {D.}~\bibnamefont {{Eisenstein}}},\ and\ \bibinfo {author} {\bibfnamefont {S.}~\bibnamefont {{Bose}}},\ }\bibfield  {title} {\bibinfo {title} {{The halo light-cone catalogues of ABACUSSUMMIT}},\ }\href {https://doi.org/10.1093/mnras/stab3066} {\bibfield  {journal} {\bibinfo  {journal} {\mnras}\ }\textbf {\bibinfo {volume} {509}},\ \bibinfo {pages} {2194} (\bibinfo {year} {2022}{\natexlab{a}})},\ \Eprint {https://arxiv.org/abs/2110.11413} {arXiv:2110.11413 [astro-ph.CO]} \BibitemShut {NoStop}%
\bibitem [{\citenamefont {{Garrison}}\ \emph {et~al.}(2019)\citenamefont {{Garrison}}, \citenamefont {{Eisenstein}},\ and\ \citenamefont {{Pinto}}}]{2019MNRAS.485.3370G}%
  \BibitemOpen
  \bibfield  {author} {\bibinfo {author} {\bibfnamefont {L.~H.}\ \bibnamefont {{Garrison}}}, \bibinfo {author} {\bibfnamefont {D.~J.}\ \bibnamefont {{Eisenstein}}},\ and\ \bibinfo {author} {\bibfnamefont {P.~A.}\ \bibnamefont {{Pinto}}},\ }\bibfield  {title} {\bibinfo {title} {{A high-fidelity realization of the Euclid code comparison N-body simulation with ABACUS}},\ }\href {https://doi.org/10.1093/mnras/stz634} {\bibfield  {journal} {\bibinfo  {journal} {\mnras}\ }\textbf {\bibinfo {volume} {485}},\ \bibinfo {pages} {3370} (\bibinfo {year} {2019})},\ \Eprint {https://arxiv.org/abs/1810.02916} {arXiv:1810.02916 [astro-ph.CO]} \BibitemShut {NoStop}%
\bibitem [{\citenamefont {{Garrison}}\ \emph {et~al.}(2021)\citenamefont {{Garrison}}, \citenamefont {{Eisenstein}}, \citenamefont {{Ferrer}}, \citenamefont {{Maksimova}},\ and\ \citenamefont {{Pinto}}}]{2021MNRAS.508..575G}%
  \BibitemOpen
  \bibfield  {author} {\bibinfo {author} {\bibfnamefont {L.~H.}\ \bibnamefont {{Garrison}}}, \bibinfo {author} {\bibfnamefont {D.~J.}\ \bibnamefont {{Eisenstein}}}, \bibinfo {author} {\bibfnamefont {D.}~\bibnamefont {{Ferrer}}}, \bibinfo {author} {\bibfnamefont {N.~A.}\ \bibnamefont {{Maksimova}}},\ and\ \bibinfo {author} {\bibfnamefont {P.~A.}\ \bibnamefont {{Pinto}}},\ }\bibfield  {title} {\bibinfo {title} {{The ABACUS cosmological N-body code}},\ }\href {https://doi.org/10.1093/mnras/stab2482} {\bibfield  {journal} {\bibinfo  {journal} {\mnras}\ }\textbf {\bibinfo {volume} {508}},\ \bibinfo {pages} {575} (\bibinfo {year} {2021})},\ \Eprint {https://arxiv.org/abs/2110.11392} {arXiv:2110.11392 [astro-ph.CO]} \BibitemShut {NoStop}%
\bibitem [{\citenamefont {{Hadzhiyska}}\ \emph {et~al.}(2022{\natexlab{b}})\citenamefont {{Hadzhiyska}}, \citenamefont {{Eisenstein}}, \citenamefont {{Bose}}, \citenamefont {{Garrison}},\ and\ \citenamefont {{Maksimova}}}]{2022MNRAS.509..501H}%
  \BibitemOpen
  \bibfield  {author} {\bibinfo {author} {\bibfnamefont {B.}~\bibnamefont {{Hadzhiyska}}}, \bibinfo {author} {\bibfnamefont {D.}~\bibnamefont {{Eisenstein}}}, \bibinfo {author} {\bibfnamefont {S.}~\bibnamefont {{Bose}}}, \bibinfo {author} {\bibfnamefont {L.~H.}\ \bibnamefont {{Garrison}}},\ and\ \bibinfo {author} {\bibfnamefont {N.}~\bibnamefont {{Maksimova}}},\ }\bibfield  {title} {\bibinfo {title} {{COMPASO: A new halo finder for competitive assignment to spherical overdensities}},\ }\href {https://doi.org/10.1093/mnras/stab2980} {\bibfield  {journal} {\bibinfo  {journal} {\mnras}\ }\textbf {\bibinfo {volume} {509}},\ \bibinfo {pages} {501} (\bibinfo {year} {2022}{\natexlab{b}})},\ \Eprint {https://arxiv.org/abs/2110.11408} {arXiv:2110.11408 [astro-ph.CO]} \BibitemShut {NoStop}%
\bibitem [{\citenamefont {{Bryan}}\ and\ \citenamefont {{Norman}}(1998)}]{1998ApJ...495...80B}%
  \BibitemOpen
  \bibfield  {author} {\bibinfo {author} {\bibfnamefont {G.~L.}\ \bibnamefont {{Bryan}}}\ and\ \bibinfo {author} {\bibfnamefont {M.~L.}\ \bibnamefont {{Norman}}},\ }\bibfield  {title} {\bibinfo {title} {{Statistical Properties of X-Ray Clusters: Analytic and Numerical Comparisons}},\ }\href {https://doi.org/10.1086/305262} {\bibfield  {journal} {\bibinfo  {journal} {\apj}\ }\textbf {\bibinfo {volume} {495}},\ \bibinfo {pages} {80} (\bibinfo {year} {1998})},\ \Eprint {https://arxiv.org/abs/astro-ph/9710107} {arXiv:astro-ph/9710107 [astro-ph]} \BibitemShut {NoStop}%
\bibitem [{\citenamefont {Zheng}\ \emph {et~al.}(2005)\citenamefont {Zheng}, \citenamefont {Berlind}, \citenamefont {Weinberg}, \citenamefont {Benson}, \citenamefont {Baugh}, \citenamefont {Cole}, \citenamefont {Dave}, \citenamefont {Frenk}, \citenamefont {Katz},\ and\ \citenamefont {Lacey}}]{Zheng:2004id}%
  \BibitemOpen
  \bibfield  {author} {\bibinfo {author} {\bibfnamefont {Z.}~\bibnamefont {Zheng}}, \bibinfo {author} {\bibfnamefont {A.~A.}\ \bibnamefont {Berlind}}, \bibinfo {author} {\bibfnamefont {D.~H.}\ \bibnamefont {Weinberg}}, \bibinfo {author} {\bibfnamefont {A.~J.}\ \bibnamefont {Benson}}, \bibinfo {author} {\bibfnamefont {C.~M.}\ \bibnamefont {Baugh}}, \bibinfo {author} {\bibfnamefont {S.}~\bibnamefont {Cole}}, \bibinfo {author} {\bibfnamefont {R.}~\bibnamefont {Dave}}, \bibinfo {author} {\bibfnamefont {C.~S.}\ \bibnamefont {Frenk}}, \bibinfo {author} {\bibfnamefont {N.}~\bibnamefont {Katz}},\ and\ \bibinfo {author} {\bibfnamefont {C.~G.}\ \bibnamefont {Lacey}},\ }\bibfield  {title} {\bibinfo {title} {{Theoretical models of the halo occupation distribution: Separating central and satellite galaxies}},\ }\href {https://doi.org/10.1086/466510} {\bibfield  {journal} {\bibinfo  {journal} {Astrophys. J.}\ }\textbf {\bibinfo {volume} {633}},\ \bibinfo {pages} {791} (\bibinfo {year} {2005})},\ \Eprint
  {https://arxiv.org/abs/astro-ph/0408564} {arXiv:astro-ph/0408564} \BibitemShut {NoStop}%
\bibitem [{Note2()}]{Note2}%
  \BibitemOpen
  \bibinfo {note} {\protect \url {https://github.com/abacusorg/abacusutils}}\BibitemShut {NoStop}%
\bibitem [{Note3()}]{Note3}%
  \BibitemOpen
  \bibinfo {note} {\protect \url {https://camb.readthedocs.io/en/latest/}}\BibitemShut {NoStop}%
\bibitem [{\citenamefont {{Hadzhiyska}}\ \emph {et~al.}(2023{\natexlab{a}})\citenamefont {{Hadzhiyska}}, \citenamefont {{Yuan}}, \citenamefont {{Blake}}, \citenamefont {{Eisenstein}}, \citenamefont {{Aguilar}}, \citenamefont {{Ahlen}}, \citenamefont {{Brooks}}, \citenamefont {{Claybaugh}}, \citenamefont {{de la Macorra}}, \citenamefont {{Doel}}, \citenamefont {{Emas}}, \citenamefont {{Forero-Romero}}, \citenamefont {{Garcia-Quintero}}, \citenamefont {{Ishak}}, \citenamefont {{Joudaki}}, \citenamefont {{Jullo}}, \citenamefont {{Kehoe}}, \citenamefont {{Kisner}}, \citenamefont {{Kremin}}, \citenamefont {{Krolewski}}, \citenamefont {{Landriau}}, \citenamefont {{Lange}}, \citenamefont {{Manera}}, \citenamefont {{Miquel}}, \citenamefont {{Nie}}, \citenamefont {{Poppett}}, \citenamefont {{Porredon}}, \citenamefont {{Rossi}}, \citenamefont {{Ruggeri}}, \citenamefont {{Saulder}}, \citenamefont {{Schubnell}}, \citenamefont {{Tarl{\'e}}}, \citenamefont {{Weaver}}, \citenamefont {{Xhakaj}},\ and\ \citenamefont
  {{Zhou}}}]{2023MNRAS.525.4367H}%
  \BibitemOpen
  \bibfield  {author} {\bibinfo {author} {\bibfnamefont {B.}~\bibnamefont {{Hadzhiyska}}}, \bibinfo {author} {\bibfnamefont {S.}~\bibnamefont {{Yuan}}}, \bibinfo {author} {\bibfnamefont {C.}~\bibnamefont {{Blake}}}, \bibinfo {author} {\bibfnamefont {D.~J.}\ \bibnamefont {{Eisenstein}}}, \bibinfo {author} {\bibfnamefont {J.}~\bibnamefont {{Aguilar}}}, \bibinfo {author} {\bibfnamefont {S.}~\bibnamefont {{Ahlen}}}, \bibinfo {author} {\bibfnamefont {D.}~\bibnamefont {{Brooks}}}, \bibinfo {author} {\bibfnamefont {T.}~\bibnamefont {{Claybaugh}}}, \bibinfo {author} {\bibfnamefont {A.}~\bibnamefont {{de la Macorra}}}, \bibinfo {author} {\bibfnamefont {P.}~\bibnamefont {{Doel}}}, \bibinfo {author} {\bibfnamefont {N.}~\bibnamefont {{Emas}}}, \bibinfo {author} {\bibfnamefont {J.~E.}\ \bibnamefont {{Forero-Romero}}}, \bibinfo {author} {\bibfnamefont {C.}~\bibnamefont {{Garcia-Quintero}}}, \bibinfo {author} {\bibfnamefont {M.}~\bibnamefont {{Ishak}}}, \bibinfo {author} {\bibfnamefont {S.}~\bibnamefont {{Joudaki}}},
  \bibinfo {author} {\bibfnamefont {E.}~\bibnamefont {{Jullo}}}, \bibinfo {author} {\bibfnamefont {R.}~\bibnamefont {{Kehoe}}}, \bibinfo {author} {\bibfnamefont {T.}~\bibnamefont {{Kisner}}}, \bibinfo {author} {\bibfnamefont {A.}~\bibnamefont {{Kremin}}}, \bibinfo {author} {\bibfnamefont {A.}~\bibnamefont {{Krolewski}}}, \bibinfo {author} {\bibfnamefont {M.}~\bibnamefont {{Landriau}}}, \bibinfo {author} {\bibfnamefont {J.~U.}\ \bibnamefont {{Lange}}}, \bibinfo {author} {\bibfnamefont {M.}~\bibnamefont {{Manera}}}, \bibinfo {author} {\bibfnamefont {R.}~\bibnamefont {{Miquel}}}, \bibinfo {author} {\bibfnamefont {J.}~\bibnamefont {{Nie}}}, \bibinfo {author} {\bibfnamefont {C.}~\bibnamefont {{Poppett}}}, \bibinfo {author} {\bibfnamefont {A.}~\bibnamefont {{Porredon}}}, \bibinfo {author} {\bibfnamefont {G.}~\bibnamefont {{Rossi}}}, \bibinfo {author} {\bibfnamefont {R.}~\bibnamefont {{Ruggeri}}}, \bibinfo {author} {\bibfnamefont {C.}~\bibnamefont {{Saulder}}}, \bibinfo {author} {\bibfnamefont {M.}~\bibnamefont
  {{Schubnell}}}, \bibinfo {author} {\bibfnamefont {G.}~\bibnamefont {{Tarl{\'e}}}}, \bibinfo {author} {\bibfnamefont {B.~A.}\ \bibnamefont {{Weaver}}}, \bibinfo {author} {\bibfnamefont {E.}~\bibnamefont {{Xhakaj}}},\ and\ \bibinfo {author} {\bibfnamefont {Z.}~\bibnamefont {{Zhou}}},\ }\bibfield  {title} {\bibinfo {title} {{Synthetic light-cone catalogues of modern redshift and weak lensing surveys waith ABACUSSUMMIT}},\ }\href {https://doi.org/10.1093/mnras/stad2563} {\bibfield  {journal} {\bibinfo  {journal} {\mnras}\ }\textbf {\bibinfo {volume} {525}},\ \bibinfo {pages} {4367} (\bibinfo {year} {2023}{\natexlab{a}})},\ \Eprint {https://arxiv.org/abs/2305.11935} {arXiv:2305.11935 [astro-ph.CO]} \BibitemShut {NoStop}%
\bibitem [{Note4()}]{Note4}%
  \BibitemOpen
  \bibinfo {note} {\protect \url {https://github.com/simonsobs/pixell}}\BibitemShut {NoStop}%
\bibitem [{Note5()}]{Note5}%
  \BibitemOpen
  \bibinfo {note} {\protect \url {https://github.com/simonsobs/so-lenspipe}}\BibitemShut {NoStop}%
\bibitem [{\citenamefont {Hu}\ \emph {et~al.}(2007)\citenamefont {Hu}, \citenamefont {DeDeo},\ and\ \citenamefont {Vale}}]{Hu:2007bt}%
  \BibitemOpen
  \bibfield  {author} {\bibinfo {author} {\bibfnamefont {W.}~\bibnamefont {Hu}}, \bibinfo {author} {\bibfnamefont {S.}~\bibnamefont {DeDeo}},\ and\ \bibinfo {author} {\bibfnamefont {C.}~\bibnamefont {Vale}},\ }\bibfield  {title} {\bibinfo {title} {{Cluster Mass Estimators from CMB Temperature and Polarization Lensing}},\ }\href {https://doi.org/10.1088/1367-2630/9/12/441} {\bibfield  {journal} {\bibinfo  {journal} {New J. Phys.}\ }\textbf {\bibinfo {volume} {9}},\ \bibinfo {pages} {441} (\bibinfo {year} {2007})},\ \Eprint {https://arxiv.org/abs/astro-ph/0701276} {arXiv:astro-ph/0701276} \BibitemShut {NoStop}%
\bibitem [{\citenamefont {Madhavacheril}\ \emph {et~al.}(2020)\citenamefont {Madhavacheril} \emph {et~al.}}]{ACT:2020izl}%
  \BibitemOpen
  \bibfield  {author} {\bibinfo {author} {\bibfnamefont {M.~S.}\ \bibnamefont {Madhavacheril}} \emph {et~al.} (\bibinfo {collaboration} {ACT}),\ }\bibfield  {title} {\bibinfo {title} {{The Atacama Cosmology Telescope: Weighing Distant Clusters with the Most Ancient Light}},\ }\href {https://doi.org/10.3847/2041-8213/abbccb} {\bibfield  {journal} {\bibinfo  {journal} {Astrophys. J. Lett.}\ }\textbf {\bibinfo {volume} {903}},\ \bibinfo {pages} {L13} (\bibinfo {year} {2020})},\ \Eprint {https://arxiv.org/abs/2009.07772} {arXiv:2009.07772 [astro-ph.CO]} \BibitemShut {NoStop}%
\bibitem [{\citenamefont {Baxter}\ \emph {et~al.}(2015)\citenamefont {Baxter} \emph {et~al.}}]{Baxter:2014frs}%
  \BibitemOpen
  \bibfield  {author} {\bibinfo {author} {\bibfnamefont {E.~J.}\ \bibnamefont {Baxter}} \emph {et~al.},\ }\bibfield  {title} {\bibinfo {title} {{A Measurement of Gravitational Lensing of the Cosmic Microwave Background by Galaxy Clusters Using Data from the South Pole Telescope}},\ }\href {https://doi.org/10.1088/0004-637X/806/2/247} {\bibfield  {journal} {\bibinfo  {journal} {Astrophys. J.}\ }\textbf {\bibinfo {volume} {806}},\ \bibinfo {pages} {247} (\bibinfo {year} {2015})},\ \Eprint {https://arxiv.org/abs/1412.7521} {arXiv:1412.7521 [astro-ph.CO]} \BibitemShut {NoStop}%
\bibitem [{\citenamefont {Raghunathan}\ \emph {et~al.}(2017)\citenamefont {Raghunathan}, \citenamefont {Patil}, \citenamefont {Baxter}, \citenamefont {Bianchini}, \citenamefont {Bleem}, \citenamefont {Crawford}, \citenamefont {Holder}, \citenamefont {Manzotti},\ and\ \citenamefont {Reichardt}}]{Raghunathan:2017cle}%
  \BibitemOpen
  \bibfield  {author} {\bibinfo {author} {\bibfnamefont {S.}~\bibnamefont {Raghunathan}}, \bibinfo {author} {\bibfnamefont {S.}~\bibnamefont {Patil}}, \bibinfo {author} {\bibfnamefont {E.~J.}\ \bibnamefont {Baxter}}, \bibinfo {author} {\bibfnamefont {F.}~\bibnamefont {Bianchini}}, \bibinfo {author} {\bibfnamefont {L.~E.}\ \bibnamefont {Bleem}}, \bibinfo {author} {\bibfnamefont {T.~M.}\ \bibnamefont {Crawford}}, \bibinfo {author} {\bibfnamefont {G.~P.}\ \bibnamefont {Holder}}, \bibinfo {author} {\bibfnamefont {A.}~\bibnamefont {Manzotti}},\ and\ \bibinfo {author} {\bibfnamefont {C.~L.}\ \bibnamefont {Reichardt}},\ }\bibfield  {title} {\bibinfo {title} {{Measuring galaxy cluster masses with CMB lensing using a Maximum Likelihood estimator: Statistical and systematic error budgets for future experiments}},\ }\href {https://doi.org/10.1088/1475-7516/2017/08/030} {\bibfield  {journal} {\bibinfo  {journal} {JCAP}\ }\textbf {\bibinfo {volume} {08}}\bibfield  {number} {\bibinfo  {number} { (8)},\ \bibinfo {pages}
  {030}},\ }\Eprint {https://arxiv.org/abs/1705.00411} {arXiv:1705.00411 [astro-ph.CO]} \BibitemShut {NoStop}%
\bibitem [{\citenamefont {Horowitz}\ \emph {et~al.}(2019)\citenamefont {Horowitz}, \citenamefont {Ferraro},\ and\ \citenamefont {Sherwin}}]{Horowitz:2017iql}%
  \BibitemOpen
  \bibfield  {author} {\bibinfo {author} {\bibfnamefont {B.}~\bibnamefont {Horowitz}}, \bibinfo {author} {\bibfnamefont {S.}~\bibnamefont {Ferraro}},\ and\ \bibinfo {author} {\bibfnamefont {B.~D.}\ \bibnamefont {Sherwin}},\ }\bibfield  {title} {\bibinfo {title} {{Reconstructing Small Scale Lenses from the Cosmic Microwave Background Temperature Fluctuations}},\ }\href {https://doi.org/10.1093/mnras/stz566} {\bibfield  {journal} {\bibinfo  {journal} {Mon. Not. Roy. Astron. Soc.}\ }\textbf {\bibinfo {volume} {485}},\ \bibinfo {pages} {3919} (\bibinfo {year} {2019})},\ \Eprint {https://arxiv.org/abs/1710.10236} {arXiv:1710.10236 [astro-ph.CO]} \BibitemShut {NoStop}%
\bibitem [{\citenamefont {Levy}\ \emph {et~al.}(2023)\citenamefont {Levy}, \citenamefont {Raghunathan},\ and\ \citenamefont {Basu}}]{Levy:2023moy}%
  \BibitemOpen
  \bibfield  {author} {\bibinfo {author} {\bibfnamefont {K.}~\bibnamefont {Levy}}, \bibinfo {author} {\bibfnamefont {S.}~\bibnamefont {Raghunathan}},\ and\ \bibinfo {author} {\bibfnamefont {K.}~\bibnamefont {Basu}},\ }\bibfield  {title} {\bibinfo {title} {{A foreground-immune CMB-cluster lensing estimator}},\ }\href {https://doi.org/10.1088/1475-7516/2023/08/020} {\bibfield  {journal} {\bibinfo  {journal} {JCAP}\ }\textbf {\bibinfo {volume} {08}}\bibfield  {number} {\bibinfo  {number} { (8)},\ \bibinfo {pages} {020}},\ }\Eprint {https://arxiv.org/abs/2305.06326} {arXiv:2305.06326 [astro-ph.CO]} \BibitemShut {NoStop}%
\bibitem [{\citenamefont {Stein}\ \emph {et~al.}(2020)\citenamefont {Stein}, \citenamefont {Alvarez}, \citenamefont {Bond}, \citenamefont {van Engelen},\ and\ \citenamefont {Battaglia}}]{Stein:2020its}%
  \BibitemOpen
  \bibfield  {author} {\bibinfo {author} {\bibfnamefont {G.}~\bibnamefont {Stein}}, \bibinfo {author} {\bibfnamefont {M.~A.}\ \bibnamefont {Alvarez}}, \bibinfo {author} {\bibfnamefont {J.~R.}\ \bibnamefont {Bond}}, \bibinfo {author} {\bibfnamefont {A.}~\bibnamefont {van Engelen}},\ and\ \bibinfo {author} {\bibfnamefont {N.}~\bibnamefont {Battaglia}},\ }\bibfield  {title} {\bibinfo {title} {{The Websky Extragalactic CMB Simulations}},\ }\href {https://doi.org/10.1088/1475-7516/2020/10/012} {\bibfield  {journal} {\bibinfo  {journal} {JCAP}\ }\textbf {\bibinfo {volume} {10}}\bibfield  {number} {\bibinfo  {number} { (10)},\ \bibinfo {pages} {012}},\ }\Eprint {https://arxiv.org/abs/2001.08787} {arXiv:2001.08787 [astro-ph.CO]} \BibitemShut {NoStop}%
\bibitem [{\citenamefont {Schaan}\ and\ \citenamefont {Ferraro}(2019)}]{Schaan:2018tup}%
  \BibitemOpen
  \bibfield  {author} {\bibinfo {author} {\bibfnamefont {E.}~\bibnamefont {Schaan}}\ and\ \bibinfo {author} {\bibfnamefont {S.}~\bibnamefont {Ferraro}},\ }\bibfield  {title} {\bibinfo {title} {{Foreground-Immune Cosmic Microwave Background Lensing with Shear-Only Reconstruction}},\ }\href {https://doi.org/10.1103/PhysRevLett.122.181301} {\bibfield  {journal} {\bibinfo  {journal} {Phys. Rev. Lett.}\ }\textbf {\bibinfo {volume} {122}},\ \bibinfo {pages} {181301} (\bibinfo {year} {2019})},\ \Eprint {https://arxiv.org/abs/1804.06403} {arXiv:1804.06403 [astro-ph.CO]} \BibitemShut {NoStop}%
\bibitem [{\citenamefont {Qu}\ \emph {et~al.}(2023)\citenamefont {Qu}, \citenamefont {Challinor},\ and\ \citenamefont {Sherwin}}]{Qu:2022qie}%
  \BibitemOpen
  \bibfield  {author} {\bibinfo {author} {\bibfnamefont {F.~J.}\ \bibnamefont {Qu}}, \bibinfo {author} {\bibfnamefont {A.}~\bibnamefont {Challinor}},\ and\ \bibinfo {author} {\bibfnamefont {B.~D.}\ \bibnamefont {Sherwin}},\ }\bibfield  {title} {\bibinfo {title} {{CMB lensing with shear-only reconstruction on the full sky}},\ }\href {https://doi.org/10.1103/PhysRevD.108.063518} {\bibfield  {journal} {\bibinfo  {journal} {Phys. Rev. D}\ }\textbf {\bibinfo {volume} {108}},\ \bibinfo {pages} {063518} (\bibinfo {year} {2023})},\ \Eprint {https://arxiv.org/abs/2208.14988} {arXiv:2208.14988 [astro-ph.CO]} \BibitemShut {NoStop}%
\bibitem [{\citenamefont {Madhavacheril}\ and\ \citenamefont {Hill}(2018)}]{Madhavacheril:2018bxi}%
  \BibitemOpen
  \bibfield  {author} {\bibinfo {author} {\bibfnamefont {M.~S.}\ \bibnamefont {Madhavacheril}}\ and\ \bibinfo {author} {\bibfnamefont {J.~C.}\ \bibnamefont {Hill}},\ }\bibfield  {title} {\bibinfo {title} {{Mitigating Foreground Biases in CMB Lensing Reconstruction Using Cleaned Gradients}},\ }\href {https://doi.org/10.1103/PhysRevD.98.023534} {\bibfield  {journal} {\bibinfo  {journal} {Phys. Rev. D}\ }\textbf {\bibinfo {volume} {98}},\ \bibinfo {pages} {023534} (\bibinfo {year} {2018})},\ \Eprint {https://arxiv.org/abs/1802.08230} {arXiv:1802.08230 [astro-ph.CO]} \BibitemShut {NoStop}%
\bibitem [{\citenamefont {Sailer}\ \emph {et~al.}(2023)\citenamefont {Sailer}, \citenamefont {Ferraro},\ and\ \citenamefont {Schaan}}]{Sailer:2022jwt}%
  \BibitemOpen
  \bibfield  {author} {\bibinfo {author} {\bibfnamefont {N.}~\bibnamefont {Sailer}}, \bibinfo {author} {\bibfnamefont {S.}~\bibnamefont {Ferraro}},\ and\ \bibinfo {author} {\bibfnamefont {E.}~\bibnamefont {Schaan}},\ }\bibfield  {title} {\bibinfo {title} {{Foreground-immune CMB lensing reconstruction with polarization}},\ }\href {https://doi.org/10.1103/PhysRevD.107.023504} {\bibfield  {journal} {\bibinfo  {journal} {Phys. Rev. D}\ }\textbf {\bibinfo {volume} {107}},\ \bibinfo {pages} {023504} (\bibinfo {year} {2023})},\ \Eprint {https://arxiv.org/abs/2211.03786} {arXiv:2211.03786 [astro-ph.CO]} \BibitemShut {NoStop}%
\bibitem [{\citenamefont {{Darwish}}\ \emph {et~al.}(2021)\citenamefont {{Darwish}}, \citenamefont {{Sherwin}}, \citenamefont {{Sailer}}, \citenamefont {{Schaan}},\ and\ \citenamefont {{Ferraro}}}]{Darwish_2021}%
  \BibitemOpen
  \bibfield  {author} {\bibinfo {author} {\bibfnamefont {O.}~\bibnamefont {{Darwish}}}, \bibinfo {author} {\bibfnamefont {B.~D.}\ \bibnamefont {{Sherwin}}}, \bibinfo {author} {\bibfnamefont {N.}~\bibnamefont {{Sailer}}}, \bibinfo {author} {\bibfnamefont {E.}~\bibnamefont {{Schaan}}},\ and\ \bibinfo {author} {\bibfnamefont {S.}~\bibnamefont {{Ferraro}}},\ }\bibfield  {title} {\bibinfo {title} {{Optimizing foreground mitigation for CMB lensing with combined multifrequency and geometric methods}},\ }\href {https://doi.org/10.48550/arXiv.2111.00462} {\bibfield  {journal} {\bibinfo  {journal} {arXiv e-prints}\ ,\ \bibinfo {eid} {arXiv:2111.00462}} (\bibinfo {year} {2021})},\ \Eprint {https://arxiv.org/abs/2111.00462} {arXiv:2111.00462 [astro-ph.CO]} \BibitemShut {NoStop}%
\bibitem [{\citenamefont {{Sailer}}\ \emph {et~al.}(2021)\citenamefont {{Sailer}}, \citenamefont {{Schaan}}, \citenamefont {{Ferraro}}, \citenamefont {{Darwish}},\ and\ \citenamefont {{Sherwin}}}]{Sailer_2021}%
  \BibitemOpen
  \bibfield  {author} {\bibinfo {author} {\bibfnamefont {N.}~\bibnamefont {{Sailer}}}, \bibinfo {author} {\bibfnamefont {E.}~\bibnamefont {{Schaan}}}, \bibinfo {author} {\bibfnamefont {S.}~\bibnamefont {{Ferraro}}}, \bibinfo {author} {\bibfnamefont {O.}~\bibnamefont {{Darwish}}},\ and\ \bibinfo {author} {\bibfnamefont {B.}~\bibnamefont {{Sherwin}}},\ }\bibfield  {title} {\bibinfo {title} {{Optimal multifrequency weighting for CMB lensing}},\ }\href {https://doi.org/10.1103/PhysRevD.104.123514} {\bibfield  {journal} {\bibinfo  {journal} {\prd}\ }\textbf {\bibinfo {volume} {104}},\ \bibinfo {eid} {123514} (\bibinfo {year} {2021})},\ \Eprint {https://arxiv.org/abs/2108.01663} {arXiv:2108.01663 [astro-ph.CO]} \BibitemShut {NoStop}%
\bibitem [{\citenamefont {{Coulton}}\ \emph {et~al.}(2024)\citenamefont {{Coulton}}, \citenamefont {{Schutt}}, \citenamefont {{Maniyar}}, \citenamefont {{Schaan}}, \citenamefont {{An}}, \citenamefont {{Atkins}}, \citenamefont {{Battaglia}}, \citenamefont {{Bond}}, \citenamefont {{Calabrese}}, \citenamefont {{Choi}}, \citenamefont {{Devlin}}, \citenamefont {{Duivenvoorden}}, \citenamefont {{Dunkley}}, \citenamefont {{Ferraro}}, \citenamefont {{Gluscevic}}, \citenamefont {{Hill}}, \citenamefont {{Hilton}}, \citenamefont {{Hincks}}, \citenamefont {{Kosowsky}}, \citenamefont {{Kramer}}, \citenamefont {{Kusiak}}, \citenamefont {{La Posta}}, \citenamefont {{Louis}}, \citenamefont {{Madhavacheril}}, \citenamefont {{Marques}}, \citenamefont {{McCarthy}}, \citenamefont {{McMahon}}, \citenamefont {{Moodley}}, \citenamefont {{Naess}}, \citenamefont {{Page}}, \citenamefont {{Partridge}}, \citenamefont {{Qu}}, \citenamefont {{Sehgal}}, \citenamefont {{Sherwin}}, \citenamefont {{Sif{\'o}n}}, \citenamefont {{Spergel}},
  \citenamefont {{Staggs}}, \citenamefont {{Van Engelen}}, \citenamefont {{Vargas}},\ and\ \citenamefont {{Wollack}}}]{2024arXiv240113033C}%
  \BibitemOpen
  \bibfield  {author} {\bibinfo {author} {\bibfnamefont {W.~R.}\ \bibnamefont {{Coulton}}}, \bibinfo {author} {\bibfnamefont {T.}~\bibnamefont {{Schutt}}}, \bibinfo {author} {\bibfnamefont {A.~S.}\ \bibnamefont {{Maniyar}}}, \bibinfo {author} {\bibfnamefont {E.}~\bibnamefont {{Schaan}}}, \bibinfo {author} {\bibfnamefont {R.}~\bibnamefont {{An}}}, \bibinfo {author} {\bibfnamefont {Z.}~\bibnamefont {{Atkins}}}, \bibinfo {author} {\bibfnamefont {N.}~\bibnamefont {{Battaglia}}}, \bibinfo {author} {\bibfnamefont {J.~R.}\ \bibnamefont {{Bond}}}, \bibinfo {author} {\bibfnamefont {E.}~\bibnamefont {{Calabrese}}}, \bibinfo {author} {\bibfnamefont {S.~K.}\ \bibnamefont {{Choi}}}, \bibinfo {author} {\bibfnamefont {M.~J.}\ \bibnamefont {{Devlin}}}, \bibinfo {author} {\bibfnamefont {A.~J.}\ \bibnamefont {{Duivenvoorden}}}, \bibinfo {author} {\bibfnamefont {J.}~\bibnamefont {{Dunkley}}}, \bibinfo {author} {\bibfnamefont {S.}~\bibnamefont {{Ferraro}}}, \bibinfo {author} {\bibfnamefont {V.}~\bibnamefont {{Gluscevic}}},
  \bibinfo {author} {\bibfnamefont {J.~C.}\ \bibnamefont {{Hill}}}, \bibinfo {author} {\bibfnamefont {M.}~\bibnamefont {{Hilton}}}, \bibinfo {author} {\bibfnamefont {A.~D.}\ \bibnamefont {{Hincks}}}, \bibinfo {author} {\bibfnamefont {A.}~\bibnamefont {{Kosowsky}}}, \bibinfo {author} {\bibfnamefont {D.}~\bibnamefont {{Kramer}}}, \bibinfo {author} {\bibfnamefont {A.}~\bibnamefont {{Kusiak}}}, \bibinfo {author} {\bibfnamefont {A.}~\bibnamefont {{La Posta}}}, \bibinfo {author} {\bibfnamefont {T.}~\bibnamefont {{Louis}}}, \bibinfo {author} {\bibfnamefont {M.~S.}\ \bibnamefont {{Madhavacheril}}}, \bibinfo {author} {\bibfnamefont {G.~A.}\ \bibnamefont {{Marques}}}, \bibinfo {author} {\bibfnamefont {F.}~\bibnamefont {{McCarthy}}}, \bibinfo {author} {\bibfnamefont {J.}~\bibnamefont {{McMahon}}}, \bibinfo {author} {\bibfnamefont {K.}~\bibnamefont {{Moodley}}}, \bibinfo {author} {\bibfnamefont {S.}~\bibnamefont {{Naess}}}, \bibinfo {author} {\bibfnamefont {L.~A.}\ \bibnamefont {{Page}}}, \bibinfo {author} {\bibfnamefont
  {B.}~\bibnamefont {{Partridge}}}, \bibinfo {author} {\bibfnamefont {F.~J.}\ \bibnamefont {{Qu}}}, \bibinfo {author} {\bibfnamefont {N.}~\bibnamefont {{Sehgal}}}, \bibinfo {author} {\bibfnamefont {B.~D.}\ \bibnamefont {{Sherwin}}}, \bibinfo {author} {\bibfnamefont {C.}~\bibnamefont {{Sif{\'o}n}}}, \bibinfo {author} {\bibfnamefont {D.~N.}\ \bibnamefont {{Spergel}}}, \bibinfo {author} {\bibfnamefont {S.~T.}\ \bibnamefont {{Staggs}}}, \bibinfo {author} {\bibfnamefont {A.}~\bibnamefont {{Van Engelen}}}, \bibinfo {author} {\bibfnamefont {C.}~\bibnamefont {{Vargas}}},\ and\ \bibinfo {author} {\bibfnamefont {E.~J.}\ \bibnamefont {{Wollack}}},\ }\bibfield  {title} {\bibinfo {title} {{The Atacama Cosmology Telescope: Detection of Patchy Screening of the Cosmic Microwave Background}},\ }\href {https://doi.org/10.48550/arXiv.2401.13033} {\bibfield  {journal} {\bibinfo  {journal} {arXiv e-prints}\ ,\ \bibinfo {eid} {arXiv:2401.13033}} (\bibinfo {year} {2024})},\ \Eprint {https://arxiv.org/abs/2401.13033}
  {arXiv:2401.13033 [astro-ph.CO]} \BibitemShut {NoStop}%
\bibitem [{\citenamefont {{Speagle}}(2020)}]{2020MNRAS.493.3132S}%
  \BibitemOpen
  \bibfield  {author} {\bibinfo {author} {\bibfnamefont {J.~S.}\ \bibnamefont {{Speagle}}},\ }\bibfield  {title} {\bibinfo {title} {{DYNESTY: a dynamic nested sampling package for estimating Bayesian posteriors and evidences}},\ }\href {https://doi.org/10.1093/mnras/staa278} {\bibfield  {journal} {\bibinfo  {journal} {\mnras}\ }\textbf {\bibinfo {volume} {493}},\ \bibinfo {pages} {3132} (\bibinfo {year} {2020})},\ \Eprint {https://arxiv.org/abs/1904.02180} {arXiv:1904.02180 [astro-ph.IM]} \BibitemShut {NoStop}%
\bibitem [{\citenamefont {Chung}\ \emph {et~al.}(2020)\citenamefont {Chung}, \citenamefont {Foreman},\ and\ \citenamefont {van Engelen}}]{Chung:2019bsk}%
  \BibitemOpen
  \bibfield  {author} {\bibinfo {author} {\bibfnamefont {E.}~\bibnamefont {Chung}}, \bibinfo {author} {\bibfnamefont {S.}~\bibnamefont {Foreman}},\ and\ \bibinfo {author} {\bibfnamefont {A.}~\bibnamefont {van Engelen}},\ }\bibfield  {title} {\bibinfo {title} {{Baryonic effects on CMB lensing and neutrino mass constraints}},\ }\href {https://doi.org/10.1103/PhysRevD.101.063534} {\bibfield  {journal} {\bibinfo  {journal} {Phys. Rev. D}\ }\textbf {\bibinfo {volume} {101}},\ \bibinfo {pages} {063534} (\bibinfo {year} {2020})},\ \bibinfo {note} {[Erratum: Phys.Rev.D 102, 109903 (2020)]},\ \Eprint {https://arxiv.org/abs/1910.09565} {arXiv:1910.09565 [astro-ph.CO]} \BibitemShut {NoStop}%
\bibitem [{Note6()}]{Note6}%
  \BibitemOpen
  \bibinfo {note} {To avoid potential bias from upward fluctuations, we do not select the bin with the highest SNR in the data, but instead use simulations. We also note that the first radial bin in the $\kappa $ measurements correlates best with the one-halo term amplitude, as it is sensitive to smaller scales.}\BibitemShut {Stop}%
\bibitem [{\citenamefont {{Schaan}}\ \emph {et~al.}(2021{\natexlab{b}})\citenamefont {{Schaan}}, \citenamefont {{Ferraro}}, \citenamefont {{Amodeo}}, \citenamefont {{Battaglia}}, \citenamefont {{Aiola}}, \citenamefont {{Austermann}}, \citenamefont {{Beall}}, \citenamefont {{Bean}}, \citenamefont {{Becker}}, \citenamefont {{Bond}}, \citenamefont {{Calabrese}}, \citenamefont {{Calafut}}, \citenamefont {{Choi}}, \citenamefont {{Denison}}, \citenamefont {{Devlin}}, \citenamefont {{Duff}}, \citenamefont {{Duivenvoorden}}, \citenamefont {{Dunkley}}, \citenamefont {{D{\"u}nner}}, \citenamefont {{Gallardo}}, \citenamefont {{Guan}}, \citenamefont {{Han}}, \citenamefont {{Hill}}, \citenamefont {{Hilton}}, \citenamefont {{Hilton}}, \citenamefont {{Hlo{\v{z}}ek}}, \citenamefont {{Hubmayr}}, \citenamefont {{Huffenberger}}, \citenamefont {{Hughes}}, \citenamefont {{Koopman}}, \citenamefont {{MacInnis}}, \citenamefont {{McMahon}}, \citenamefont {{Madhavacheril}}, \citenamefont {{Moodley}}, \citenamefont {{Mroczkowski}},
  \citenamefont {{Naess}}, \citenamefont {{Nati}}, \citenamefont {{Newburgh}}, \citenamefont {{Niemack}}, \citenamefont {{Page}}, \citenamefont {{Partridge}}, \citenamefont {{Salatino}}, \citenamefont {{Sehgal}}, \citenamefont {{Schillaci}}, \citenamefont {{Sif{\'o}n}}, \citenamefont {{Smith}}, \citenamefont {{Spergel}}, \citenamefont {{Staggs}}, \citenamefont {{Storer}}, \citenamefont {{Trac}}, \citenamefont {{Ullom}}, \citenamefont {{Van Lanen}}, \citenamefont {{Vale}}, \citenamefont {{van Engelen}}, \citenamefont {{Maga{\~n}a}}, \citenamefont {{Vavagiakis}}, \citenamefont {{Wollack}}, \citenamefont {{Xu}},\ and\ \citenamefont {{Atacama Cosmology Telescope Collaboration}}}]{Schaan21}%
  \BibitemOpen
  \bibfield  {author} {\bibinfo {author} {\bibfnamefont {E.}~\bibnamefont {{Schaan}}}, \bibinfo {author} {\bibfnamefont {S.}~\bibnamefont {{Ferraro}}}, \bibinfo {author} {\bibfnamefont {S.}~\bibnamefont {{Amodeo}}}, \bibinfo {author} {\bibfnamefont {N.}~\bibnamefont {{Battaglia}}}, \bibinfo {author} {\bibfnamefont {S.}~\bibnamefont {{Aiola}}}, \bibinfo {author} {\bibfnamefont {J.~E.}\ \bibnamefont {{Austermann}}}, \bibinfo {author} {\bibfnamefont {J.~A.}\ \bibnamefont {{Beall}}}, \bibinfo {author} {\bibfnamefont {R.}~\bibnamefont {{Bean}}}, \bibinfo {author} {\bibfnamefont {D.~T.}\ \bibnamefont {{Becker}}}, \bibinfo {author} {\bibfnamefont {R.~J.}\ \bibnamefont {{Bond}}}, \bibinfo {author} {\bibfnamefont {E.}~\bibnamefont {{Calabrese}}}, \bibinfo {author} {\bibfnamefont {V.}~\bibnamefont {{Calafut}}}, \bibinfo {author} {\bibfnamefont {S.~K.}\ \bibnamefont {{Choi}}}, \bibinfo {author} {\bibfnamefont {E.~V.}\ \bibnamefont {{Denison}}}, \bibinfo {author} {\bibfnamefont {M.~J.}\ \bibnamefont {{Devlin}}}, \bibinfo
  {author} {\bibfnamefont {S.~M.}\ \bibnamefont {{Duff}}}, \bibinfo {author} {\bibfnamefont {A.~J.}\ \bibnamefont {{Duivenvoorden}}}, \bibinfo {author} {\bibfnamefont {J.}~\bibnamefont {{Dunkley}}}, \bibinfo {author} {\bibfnamefont {R.}~\bibnamefont {{D{\"u}nner}}}, \bibinfo {author} {\bibfnamefont {P.~A.}\ \bibnamefont {{Gallardo}}}, \bibinfo {author} {\bibfnamefont {Y.}~\bibnamefont {{Guan}}}, \bibinfo {author} {\bibfnamefont {D.}~\bibnamefont {{Han}}}, \bibinfo {author} {\bibfnamefont {J.~C.}\ \bibnamefont {{Hill}}}, \bibinfo {author} {\bibfnamefont {G.~C.}\ \bibnamefont {{Hilton}}}, \bibinfo {author} {\bibfnamefont {M.}~\bibnamefont {{Hilton}}}, \bibinfo {author} {\bibfnamefont {R.}~\bibnamefont {{Hlo{\v{z}}ek}}}, \bibinfo {author} {\bibfnamefont {J.}~\bibnamefont {{Hubmayr}}}, \bibinfo {author} {\bibfnamefont {K.~M.}\ \bibnamefont {{Huffenberger}}}, \bibinfo {author} {\bibfnamefont {J.~P.}\ \bibnamefont {{Hughes}}}, \bibinfo {author} {\bibfnamefont {B.~J.}\ \bibnamefont {{Koopman}}}, \bibinfo {author}
  {\bibfnamefont {A.}~\bibnamefont {{MacInnis}}}, \bibinfo {author} {\bibfnamefont {J.}~\bibnamefont {{McMahon}}}, \bibinfo {author} {\bibfnamefont {M.~S.}\ \bibnamefont {{Madhavacheril}}}, \bibinfo {author} {\bibfnamefont {K.}~\bibnamefont {{Moodley}}}, \bibinfo {author} {\bibfnamefont {T.}~\bibnamefont {{Mroczkowski}}}, \bibinfo {author} {\bibfnamefont {S.}~\bibnamefont {{Naess}}}, \bibinfo {author} {\bibfnamefont {F.}~\bibnamefont {{Nati}}}, \bibinfo {author} {\bibfnamefont {L.~B.}\ \bibnamefont {{Newburgh}}}, \bibinfo {author} {\bibfnamefont {M.~D.}\ \bibnamefont {{Niemack}}}, \bibinfo {author} {\bibfnamefont {L.~A.}\ \bibnamefont {{Page}}}, \bibinfo {author} {\bibfnamefont {B.}~\bibnamefont {{Partridge}}}, \bibinfo {author} {\bibfnamefont {M.}~\bibnamefont {{Salatino}}}, \bibinfo {author} {\bibfnamefont {N.}~\bibnamefont {{Sehgal}}}, \bibinfo {author} {\bibfnamefont {A.}~\bibnamefont {{Schillaci}}}, \bibinfo {author} {\bibfnamefont {C.}~\bibnamefont {{Sif{\'o}n}}}, \bibinfo {author} {\bibfnamefont
  {K.~M.}\ \bibnamefont {{Smith}}}, \bibinfo {author} {\bibfnamefont {D.~N.}\ \bibnamefont {{Spergel}}}, \bibinfo {author} {\bibfnamefont {S.}~\bibnamefont {{Staggs}}}, \bibinfo {author} {\bibfnamefont {E.~R.}\ \bibnamefont {{Storer}}}, \bibinfo {author} {\bibfnamefont {H.}~\bibnamefont {{Trac}}}, \bibinfo {author} {\bibfnamefont {J.~N.}\ \bibnamefont {{Ullom}}}, \bibinfo {author} {\bibfnamefont {J.}~\bibnamefont {{Van Lanen}}}, \bibinfo {author} {\bibfnamefont {L.~R.}\ \bibnamefont {{Vale}}}, \bibinfo {author} {\bibfnamefont {A.}~\bibnamefont {{van Engelen}}}, \bibinfo {author} {\bibfnamefont {M.~V.}\ \bibnamefont {{Maga{\~n}a}}}, \bibinfo {author} {\bibfnamefont {E.~M.}\ \bibnamefont {{Vavagiakis}}}, \bibinfo {author} {\bibfnamefont {E.~J.}\ \bibnamefont {{Wollack}}}, \bibinfo {author} {\bibfnamefont {Z.}~\bibnamefont {{Xu}}},\ and\ \bibinfo {author} {\bibnamefont {{Atacama Cosmology Telescope Collaboration}}},\ }\bibfield  {title} {\bibinfo {title} {{Atacama Cosmology Telescope: Combined kinematic and
  thermal Sunyaev-Zel'dovich measurements from BOSS CMASS and LOWZ halos}},\ }\href {https://doi.org/10.1103/PhysRevD.103.063513} {\bibfield  {journal} {\bibinfo  {journal} {\prd}\ }\textbf {\bibinfo {volume} {103}},\ \bibinfo {eid} {063513} (\bibinfo {year} {2021}{\natexlab{b}})},\ \Eprint {https://arxiv.org/abs/2009.05557} {arXiv:2009.05557 [astro-ph.CO]} \BibitemShut {NoStop}%
\bibitem [{\citenamefont {{Popesso}}\ \emph {et~al.}(2024)\citenamefont {{Popesso}}, \citenamefont {{Biviano}}, \citenamefont {{Marini}}, \citenamefont {{Dolag}}, \citenamefont {{Vladutescu-Zopp}}, \citenamefont {{Csizi}}, \citenamefont {{Biffi}}, \citenamefont {{Lamer}}, \citenamefont {{Robothan}}, \citenamefont {{Bravo}}, \citenamefont {{Lovisari}}, \citenamefont {{Ettori}}, \citenamefont {{Angelinelli}}, \citenamefont {{Driver}}, \citenamefont {{Toptun}}, \citenamefont {{Dev}}, \citenamefont {{Mazengo}}, \citenamefont {{Merloni}}, \citenamefont {{Comparat}}, \citenamefont {{Ponti}}, \citenamefont {{Mroczkowski}}, \citenamefont {{Bulbul}}, \citenamefont {{Grandis}},\ and\ \citenamefont {{Bahar}}}]{2024arXiv241116555P}%
  \BibitemOpen
  \bibfield  {author} {\bibinfo {author} {\bibfnamefont {P.}~\bibnamefont {{Popesso}}}, \bibinfo {author} {\bibfnamefont {A.}~\bibnamefont {{Biviano}}}, \bibinfo {author} {\bibfnamefont {I.}~\bibnamefont {{Marini}}}, \bibinfo {author} {\bibfnamefont {K.}~\bibnamefont {{Dolag}}}, \bibinfo {author} {\bibfnamefont {S.}~\bibnamefont {{Vladutescu-Zopp}}}, \bibinfo {author} {\bibfnamefont {B.}~\bibnamefont {{Csizi}}}, \bibinfo {author} {\bibfnamefont {V.}~\bibnamefont {{Biffi}}}, \bibinfo {author} {\bibfnamefont {G.}~\bibnamefont {{Lamer}}}, \bibinfo {author} {\bibfnamefont {A.}~\bibnamefont {{Robothan}}}, \bibinfo {author} {\bibfnamefont {M.}~\bibnamefont {{Bravo}}}, \bibinfo {author} {\bibfnamefont {L.}~\bibnamefont {{Lovisari}}}, \bibinfo {author} {\bibfnamefont {S.}~\bibnamefont {{Ettori}}}, \bibinfo {author} {\bibfnamefont {M.}~\bibnamefont {{Angelinelli}}}, \bibinfo {author} {\bibfnamefont {S.}~\bibnamefont {{Driver}}}, \bibinfo {author} {\bibfnamefont {V.}~\bibnamefont {{Toptun}}}, \bibinfo {author}
  {\bibfnamefont {A.}~\bibnamefont {{Dev}}}, \bibinfo {author} {\bibfnamefont {D.}~\bibnamefont {{Mazengo}}}, \bibinfo {author} {\bibfnamefont {A.}~\bibnamefont {{Merloni}}}, \bibinfo {author} {\bibfnamefont {J.}~\bibnamefont {{Comparat}}}, \bibinfo {author} {\bibfnamefont {G.}~\bibnamefont {{Ponti}}}, \bibinfo {author} {\bibfnamefont {T.}~\bibnamefont {{Mroczkowski}}}, \bibinfo {author} {\bibfnamefont {E.}~\bibnamefont {{Bulbul}}}, \bibinfo {author} {\bibfnamefont {S.}~\bibnamefont {{Grandis}}},\ and\ \bibinfo {author} {\bibfnamefont {E.}~\bibnamefont {{Bahar}}},\ }\bibfield  {title} {\bibinfo {title} {{The hot gas mass fraction in halos. From Milky Way-like groups to massive clusters}},\ }\href {https://doi.org/10.48550/arXiv.2411.16555} {\bibfield  {journal} {\bibinfo  {journal} {arXiv e-prints}\ ,\ \bibinfo {eid} {arXiv:2411.16555}} (\bibinfo {year} {2024})},\ \Eprint {https://arxiv.org/abs/2411.16555} {arXiv:2411.16555 [astro-ph.GA]} \BibitemShut {NoStop}%
\bibitem [{\citenamefont {{Lucie-Smith}}\ \emph {et~al.}(2025)\citenamefont {{Lucie-Smith}}, \citenamefont {{Peiris}}, \citenamefont {{Pontzen}}, \citenamefont {{Halder}}, \citenamefont {{Schaye}}, \citenamefont {{Schaller}}, \citenamefont {{Helly}}, \citenamefont {{McGibbon}},\ and\ \citenamefont {{Elbers}}}]{Lucie-Smith:2025hgj}%
  \BibitemOpen
  \bibfield  {author} {\bibinfo {author} {\bibfnamefont {L.}~\bibnamefont {{Lucie-Smith}}}, \bibinfo {author} {\bibfnamefont {H.~V.}\ \bibnamefont {{Peiris}}}, \bibinfo {author} {\bibfnamefont {A.}~\bibnamefont {{Pontzen}}}, \bibinfo {author} {\bibfnamefont {A.}~\bibnamefont {{Halder}}}, \bibinfo {author} {\bibfnamefont {J.}~\bibnamefont {{Schaye}}}, \bibinfo {author} {\bibfnamefont {M.}~\bibnamefont {{Schaller}}}, \bibinfo {author} {\bibfnamefont {J.}~\bibnamefont {{Helly}}}, \bibinfo {author} {\bibfnamefont {R.~J.}\ \bibnamefont {{McGibbon}}},\ and\ \bibinfo {author} {\bibfnamefont {W.}~\bibnamefont {{Elbers}}},\ }\bibfield  {title} {\bibinfo {title} {{Cosmological feedback from a halo assembly perspective}},\ }\href {https://doi.org/10.48550/arXiv.2505.18258} {\bibfield  {journal} {\bibinfo  {journal} {arXiv e-prints}\ ,\ \bibinfo {eid} {arXiv:2505.18258}} (\bibinfo {year} {2025})},\ \Eprint {https://arxiv.org/abs/2505.18258} {arXiv:2505.18258 [astro-ph.CO]} \BibitemShut {NoStop}%
\bibitem [{\citenamefont {{Kova{\v{c}}}}\ \emph {et~al.}(2025)\citenamefont {{Kova{\v{c}}}}, \citenamefont {{Nicola}}, \citenamefont {{Bucko}}, \citenamefont {{Schneider}}, \citenamefont {{Reischke}}, \citenamefont {{Giri}}, \citenamefont {{Teyssier}}, \citenamefont {{Schaller}},\ and\ \citenamefont {{Schaye}}}]{Kovac:2025zqy}%
  \BibitemOpen
  \bibfield  {author} {\bibinfo {author} {\bibfnamefont {M.}~\bibnamefont {{Kova{\v{c}}}}}, \bibinfo {author} {\bibfnamefont {A.}~\bibnamefont {{Nicola}}}, \bibinfo {author} {\bibfnamefont {J.}~\bibnamefont {{Bucko}}}, \bibinfo {author} {\bibfnamefont {A.}~\bibnamefont {{Schneider}}}, \bibinfo {author} {\bibfnamefont {R.}~\bibnamefont {{Reischke}}}, \bibinfo {author} {\bibfnamefont {S.~K.}\ \bibnamefont {{Giri}}}, \bibinfo {author} {\bibfnamefont {R.}~\bibnamefont {{Teyssier}}}, \bibinfo {author} {\bibfnamefont {M.}~\bibnamefont {{Schaller}}},\ and\ \bibinfo {author} {\bibfnamefont {J.}~\bibnamefont {{Schaye}}},\ }\bibfield  {title} {\bibinfo {title} {{Baryonification II: Constraining feedback with X-ray and kinematic Sunyaev-Zel'dovich observations}},\ }\href {https://doi.org/10.48550/arXiv.2507.07991} {\bibfield  {journal} {\bibinfo  {journal} {arXiv e-prints}\ ,\ \bibinfo {eid} {arXiv:2507.07991}} (\bibinfo {year} {2025})},\ \Eprint {https://arxiv.org/abs/2507.07991} {arXiv:2507.07991 [astro-ph.CO]}
  \BibitemShut {NoStop}%
\bibitem [{\citenamefont {{Nelson}}\ \emph {et~al.}(2019)\citenamefont {{Nelson}}, \citenamefont {{Springel}}, \citenamefont {{Pillepich}}, \citenamefont {{Rodriguez-Gomez}}, \citenamefont {{Torrey}}, \citenamefont {{Genel}}, \citenamefont {{Vogelsberger}}, \citenamefont {{Pakmor}}, \citenamefont {{Marinacci}}, \citenamefont {{Weinberger}}, \citenamefont {{Kelley}}, \citenamefont {{Lovell}}, \citenamefont {{Diemer}},\ and\ \citenamefont {{Hernquist}}}]{2019ComAC...6....2N}%
  \BibitemOpen
  \bibfield  {author} {\bibinfo {author} {\bibfnamefont {D.}~\bibnamefont {{Nelson}}}, \bibinfo {author} {\bibfnamefont {V.}~\bibnamefont {{Springel}}}, \bibinfo {author} {\bibfnamefont {A.}~\bibnamefont {{Pillepich}}}, \bibinfo {author} {\bibfnamefont {V.}~\bibnamefont {{Rodriguez-Gomez}}}, \bibinfo {author} {\bibfnamefont {P.}~\bibnamefont {{Torrey}}}, \bibinfo {author} {\bibfnamefont {S.}~\bibnamefont {{Genel}}}, \bibinfo {author} {\bibfnamefont {M.}~\bibnamefont {{Vogelsberger}}}, \bibinfo {author} {\bibfnamefont {R.}~\bibnamefont {{Pakmor}}}, \bibinfo {author} {\bibfnamefont {F.}~\bibnamefont {{Marinacci}}}, \bibinfo {author} {\bibfnamefont {R.}~\bibnamefont {{Weinberger}}}, \bibinfo {author} {\bibfnamefont {L.}~\bibnamefont {{Kelley}}}, \bibinfo {author} {\bibfnamefont {M.}~\bibnamefont {{Lovell}}}, \bibinfo {author} {\bibfnamefont {B.}~\bibnamefont {{Diemer}}},\ and\ \bibinfo {author} {\bibfnamefont {L.}~\bibnamefont {{Hernquist}}},\ }\bibfield  {title} {\bibinfo {title} {{The IllustrisTNG simulations:
  public data release}},\ }\href {https://doi.org/10.1186/s40668-019-0028-x} {\bibfield  {journal} {\bibinfo  {journal} {Computational Astrophysics and Cosmology}\ }\textbf {\bibinfo {volume} {6}},\ \bibinfo {eid} {2} (\bibinfo {year} {2019})},\ \Eprint {https://arxiv.org/abs/1812.05609} {arXiv:1812.05609 [astro-ph.GA]} \BibitemShut {NoStop}%
\bibitem [{\citenamefont {{Zhou}}\ \emph {et~al.}(2021)\citenamefont {{Zhou}}, \citenamefont {{Newman}}, \citenamefont {{Mao}}, \citenamefont {{Meisner}}, \citenamefont {{Moustakas}}, \citenamefont {{Myers}}, \citenamefont {{Prakash}}, \citenamefont {{Zentner}}, \citenamefont {{Brooks}}, \citenamefont {{Duan}}, \citenamefont {{Landriau}}, \citenamefont {{Levi}}, \citenamefont {{Prada}},\ and\ \citenamefont {{Tarle}}}]{2021MNRAS.501.3309Z}%
  \BibitemOpen
  \bibfield  {author} {\bibinfo {author} {\bibfnamefont {R.}~\bibnamefont {{Zhou}}}, \bibinfo {author} {\bibfnamefont {J.~A.}\ \bibnamefont {{Newman}}}, \bibinfo {author} {\bibfnamefont {Y.-Y.}\ \bibnamefont {{Mao}}}, \bibinfo {author} {\bibfnamefont {A.}~\bibnamefont {{Meisner}}}, \bibinfo {author} {\bibfnamefont {J.}~\bibnamefont {{Moustakas}}}, \bibinfo {author} {\bibfnamefont {A.~D.}\ \bibnamefont {{Myers}}}, \bibinfo {author} {\bibfnamefont {A.}~\bibnamefont {{Prakash}}}, \bibinfo {author} {\bibfnamefont {A.~R.}\ \bibnamefont {{Zentner}}}, \bibinfo {author} {\bibfnamefont {D.}~\bibnamefont {{Brooks}}}, \bibinfo {author} {\bibfnamefont {Y.}~\bibnamefont {{Duan}}}, \bibinfo {author} {\bibfnamefont {M.}~\bibnamefont {{Landriau}}}, \bibinfo {author} {\bibfnamefont {M.~E.}\ \bibnamefont {{Levi}}}, \bibinfo {author} {\bibfnamefont {F.}~\bibnamefont {{Prada}}},\ and\ \bibinfo {author} {\bibfnamefont {G.}~\bibnamefont {{Tarle}}},\ }\bibfield  {title} {\bibinfo {title} {{The clustering of DESI-like luminous red
  galaxies using photometric redshifts}},\ }\href {https://doi.org/10.1093/mnras/staa3764} {\bibfield  {journal} {\bibinfo  {journal} {\mnras}\ }\textbf {\bibinfo {volume} {501}},\ \bibinfo {pages} {3309} (\bibinfo {year} {2021})},\ \Eprint {https://arxiv.org/abs/2001.06018} {arXiv:2001.06018 [astro-ph.CO]} \BibitemShut {NoStop}%
\bibitem [{\citenamefont {Pillepich}\ \emph {et~al.}(2018)\citenamefont {Pillepich} \emph {et~al.}}]{Pillepich:2017fcc}%
  \BibitemOpen
  \bibfield  {author} {\bibinfo {author} {\bibfnamefont {A.}~\bibnamefont {Pillepich}} \emph {et~al.},\ }\bibfield  {title} {\bibinfo {title} {{First results from the IllustrisTNG simulations: the stellar mass content of groups and clusters of galaxies}},\ }\href {https://doi.org/10.1093/mnras/stx3112} {\bibfield  {journal} {\bibinfo  {journal} {Mon. Not. Roy. Astron. Soc.}\ }\textbf {\bibinfo {volume} {475}},\ \bibinfo {pages} {648} (\bibinfo {year} {2018})},\ \Eprint {https://arxiv.org/abs/1707.03406} {arXiv:1707.03406 [astro-ph.GA]} \BibitemShut {NoStop}%
\bibitem [{\citenamefont {{Hadzhiyska}}\ \emph {et~al.}(2021)\citenamefont {{Hadzhiyska}}, \citenamefont {{Tacchella}}, \citenamefont {{Bose}},\ and\ \citenamefont {{Eisenstein}}}]{2021MNRAS.502.3599H}%
  \BibitemOpen
  \bibfield  {author} {\bibinfo {author} {\bibfnamefont {B.}~\bibnamefont {{Hadzhiyska}}}, \bibinfo {author} {\bibfnamefont {S.}~\bibnamefont {{Tacchella}}}, \bibinfo {author} {\bibfnamefont {S.}~\bibnamefont {{Bose}}},\ and\ \bibinfo {author} {\bibfnamefont {D.~J.}\ \bibnamefont {{Eisenstein}}},\ }\bibfield  {title} {\bibinfo {title} {{The galaxy-halo connection of emission-line galaxies in IllustrisTNG}},\ }\href {https://doi.org/10.1093/mnras/stab243} {\bibfield  {journal} {\bibinfo  {journal} {\mnras}\ }\textbf {\bibinfo {volume} {502}},\ \bibinfo {pages} {3599} (\bibinfo {year} {2021})},\ \Eprint {https://arxiv.org/abs/2011.05331} {arXiv:2011.05331 [astro-ph.GA]} \BibitemShut {NoStop}%
\bibitem [{\citenamefont {{Hadzhiyska}}\ \emph {et~al.}(2023{\natexlab{b}})\citenamefont {{Hadzhiyska}}, \citenamefont {{Hernquist}}, \citenamefont {{Eisenstein}}, \citenamefont {{Delgado}}, \citenamefont {{Bose}}, \citenamefont {{Kannan}}, \citenamefont {{Pakmor}}, \citenamefont {{Springel}}, \citenamefont {{Contreras}}, \citenamefont {{Barrera}}, \citenamefont {{Ferlito}}, \citenamefont {{Hern{\'a}ndez-Aguayo}}, \citenamefont {{White}},\ and\ \citenamefont {{Frenk}}}]{2023MNRAS.524.2524H}%
  \BibitemOpen
  \bibfield  {author} {\bibinfo {author} {\bibfnamefont {B.}~\bibnamefont {{Hadzhiyska}}}, \bibinfo {author} {\bibfnamefont {L.}~\bibnamefont {{Hernquist}}}, \bibinfo {author} {\bibfnamefont {D.}~\bibnamefont {{Eisenstein}}}, \bibinfo {author} {\bibfnamefont {A.~M.}\ \bibnamefont {{Delgado}}}, \bibinfo {author} {\bibfnamefont {S.}~\bibnamefont {{Bose}}}, \bibinfo {author} {\bibfnamefont {R.}~\bibnamefont {{Kannan}}}, \bibinfo {author} {\bibfnamefont {R.}~\bibnamefont {{Pakmor}}}, \bibinfo {author} {\bibfnamefont {V.}~\bibnamefont {{Springel}}}, \bibinfo {author} {\bibfnamefont {S.}~\bibnamefont {{Contreras}}}, \bibinfo {author} {\bibfnamefont {M.}~\bibnamefont {{Barrera}}}, \bibinfo {author} {\bibfnamefont {F.}~\bibnamefont {{Ferlito}}}, \bibinfo {author} {\bibfnamefont {C.}~\bibnamefont {{Hern{\'a}ndez-Aguayo}}}, \bibinfo {author} {\bibfnamefont {S.~D.~M.}\ \bibnamefont {{White}}},\ and\ \bibinfo {author} {\bibfnamefont {C.}~\bibnamefont {{Frenk}}},\ }\bibfield  {title} {\bibinfo {title} {{The MillenniumTNG
  Project: refining the one-halo model of red and blue galaxies at different redshifts}},\ }\href {https://doi.org/10.1093/mnras/stad279} {\bibfield  {journal} {\bibinfo  {journal} {\mnras}\ }\textbf {\bibinfo {volume} {524}},\ \bibinfo {pages} {2524} (\bibinfo {year} {2023}{\natexlab{b}})},\ \Eprint {https://arxiv.org/abs/2210.10068} {arXiv:2210.10068 [astro-ph.CO]} \BibitemShut {NoStop}%
\end{thebibliography}%

\end{document}